%% file: thesis.tex
\documentclass[reqno,12pt,oneside]{report} 

\usepackage{rac}         
\usepackage{aas_macros}  
\usepackage{amsmath} 
\usepackage{amsxtra}     
\usepackage{amsthm}
\usepackage{amssymb}
\usepackage{amsfonts}
\usepackage{graphicx}    
\usepackage{rotating}
\usepackage{color}
\usepackage{epsfig}
\usepackage{subfigure}   
\usepackage{verbatim}
\usepackage[square,numbers,sort&compress]{natbib}      
\usepackage[printonlyused]{acronym} 
\usepackage{setspace}    
\usepackage{physics}
\usepackage[utf8]{inputenc}
\usepackage{tikz}
\usepackage[hidelinks]{hyperref}
\usepackage{xcolor}
\usepackage{graphicx}
\usepackage{float}
\usepackage{bbm}

\newcommand{\fft}[2]{\frac{#1}{#2}}
\newcommand{\ft}[2]{{\textstyle{\frac{#1}{#2}}}}
\newcommand{\nn}{\nonumber}
\newcommand{\Ai}{\operatorname{Ai}}
\renewcommand{\Re}{\operatorname{Re}}
\renewcommand{\Im}{\operatorname{Im}}
\newcommand{\Li}{\operatorname{Li}}
\newcommand{\Cl}{\operatorname{Cl}}
\newcommand{\ol}[1]{\overline{#1}}
\newcommand{\cN}{\mathcal{N}}
\newcommand{\cL}{\mathcal{L}}

\newcommand{\cO}{\mathcal{O}}
\newcommand{\cC}{\mathcal{C}}

\newcommand{\cF}{\mathcal{F}}
\newcommand{\cS}{\mathcal{S}}
\newcommand{\bR}{\mathbb{R}}
\newcommand{\cR}{\mathcal{R}}
\newcommand{\ul}[1]{\underline{#1}}

\DeclareFontEncoding{LS1}{}{}
\DeclareFontSubstitution{LS1}{stix}{m}{n}
\newcommand*\kay{%
  \text{%
  \fontencoding{LS1}%
  \fontfamily{stixscr}%
  \fontseries{\textmathversion}%
  \fontshape{n}%
  \selectfont\symbol{"6B}}}
\makeatletter
  \newcommand*\textmathversion{\csname textmv@\math@version\endcsname}
  \newcommand*\textmv@normal{m}
  \newcommand*\textmv@bold{b}
\makeatother

\numberwithin{equation}{section}

\theoremstyle{plain}

\theoremstyle{definition}

\theoremstyle{remark}

\numberwithin{theorem}{chapter}     

\makeatletter
\def\cleardoublepage{\clearpage\if@twoside \ifodd\c@page\else
\hbox{}
\thispagestyle{empty}
\newpage
\if@twocolumn\hbox{}\newpage\fi\fi\fi}
\makeatother 

\begin{document}

\bibliographystyle{JHEP}    

\titlepage{Explorations in Precision Holography and Higher-derivative Supergravity}{Robert J. Saskowski}{Doctor of Philosophy}
{Physics}{2024}
{ Professor James T. Liu, Chair\\
  Professor Christine A. Aidala\\
  Professor Leopoldo A. Pando Zayas\\
  Professor Aaron Pierce\\
  Professor Laura Ruetsche}

\initializefrontsections

\copyrightpage{Robert J. Saskowski}

\makeatletter
\if@twoside \setcounter{page}{4} \else \setcounter{page}{1} \fi
\makeatother
 

\startacknowledgementspage
\input{Intro/Acknowledgements}
\label{Acknowledgements}


\tableofcontents     
\listoffigures       
\listofappendices    
\listofabbreviations 

\startabstractpage
\input{Abstract/Abstract}
\label{Abstract}

\startthechapters 

 \chapter{Introduction}
 \label{chap:Intro}
 \input{Intro/Intro}

 \chapter{Subleading Corrections in $\mathcal N=3$ Gaiotto-Tomasiello Theory}
 \label{chap:GT}
 \input{Chapters/chap2}

 \chapter{Four-derivative Corrections to Minimal Gauged
Supergravity in Five Dimensions}
 \label{chap:UE}
 \input{Chapters/chap3}

 \chapter{$c$-functions in Higher-derivative Flows Across Dimensions}
 \label{chap:RG}
 \input{Chapters/chap4}

 \chapter{Consistent Truncations in Higher-derivative
Supergravity I: The Torus}
 \label{chap:torus}
 \input{Chapters/chap5}

 \chapter{Consistent Truncations in Higher-derivative
Supergravity II: The Sphere}
 \label{chap:sphere}
 \input{Chapters/chap6}

 \chapter{Concluding Remarks}
 \label{chap:conc}
 \input{Chapters/chap7}

\startappendices
\addtocontents{toc}{\protect\setcounter{tocdepth}{0}}
 \appendix{GT Endpoints and Free Energy}
 \label{app:GT}
 \input{Appendices/Appendix_GT}
 \appendix{Technical Details for Minimal $D=5$ Supergravity}
 \label{app:UnreasonableEffectiveness}
 \input{Appendices/Appendix_Unreasonable_Effectiveness}
 \appendix{Supplemental Computations for RG Flows}
 \label{app:RG}
 \input{Appendices/Appendix_RG}
  \appendix{Torsionful Tensors}
 \label{app:torus}
 \input{Appendices/Appendix_torus}
 \addtocontents{toc}{\protect\setcounter{tocdepth}{2}}
 
\startbibliography
 \begin{singlespace} 
  \bibliography{References}   
 \end{singlespace}


\end{document}

%% file: Intro/Acknowledgements.tex
I want to start by expressing my gratitude to Jim Liu, my advisor, for all of his help and support during my graduate studies. His insights and explanations have helped mold me as a physicist and, by extension, as a scientist. His advice, encouragement, and wisdom are a large part of what made the last five years so exciting and fulfilling. Much of the review sections of this dissertation are filled with physics which he explained to me many times. 

Additionally, I would like to thank Professor Leo Pando Zayas for his very fruitful collaboration with me, as well as for teaching three of the courses I took in grad school: string theory, gauge/gravity duality, and modern relativity. I would also like to thank Professor Henning Samtleben for taking the time to mentor and advise me as part of the String Theory Mentorship Program.

I would like to express my appreciation to Sabare Jayaprakash, Yide Cai, and Evan Deddo for their invaluable collaboration; without them, my graduate career would have been much less fruitful.

In addition, I am grateful to the following professors: Finn Larsen, Henriette Elvang, Aaron Pierce, James Wells, Ratindranath Akhoury, Lydia Bieiri, Andrew Snowden, Alejandro Uribe Ahuamada, Laura Ruetsche, Gordon Belot, Dave Baker, and Christine Aidala for the physics, mathematics, and philosophy that I have learned from them both through conversation and in various courses. I would also like to thank Gabriele Carcassi for many interesting discussions.

I also wish to thank my undergraduate advisor Professor Tom Kephart for accepting me as a young researcher and introducing me to the field of theoretical physics.

I also want to express my gratitude to all of the postdocs—both current and past—that I have had the pleasure of interacting with, including Christoph Uhlemann, Prudhvi Bhattiprolu, Jose de la Cruz Moreno, Nick Geiser, and Yingchun Zhang.

Thanks to all of the graduate students at Michigan. In particular, I wish to acknowledge interactions, both physics and recreational, with Brian McPeak, Callum Jones, Shruti Paranjape, Junho Hong, Sangmin Choi, Marina David, Ben Sheff, Zach Johnson, Aidan Herderschee, Nizar Ezroura, Siyul Lee, Alan Chen, Max Jerdee, Leia Barrowes, Evan Deddo, Justin Berman, Evan Petrosky, Sabare Jayaprakash, Daniel Sela, Loki Lin, Shaghayegh Emami, Alex Takla, Ilia Nekrasov, and Francisco Calderon, as well as the various visiting students we have had: Gabriel Larios, Jack Hudson, and Jingchao Zhang.

I would also like to thank Karen O'Donnovan for many useful conversations and much administrative help. I would also like to thank Harvey for much emotional support.

I would especially like to thank all my friends and family for their unending and unconditional support, especially my partner Ari Wright. I love you all.

Finally, thanks to whoever put together this nice \LaTeX\, template for me to use.

%% file: Abstract/Abstract.tex
This thesis explores topics related to the study of quantum gravity, with a focus on precision holography and higher-derivative supergravity. First, we study subleading corrections to the free energy of a particular 3D $\mathcal N=3$ Chern-Simons-matter theory found by Gaiotto and Tomasiello, which is given by a matrix model after supersymmetric localization. This theory is dual to massive IIA supergravity on $\mathrm{AdS}_4\times\mathbb{CP}^3$, and consequently, the structure of subleading corrections to the field theory naturally elucidates the higher-derivative corrections to the gravity dual. We extract the first order of corrections to the free energy using resolvent methods, and our results imply that particular terms in the supergravity action should vanish on-shell.

Next, we consider the ``unreasonable effectiveness'' of five-dimensional minimal gauged supergravity. There are three independent supersymmetric four-derivative terms that one can add to the action; nevertheless, after going on-shell (or, equivalently, after a field redefinition that pushes the off-shell discrepancies to six-derivative order), there is a unique supersymmetric invariant.

Third, we consider the effect of higher-derivative corrections in holographic renormalization group flows across dimensions. In particular, we construct a local holographic $c$-function out of metric functions and show its monotonicity via the Null Energy Condition. We also construct a $c$-function from the entanglement entropy for flows with a $\mathrm{CFT}_2$ IR fixed point, and we show that such flows are monotonic.

Finally, we consider consistent truncations of four-derivative heterotic supergravity. In particular, we show that reducing both on an $n$-dimensional torus $T^n$ or on $S^3$ and truncating the vector multiplets is indeed a consistent truncation at the four-derivative level. Moreover, we find examples of two-derivative consistent truncations which fail to extend to four-derivative ones.

%% file: Intro/Intro.tex
Einstein's theory of \ac{GR} is wildly successful, encapsulating much of the dynamics of classical gravity in a single, short line 
\begin{equation}
    R_{\mu\nu}-\frac{1}{2}Rg_{\mu\nu}-\Lambda g_{\mu\nu}=\frac{8\pi G_N}{c^4}T_{\mu\nu}.
\end{equation}
The left-hand side captures the geometry of spacetime via the Ricci tensor $R_{\mu\nu}$, the Ricci scalar $R$, the metric $g_{\mu\nu}$, and the cosmological constant $\Lambda$. In contrast, the right-hand side,  proportional to the stress-energy tensor $T_{\mu\nu}$, is a function of the matter and energy distribution. Thus, the shape of spacetime is determined by the matter we put in it, and the curvature of spacetime controls the motion of matter. There have been many experimental verifications of GR's predictions, which include the precession of the perihelion of Mercury's orbit, the bending of light, the gravitational redshift of light, Shapiro time delay, frame dragging, and gravitational waves.

However, we know the world is quantum, so we seek to upgrade general relativity to be a theory of \emph{quantum gravity}. In principle, we interpret the Einstein equation as the classical equations of motion corresponding to a field theory with action\footnote{Technically, we should also add a Gibbons-Hawking term to make the variational principle well-defined.}
\begin{equation}
    S_\mathrm{grav}=\int\dd[d]{x}\sqrt{-g}\,\qty(\frac{c^3}{16\pi G_N}R+\mathcal L_\mathrm{matter}),
\end{equation}
put it into a path integral
\begin{equation}
    Z_\mathrm{grav}=\int\mathcal Dg\,\mathcal D\phi \,e^{iS_\mathrm{grav}/\hbar},
\end{equation}
and we have quantum gravity. Unfortunately, this does not work. Unlike the Standard Model, if we try to quantize the perturbative (Fierz-Pauli) expansion of GR, we find that the coupling constant (the Planck mass) is dimensionful, and we expect to obtain a non-renormalizable theory. In other words, general relativity simply fails to accurately describe the world at sufficiently high energies.

But even classical general relativity contains hints of its non-fundamentality. Consider the simplest ($d$-dimensional) Schwarzschild black hole solution of mass $M$ with metric
\begin{equation}
    \dd s^2=\qty(1-\frac{2G_NM}{c^2r^{d-3}})c^2\dd t^2+\qty(1-\frac{2G_NM}{c^2r^{d-3}})^{-1}\dd r^2+r^2\dd\Omega_{d-2}^2.
\end{equation}
This becomes singular as $r\to 0$. Being a gauge-dependent object, the metric is not always a good indicator of singular behavior; there is also seemingly a singularity at the event horizon $r_s^{d-3}=2G_NM/c^2$, but this turns out to just be an artifact of the coordinate system. However, one can check that invariant objects such as the Kretschmann scalar $(R_{\mu\nu\rho\sigma})^2$ will diverge as $r\to 0$ (and are well-behaved around the horizon).\footnote{Although the singularity is hidden behind a horizon, GR still predicts its existence and an infalling observer could reach it. A fundamental theory should not have such singularities.} Since infinities usually signal the existence of phenomena taking place at new scales, \emph{something} about GR must be modified at sufficiently short distances. Heuristically, given some localized matter distribution, we expect that quantum effects should begin to dominate when the Compton wavelength becomes of order the Schwarzschild radius; this happens when the mass becomes of order the Planck mass, so we expect the Planck scale to be the energy scale at which the theory breaks down.

Now, given the enormity of the Planck scale, one might na\"ively believe that this is a problem that only applies to phenomena beyond our capacity of observation such as black hole interiors and solar system-sized\footnote{To be precise, it may be possible to probe the Planck length with a collider sized as small as $10^{10}$ m, which, for context, is a tenth of the distance between earth and sun \cite{Zimmermann:2018koi}.} super-colliders, and that semiclassical\footnote{By semiclassical, we mean classical gravity coupled to quantum matter.} gravity should be empirically adequate for everything we can observe; but this is too hasty. Indeed, it has long been known that black holes are thermodynamic objects (see \emph{e.g.} \cite{Bekenstein:1972tm,Bardeen:1973gs,Bekenstein:1973ur,Bekenstein:1974ax,Wald:1975kc,Hawking:1975vcx,Bekenstein:1975tw,Hawking:1976de}), even semiclassically. There is a simple thought experiment to demonstrate this: suppose black holes have no entropy, toss in some (entropy-rich) matter, and let the system evolve into a new black hole. By assumption, this new black hole would still have no entropy, and thus the total entropy would have decreased, violating the second law of thermodynamics. Hence, one concludes that black holes must have entropy. 

In particular, it can be shown that a black hole has a temperature $T$ related to its surface gravity $\kappa$ by
\begin{equation}
    T=\frac{\hbar}{ck_B}\frac{\kappa}{2\pi},
\end{equation}
and an entropy $\mathcal S$ given by one-quarter of its horizon area $\mathcal A_H$ in Planck units
\begin{equation}
    \mathcal S=\frac{k_B c^3}{G_{N}\hbar}\,\frac{\mathcal A_H}{4}.
\end{equation}
Remarkably, this brings together the various domains of physics: $k_B$ refers to statistical physics, $G_N$ to gravity, $c$ to special relativity, and $\hbar$ to quantum mechanics. Another surprising feature is that the entropy scales with the area of the black hole, rather than the volume as we would have expected. There is thus a sense in which the degrees of freedom live in one lower dimension. 

Being thermodynamic objects, black holes turn out to satisfy the four laws of thermodynamics
\begin{enumerate}\setcounter{enumi}{-1}
    \item The surface gravity (temperature) is constant over its event horizon
    \item Conservation of energy\footnote{The mass may be thought of as the internal energy for asymptotically flat black holes, but it should be interpreted as the enthalpy for asymptotically AdS black holes, for which the cosmological constant functions as a thermodynamic pressure \cite{Kubiznak:2014zwa}.}
    \begin{equation}
        \dd M =T\,\dd\mathcal S+\Omega\,\dd J+\Phi\,\dd Q,
    \end{equation}
    where $\Omega$ is the angular velocity, $J$ is the angular momentum, $\Phi$ is the electric potential, and $Q$ is the charge
    \item  The surface area (entropy) of a black hole never decreases\footnote{This is true classically. Quantum effects cause the black hole to emit Hawking radiation. However, a \emph{generalized second law} still holds, in the sense that the total entropy of the black hole-radiation system still increases.}
    \item The entropy of a black hole goes to a constant as the temperature vanishes
\end{enumerate}
Now, the fact that black holes have entropy,\footnote{For context, a solar mass black hole would have an enormous entropy, on the order of $10^{77}\cdot k_B$. That is about twenty orders of magnitude larger than the entropy of the sun itself.} by Boltzmann's famous equation
\begin{equation}
    \mathcal S=k_B\log \Omega,
\end{equation}
implies that there must be some quantum microstates, but semiclassical gravity offers no hints of their identity. 

All this is to say that there are many open questions regarding quantum gravity. One approach to address such fundamental questions is to detour through string theory, which is the only known UV-complete theory of gravity.\footnote{In principle, another candidate would be loop quantum gravity, which is known to be UV-complete, but it is not known how to recover gravity from the low-energy limit.} This will lead us to study holographic duality, which is the theme of this dissertation.

\subsubsection*{Notation}
Having left some unitful quantities for clarity in the preceding discussion, we will now set $\hbar=c=k_B=1$ and $G_N=1/16\pi$ for the rest of this dissertation, although we will occasionally leave $G_N$ explicit for emphasis.

\section{Holography}
String theory originated as a theory of the strong nuclear force: In the late 1960s, it was found that hadrons arranged themselves into Regge trajectories, with squared energy proportional to their angular momentum, and theorists showed that such a relationship emerged naturally from a rotating relativistic string. Seemingly unfortunately, attempts to model hadrons as strings came with unwanted massless spin-2 excitations (whereas no such particle appears in usual hadronic physics). However, such a particle must necessarily mediate a force with the properties of gravity \cite{Weinberg:1980kq}. So, in 1974, Scherk and Schwarz suggested that string theory was not a theory of nuclear physics but a theory of quantum gravity \cite{Scherk:1974ca}. Around that time, it was realized that hadrons are composed of quarks and hadronic string theory was abandoned in favor of quantum chromodynamics.\footnote{It is worth noting that the the string theory approach to non-perturbative quantum chromodynamics has since been revived, see \emph{e.g.} \cite{Makeenko:2010dq}.} However, string theory continued to develop independently as a theory of quantum gravity. The two theories became intertwined once again in 1974, when 't Hooft considered the large-$N$ limit of Yang-Mills theories and argued that certain calculations in quantum field theory resemble calculations in string theory in this limit \cite{tHooft:1973alw}.

Later, in 1986, Brown and Henneaux investigated the asymptotic symmetries of three-dimensional \ac{AdS} space \cite{Brown:1986nw}. If one considers diffeomorphisms that leave the asymptotic behavior of the metric unchanged, then the algebra of these symmetries is precisely the Virasoro algebra with central charge
\begin{equation}
    c=\frac{3L}{2G_N},
\end{equation}
where $L$ is the AdS radius. This then hinted at a connection between $\mathrm{AdS}_3$ and two-dimensional \ac{CFT}, which also has a Virasoro symmetry algebra. In 1995, Henneaux, along with Coussaert and van Driel, elucidated this connection by suggesting that 3D gravity in AdS is equivalent to Liouville field theory~\cite{Coussaert:1995zp}.

Meanwhile, in 1993, 't Hooft wrote a paper revisiting black hole thermodynamics and concluded that the total number of degrees of freedom in a region of spacetime surrounding a black hole must be proportional to the surface area of the horizon \cite{tHooft:1993dmi}. This \emph{holographic principle} was subsequently expanded upon by Susskind in \cite{Susskind:1994vu}.

Finally, in 1998, Maldacena published his landmark paper that initiated the study of AdS/CFT \cite{Maldacena:1997re}. The essential observation was that there are two descriptions of branes: One is the string/M-theoretic description of D/M-branes, and the other is the supergravity description. The string/M-theoretic description gives a conformal field theory living on the worldvolume traced out by the brane. On the other hand, the supergravity description generally has a near-horizon limit that resembles AdS. Hence, Maldacena conjectured that the two descriptions are, in fact, the same.

Specifically, we get a ``holographic dictionary'' that relates gravity and CFT quantities. This dictionary generally depends on the precise correspondence under consideration, but as an example, we may consider the duality obtained from a stack of $N$ coincident D3-branes in IIB string theory. There is both an open string description as well as a supergravity description of these branes, which leads to the duality. On the gravity side of this correspondence, we have IIB supergravity on $\mathrm{AdS}_5\times S^5$ with $N$ units of flux through $S^5$, string coupling $g_s$, AdS radius $L$, and string length $\alpha'=\ell_s^2$. On the CFT side, we have $d=4$, $\mathcal{N}=4$ super-Yang-Mills with gauge group $SU(N)$ and coupling $g_\text{YM}$. These parameters are then related by
\begin{equation}
    g_\text{YM}^2=2\pi g_s,\qquad 2g_\text{YM}^2N=\qty(\frac{L^2}{\alpha'})^2.\label{eq:IIBdict}
\end{equation}
This dictionary tells us quite a bit about the dual field theory. In particular, the supergravity description is valid for $g_s\ll 1$ and $\alpha'/L^2\ll 1$. This tells us that we must have $g_\text{YM}\ll 1$ and $N\gg 1$ such that the effective coupling $\lambda=g_\text{YM}^2N\gg 1$. That is, while the gravity theory is weakly coupled, the dual field theory is \emph{strongly coupled}. 

We also see that the symmetries match on both sides of the correspondence. The AdS$_5$ isometry group and the 4D conformal group are both $SO(4,2)$; similarly, $S^5$ has isometry group $SO(6)$ and the $R$-symmetry group of $\mathcal N=4$ SYM is $SU(4)\sim SO(6)$, so we see that the ``internal''\footnote{It is a bit misleading to refer to the $S^5$ as ``internal'' since the $S^5$ radius and the $\mathrm{AdS}_5$ radius are equal. However, one may still consider a consistent truncation to $\mathrm{AdS}_5$.} dimensions geometrize the $R$-symmetry. This is only the bosonic symmetries, but a more careful analysis involving fermions reveals that the full $PSU(2,2\mid 4)$ symmetry matches.

This correspondence was subsequently fleshed out in papers by Gubser, Klebanov, and Polyakov \cite{Gubser:1998bc} and Witten \cite{Witten:1998qj}. In Poincar\'e coordinates, the boundary of AdS lies at $z=0$, and bulk supergravity fields have boundary behavior
\begin{equation}
    \phi(x,z)\overset{z\to0}{\sim}z^\Delta\phi_{(0)}(x)\label{eq:bc},
\end{equation}
where $\phi$ generically refers to any field and $\Delta$ corresponds to the conformal scaling dimension. $\phi_{(0)}$ is then interpreted as the source of the CFT operator $\mathcal O$ of dimension $d-\Delta$, where $d$ is the dimension of the dual CFT. This then means that we may equate the bulk partition function of supergravity subject to the boundary condition \eqref{eq:bc} with the generating functional of the CFT
\begin{equation}
    Z_\text{sugra}\big\vert_{\phi\to\phi_{(0)}}=Z_\text{CFT}[\phi_{(0)}].
\end{equation}
This implicitly gives us $n$-point functions of $\mathcal O$ and formalizes the notion that ``the CFT lives on the boundary.'' Making use of the semiclassical limit of supergravity, we may use the stationary phase approximation
\begin{equation}
    Z_\text{sugra}\approx e^{iI_\text{sugra}},
\end{equation}
where we have used $I$ to denote the on-shell action.\footnote{This is in contrast to the ``off-shell'' action $S[g,\phi]$, which is a functional that may take as input any choice of field configuration, whereas the on-shell action $I$ is a function of the boundary data obtained by plugging a saddle-point configuration into $S$.} Notably, this gives us a tractable way to do computations for a strongly coupled field theory.

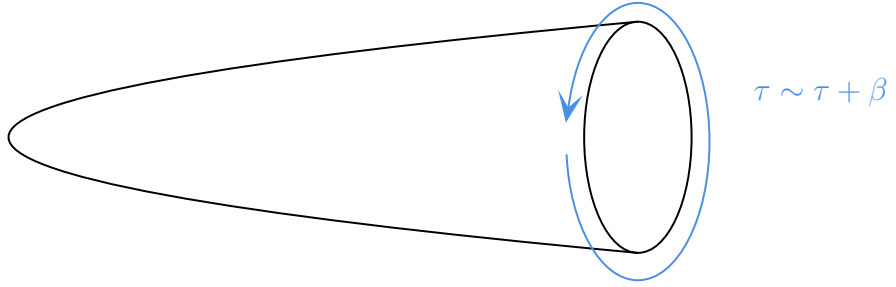
\begin{figure}[t]
    \centering
    \tikzset{every picture/.style={line width=0.75pt}} 

    \begin{tikzpicture}[x=0.75pt,y=0.75pt,yscale=-1,xscale=1]
    
    \draw   (427.41,25.1) .. controls (4.19,63.99) and (4.19,102.88) .. (427.41,141.77) ;
    \draw   (400.32,83.43) .. controls (400.32,51.22) and (412.45,25.1) .. (427.41,25.1) .. controls (442.37,25.1) and (454.5,51.22) .. (454.5,83.43) .. controls (454.5,115.65) and (442.37,141.77) .. (427.41,141.77) .. controls (412.45,141.77) and (400.32,115.65) .. (400.32,83.43) -- cycle ;
    \draw  [draw opacity=0] (392.25,69.49) .. controls (396.01,38.6) and (410.33,15.6) .. (427.41,15.6) .. controls (447.36,15.6) and (463.53,46.94) .. (463.53,85.6) .. controls (463.53,124.26) and (447.36,155.6) .. (427.41,155.6) .. controls (408.58,155.6) and (393.12,127.67) .. (391.44,92.02) -- (427.41,85.6) -- cycle ; \draw  [color={rgb, 255:red, 74; green, 144; blue, 226 }  ,draw opacity=1 ] (392.25,69.49) .. controls (396.01,38.6) and (410.33,15.6) .. (427.41,15.6) .. controls (447.36,15.6) and (463.53,46.94) .. (463.53,85.6) .. controls (463.53,124.26) and (447.36,155.6) .. (427.41,155.6) .. controls (408.58,155.6) and (393.12,127.67) .. (391.44,92.02) ;  
    \draw  [color={rgb, 255:red, 74; green, 144; blue, 226 }  ,draw opacity=1 ][fill={rgb, 255:red, 74; green, 144; blue, 226 }  ,fill opacity=1 ] (397.56,64.6) -- (391.31,74.77) -- (388.5,62.73) -- (392.17,69.22) -- cycle ;

    \draw (484.43,52) node [anchor=north west][inner sep=0.75pt]    {$\textcolor[rgb]{0.29,0.56,0.89}{{\textstyle \tau \sim \tau +\beta }}$};

    \end{tikzpicture}
    
    \caption[Cigar geometry of Schwarzschild black hole]{The Schwarzschild geometry in Euclidean signature. Here the angular directions are suppressed, which should be interpreted as a non-contractible $S^{d-2}$ at each point. The geometry smoothly caps off as we go to the horizon (hence the time circle is contractible), while as $r\to\infty$, we have periodicity $\beta$.}
    \label{fig:cigar}
\end{figure}

Moreover, this correspondence naturally incorporates temperature. In particular, one may consider asymptotically AdS supergravity solutions, such as a black hole. In Euclidean signature, the black hole temperature $T$ causes time to have an asymptotic periodicity $\beta=1/T$. For the case of the Schwarzschild black hole, this results in a cigar geometry, as shown in Figure \ref{fig:cigar}.
Notably, the boundary time is identified with the CFT time, so this periodicity extends to the CFT. Thus, we see that the CFT is also at temperature $T$.

Of course, the idea of gauge/gravity duality is broader than the original AdS/CFT correspondence. In particular, there are generalizations away from conformal symmetry to more general \ac{QFT} \cite{Glazek:2013jba}, as well as generalizations away from AdS to Minkowski \cite{Pasterski:2016qvg} and dS \cite{Strominger:2001pn}. There are also bottom-up constructions that do not \emph{a priori} originate from string theory or even require supersymmetry, such as the AdS$_3$/CFT$_2$ \cite{Coussaert:1995zp,Witten:2007kt}, JT/SYK\footnote{Although it will not be relevant for our present purposes, the JT/SYK correspondence provides a duality between a near-AdS$_2$ dilaton gravity due to Jackiw and Teitelbohm \cite{Jackiw:1984je,Teitelboim:1983ux} and a particular conformal quantum mechanics due to Sachdev, Ye, and Kitaev \cite{Sachdev:1992fk,Sachdev:2015efa,Kitaev:2015}.} \cite{Sarosi:2017ykf}, Kerr/CFT \cite{Guica:2008mu}, and higher-spin/$O(N)$ \cite{Giombi:2012ms} correspondences. This hints that holography is a general property of gravity, and hence understanding holography implies understanding (at least some part of) gravity. In particular, because the field theory side is generally well-defined for all parameter regimes, holography provides a non-perturbative definition of quantum gravity.

\section{Supergravity}
Supergravity has a long and rich history starting with its discovery in 1976 \cite{Freedman:1976xh,Deser:1976eh}, and much effort has been devoted to its study (see \emph{e.g.} \cite{Freedman:2012zz,Sezgin:2023hkc}). However, for our purposes, it boils down to general relativity coupled to appropriate choices of matter (gauge fields, scalars, spinors, \emph{etc}.) in such a way as to also respect supersymmetry. Equivalently, it is a theory of local supersymmetry \cite{Nath:1975nj}; this follows from the fact that the supercharges generically anticommute to the momentum operator, which means that to gauge diffeomorphisms, we must gauge supersymmetry and \emph{vice versa}.

Motivated by the preceding discussion of holography, we are predominantly interested in classical supergravity solutions since we work in the large-$N$, large-$\lambda$ limit, where $G_N\hbar\ll 1$. In general, any non-vanishing fermion vacuum expectation value would violate our spacetime isometries, so the fermions must be set to zero. Since we want to remove the fermions anyway, we just work with the bosonic terms in the action.\footnote{The unwary reader may na\"ively think that the fermion terms ``can't be that bad.'' Said reader is directed to look at the 227 pages of fermion terms in \cite{Gold:2023ykx}.} Imposing supersymmetry then reduces to checking the vanishing of the fermion supersymmetry variations, giving us relations that we will refer to as the \emph{Killing spinor equations}.

\subsection{Higher derivatives}
Many quantum field theories, especially non-renormalizable ones, are now generally thought of in the framework of \ac{EFT}; in particular, Einstein gravity may be viewed as a low energy EFT of some UV complete theory of quantum gravity \cite{Donoghue:1994dn}. From this perspective, one can be agnostic about what precise theory the UV completion entails so long as we stay within appropriate kinematic regions. It has long been known that the one-loop renormalization of Einstein gravity contains higher-derivative corrections \cite{tHooft:1974toh,Deser:1974cy,Deser:1974cz,Deser:1974xq,Barvinsky:1981rw} (see also \cite{Goroff:1985sz} for the case of two loops). So, from a Wilsonian RG perspective, it is natural to include higher-derivative corrections as irrelevant operators modifying the EFT at high energy. Several methods for constructing such higher-derivative corrections are the Noether procedure \cite{VanNieuwenhuizen:1981ae}, superconformal tensor calculus \cite{Hanaki:2006pj,Ozkan:2013uk,Ozkan:2013nwa,Gold:2023ymc,Gold:2023ykx}, ordinary superspace \cite{Gates:1983nr,Howe:1995md,Kuzenko:1998tsq,Cederwall:2004cg,Gold:2023dfe}, harmonic superspace \cite{deHaro:2002vk}, the superform method (ectoplasm) \cite{Gates:1997ag,Gates:1997kr,Gates:1998hy}, and holography \cite{Chester:2018aca,Binder:2018yvd}. See \cite{Ozkan:2024euj} for a comprehensive review.

Conversely, if we consider our UV completion to be a (super)string theory, then the string has some finite size characterized by $\alpha'=\ell_s^2$. So, if we take the relevant curvature scale $L$ to be much larger than the string length, $L^2/\alpha'\gg 1$, it is natural to expand the action in powers of $\alpha'$. Since $\alpha'$ has units of (Length)$^2$, we also need a couple of powers of inverse length that naturally come in the form of derivatives. This leads to the well-known string effective actions, with the zero-slope limit being precisely two-derivative (super)gravity. In particular, string theories have massless modes (which we shall schematically denote $\phi_0$) in addition to an infinite tower of massive modes (which we shall schematically denote $\phi_n$), and the effective actions are obtained by integrating out these massive modes,
\begin{equation}
    e^{iS_\mathrm{eff}[\phi_0]}=\int\mathcal D\phi_n\,e^{iS[\phi_0,\phi_n]},
\end{equation}
whose influence is encoded in derivatives of the massless fields, containing both classical and loop contributions. In practice, these are generally obtained from S-matrix elements \cite{Gross:1986iv,Gross:1986mw}, sigma model $\beta$-functions \cite{Grisaru:1986vi,Freeman:1986zh}, imposing local supersymmetry \cite{Becker:2006dvp,Gates:1986dm,Gates:1985wh,Bergshoeff:1986wc}, imposing $T$-duality \cite{Liu:2013dna,Liu:2019ses,Garousi:2019mca,Garousi:2020gio,Garousi:2023qwj}, using Double Field Theory \cite{Hohm:2013jaa,Hohm:2014xsa,Bedoya:2014pma,Marques:2015vua,Lee:2015kba,Baron:2017dvb}, or using string field theory \cite{Sen:2016qap,Maccaferri:2018vwo,Erbin:2019spp}.

One might be interested in such higher-derivative corrections because they correspond to subleading (in $N$) corrections in holography. As an example, we may revisit the example of IIB supergravity on $\mathrm{AdS}_5\times S^5$, with holographic dictionary given in \eqref{eq:IIBdict}. Higher-derivative corrections give us extra powers of $\alpha'$, corresponding to subleading corrections with respect to $N$. This pushes us away from the strict $N=\infty$ regime and allows us to do \emph{precision holography}.

Now, a word of caution is in order. We must treat these higher-derivative terms as \emph{corrections}, or else we suffer from the infamous Ostrogradsky instability \cite{Ostrogradsky:1850fid}. The essential point is that a higher-derivative action, which is not thought of as corrections to the two-derivative theory, will generically have a Hamiltonian not bounded from below. At the quantum level, such a theory suffers from pathologies such as position and momentum commuting, as well as ghost modes. To build some intuition, consider the example of a massless scalar in flat space (see \cite{Myers:2010tj}). We may consider a higher-derivative modified Klein-Gordon equation
\begin{equation}
    \qty(\Box+\frac{a}{M^2}\Box^2)\phi=0,
\end{equation}
where $M^2$ is some high-energy scale analogous to $1/\alpha'$ (and required for the dimensions to work out). The propagator is thus
\begin{equation}
    \Delta(p)=\frac{1}{p^2\qty(1-\frac{a}{M^2}p^2)}=\frac{1}{p^2}-\frac{1}{p^2-\frac{M^2}{a}}.
\end{equation}
The $1/p^2$ pole is associated with the usual massless scalar, whereas the second pole is associated with additional massive modes. For $a<0$, these modes are tachyonic, but for any choice of $a$ they are ghosts. This is because their contribution to the propagator has the wrong sign and must correspond to a negative-norm state. One might hope this is a pathological example, but it is a general feature of higher-derivative theories.

The problem is that the equations of motion become higher-order in derivatives, which require more initial data to specify a solution. So we are forced to view the system as having additional degrees of freedom. If we view the higher derivatives simply as corrections and expand perturbatively around the two-derivative solutions, then we have the same degrees of freedom as the two-derivative theory, and all our issues with instabilities and ghosts go away \cite{Jaen:1986iz,Simon:1990ic,Simon:1990jn,Cheng:2001du}. Effectively, this discards solutions that are not continuously connected to the two-derivative solutions, which gives us a decrease in degrees of freedom. In the above example, we see that the ghost term in the propagator diverges as $a/M^2\to 0$, so it would be discarded. This prescription makes sense since we would have gotten the same result by taking the full string effective action\footnote{If we pretend for the moment that we knew the string effective action to all orders in $\alpha'$.} and expanding in $\alpha'$; requiring convergence to the original string action automatically discards the spuriously divergent solutions. Said another way, the spirit of effective field theory is to use the IR degrees of freedom and compute corrections perturbatively, and we are safe so long as we do that.

However, there is a notable exception to this instability, which is when the higher-derivative terms still lead to second-order equations of motion: This happens for the so-called Lovelock terms \cite{Lovelock:1971yv}
\begin{align}
    e^{-1}\mathcal L_0&=\Lambda,\nn\\
    e^{-1}\mathcal L_1&=R,\nn\\
    e^{-1}\mathcal L_2&=R_{\mu\nu\rho\sigma}R^{\mu\nu\rho\sigma}-4R_{\mu\nu}R^{\mu\nu}+R^2,\nn\\
    \vdots&\nn\\
    e^{-1}\mathcal L_n&=\frac{1}{2^n}\delta^{\mu_1\nu_1...\mu_n\nu_n}_{\alpha_1\beta_1...\alpha_n\beta_n}\prod_{r=1}^n R^{\alpha_r\beta_r}{}_{\mu_r\nu_r},
\end{align}
where the $n$-th Lovelock term is the Euler density in $2n$ dimensions, with the zeroth Lovelock term being a cosmological constant and the first being the usual Einstein-Hilbert term. Since the Lovelock terms always lead to second-order equations of motion,\footnote{Such a higher-derivative term in the action is called \emph{quasi-topological}.} they may be treated as a proper theory without viewing them as corrections. That being said, in this dissertation, we will still treat them as corrections.

\subsection{Compactifications and consistent truncations}
The idea of dimensional reduction dates back more than a century to the 1914 work of Nordstr\"om \cite{Nordstrom:1914fn}, who formulated a unified theory of electromagnetism and scalar gravity starting from five-dimensional Maxwell theory, work which predates Einstein's 1915 theory of general relativity \cite{Einstein:1914bt}. It was Kaluza \cite{Kaluza:1921tu} who showed in 1921 that reducing general relativity from five to four dimensions yields gravity coupled to electromagnetism and a massless scalar, which was subsequently often set to zero. Klein \cite{Klein:1926tv} later came up with the idea of compactifying this fifth dimension on a circle, which led to an expansion in Fourier modes, known as the \ac{KK} tower. By identifying the first Fourier mode with the electric charge, he was able to compute the radius of the compact dimension, which turned out to be of order the Planck length, and gave a geometric explanation for the quantization of electric charge. On the other hand, it predicted that the first massive mode was of order the Planck mass. 

It was not until the work of Jordan in 1947 \cite{Jordan:1947} and Thiry in 1948 \cite{Thiry:1948ltk} that it was appreciated that it is inconsistent to set the scalar to zero unless the Maxwell field is also set to zero. Later, in 1969, Hawking \cite{Hawking:1968zw} considered the consistency of dimensional reductions in the context of Bianchi cosmologies obtained by reducing GR on a three-dimensional Lie group. He found that if the structure constants were not traceless (\emph{i.e.}, if the group was not unimodular), then the reduction of the equations of motion failed to match the equations of motion of the reduced Lagrangian. This requirement was pointed out in more general reductions by Scherk and Schwarz in 1979 \cite{Scherk:1979zr}. 

Let us now review the circle KK reduction to illustrate several important points. We start with a metric in $D+1$ dimensions and decompose it as
\begin{equation}
    \dd\hat s^2\equiv \hat g_{MN}\dd \hat x^M\dd \hat x^N=g_{\mu\nu}\dd x^\mu\dd x^\nu+e^{2\phi}\qty(\dd z+A_\mu\dd x^\mu)^2,
\end{equation}
where hats denote $(D+1)$-dimensional quantities and we have split the coordinates as $\hat x^M=(x^\mu, z)$. If we allow $g$, $A$, and $\phi$ to depend on both $x$ and $z$ then this is simply a rewriting of the original metric in a particular gauge. However, we are usually interested in getting a $D$-dimensional theory out, so we further impose the cylinder condition
\begin{equation}
    \partial_z=0,
\end{equation}
which is the statement that nothing depends on the internal coordinate $z$. This lets one view the $(D+1)$-dimensional spacetime manifold as a circle fibration over a $D$-dimensional base.\footnote{It is very common to refer to putting a theory on $M\times X$. This is almost always a warped product or just a general fibration, and rarely ever a direct product as the symbol $\times$ implies.} Moreover, the $U(1)$ subset of $(D+1)$-dimensional diffeomorphisms
\begin{equation}
    z\to z-\lambda(x),
\end{equation}
leads to a gauge symmetry
\begin{equation}
    A\to A+\dd\lambda,
\end{equation}
and so $A$ is naturally interpreted as a $U(1)$ gauge field. This geometrizes the Maxwell gauge symmetry.

There are two distinct but often confused philosophies of dimensional reduction that we wish to distinguish: \emph{compactification} and \emph{consistent truncation}. In a compactification, one views the resulting theory as merely an effective description of a truly $(D+1)$-dimensional theory. Hence, the fields, which we will schematically denote as $\hat\Psi(x,z)$, may be expanded into Fourier modes as
\begin{equation}
    \hat\Psi(x,z)=\sum_{n\in\mathbb Z} \psi_n(x) e^{inz/R},
\end{equation}
where $R$ is the radius of the circle. We may view this as rewriting each $(D+1)$-dimensional field as an infinite tower of $D$-dimensional fields. The inverse is given by
\begin{equation}
    \psi_n(x)=\int_{0}^{2\pi}\dd z \, e^{-inz/R}\hat\Psi(x,z).
\end{equation}
Returning to our earlier example of a circle reduction, if we set $A=0=\phi$ for simplicity\footnote{Otherwise, we would get a mess of interaction terms that obscure the point. This is, of course, how one sees the appearance of a gauge-covariant derivative.} and consider the Klein-Gordon equation for the dilaton $\phi$, we see
\begin{equation}
    \hat\Box\phi(x,z)=\sum_{n=-\infty}^\infty \qty(\Box-\frac{n^2}{R^2})\phi_n(x) e^{inz/R}.
\end{equation}
That is, the $n$-th mode has an effective $D$-dimensional mass of $|n|/R$. So if the compactified dimension is very small (compared to all other relevant length scales), then the mass is very large, and all the modes decouple. So we may effectively only consider the zero modes $\phi_0$, but with the idea that if we went to high enough energies, we would see those massive modes appearing. We will take this perspective in Chapter \ref{chap:RG}.

On the other hand, the idea of a consistent truncation is simply to view the reduced theory as a solution to the equations of motion. That is, the cylinder condition is consistent if the equations of motion that we get from the reduced Lagrangian match the reduced equations of motion from the original Lagrangian, \emph{i.e.}, we want the diagram 
\begin{center}
    \begin{tikzpicture}
    \node[] at (0,0) (L){$\mathcal{L}$};
    \node[] at (0,-2) (Lp) {$\mathcal{L}'$};
    \node[] at (2,0) (E) {$\mathcal{E}$};
    \node[] at (2,-2) (Ep) {$\mathcal{E}'$};

    \draw[->,dashed] (L) -- (Lp) node[midway,left,black]{\small $\partial_z=0$};
    \draw[->] (L) -- (E) node[midway,above,black]{$\delta$};
    \draw[->] (Lp) -- (Ep) node[midway,above,black]{$\delta$};
    \draw[->,dashed] (E) -- (Ep) node[midway,right,black]{\small $\partial_z=0$};
\end{tikzpicture}
\end{center}
to commute, where $\mathcal E$ ($\mathcal E'$) denotes the equations of motion obtained by varying ($\delta$) the Lagrangian $\mathcal L$ ($\mathcal L'$). This ensures that solutions of the reduced theory uplift to solutions of the original theory. In this sense, consistent truncations generate a new (generally simpler) theory from one we already have, and the higher-dimensional theory functions as an intermediate device. We will take this perspective in Chapters \ref{chap:torus} and \ref{chap:sphere}. This perspective also allows us to easily see why we cannot set the scalar to zero:\footnote{\emph{i.e.}, the further truncation $\phi=0$ is generically inconsistent in the above sense.} It is sourced by the Maxwell field
\begin{equation}
    \Box\phi\propto F^2,
\end{equation}
where $F=\dd A$ locally, and so the truncation would only be consistent if we set $F=0$ too.

As it turns out, the cylinder condition always results in a consistent truncation on a circle. This has to do with group theory. The situation that one worries about, as with the dilaton above, is a field equation of the form
\begin{equation}
    \Box H=L^2,
\end{equation}
where $H$ is a field we wish to truncate and $L$ is a field we wish to keep. However, for the cylinder condition, $L$ must necessarily be a singlet under the $U(1)$ isometry of $S^1$, while $H$ is necessarily charged under $U(1)$. But no product of non-singlets can produce a singlet on $S^1$, so nothing can ever go wrong. Likewise, it is straightforward to generalize this procedure to the torus, which we will have more to say about in Chapter \ref{chap:torus}.

\section{Overview of the dissertation}
This dissertation explores topics in holography and supergravity in the pursuit of further advancing
our understanding of quantum gravity. The work presented is based on articles written with
my advisor Professor James T. Liu and collaborators, Professor Leopoldo Pando Zayas and Evan Deddo. In the following, we give an overview of the structure of the remaining parts of the thesis. A more extensive introduction to these topics will be given in each chapter.

\subsection*{Chapter II}
In this chapter, we study subleading corrections to the genus-zero free energy of the $\mathcal{N}=3$ Gaiotto-Tomasiello theory. In general, we obtain the endpoints and free energy as a set of parametric equations via contour integrals of the planar resolvent up to exponentially suppressed corrections. In the case that the two gauge groups in the quiver are of equal rank, we find an explicit (perturbative) expansion for the free energy. If, additionally, both groups have equal levels, then we find the full expression for the genus-zero free energy, modulo exponentially suppressed corrections. We also verify our results numerically.

This chapter is based on \cite{Liu:2021njm}.

\subsection*{Chapter III}
In this chapter, we study four-derivative corrections to pure $\cN=2$, $D=5$ gauged supergravity. In particular, we find that, up to field redefinitions, there is a single four-derivative superinvariant that one can add to the action, up to factors of the two-derivative action. Consequently, this selects a unique set of coefficients for the four-derivative corrections. We confirm these coefficients (in the ungauged limit) on the BMPV solution.

This chapter is based on \cite{Liu:2022sew}.

\subsection*{Chapter IV}
In this chapter, we study the role of higher-derivative corrections to Einstein gravity in the context of gravitational theories describing renormalization group flows across dimensions via AdS/CFT. We use the Null Energy Condition to derive monotonicity properties of candidate holographic central charges formed by combinations of metric functions. We also implement an entropic approach to the characterization of the four-derivative flows using the Jacobson-Myers functional and demonstrate, under reasonable conditions, the monotonicity of certain terms in the entanglement entropy via the appropriate generalization of the Ryu-Takayanagi prescription. In particular, we show that any flow from a higher dimensional theory to a holographic CFT$_2$ satisfies a type of monotonicity. We also uncover direct relations between NEC-motivated and entropic central charges. 

This chapter is based on \cite{Liu:2023fqq}.

\subsection*{Chapter V}
In this chapter, we consider the torus reduction of heterotic supergravity in the presence of four-derivative corrections. In particular, the reduction on $T^n$ generically leads to a half-maximal supergravity coupled to $n$ vector multiplets, and we show that it is consistent to truncate out said vector multiplets. This is done by analyzing both the bosonic equations of motion and the Killing spinor equations.  As an application of the consistent truncation, we examine the four-derivative corrected BPS black string that reduces to a black hole in minimal nine-dimensional supergravity.

This chapter is based on \cite{Deddo:2023pid}.

\subsection*{Chapter VI}
At the two-derivative order, the group manifold reduction of heterotic supergravity on $S^3$ results in a half-maximal 7D gauged supergravity coupled to three vector multiplets, and a further truncation can be taken to remove the vector multiplets. In this chapter, we demonstrate that this truncation remains consistent at the four-derivative level; we do so both by analysis of the equations of motion and the supersymmetry variations. 

This chapter is based on \cite{Liu:2023fmv}.

%% file: Chapters/chap2.tex
As discussed in Chapter \ref{chap:Intro}, the AdS/CFT correspondence conjectures a remarkable equivalence between large-$N$ gauge theories and string/M-theory on asymptotically AdS backgrounds.  In this context, Chern-Simons-matter theories are of particular interest in regards to the dynamics of M2-branes \cite{Bagger:2006sk,Bagger:2007jr,Bagger:2007vi,Gustavsson:2007vu,Gustavsson:2008dy,VanRaamsdonk:2008ft,Aharony:2008ug}. In particular, the worldvolume theory of $N$ coincident M2-branes probing the singularity of a $\mathbb{C}^4/\mathbb{Z}_k$ orbifold was constructed in \cite{Aharony:2008ug} and is known as the \ac{ABJM} theory. ABJM theory is an $\mathcal N=6$, $U(N)_k\times U(N)_{-k}$ Chern-Simons-matter theory, and in the large-$N$ limit is dual to either M-theory on AdS$_4\times S^7/\mathbb{Z}_k$ or IIA string theory on AdS$_4\times\mathbb{CP}^3$, depending on the limit taken.

ABJM theory and its holographic dual provide an excellent opportunity to probe the dynamics of string/M-theory as well as quantum gravity and AdS$_4$ black holes.  However, as AdS/CFT is a strong/weak coupling duality, it is highly non-trivial to directly compare both sides of the duality.  Nevertheless, certain path integrals in superconformal Chern-Simons-matter theories reduce to matrix models via supersymmetric localization \cite{Pestun:2007rz,Kapustin:2009kz}. Such localization techniques have long been studied in the context of supersymmetric and topological QFTs, and the application of \cite{Pestun:2007rz,Kapustin:2009kz} to superconformal field theories have proven a powerful technique to analyze observables via matrix models. In particular, ABJM theory can be localized to a two-matrix model \cite{Kapustin:2009kz}, which can then be studied via standard methods of random matrix theory or by novel methods such as the ideal Fermi gas approach \cite{Marino:2011eh}.

Many important results have been obtained for the supersymmetric partition function and Wilson loop observables in ABJM theory \cite{Marino:2009jd,Marino:2011eh,Hatsuda:2012hm,Putrov:2012zi} and the ABJ generalization \cite{Awata:2012jb,Honda:2014npa,Hatsuda:2016rmv,Cavaglia:2016ide}.  In particular, the $S^3$ partition function at fixed Chern-Simons levels $k$ and $-k$ was shown to have the form of an Airy function
\begin{equation}
    Z_{\mathrm{ABJM}}^{S^3}=\left(\fft2{\pi^2k}\right)^{-1/3}e^{A(k)}\Ai\left[\left(\fft2{\pi^2k}\right)^{-1/3}\left(N-\fft1{3k}-\fft{k}{24}\right)\right]+Z_{\mathrm{np}},
\end{equation}
where $A(k)$ encodes certain quantum corrections and $Z_{\mathrm{np}}$ is a non-perturbative contribution.  Taking $F=\log Z$ then leads to a fixed $k$ expansion of the free energy as
\begin{equation}
    F_{\mathrm{ABJM}}=\fft{\pi\sqrt2}3k^{1/2}N^{3/2}-\fft\pi{\sqrt{2k}}\left(\fft{k^2}{24}+\fft13\right)N^{1/2}+\fft14\log N+\mathcal O(1).
\end{equation}
In the M-theory dual, the $N^{3/2}$ term can be matched to the on-shell classical supergravity action, while the $N^{1/2}$ term is related to eight-derivative couplings in M-theory \cite{Bergman:2009zh,Aharony:2009fc} which reduce to four-derivative couplings in AdS$_4$ supergravity \cite{Bobev:2020egg,Bobev:2021oku}.  The Airy function form of the partition function holds for a wide range of Chern-Simons-matter theories beyond ABJM theory.  Then, by expanding the Airy function at large $N$, one can see that the $\fft14\log N$ term is universal to this full set of theories.  As an important test of quantum gravity, this log term has been reproduced successfully by a one-loop calculation in eleven-dimensional supergravity on AdS$_4\times X^7$ \cite{Bhattacharyya:2012ye}.

Given the remarkable successes of precision tests of ABJM holography, we wish to extend such investigations to the \ac{GT} case \cite{Gaiotto:2009mv}.  The GT theory is an $\mathcal{N}=3$ Chern-Simons-matter theory and can be thought of as a generalization of the ABJM theory to arbitrary Chern-Simons levels, $k_1$ and $k_2$, with $F_0=k_1+k_2\ne0$.  This model is dual to massive IIA supergravity with $F_0$ playing the role of the Romans mass \cite{Romans:1985tz}.  The leading order behavior of GT free energy is \cite{Suyama:2010hr,Suyama:2011yz,Jafferis:2011zi}
\begin{equation}
    F_{\mathrm{GT}}=\fft{3^{5/3}\pi }{5\cdot2^{4/3}}e^{-i\pi/6}(k_1+k_2)^{1/3}N^{5/3}+\cdots.
\label{eq:FGT0}
\end{equation}
The $N^{5/3}k^{1/3}$ scaling is in contrast to the $N^{3/2}k^{1/2}$ scaling of the ABJM free energy and has confirmed on the supergravity side \cite{Aharony:2010af}.  While this leading-order behavior is well established and generalizes to a large class of $\mathcal N=3$ necklace quiver models with $F_0\ne0$, less is known about its subleading corrections, which is the focus of this chapter.

Although the partition function for GT theory can also be mapped to a corresponding ideal Fermi gas system, unlike for the ABJM model, the resulting expression does not take the form of an Airy function \cite{Marino:2011eh,Hong:2021bsb}.  Furthermore, the mapping to the quantum Fermi gas system promoted in \cite{Marino:2011eh} involves taking
\begin{equation}
    \fft{4\pi}\hbar=\fft1{k_1}-\fft1{k_2}.
\end{equation}
This demonstrates that a small $\hbar$ expansion is in tension with taking $k_1\approx k_2$, the natural realm for exploring the free energy in (\ref{eq:FGT0}).  We thus find it more natural to work directly with the GT theory partition function written as a two-matrix model.  While a saddle point analysis was performed in \cite{Hong:2021bsb}, here we use a standard resolvent approach and compute the genus-zero partition function as an expansion in inverse powers of the 't~Hooft parameter $t=g_sN$ with $g_s=2\pi i/k$ where $k$ is an effective overall Chern-Simons level.  For equal levels, $k=k_1=k_2$, we find (at genus zero)
\begin{equation}
    F_{\mathrm{GT}}^{k_1=k_2}=\fft1{g_s^2}\left[\fft35\left(\fft{3\pi^2}2\right)^{2/3}\left(t+\fft{\zeta(3)}{2\pi^2}\right)^{5/3}-\fft{\pi^2}{12}t+\mbox{const.}\right],
\label{eq:FGT=}
\end{equation}
up to exponentially small corrections in the large $|t|$ limit.

To highlight the first subleading corrections to the planar free energy, we  substitute $t=2\pi iN/k$ into (\ref{eq:FGT=}) and expand to obtain
\begin{equation}
    F_{\mathrm{GT}}^{k_1=k_2}=\fft{3^{5/3}\pi}{10}e^{-i\pi/6}k^{1/3}N^{5/3}+\fft{i\pi}{24}kN+\fft{3^{2/3}}{8\pi^2}e^{-2\pi i/3}\zeta(3)k^{4/3}N^{2/3}+\mathcal O(1).
\end{equation}
The leading order $N^{5/3}$ term matches (\ref{eq:FGT0}), while the linear-$N$ term was previously obtained in \cite{Hong:2021bsb}, and is pure imaginary for real Chern-Simons levels.  At the next order, we find a $N^{2/3}$ term with a coefficient proportional to $\zeta(3)$.  This term is of $\mathcal O(1/t)$ compared to the leading order and has a natural interpretation in the massive IIA supergravity dual as originating from a tree-level $\alpha'^3R^4$ coupling.

The rest of this chapter is organized as follows. In Section~\ref{sec:GTreview}, we predominantly follow \cite{Suyama:2010hr} in summarizing important results about the planar limit and the resolvent in GT theory.  We then proceed in Section~\ref{sec:GTfree} to obtain the planar free energy from the resolvent in the limit of large 't~Hooft coupling and further check our results against numerical data.  Finally, we conclude in Section~\ref{sec:disc} with some open questions.  Some of the more technical calculations are relegated to two appendices.

\section{GT theory and the planar resolvent}
\label{sec:GTreview}

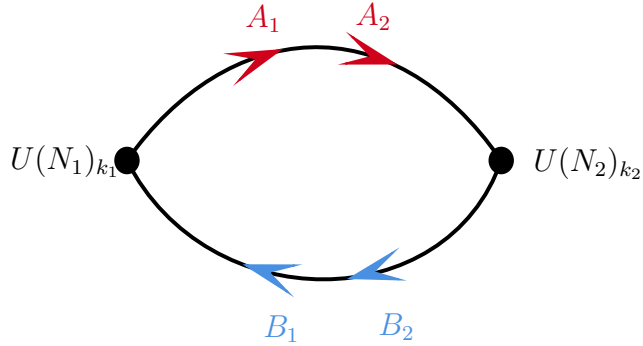
\begin{figure}[t]
    \centering
    \tikzset{every picture/.style={line width=0.75pt}} 
    
    \begin{tikzpicture}[x=0.75pt,y=0.75pt,yscale=-.75,xscale=.75]
    
    \draw  [fill={rgb, 255:red, 0; green, 0; blue, 0 }  ,fill opacity=1 ] (149.51,129.68) .. controls (149.51,124.93) and (153.03,121.07) .. (157.38,121.07) .. controls (161.73,121.07) and (165.25,124.93) .. (165.25,129.68) .. controls (165.25,134.44) and (161.73,138.3) .. (157.38,138.3) .. controls (153.03,138.3) and (149.51,134.44) .. (149.51,129.68) -- cycle ;
    \draw  [fill={rgb, 255:red, 0; green, 0; blue, 0 }  ,fill opacity=1 ] (401.18,129.68) .. controls (401.18,124.93) and (404.7,121.07) .. (409.05,121.07) .. controls (413.4,121.07) and (416.92,124.93) .. (416.92,129.68) .. controls (416.92,134.44) and (413.4,138.3) .. (409.05,138.3) .. controls (404.7,138.3) and (401.18,134.44) .. (401.18,129.68) -- cycle ;
    \draw [color={rgb, 255:red, 0; green, 0; blue, 0 }  ,draw opacity=1 ][line width=1.5]    (157.38,129.68) .. controls (211.53,236.99) and (372.55,235.68) .. (409.05,129.68) ;
    \draw [color={rgb, 255:red, 0; green, 0; blue, 0 }  ,draw opacity=1 ][line width=1.5]    (157.38,129.68) .. controls (224.65,34.65) and (334.38,21.6) .. (409.05,129.68) ;
    \draw  [color={rgb, 255:red, 208; green, 2; blue, 27 }  ,draw opacity=1 ][fill={rgb, 255:red, 208; green, 2; blue, 27 }  ,fill opacity=1 ] (227.29,57.97) -- (259.97,55.51) -- (232.87,75.63) -- (245.03,61.16) -- cycle ;
    \draw  [color={rgb, 255:red, 208; green, 2; blue, 27 }  ,draw opacity=1 ][fill={rgb, 255:red, 208; green, 2; blue, 27 }  ,fill opacity=1 ] (308.8,46.42) -- (337.13,64.42) -- (304.37,64.47) -- (321.86,59.93) -- cycle ;
    \draw  [color={rgb, 255:red, 74; green, 144; blue, 226 }  ,draw opacity=1 ][fill={rgb, 255:red, 74; green, 144; blue, 226 }  ,fill opacity=1 ] (265.61,218.32) -- (237.35,200.19) -- (270.11,200.29) -- (252.6,204.75) -- cycle ;
    \draw  [color={rgb, 255:red, 74; green, 144; blue, 226 }  ,draw opacity=1 ][fill={rgb, 255:red, 74; green, 144; blue, 226 }  ,fill opacity=1 ] (340.36,209.73) -- (307.62,208.85) -- (336.32,191.57) -- (322.98,204.75) -- cycle ;
    
    \draw (235.16,23.26) node [anchor=north west][inner sep=0.75pt]  [color={rgb, 255:red, 208; green, 2; blue, 27 }  ,opacity=1 ]  {$A_{1}$};
    \draw (309.11,21.96) node [anchor=north west][inner sep=0.75pt]  [color={rgb, 255:red, 208; green, 2; blue, 27 }  ,opacity=1 ]  {$A_{2}$};
    \draw (247.09,232.04) node [anchor=north west][inner sep=0.75pt]  [color={rgb, 255:red, 74; green, 144; blue, 226 }  ,opacity=1 ]  {$B_{1}$};
    \draw (324.62,230.74) node [anchor=north west][inner sep=0.75pt]  [color={rgb, 255:red, 74; green, 144; blue, 226 }  ,opacity=1 ]  {$B_{2}$};
    \draw (76.8,117.47) node [anchor=north west][inner sep=0.75pt]    {$U( N_{1})_{k_{1}}$};
    \draw (428.67,118.78) node [anchor=north west][inner sep=0.75pt]    {$U( N_{2})_{k_{2}}$};
    \end{tikzpicture}

    \caption[GT quiver diagram]{The $\cN=3$ GT quiver diagram. $A_1$ and $A_2$ are bifundamental hypermultiplets and $B_1$ and $B_2$ are anti-bifundamental hypermultiplets coupling the nodes of the quiver.}
    \label{fig:quiver}
\end{figure}

GT theory is an $\mathcal{N}=3$ superconformal Chern-Simons-matter theory with $U(N_1)_{k_1}\times U(N_2)_{k_2}$ gauge group and quiver diagram given in Figure~\ref{fig:quiver}.  It was originally constructed as a deformation of ABJM theory in \cite{Gaiotto:2009mv} by allowing the two $U(N)$ quivers to take on arbitrary ranks and levels, which in turn knocks the supersymmetry down from $\mathcal{N}=6$ to $\mathcal{N}=3$. On the dual gravity side, which was constructed to first order in perturbation theory in \cite{Gaiotto:2009yz}, this corresponds to turning on a nonzero Romans mass $F_0=k_1+k_2$, which is a 0-form R-R flux sourced by D8-branes.  The supergravity description then corresponds to the massive IIA theory where the 2-form NS-NS $B$-field acquires a mass precisely equal to $F_0$ by ``eating'' the 1-form gauge field in a Higgs-like mechanism \cite{Romans:1985tz}. It is generally believed that there is no M-theory limit \cite{Aharony:2010af} when this mass is non-vanishing.

Since GT theory still retains $\mathcal{N}=3$ supersymmetry, its partition function can be localized following \cite{Kapustin:2009kz}, just as in the AJBM case.  The resulting matrix model takes the form
\begin{equation}
    Z=\fft1{N_1!N_2!}\int\prod_{i=1}^{N_1}\frac{\dd{u_i}}{2\pi}\prod_{j=1}^{N_2}\frac{\dd{v_j}}{2\pi}e^{-S(u_i,v_j)},
\end{equation}
where the action is given by
\begin{equation}
    e^{-S}=\exp\qty[\frac{ik_1}{4\pi}\sum_{i=1}^{N_1}u_i^2+\frac{ik_2}{4\pi}\sum_{i=1}^{N_2}v_i^2]\frac{\prod_{i<j}^{N_1}\sinh^2\qty(\frac{u_i-u_j}{2})\prod_{i<j}^{N_2}\sinh^2\qty(\frac{v_i-v_j}{2})}{\prod_{i=1}^{N_1}\prod_{j=1}^{N_2}\cosh^2\qty(\frac{u_i-v_j}{2})}.
\end{equation}
Since there are two independent Chern-Simons levels, $k_1$ and $k_2$, we can define two 't~Hooft couplings, $\lambda_1=N_1/k_1$ and $\lambda_2=N_2/k_2$.  However, to highlight the planar limit, we find it more convenient to follow \cite{Suyama:2010hr} by introducing an auxiliary parameter $k$ and defining
\begin{equation}
    t_1=\fft{2\pi iN_1}k,\qquad t_2=\fft{2\pi iN_2}k,\qquad\kappa_1=\fft{k_1}k,\qquad\kappa_2=\fft{k_2}k.
\end{equation}
The planar limit is then taken by sending $k\to\infty$ while holding $t_i$ and $\kappa_i$ fixed.

Written in terms of the above quantities, the action now takes the form%
\footnote{Note that this choice of parameters differs from that of \cite{Suyama:2010hr} in the choice of sign of $t_2$ and $\kappa_2$.  In particular, $(t_2)_{\mathrm{there}}=(-t_2)_{\mathrm{here}}$ and $(\kappa_2)_{\mathrm{there}}=(-\kappa_2)_{\mathrm{here}}$.}
\begin{align}
    S&=\fft1{g_s^2}\Biggl[\fft{\kappa_1t_1}{2N_1}\sum_{i=1}^{N_1}u_i^2+\fft{\kappa_2t_2}{2N_2}\sum_{i=1}^{N_2}v_i^2-\fft{t_1^2}{N_1^2}\sum_{i<j}^{N_1}\log\sinh^2\fft{u_i-u_j}2\nn\\
    &\qquad\qquad-\fft{t_2^2}{N_2^2}\sum_{i<j}^{N_2}\log\sinh^2\fft{v_i-v_j}2+\fft{t_1t_2}{N_1N_2}\sum_{i=1}^{N_1}\sum_{j=1}^{N_2}\log\cosh^2\fft{u_i-v_j}2\Biggr],
\label{eq:action}
\end{align}
where we have introduced $g_s=2\pi i/k$.  While the physical Chern-Simons levels $k_1$ and $k_2$ are real, below we will analytically continue to imaginary levels such that the couplings $t_i$ and $\kappa_i$ are real.  This will allow us to work with a real action and corresponding real saddle point equations.  In particular, varying the action, (\ref{eq:action}), with respect to $u_i$ and $v_j$ gives the saddle-point equations
\begin{subequations}
\begin{align}
     \kappa_{1} u_{i} &=\frac{t_{1}}{N_{1}} \sum_{j \neq i}^{N_{1}} \operatorname{coth} \frac{u_{i}-u_{j}}{2}-\frac{t_{2}}{N_{2}} \sum_{j=1}^{N_{2}} \tanh \frac{u_{i}-v_{j}}{2}, \\
    \kappa_{2} v_{i} &=\frac{t_{2}}{N_{2}} \sum_{j \neq i}^{N_{2}} \operatorname{coth} \frac{v_{i}-v_{j}}{2}-\frac{t_{1}}{N_{1}} \sum_{j=1}^{N_{1}} \tanh \frac{v_{i}-u_{j}}{2}.
\end{align}
\label{eq:unexponentatedSPE}
\end{subequations}
At this stage, it is convenient to switch to exponentiated coordinates
\begin{equation}
    z_i:=e^{u_i},\qquad w_i:=-e^{v_i}.
\end{equation}
Making note of the sign in the definition of the $\{w_i\}$, the saddle-point equations then take the form
\begin{subequations}
\begin{align}
    \kappa_{1} \log z_{i} &=\frac{t_{1}}{N_{1}} \sum_{j \neq i}^{N_{1}} \frac{z_i+z_{j}}{z_{i}-z_{j}}-\frac{t_{2}}{N_{2}} \sum_{j=1}^{N_{2}} \frac{z_i+w_{j}}{z_{i}-w_{j}}, \\
    \kappa_{2} \log(- w_{i}) &=\frac{t_{2}}{N_{2}} \sum_{j \neq i}^{N_{2}} \frac{w_i+w_{j}}{w_{i}-w_{j}}-\frac{t_{1}}{N_{1}} \sum_{j=1}^{N_{1}} \frac{w_i+z_{j}}{w_{i}-z_{j}}.
\end{align}
\label{eq:mmspe}
\end{subequations}

We now define the planar resolvent in terms of the exponentiated variables
\begin{equation}
    v(z):=v_1(z)-v_2(z)=\fft{t_1}{N_1}\sum_{i=1}^{N_1}\fft{z+z_i}{z-z_i}-\fft{t_2}{N_2}\sum_{i=1}^{N_2}\fft{z+w_i}{z-w_i},
\label{eq:resolventi}
\end{equation}
where the eigenvalues $\left\{z_i\right\}_{i=1}^{N_1}$ and $\left\{w_i\right\}_{i=1}^{N_2}$ solve the saddle-point equations \eqref{eq:mmspe}. In the planar limit, $k\to\infty$, we expect the eigenvalue distributions $\{z_i\}_{i=1}^{N_1}$ to localize to a cut $[c,d]\subset\mathbb{R}^+$ and 
$\{w_i\}_{i=1}^{N_2}$ to localize to a cut $[a,b]\subset\mathbb{R}^-$.  We thus introduce eigenvalue densities $\rho(x)$ and $\tilde\rho(x)$ and write the planar resolvent as
\begin{equation}
    v(z):=t_1\int_c^d\dd{x}\rho(x)\frac{z+x}{z-x}-t_2\int_a^b\dd{x}\tilde{\rho}(x)\frac{z+x}{z-x}.
\label{eq:expResolvent}
\end{equation}
Note that $v(z)$ has branch-cut discontinuities along $[a,b]$ and $[c,d]$ where the eigenvalues condense.  In terms of this resolvent, we can rewrite the saddle-point equations quite simply as
\begin{subequations}
\begin{align}
    \kappa_{1} \log z& =\ft12[v(z+i 0)+v(z-i 0)],\qquad y\in[c,d], \\
    -\kappa_{2} \log (-z) & =\ft12[v(z+i 0)+v(z-i 0)],\qquad y\in[a,b].
\end{align}\label{eq:expSPE}
\end{subequations}
These equations can be solved by standard methods that have been developed in random matrix theory (see \textit{e.g.}~\cite{Marino:2011nm}).

The planar resolvent for GT theory was already worked out in \cite{Suyama:2010hr} by solving the Riemann-Hilbert problem.  The idea is to convert the saddle-point equations (\ref{eq:expSPE}), which correspond to the principal value of the resolvent along the two cuts, into corresponding discontinuity equations by introducing
\begin{equation}
    f(z)=\fft{v(z)}{\sqrt{(z-a)(z-b)(z-c)(z-d)}}.
\end{equation}
We then use Cauchy's theorem to write
\begin{equation}
    f(z)=\oint\fft{d\zeta}{2\pi i}\fft{f(\zeta)}{\zeta-z},
\end{equation}
where the contour is a small circle surrounding $z$.  By deforming the contour to go around the two cuts and using the saddle-point equations, we can obtain an integral expression for $f(z)$.  Converted back to the resolvent, $v(z)$, we finally obtain \cite{Suyama:2010hr}
\begin{align}
    v(z)&=\frac{\kappa_1}{\pi}\int_c^d\dd{x}\frac{\log\qty(x)}{z-x}\frac{\sqrt{(z-a)(z-b)(z-c)(z-d)}}{\sqrt{|(x-a)(x-b)(x-c)(x-d)|}}\nonumber \\
    &\quad+\frac{\kappa_2}{\pi}\int_a^b\dd{x}\frac{\log\qty(-x)}{z-x}\frac{\sqrt{(z-a)(z-b)(z-c)(z-d)}}{\sqrt{|(x-a)(x-b)(x-c)(x-d)|}}.
\label{eq:vzint}
\end{align}
This is the starting point for the subsequent analysis.

\subsection{Fixing the endpoints}

While the GT theory is parametrized by the couplings $t_1$ and $t_2$, the expression (\ref{eq:vzint}) for the resolvent is instead parametrized by the endpoints $a,b,c,d$ of the two cuts.  We thus want to relate these two sets of parameters.  The problem can be simplified by noticing that the saddle-point equations, \eqref{eq:mmspe}, are invariant under $z\to z^{-1}$ and $w\to w^{-1}$. This suggests that the eigenvalue distributions should also be invariant under this map, which leads to an ansatz
\begin{equation}
    ab=1,\qquad cd=1.\label{eq:endpointAnsatz}
\end{equation}
It was shown in \cite{Suyama:2010hr} that this ansatz is consistent with the constraints imposed by the asymptotic behavior of the resolvent $v(z)$ in the limits $z\to\infty$ and $z\to0$.  We must still relate the two undetermined parameters (say $a$ and $d$) to the couplings $t_1$ and $t_2$.  This can be done using the relations
\begin{subequations}
\begin{align}
    t_1&=\fft1{4\pi i}\oint_{\mathcal{C}_1} {\dd{z}}\,\frac{v(z)}{z},\\
    t_2&=\fft1{4\pi i}\oint_{\mathcal{C}_2} {\dd{z}}\,\frac{v(z)}{z},
\end{align}
\label{eq:endpointContour}%
\end{subequations}
which can be derived directly from the expression \eqref{eq:expResolvent} for the resolvent.  Here $\mathcal{C}_1$ and $\mathcal{C}_2$ are contours enclosing the branch cuts $[c,d]$ and $[a,b]$, respectively.

While the resolvent, (\ref{eq:vzint}), does not appear to admit a simple analytic form, we can work with it as an integral expression.  This is facilitated in the strong coupling limit $t_1,t_2\gg1$, where it was shown in \cite{Suyama:2011yz} that the endpoints of the two cuts scale uniformly when $t_1\approx t_2\to\infty$.  In particular, making note of (\ref{eq:endpointAnsatz}), we let
\begin{equation}
    a=-e^\alpha,\qquad b=-e^{-\alpha},\qquad c=e^{-\beta},\qquad d=e^\beta.
\label{eq:abcd}
\end{equation}
Since the strong coupling limit is taken with $\alpha\approx\beta$, we find it convenient to further parametrize the endpoints by
\begin{equation}
    \alpha=\gamma+\delta,\qquad\beta=\gamma-\delta.
\label{eq:abgd}
\end{equation}
The symmetric case, $t_1=t_2$ and $\kappa_1=\kappa_2$, corresponds to $\delta=0$ and
\begin{equation}
    t_1=t_2\sim\frac{\kappa_1+\kappa_2}{3\pi^2}\gamma^3,
\label{eq:endptIntermed}
\end{equation}
at least to leading order \cite{Suyama:2011yz}.  More generally, the scaling $t_i\sim\gamma^3$ continues to hold, while $\delta$ is of subleading order compared with $\gamma$.  The relation between $\{\gamma,\delta\}$ and $\{t_1,t_2\}$ will be worked out in more detail below.

\subsection{Computing the free energy}\label{subsec:freeEnergy}

\begin{figure}[t]
    \centering
    \tikzset{every picture/.style={line width=0.75pt}} 
    
    \begin{tikzpicture}[x=0.75pt,y=0.75pt,yscale=-1,xscale=1]
    
    \draw   (455.42,71.6) .. controls (460.12,71.6) and (463.92,83.89) .. (463.92,99.06) .. controls (463.92,114.22) and (460.12,126.52) .. (455.42,126.52) .. controls (450.73,126.52) and (446.92,114.22) .. (446.92,99.06) .. controls (446.92,83.89) and (450.73,71.6) .. (455.42,71.6) -- cycle ;
    \draw   (455.42,126.52) .. controls (324.74,160.97) and (324.74,195.43) .. (455.42,229.89) ;
    \draw   (106.47,226.12) .. controls (237.15,191.66) and (237.15,157.21) .. (106.47,122.75) ;
    \draw [color={rgb, 255:red, 208; green, 2; blue, 27 }  ,draw opacity=1 ] [dash pattern={on 4.5pt off 4.5pt}]  (357.41,178.2) .. controls (313.38,159.27) and (246.42,152.49) .. (204.48,174.43) ;
    \draw [shift={(280.98,161.03)}, rotate = 363.69] [fill={rgb, 255:red, 208; green, 2; blue, 27 }  ,fill opacity=1 ][line width=0.08]  [draw opacity=0] (8.93,-4.29) -- (0,0) -- (8.93,4.29) -- cycle    ;
    \draw [color={rgb, 255:red, 74; green, 144; blue, 226 }  ,draw opacity=1 ]   (458.8,122.6) .. controls (319.3,136.57) and (282.18,208.31) .. (457.8,234.6) ;
    \draw [color={rgb, 255:red, 208; green, 2; blue, 27 }  ,draw opacity=1 ]   (204.48,174.43) .. controls (262.88,206.55) and (314.69,198.93) .. (357.41,178.2) ;
    \draw [shift={(280.64,196.12)}, rotate = 182.38] [fill={rgb, 255:red, 208; green, 2; blue, 27 }  ,fill opacity=1 ][line width=0.08]  [draw opacity=0] (8.93,-4.29) -- (0,0) -- (8.93,4.29) -- cycle    ;
    \draw  [color={rgb, 255:red, 74; green, 144; blue, 226 }  ,draw opacity=1 ][fill={rgb, 255:red, 74; green, 144; blue, 226 }  ,fill opacity=1 ] (381.64,221.32) -- (374.73,213) -- (385.96,211.5) -- (379.26,214.7) -- cycle ;
    \draw  [color={rgb, 255:red, 74; green, 144; blue, 226 }  ,draw opacity=1 ][fill={rgb, 255:red, 74; green, 144; blue, 226 }  ,fill opacity=1 ] (376.62,136.42) -- (387.43,136.24) -- (381.57,145.93) -- (383.26,138.7) -- cycle ;
    \draw   (455.42,229.89) .. controls (460.12,229.89) and (463.92,242.18) .. (463.92,257.34) .. controls (463.92,272.51) and (460.12,284.8) .. (455.42,284.8) .. controls (450.73,284.8) and (446.92,272.51) .. (446.92,257.34) .. controls (446.92,242.18) and (450.73,229.89) .. (455.42,229.89) -- cycle ;
    \draw   (106.47,226.12) .. controls (111.16,226.12) and (114.97,238.41) .. (114.97,253.57) .. controls (114.97,268.74) and (111.16,281.03) .. (106.47,281.03) .. controls (101.77,281.03) and (97.97,268.74) .. (97.97,253.57) .. controls (97.97,238.41) and (101.77,226.12) .. (106.47,226.12) -- cycle ;
    \draw   (106.47,67.83) .. controls (111.16,67.83) and (114.97,80.12) .. (114.97,95.29) .. controls (114.97,110.45) and (111.16,122.75) .. (106.47,122.75) .. controls (101.77,122.75) and (97.97,110.45) .. (97.97,95.29) .. controls (97.97,80.12) and (101.77,67.83) .. (106.47,67.83) -- cycle ;
    \draw [color={rgb, 255:red, 155; green, 155; blue, 155 }  ,draw opacity=1 ] [dash pattern={on 0.84pt off 2.51pt}]  (279.8,94.6) .. controls (298.8,126.6) and (309.8,219.6) .. (277.8,258.6) ;
    \draw    (106.47,67.83) .. controls (294.8,102.6) and (275.8,102.6) .. (455.42,71.6) ;
    \draw    (106.47,281.03) .. controls (281.8,247.6) and (285.8,254.6) .. (455.42,284.8) ;
    
    \draw (392,111.4) node [anchor=north west][inner sep=0.75pt]    {$\textcolor[rgb]{0.29,0.56,0.89}{B}$};
    \draw (231,196.4) node [anchor=north west][inner sep=0.75pt]    {$\textcolor[rgb]{0.82,0.01,0.11}{A}$};
    \draw (271,269.4) node [anchor=north west][inner sep=0.75pt]  [font=\scriptsize]  {$-i\pi $};
    \draw (274,72.4) node [anchor=north west][inner sep=0.75pt]  [font=\scriptsize]  {$i\pi $};

    \end{tikzpicture}

    \caption[Chern-Simons integration contours]{The $A$ and $B$-cycle contours for Chern-Simons theory. Note that the Riemann sheets are curled up due to the $2\pi i$ periodicity of the resolvent.}
    \label{fig:Bcycle}
\end{figure}
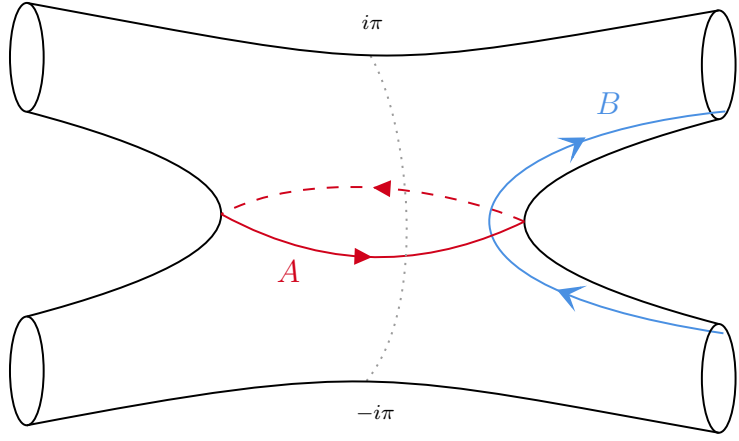

While the leading order free energy, (\ref{eq:FGT0}), can be obtained directly from a large-$N$ saddle point solution \cite{Jafferis:2011zi}, since we are interested in subleading corrections, we will instead work with the resolvent, following \cite{Suyama:2010hr,Suyama:2011yz}.  In particular, making the identification $g_s={2\pi i}/{k}$, the free energy can be written in the form of a genus expansion
\begin{equation}
    F=\sum_{g=0}^\infty g_s^{2g-2}F_g(t).
\end{equation}
It has long been known that the genus-zero free energy, $F_0(t)$, for such matrix models can be written as an integral of the planar resolvent over a particular contour \cite{Dijkgraaf:2002fc,Halmagyi:2003fy,Halmagyi:2003ze}.%
\footnote{Note that our convention for the free energy differs from that in \cite{Halmagyi:2003fy,Halmagyi:2003ze} in that we take the free energy to be $F=-\log Z$.}
The basic idea is to look at the change in the leading order free energy from adding one eigenvalue to the branch cut and use this to deduce the derivative of the genus-zero free energy with respect to the 't~Hooft parameter. The resulting expression can then be shown to be an integral of the resolvent around the $B$-cycle, a contour that starts at infinity on one Riemann sheet, passes through the branch cut, and goes off to infinity on the other Riemann sheet. This results in a beautiful geometric picture, where the $A$-cycle determines the endpoints and the $B$-cycle determines the free energy; this is depicted in Figure \ref{fig:Bcycle} for the Chern-Simons matrix model.  This is the strategy we will employ for GT theory.

\begin{figure}[t]
    \centering
    \tikzset{every picture/.style={line width=0.75pt}} 
    
    \begin{tikzpicture}[x=0.75pt,y=0.75pt,yscale=-1,xscale=1]
    
    \draw   (274.46,156.34) .. controls (274.46,128.56) and (298.36,106.04) .. (327.85,106.04) .. controls (357.33,106.04) and (381.23,128.56) .. (381.23,156.34) .. controls (381.23,184.12) and (357.33,206.64) .. (327.85,206.64) .. controls (298.36,206.64) and (274.46,184.12) .. (274.46,156.34) -- cycle ;
    \draw [color={rgb, 255:red, 208; green, 2; blue, 27 }  ,draw opacity=1 ]   (178.28,152.81) .. controls (195.55,163.46) and (247.35,169.2) .. (274.46,152.45) ;
    \draw [shift={(226.58,162.99)}, rotate = 181.65] [fill={rgb, 255:red, 208; green, 2; blue, 27 }  ,fill opacity=1 ][line width=0.08]  [draw opacity=0] (8.93,-4.29) -- (0,0) -- (8.93,4.29) -- cycle    ;
    \draw [color={rgb, 255:red, 208; green, 2; blue, 27 }  ,draw opacity=1 ]   (381.23,154.74) .. controls (398.49,165.39) and (450.29,171.13) .. (477.41,154.38) ;
    \draw [shift={(429.52,164.93)}, rotate = 181.65] [fill={rgb, 255:red, 208; green, 2; blue, 27 }  ,fill opacity=1 ][line width=0.08]  [draw opacity=0] (8.93,-4.29) -- (0,0) -- (8.93,4.29) -- cycle    ;
    \draw  [color={rgb, 255:red, 74; green, 144; blue, 226 }  ,draw opacity=1 ][fill={rgb, 255:red, 74; green, 144; blue, 226 }  ,fill opacity=1 ] (152.56,223.76) -- (163.79,222.39) -- (159.27,232.17) -- (159.85,225.18) -- cycle ;
    \draw  [color={rgb, 255:red, 74; green, 144; blue, 226 }  ,draw opacity=1 ][fill={rgb, 255:red, 74; green, 144; blue, 226 }  ,fill opacity=1 ] (171.55,98.21) -- (168.56,87.92) -- (179.47,90.79) -- (172.04,91.21) -- cycle ;
    \draw  [color={rgb, 255:red, 74; green, 144; blue, 226 }  ,draw opacity=1 ][fill={rgb, 255:red, 74; green, 144; blue, 226 }  ,fill opacity=1 ] (499.41,72.01) -- (510.67,73.06) -- (503.93,81.63) -- (506.17,74.94) -- cycle ;
    \draw  [color={rgb, 255:red, 74; green, 144; blue, 226 }  ,draw opacity=1 ][fill={rgb, 255:red, 74; green, 144; blue, 226 }  ,fill opacity=1 ] (513.35,242.65) -- (508.13,233.18) -- (519.44,233.82) -- (512.26,235.7) -- cycle ;
    \draw    (27.5,45.04) .. controls (220.94,44.99) and (238.2,260.3) .. (26.48,260.58) ;
    \draw [color={rgb, 255:red, 74; green, 144; blue, 226 }  ,draw opacity=1 ]   (27.5,45.04) .. controls (258.52,45.95) and (263.59,262.21) .. (26.48,260.58) ;
    \draw    (629.21,47.57) .. controls (422.03,49.77) and (430.15,258.38) .. (629.21,262.15) ;
    \draw [color={rgb, 255:red, 74; green, 144; blue, 226 }  ,draw opacity=1 ]   (629.21,47.61) .. controls (394.61,46.9) and (397.65,261.25) .. (629.21,262.19) ;
    \draw [color={rgb, 255:red, 208; green, 2; blue, 27 }  ,draw opacity=1 ] [dash pattern={on 4.5pt off 4.5pt}]  (477.41,154.38) .. controls (441.71,139.48) and (415.23,137.47) .. (381.23,154.74) ;
    \draw [shift={(429.14,142.51)}, rotate = 363] [fill={rgb, 255:red, 208; green, 2; blue, 27 }  ,fill opacity=1 ][line width=0.08]  [draw opacity=0] (8.93,-4.29) -- (0,0) -- (8.93,4.29) -- cycle    ;
    \draw [color={rgb, 255:red, 208; green, 2; blue, 27 }  ,draw opacity=1 ] [dash pattern={on 4.5pt off 4.5pt}]  (274.46,152.45) .. controls (238.76,137.55) and (212.28,135.54) .. (178.28,152.81) ;
    \draw [shift={(226.2,140.57)}, rotate = 363] [fill={rgb, 255:red, 208; green, 2; blue, 27 }  ,fill opacity=1 ][line width=0.08]  [draw opacity=0] (8.93,-4.29) -- (0,0) -- (8.93,4.29) -- cycle    ;
    
    \draw (170.97,61.23) node [anchor=north west][inner sep=0.75pt]  [color={rgb, 255:red, 74; green, 144; blue, 226 }  ,opacity=1 ]  {$B_{2}$};
    \draw (454.92,73.89) node [anchor=north west][inner sep=0.75pt]  [color={rgb, 255:red, 74; green, 144; blue, 226 }  ,opacity=1 ]  {$B_{1}$};
    \draw (416.36,118.08) node [anchor=north west][inner sep=0.75pt]  [color={rgb, 255:red, 208; green, 2; blue, 27 }  ,opacity=1 ]  {$\mathcal{C}_{1}$};
    \draw (224.91,117.38) node [anchor=north west][inner sep=0.75pt]  [color={rgb, 255:red, 208; green, 2; blue, 27 }  ,opacity=1 ]  {$\mathcal{C}_{2}$};
    \end{tikzpicture}
    \caption[GT integration contours]{The integration contours (in exponentiated coordinates) used in the derivation of the genus-zero free energy. Note that we no longer have the $2\pi i$ periodicity of the Riemann sheets because we are in exponentiated coordinates.}
    \label{fig:contours}
\end{figure}
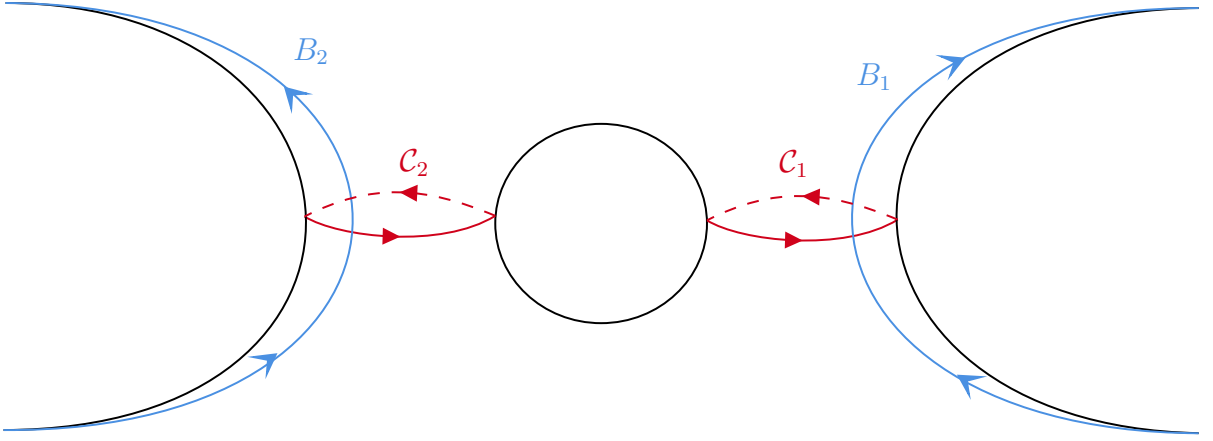

The two-node GT theory has two gauge groups whose eigenvalues condense along separate cuts in the complex plane.  As a result, there are two B-cycle integrals to consider.  We start by taking the genus-zero free energy $F_0=\left.g_s^2S\right|_{N\to\infty}$ from the effective action, (\ref{eq:action}).  For the first gauge group, we play the trick of adding one more $\hat u$ eigenvalue to the first branch cut (\textit{i.e.} we take $N_1\to N_1+1$).  The 't~Hooft parameter correspondingly changes by $\Delta t_1={2\pi}/{k}$. This gives
\begin{equation}
    \fft{\Delta F_0}{\Delta t_1}=\fft{\kappa_1}2\hat u^2-t_1\fft1{N_1}\sum_i^{N_1}\log\sinh^2\fft{\hat u-u_i}2+t_2\fft1{N_2}\sum_i^{N_2}\log\cosh^2\fft{\hat u-v_i}2.
\end{equation}
Integrating the resolvent, (\ref{eq:resolventi})
\begin{equation}
    v_1(z)=\fft{t_1}{N_1}\sum_{i=1}^{N_1}\fft{z+z_i}{z-z_i},
\end{equation}
we then obtain
\begin{equation}
    \fft{t_1}{N_1}\sum_{i=1}^{N_1}\log\sinh^2\fft{\hat u-u_i}2=-\int_{e^{\hat u}}^{e^\Lambda}v_1(z)\fft{\dd z}z+t_1(\Lambda-\log4),
\end{equation}
where $\Lambda$ is a large cutoff, and we have dropped exponentially small terms of the form $e^{-\Lambda}$.  Using this expression and a similar one for the integral of $v_2(z)$ gives, in the large-$N$ limit
\begin{equation}
    \fft{\partial F_0}{\partial t_1}=\fft{\kappa_1}2\hat u^2+\int_{e^{\hat u}}^{e^\Lambda}v(z)\fft{\dd z}z-(t_1-t_2)(\Lambda-\log4).
\end{equation}
We take the last eigenvalue $\hat u$ at the right endpoint of the cut, (\ref{eq:abcd}), and write
\begin{equation}
    \fft{\partial F_0}{\partial t_1}=\fft{\kappa_1}2\beta^2+\int_\beta^\Lambda v(e^u)\dd u-(t_1-t_2)(\Lambda-\log4).
\label{eq:dF0dt1}
\end{equation}
Geometrically, this is the $B_1$-cycle integral, which we have graphically depicted in Figure~\ref{fig:contours}.  By swapping the two gauge groups, we can obtain a similar $B_2$-cycle integral for $\partial F_0/\partial t_2$.  This integral will be worked out perturbatively in the next section.

\section{Subleading corrections to the free energy}
\label{sec:GTfree}

We now turn to an evaluation of the free energy beyond leading order.  As we have seen in (\ref{eq:vzint}), the planar resolvent for the GT model can be written down in integral form.  While the integral is challenging to perform analytically, the general expression will be sufficient when working out the free energy.

Our goal is to compute the derivative of the free energy, (\ref{eq:dF0dt1}), up to exponentially small terms in the large 't~Hooft parameter limit.  To do so, we insert the integral expression for the resolvent, (\ref{eq:vzint}), into (\ref{eq:dF0dt1}) and work out the double integral in the large $t_1$ and $t_2$ limit.  However, since this gives an expression for $\partial F_0/\partial t_1$ as a function of the endpoints of the cuts, (\ref{eq:abcd}), we additionally need to relate the endpoints to the 't~Hooft couplings using the $A$-cycle integrals (\ref{eq:endpointContour}).  We will work this out first and then return to the free energy integral.

\subsection{Correction to the endpoints}

At leading order, the endpoints of the cuts scale with the 't~Hooft couplings according to (\ref{eq:endptIntermed}).  However, this will pick up corrections, both for $t_1\ne t_2$ and subleading in the couplings.  We explicitly work out the $A$-cycle integral for $t_1$; then the $t_2$ expression follows from symmetry under $t_1\leftrightarrow t_2$ and $\kappa_1\leftrightarrow\kappa_2$ interchange.

Substituting the integral expression for the resolvent, (\ref{eq:vzint}), into (\ref{eq:endptIntermed}), then explicitly writing out the $A$-cycle integral as an integral over the discontinuity across the cut and finally interchanging the order of integration gives
\begin{equation}
    t_1=\fft{\kappa_1}{2\pi^2}J_1+\fft{\kappa_2}{2\pi^2}J_2,
\end{equation}
where
\begin{subequations}
\begin{align}
    J_1=\int_c^d\dd x\fft{\log x}{\sqrt{(x-a)(x-b)(x-c)(d-x)}}I(x),\\ J_2=\int_a^b\dd x\fft{\log(-x)}{\sqrt{(x-a)(b-x)(c-x)(d-x)}}I(x),
\end{align}
\label{eq:J1J2ints}%
\end{subequations}
with
\begin{equation}
    I(z)=\int_c^d\fft{\dd y}y\fft{\sqrt{(y-a)(y-b)(y-c)(d-y)}}{z-y}.
\end{equation}
Here the principal value of $I(x)$ has to be taken in the $J_1$ integral.  We proceed by rewriting these expressions in terms of exponentiated variables:
\begin{subequations}
\begin{align}
    J_1&=\int_{-\beta}^\beta \dd v\,\fft{vI(e^v)}{4\sqrt{\cosh\fft{\alpha+v}2\cosh\fft{\alpha-v}2\sinh\fft{\beta+v}2\sinh\fft{\beta-v}2}},\\
    J_2&=\int_{-\alpha}^\alpha \dd v\,\fft{vI(-e^v)}{4\sqrt{\sinh\fft{\alpha+v}2\sinh\fft{\alpha-v}2\cosh\fft{\beta+v}2\cosh\fft{\beta-v}2}},
\end{align}
\end{subequations}
and
\begin{equation}
    I(z)=\int_{-\beta}^\beta \dd u\,\fft{4\sqrt{\cosh\fft{\alpha+u}2\cosh\fft{\alpha-u}2\sinh\fft{\beta+u}2\sinh\fft{\beta-u}2}}{ze^{-u}-1}.
\end{equation}
Note that the $\cosh$ terms never vanish, while the $\sinh$ terms vanish at the endpoints.  Moreover, the square-root factors are all even under $v\to-v$ or $u\to-u$.  This suggests that we split up the regions of integration into half intervals and write
\begin{subequations}
\begin{align}
    J_1&=\int_0^\beta \dd v\,\fft{vI_1(v)}{4\sqrt{\cosh\fft{\alpha+v}2\cosh\fft{\alpha-v}2\sinh\fft{\beta+v}2\sinh\fft{\beta-v}2}},\\
    J_2&=\int_0^\alpha \dd v\,\fft{vI_2(v)}{4\sqrt{\sinh\fft{\alpha+v}2\sinh\fft{\alpha-v}2\cosh\fft{\beta+v}2\cosh\fft{\beta-v}2}},
\end{align}
\end{subequations}
where
\begin{subequations}
\begin{align}
    I_1(v)&=\int_0^\beta \dd u\,4\sqrt{\textstyle\cosh\fft{\alpha+u}2\cosh\fft{\alpha-u}2\sinh\fft{\beta+u}2\sinh\fft{\beta-u}2}\left(\coth\fft{v-u}2+\coth\fft{v+u}2\right),\\
    I_2(v)&=\int_0^\beta \dd u\,4\sqrt{\textstyle\cosh\fft{\alpha+u}2\cosh\fft{\alpha-u}2\sinh\fft{\beta+u}2\sinh\fft{\beta-u}2}\left(\tanh\fft{v-u}2+\tanh\fft{v+u}2\right).
\end{align}
\end{subequations}
Here we see explicitly that the integrand of $I_1$ has a pole when $v-u$ vanishes, so the principal value should be taken when evaluating the integral.

So far, these expressions are still exact, as far as the planar resolvent is concerned.  However, the integrals are not easy to evaluate.  To proceed, we now focus on the large 't~Hooft coupling limit, where $\alpha,\beta\gg1$.  Since the integrals are over half intervals, we can approximate $\alpha+v\gg1$, $\beta+v\gg1$, and similarly for $v$ replaced by $u$.  As a result, up to exponentially suppressed terms, we have
\begin{align}
    J_1=\int_0^\beta \dd v\,\fft{ve^{-\fft12(\gamma+v)}I_1(v)}{2\sqrt{\cosh\fft{\alpha-v}2\sinh\fft{\beta-v}2}},\qquad
    J_2=\int_0^\alpha \dd v\,\fft{ve^{-\fft12(\gamma+v)}I_2(v)}{2\sqrt{\sinh\fft{\alpha-v}2\cosh\fft{\beta-v}2}},
\label{eq:J1intdef}
\end{align}
with
\begin{subequations}
\begin{align}
    I_1(v)&=\int_0^\beta \dd u\,2e^{\fft12(\gamma+u)}\sqrt{\textstyle\cosh\fft{\alpha-u}2\sinh\fft{\beta-u}2}\left(\coth\fft{v-u}2+\coth\fft{v+u}2\right),\\
    I_2(v)&=\int_0^\beta \dd u\,2e^{\fft12(\gamma+u)}\sqrt{\textstyle\cosh\fft{\alpha-u}2\sinh\fft{\beta-u}2}\left(\tanh\fft{v-u}2+\tanh\fft{v+u}2\right).
\end{align}
\label{eq:I12intdef}%
\end{subequations}
Recall that we have defined $\gamma=(\alpha+\beta)/2$ and $\delta=(\alpha-\beta)/2$, following (\ref{eq:abcd}).

The $I_1$ and $I_2$ integrals can be performed explicitly and then substituted into the integrands for $J_1$ and $J_2$.  The remaining integrals are more challenging, and we have been unable to obtain a closed-form expression for $J_1$ and $J_2$.  Nevertheless, they can be reduced to polynomial expressions in $\gamma$ up to exponentially suppressed terms.  The integration is worked out in Appendix~\ref{appendix:endpoints}, and the result is a relation between the 't~Hooft couplings $t_1$ and $t_2$ and the endpoints of the cuts as parameterized by $\gamma$ and $\delta$.  After defining convenient combinations of $t_1$ and $t_2$,
\begin{equation}
    \bar t=\ft12(t_1+t_2),\qquad\Delta=\ft12(t_1-t_2),
\end{equation}
we find
\begin{subequations}
\begin{align}
    \bar t&=\fft{\kappa_1+\kappa_2}{4\pi^2}\biggl[\fft43(\gamma-\log\ft12\cosh\delta)^3+4\gamma\tan^{-1}\sinh\delta(\tan^{-1}\sinh\delta-\xi)\nn\\
    &\kern5em+\fft43\log^3(\ft12\cosh\delta)+j_{1,e}(\delta)+j_{2,e}(\delta)+\fft{2\xi}\pi j_{1,o}(\delta)\biggr],\\
    \Delta&=\fft{\kappa_1+\kappa_2}{4\pi^2}\Bigl[-2\pi\gamma(\tan^{-1}\sinh\delta-\xi)+j_{1,o}(\delta)+2\pi \xi\log\ft12\cosh\delta\Bigr],
\end{align}
\label{eq:tbD}%
\end{subequations}
where
\begin{equation}
    \xi:=\fft\pi2\fft{\kappa_1-\kappa_2}{\kappa_1+\kappa_2}=\fft\pi2\fft{k_1-k_2}{k_1+k_2},
\label{eq:xdef}
\end{equation}
is the relative difference in Chern-Simons levels.  Here $j_1(\delta)$ and $j_2(\delta)$ are particular functions explicitly defined in Appendix \ref{appendix:endpoints}, and the subscripts $e$ and $o$ denote their even and odd parts, respectively.

For the most part, we are interested in the case of equal ranks, $N_1=N_2$, where the difference $\Delta$ vanishes.  Setting $\Delta=0$ in (\ref{eq:tbD}) then gives a straightforward expression for $\gamma$ in terms of $\delta$
\begin{equation}
    2\pi\gamma(\tan^{-1}\sinh\delta-\xi)=j_{1,o}(\delta)+2\pi \xi\log\ft12\cosh\delta.
\label{eq:2pgeqn}
\end{equation}
However, we are more interested in obtaining $\delta$ in terms of $\gamma$ since we are focused on the large coupling expansion generalizing (\ref{eq:endptIntermed}) where $t_i\sim\gamma^3$ with $\delta$ being subdominant.  Working to leading order in $\gamma$, we can disregard the last two terms in the expression for $\Delta$ in (\ref{eq:tbD}) so that
\begin{equation}
    \delta\approx\sinh^{-1}\tan\left(\xi\right).
\end{equation}
However, we can do better than this. Since we assume $\gamma\gg1$, we can expand perturbatively
\begin{equation}
    \delta\approx\sinh^{-1}\tan\left(\xi\right)+\frac{\delta_1}{\gamma}+\frac{\delta_2}{\gamma^2}+\cO\qty(\frac{1}{\gamma^3}).
\label{eq:delta}
\end{equation}
Solving the $\cO(\gamma^0)$ expression in $\Delta$ gives
\begin{equation}
    \delta_1=\sec \left(\xi\right)\Cl_2\left(\pi+2\xi\right),
\end{equation}
where $\Cl_2(x)$ denotes the Clausen function
\begin{equation}
    \Cl_2(x)=\Im\Li_2(e^{ix}).
\end{equation}
The expression for $\delta_2$ is rather more involved
\makeatletter
\newcommand{\vast}{\bBigg@{4}}
\makeatother
\begin{align}
    \delta_2&=\frac{1}{2}\sec\qty(\xi)\Cl_2\qty(\pi+2\xi)
    \Bigg[\tan \left(\xi\right) \left(2\xi-2 \text{gd}\left(\sinh ^{-1}\left(\tan \left(\xi\right)\right)\right)+\Cl_2\qty(\pi+2\xi)\right)\nn\\
    &-4 \sinh ^{-1}\left(\tan \left(\xi\right)\right)-2 \log \left(\frac{8\sec\qty(\xi)}{\left(\qty(\tan \left(\xi\right)+\sec \left(\xi\right))^2+1\right){}^2}\right)\Bigg],
\end{align}
where $\text{gd}$ denotes the Gudermannian function
\begin{equation}
    \mathrm{gd}(z)=2\arctan\tanh\qty(\tfrac{1}{2}z).
\end{equation}
This is a rather messy expression, but for $\xi\ll 1$, it takes the nice perturbative form
\begin{equation}
    \delta_2\approx 2\xi \log ^2(2)+\xi^3\left(3\log ^2(2)-\fft43 \log (2)\right)+\mathcal{O}\qty(\xi^5).
\end{equation}

Having obtained $\delta$, at least perturbatively, we now proceed to relate $\gamma$ and $\bar t$.  Keeping $\Delta=0$, we first eliminate the second term in the $\bar t$ expression in (\ref{eq:tbD}) to obtain
\begin{align}
    \bar t&=\fft{\kappa_1+\kappa_2}{4\pi^2}\biggl[\fft43(\gamma-\log\ft12\cosh\delta)^3+\fft2\pi(\tan^{-1}\sinh\delta+\xi)j_{1,o}(\delta)\nn\\
    &\kern5em+4\log(\ft12\cosh\delta)\left(\fft13\log^2(\ft12\cosh\delta)+\xi\tan^{-1}\sinh\delta\right)
    +j_{1,e}(\delta)+j_{2,e}(\delta)\biggr].
\end{align}
This expression is useful since the only $\gamma$ dependence appears in the first term.  We can now substitute the perturbative expression (\ref{eq:delta}).  To the first non-trivial order, we find
\begin{align}
    \bar t&=\fft{\kappa_1+\kappa_2}{4\pi^2}\biggl[\fft43(\gamma-\log\ft12\cosh\delta)^3-4\log(2\cos \xi)\left(\fft13\log^2(2\cos \xi)+\xi^2\right)\nn\\
    &\kern5em+\fft{4\xi}\pi j_{1,o}(\delta)+j_{1,e}(\delta)+j_{2,e}(\delta)+\mathcal O(\gamma^{-1})\biggr].
\end{align}
The transcendental functions on the second line are a bit troublesome to work with.  However, by studying the series expansion of $j_1(\delta)$ and $j_2(\delta)$, we can determine empirically that
\begin{equation}
    \bar t=\fft{\kappa_1+\kappa_2}{4\pi^2}\left[\fft43(\gamma-\log\ft12\cosh\delta)^3-4(\Cl_3(\pi-2\xi)+\zeta(3))+\mathcal O(\gamma^{-1})\right],
\label{eq:tbgeqn}
\end{equation}
where
\begin{equation}
    \Cl_3(x)=\Re\Li_3(e^{ix}).
\end{equation}
We will use this expression below when computing the planar free energy.

\subsection{The free energy}

We now turn to evaluating the free energy, which can be obtained from the integral expression (\ref{eq:dF0dt1}).  The $B$-cycle integral can be evaluated similarly to the $A$-cycle integral performed above for computing the endpoint relation.  In particular, using the integral expression for the resolvent, (\ref{eq:vzint}), we can write
\begin{equation}
    \fft{\partial F_0}{\partial t_1}=\fft{\kappa_1}2\beta^2-(t_1-t_2)(\Lambda-\log4)-\fft{\kappa_1}\pi K_1-\fft{\kappa_2}\pi K_2,
\end{equation}
where
\begin{subequations}
\begin{align}
    K_1=\int_c^d\dd x\fft{\log x}{\sqrt{(x-a)(x-b)(x-c)(d-x)}}I_B(x),\\ K_2=\int_a^b\dd x\fft{\log(-x)}{\sqrt{(x-a)(b-x)(c-x)(d-x)}}I_B(x).
\end{align}
\end{subequations}
These integrals are similar to the $J_1$ and $J_2$ integrals in (\ref{eq:J1J2ints}), except that now $I_B(x)$ is a $B$-cycle integral
\begin{equation}
    I_B(z)=\int_d^{e^\Lambda}\fft{\dd y}y\fft{\sqrt{(y-a)(y-b)(y-c)(y-d)}}{z-y}.
\end{equation}
These integrals can be evaluated up to exponentially small terms in a similar manner as was done for the endpoint integrals.  Combining $\partial F_0/\partial t_1$ and the corresponding expression for $\partial F_0/\partial t_2$, we find the relatively compact expression
\begin{equation}
    \fft{\partial F_0}{\partial\bar t}=\fft{\kappa_1+\kappa_2}2\Bigl[(\gamma-\log\ft12\cosh\delta)^2+(\tan^{-1}\sinh\delta-\xi)^2-\ft1{12}\pi^2-\xi^2\Bigr].
\label{eq:dF0dtb}
\end{equation}
Details of the calculation are given in Appendix \ref{appendix:freeEnergy}.

We now have everything we need to obtain the planar free energy from the resolvent.  Since the derivative $\partial F_0/\partial\bar t$ is given in terms of the endpoint parameters $\gamma$ and $\delta$, the general procedure is to first obtain these parameters from the 't~Hooft couplings $t_1$ and $t_2$ by inverting the endpoint relations (\ref{eq:tbD}).  After doing so, it becomes straightforward to integrate (\ref{eq:dF0dtb}) to obtain the planar free energy $F_0$ up to a $\bar t$ independent constant, which remains to be fixed.

Focusing on the case $\Delta=0$, the relation (\ref{eq:2pgeqn}) demonstrates that the combination $(\tan^{-1}\sinh\delta-\xi)$ is of $\mathcal O(\gamma^{-1})$.  As a result, (\ref{eq:dF0dtb}) can be written as
\begin{equation}
     \fft{\partial F_0}{\partial\bar t}=\fft{\kappa_1+\kappa_2}2\Bigl[(\gamma-\log\ft12\cosh\delta)^2-\ft1{12}\pi^2-\xi^2+\mathcal O(\gamma^{-2})\Bigr].
\end{equation}
Making use of the $\bar t$ versus $\gamma$ relation, (\ref{eq:tbgeqn}), and integrating then gives the genus zero free energy
\begin{align}
    F_0&=\bar\kappa^2\left[\fft35\left(\fft{3\pi^2}2\right)^{2/3}\left(\fft{\bar t}{\bar\kappa}+\fft2{\pi^2}\left(\Cl_3(\pi-2\xi)+\zeta(3)\right)\right)^{5/3}-\left(\fft{\pi^2}{12}+\xi^2\right)\fft{\bar t}{\bar\kappa}+\mathcal O(\bar t^{1/3})\right.\nonumber\\
    &\qquad\quad+\mbox{const.}\Bigg],
\label{eq:f0fin}
\end{align}
where we have defined
\begin{equation}
    \bar\kappa=\fft{\kappa_1+\kappa_2}2=\fft{k_1+k_2}{2k}.
\end{equation}
Several points are now in order.  Firstly, the ``constant'' term is independent of $\bar t$ but can depend on the fractional difference of Chern-Simons levels, $\xi$.  However, it cannot be obtained directly from integrating the derivative of the free energy.%
\footnote{In contrast, the $\cO(\ol{t}^{1/3})$ part can, in principle, be obtained term-by-term from higher-order perturbation theory. We denote the $\cO(\ol{t}^{1/3})$ and constant terms separately to emphasize this distinction.}
In addition, the leading term in the large-$\bar t$ expansion of this expression matches what we expect from \eqref{eq:FGT=}.  Finally, note that the $\mathcal O(\bar t^{1/3})$ term vanishes in the $\xi=0$ limit (ie, for $k_1=k_2$).  In this case, expression (\ref{eq:f0fin}) is exact up to exponentially small terms in $\bar t$.

It is easily seen that $\delta$ vanishes in the $\xi=0$ limit.  As a result, (\ref{eq:tbgeqn}) takes on the simple relation
\begin{equation}
    t=\fft{\kappa}{2\pi^2}\left(\ft43(\gamma+\log2)^3-\zeta(3)\right),
\end{equation}
and the planar free energy, (\ref{eq:f0fin}) becomes
\begin{equation}
    F_0=\kappa^2\left[\fft35\left(\fft{3\pi^2}2\right)^{2/3}\left(\frac{t}{\kappa}+\fft{\zeta(3)}{2\pi^2}\right)^{5/3}-\fft{\pi^2}{12}\frac{t}{\kappa}+\mbox{const.}+\mathcal O(e^{-t})\right].
\label{eq:F0equal}
\end{equation}
Here we have dropped the bars on $t$ and $\kappa$ as we are considering $t_1=t_2$ and $\kappa_1=\kappa_2$.  If desired, this can be expanded in inverse powers of $t$
\begin{equation}
    F_0(t)=-\frac{\pi^2}{12} t+\frac{3\cdot6^{2/3}}{40\pi^2}\qty(2\pi^2{t})^{5/3}\sum_{n=0}^\infty \frac{\qty(\tfrac{5}{3})_n}{n!}\qty(\frac{\zeta(3)}{2\pi^2t})^n,
\label{eq:F0expand}
\end{equation}
where $(\ )_n$ denotes the Pochhammer symbol.  Since this expression holds for $k_1=k_2$, we have set $\kappa=1$ and $t=2\pi iN/k$.  Note that $t$ is imaginary when we take $N$ and $k$ to be real.  In this case, the first term, which is linear in $t$, does not contribute to the real part of the free energy.

\subsection{Numerical analysis}

Our main result is the expression, (\ref{eq:f0fin}), for the genus zero free energy $F_0(N,k_1,k_2)$ at large 't~Hooft coupling $\bar t$.  While the first term is complete, additional terms of $\mathcal O(\bar t^{1/3})$ and smaller will contribute when the Chern-Simons levels are different, as parametrized by $\xi$ defined in (\ref{eq:xdef}).  To get an idea of the size of these terms, we carried out a numerical investigation of the large-$N$ partition function.  In this limit, we solved the saddle-point equations in Mathematica for $N$ ranging from 100 to 340 at fixed (real positive) 't~Hooft coupling $\bar t$ and extrapolated $N\to \infty$, using a working precision of 50. This was done for various values of $\bar t$ and then fitted to extract the subleading coefficients $f_1(\xi)$ and $f_2(\xi)$ in the expansion
\begin{equation}
    F_0(\bar t,\bar\kappa,\xi)=\fft35\left(\fft{3\pi^2}2\right)^{2/3}\bar t^{5/3}\bar\kappa^{1/3}+f_1(\xi)\bar t+f_2(\xi)\bar t^{2/3}+\cdots,
\label{eq:F0exp...}
\end{equation}
Throughout these fits, we hold $\bar\kappa=1$ fixed since changing the value of $\bar\kappa$ is equivalent to an overall rescaling of $k$.  The coefficients $f_1(\xi)$ and $f_2(\xi)$ are then extracted from the numerical free energy for various values of $\xi=(\pi/4)(\kappa_1-\kappa_2)$. Due to the computational difficulty of this process, this was only done for five sample points corresponding to $\xi=\{0,\tfrac{\pi}{40},\tfrac{\pi}{20},\tfrac{\pi}{10},\tfrac{3\pi}{20}\}$.

\begin{figure}[t]
\includegraphics[width=0.75\textwidth]{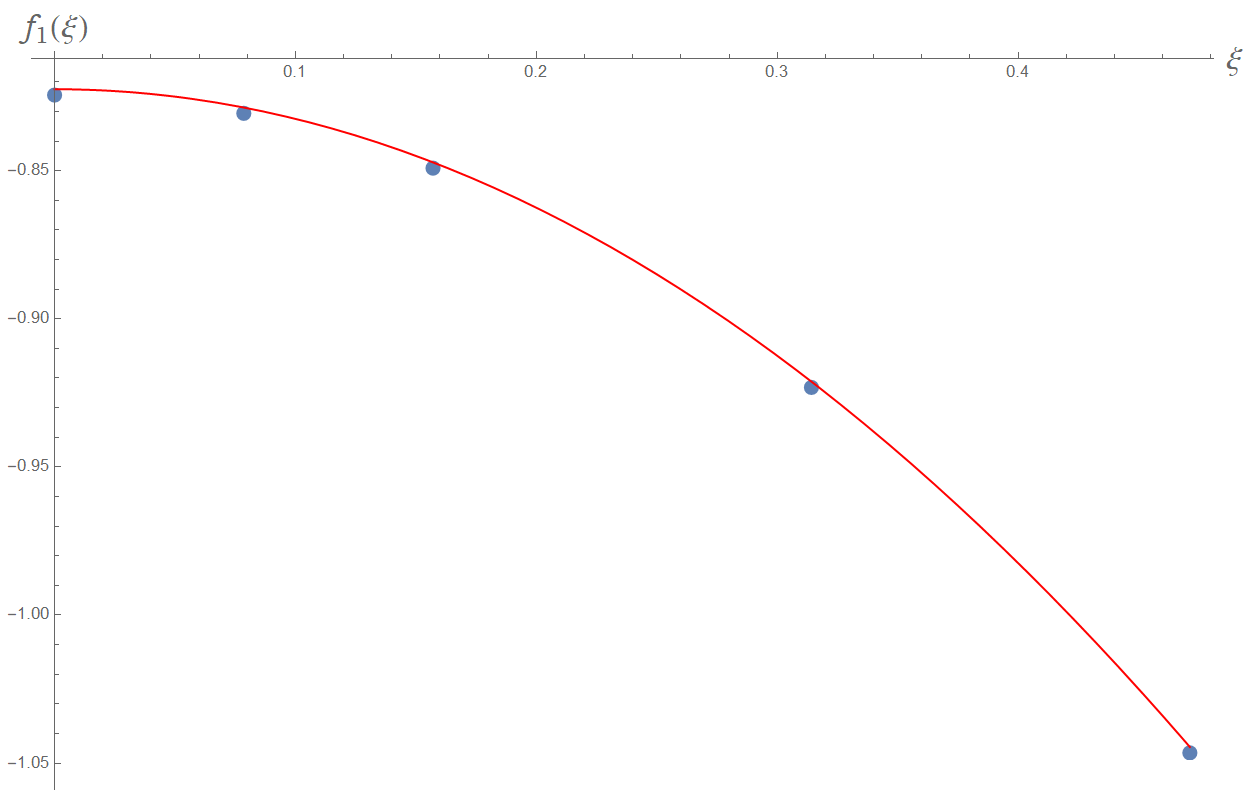}
\centering
\caption[Plot of the coefficient $f_1(\xi)$]{Plot of the coefficient $f_1(\xi)$. The red line is the analytic prediction from \eqref{eq:f0fin}, and the blue dots are sample points for numerical simulations performed in Mathematica for $\xi=0$, $\tfrac{\pi}{40}$, $\tfrac{\pi}{20}$, $\tfrac{\pi}{10}$, and $\tfrac{3\pi}{20}$.}
\label{fig:linearTerm}
\end{figure}

\begin{figure}[t]
\includegraphics[width=0.75\textwidth]{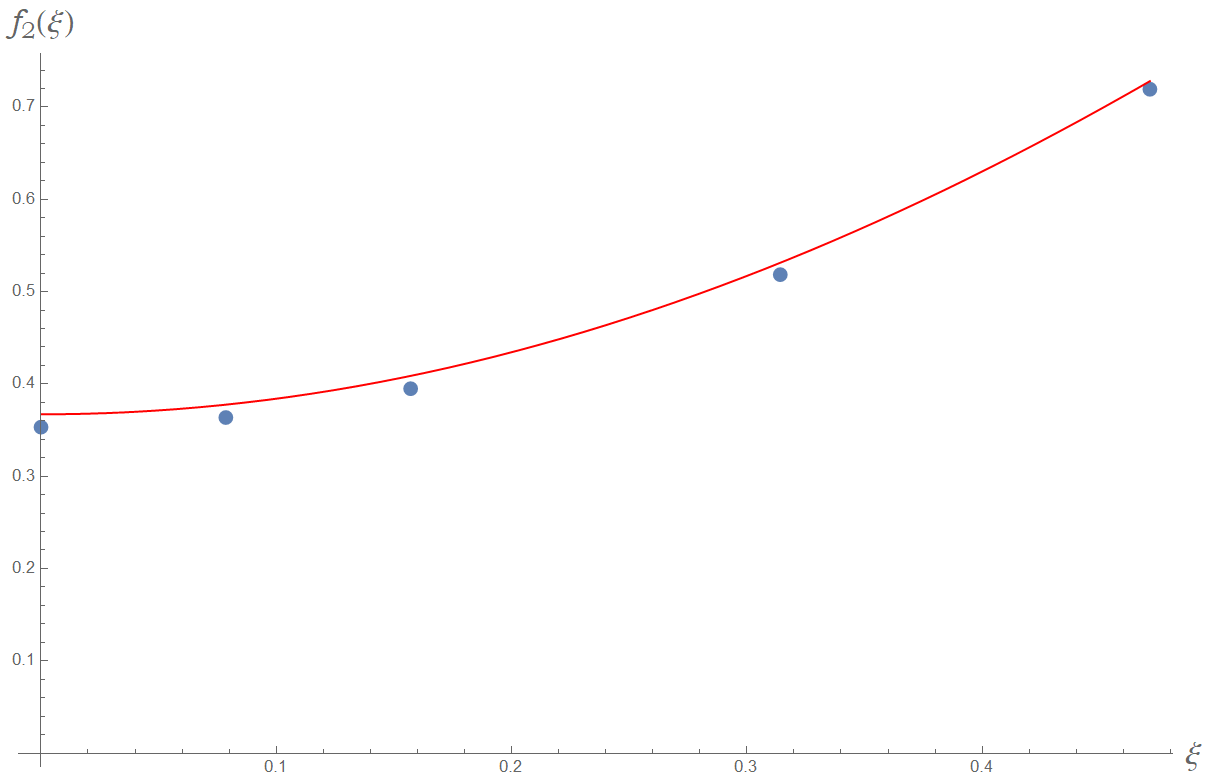}
\centering
\caption[Plot of the coefficient $f_2(\xi)$]{Plot of the coefficient $f_2(\xi)$. The red line is the analytic prediction from \eqref{eq:f0fin}, and the blue dots are sample points for numerical simulations performed in Mathematica for $\xi=0$, $\tfrac{\pi}{40}$, $\tfrac{\pi}{20}$, $\tfrac{\pi}{10}$, and $\tfrac{3\pi}{20}$.}
\label{fig:t2/3term}
\end{figure}

We have verified that the leading order term in \eqref{eq:F0exp...} is reproduced numerically to very high precision and that no term of $\mathcal O(\bar t^{4/3})$ shows up within numerical uncertainties.  As a result, we subtracted the analytic value of the leading term and fit only the subdominant coefficients.  The coefficient $f_1(\xi)$ of the linear $\ol{t}$ term shows very good agreement and is plotted in Figure~\ref{fig:linearTerm}.
We also plot the coefficient $f_2(\xi)$ of $\ol{t}^{2/3}$ in Figure~\ref{fig:t2/3term}. Here, the coefficient is slightly less numerically stable, and we cannot see the agreement quite as well.  Nonetheless, we still see fairly good agreement with the data.
%

\section{Discussion}
\label{sec:disc}

While the leading order $N^{5/3}k^{1/3}$ behavior of the free energy of GT theory was essentially known since the model was first introduced, the subleading corrections have been surprisingly difficult to obtain analytically.  The planar resolvent was constructed in \cite{Suyama:2010hr}.  However, its form did not readily lend itself to a simple expression for the free energy beyond the leading order.  Even the remarkable Fermi-gas approach to Chern-Simons-matter theories \cite{Marino:2011eh} runs into limitations when exploring higher-order corrections \cite{Hong:2021bsb}.

We obtained the planar free energy up to exponentially small corrections in the limit of large 't~Hooft coupling by working with the resolvent (\ref{eq:vzint}) in integral form.  The main technical observation is that the endpoints of the cuts can be obtained from $A$-cycle integrals of the resolvent integral while the derivative of the free energy can be obtained from $B$-cycle integrals.  The order of the resulting double integrals can then be swapped, leading to expressions that can be more readily worked with.  The key results are then the endpoint relations (\ref{eq:tbD}) and the free energy expression (\ref{eq:dF0dtb}).

The expressions (\ref{eq:tbD}) and (\ref{eq:dF0dtb}) in principle allow us to obtain the planar free energy $F_0(N_1,N_2,k_1,k_2)$ in the $\bar t\gg1$ limit directly in terms of the parameters of the model.  However, inverting the endpoint equations is generally non-trivial.  Nevertheless, for small differences in the Chern-Simons levels, $|k_1-k_2|\ll|k_1+k_2|$, these equations can be inverted perturbatively, assuming the self-consistent condition $|\delta|\ll1$ on the endpoints.  Focusing on the equal rank case $N_1=N_2$, or equivalently $\Delta=0$, 
we have found an explicit expansion of the free energy. If, in addition, the Chern-Simons levels are equal, we obtain the closed-form expression (\ref{eq:F0equal}), which is exact up to exponentially suppressed terms.

While we have focused on the equal rank case, one can work with unequal ranks if desired.  Here some care may be needed depending on how $N_1$ and $N_2$ scale in the large-$N$ limit, as there are now two independent 't~Hooft parameters.  If the difference in ranks, $N_1-N_2$, is held fixed, then $\Delta$ is a constant, and the perturbative inversion of the endpoint equations (\ref{eq:tbD}) can be worked out as usual.  However, if $\Delta$ is not fixed, then the inversion of $\{t_1,t_2\}\leftrightarrow\{\gamma,\delta\}$ becomes more involved, and the free energy as a function of two independent 't~Hooft parameters becomes less obvious.

From a technical point of view, it is possible that the way we have chosen to break the integrals into intermediate functions is not necessarily the most efficient. Many of the expressions in Appendices \ref{appendix:endpoints} and \ref{appendix:freeEnergy} are quite complicated, and one may wonder if there is a simpler parameterization that makes the formulation more elegant.  One possibility is to organize the expressions by the degree of transcendentality.  However, it is not clear if this would make them simpler.

One of the motivations for examining the subleading behavior of the free energy is to compare it with the holographic dual.  From this point of view, it is interesting to observe that the expansion (\ref{eq:F0expand}) involves powers of  $\zeta(3)/t$.  From the supergravity point of view, this is suggestive of the $\alpha'$ expansion of the tree-level closed string effective action, which starts with a term of the form $\zeta(3)\alpha'^3R^4$ \cite{Gross:1986iv}.  More generally, at higher derivative order, one expects a series of corrections of the form $\alpha'^{3(n+1)}\zeta(3)^nD^{6n}R^4$, or equivalently $\alpha'^{3(n+1)}\zeta(3)^nR^{4+3n}$, which would provide an obvious source of corrections to the dual free energy.

Of course, this is only a heuristic picture for now, as many open questions remain to be addressed before the comparison can be made rigorous.  For one thing, while the higher derivative couplings have been extensively studied for type II strings, the dual to GT theory is massive IIA supergravity, which may not receive the same corrections as ordinary type II supergravity.  Nevertheless, we expect the structure to be very similar, at least if we assume a common M-theory origin.

Perhaps more importantly, advances in computing open and closed tree-level string amplitudes have provided a clearer picture of the structure of higher derivative corrections beyond $\alpha'^3R^4$.  In particular, it is known that the $\alpha'$ expansion yields terms of the form $\alpha'^{3+n}D^{2n}R^4$ (along with counterparts such as $\alpha'^{3+n}R^{4+n}$) multiplied by various combinations of $\zeta(n)$.  Assuming the free energy can be expanded only in powers of $\zeta(3)$ then demands that these other terms not proportional to $\zeta(3)^n$ do not contribute to the free energy, and hence must vanish on-shell in the gravity dual.

Finally, the form of the planar free energy, (\ref{eq:F0equal}), where the large-$t$ expansion involves a linear function of $t$ raised to a fractional power, may hint at some underlying symmetry in the $\alpha'$ expansion.  It would be interesting to study the dual massive IIA description of GT theory and to clarify some of these questions.  One obstacle in doing so is the lack of an explicit construction of the dual supergravity background beyond the limit of 
infinitesimally small Romans mass \cite{Gaiotto:2009yz}. However, we hope that such a solution may be found in the future.

%% file: Chapters/chap3.tex

While higher-derivative corrections can be obtained directly from the underlying string theory, they can also be parameterized more generally by working directly in the supergravity theory.  Various higher derivative superinvariants have been constructed, both in the off-shell conformal supergravity approach and in the Poincar\'e frame.  The former is particularly powerful, although it is generally limited to theories with at most eight real supercharges.  In this manner, four-derivative corrections have been constructed in four-dimensional $\mathcal N=2$ \cite{deWit:1979dzm,deWit:1980lyi,deWit:1984rvr,deWit:1984wbb,LopesCardoso:2000qm,deWit:2006gn,Butter:2013lta}, five-dimensional $\mathcal N=2$ \cite{Hanaki:2006pj,Bergshoeff:2011xn,Ozkan:2013uk,Ozkan:2013nwa,Baggio:2014iv}, and six-dimensional $\mathcal N=(1,0)$ \cite{Bergshoeff:1986vy,Bergshoeff:1986wc,Bergshoeff:1987rb,Novak:2017wqc,Butter:2018wss} supergravities.  As these constructions are not yet in the Poincar\'e frame, an additional step is needed in integrating out the auxiliary fields to obtain conformally gauge-fixed superinvariants.

Supersymmetric higher derivative actions have a wide range of applications, from black holes to precision holography.  Higher-derivative corrected black holes provide a window into quantum gravity and can also shed light on the black hole weak gravity conjecture and the swampland.  Higher derivative corrections have also played an important role in holographic hydrodynamics and bounds on $\eta/s$, the ratio of the shear viscosity to the entropy density of the plasma.

Recently, there has been interesting work on four-derivative corrections in $\cN=2$, $D=4$ gauged supergravity with regards to AdS$_4$/CFT$_3$ holography \cite{Bobev:2020egg,Bobev:2020zov, Bobev:2021oku}. In particular, there are two off-shell four-derivative superinvariants that one can add to the action, namely the Weyl multiplet and the $\mathbb{T}$log multiplet
\begin{equation}
    S_\text{HD}^{4d}=S_{2\partial}+\alpha_1 S_{\text{Weyl}}+\alpha_2 S_{\mathbb{T}\text{log}},
\end{equation}
where $S_{2\partial}$ denotes the usual two-derivative action. The authors of \cite{Bobev:2020egg,Bobev:2021oku} showed that on-shell these reduce to a single four-derivative superinvariant, the Gauss-Bonnet action, as well as a term proportional to the two derivative action
\begin{equation}
    I_\text{HD}^{4d}= \qty(1+\frac{4}{L^2}(\alpha_2-\alpha_1))I_{2\partial}+\alpha_1I_{\text{GB}},
\label{eq:SHDos}
\end{equation}
where $L$ is the four-dimensional AdS radius.  Here $I$ denotes the on-shell value of the action~$S$.

At this point, a comment on our usage of off-shell and on-shell is in order, as there are perhaps two notions of on/off-shell in the supergravity literature: on-shell in the context of conformal supergravity and on-shell in the context of evaluating actions.  In the former, the off-shell action includes non-dynamical auxiliary fields needed for off-shell closure of the supersymmetry algebra, while on-shell indicates that the auxiliary fields have been integrated out.  In the latter, on-shell means the equations of motion have been used when evaluating the action, thus yielding an on-shell value that can no longer be used for dynamics.  In the interest of clarity, we will refer to on-shell in the conformal supergravity sense as the Poincar\'e frame and reserve the use of on-shell to denote computing the on-shell value of the action.%
\footnote{On-shell here should not be confused with the use of field redefinitions to transform the higher-derivative actions.  As further discussed in Section~\ref{sec:fieldredef}, we can shift the higher derivative action by terms proportional to the two-derivative equations of motion.  However, the resulting action is still a dynamical action that is physically equivalent to the original one in that it yields identical on-shell observables.}

The remarkable feature of the four-derivative on-shell action, (\ref{eq:SHDos}), as applied to AdS$_4$/CFT$_3$, is that it provides a natural split between geometrical and theory-dependent parameters.  The former arise from the on-shell values of $I_{2\partial}$ and $I_{\mathrm{GB}}$, while the latter correspond to $\alpha_1$, $\alpha_2$ and the five-dimensional AdS radius, $L$.  As shown in \cite{Bobev:2020egg}, the partition function of the dual SCFT obtained from (\ref{eq:SHDos}) thus takes the universal form
\begin{equation}
    -\log Z=\pi\cF\qty[AN^{3/2}+BN^{1/2}]-\pi(\cF-\chi)CN^{1/2}+o(N^{1/2}),
\end{equation}
where $\cF$ and $\chi$ depend on the three-dimensional manifold that the SCFT lives on while $\{A,B,C\}$ are related to $\{\alpha_1,\alpha_2,L\}$, and are specific to the SCFT in question. This motivates us to consider the analogous case for AdS$_5$/CFT$_4$, namely $\cN=2$, $D=5$ gauged supergravity.

In the five-dimensional case, three independent four-derivative off-shell terms have been constructed, corresponding to the supersymmetrization of $R_{\mu\nu\rho\sigma}^2$, $R_{\mu\nu}^2$ and $R^2$ \cite{Ozkan:2013nwa}.%
\footnote{The supersymmetrization of the 5D Weyl-squared action was performed in \cite{Hanaki:2006pj}, and the resulting corrections to black holes have been investigated in \cite{Castro:2007ci,Castro:2007hc,Castro:2007sd,Castro:2008ne,Castro:2008ys,Alishahiha:2007nn}.}
Following \cite{Ozkan:2013nwa}, we choose an equivalent basis of $C_{\mu\nu\rho\sigma}C^{\mu\nu\rho\sigma}+\frac{1}{6}R^2$, $R_{\mu\nu\rho\sigma}R^{\mu\nu\rho\sigma}$, and $R^2$.  Then
\begin{equation}
    S_\text{HD}^{5d}=S_{2\partial}+\alpha_1 S_{C^2+\frac{1}{6}R^2}+\alpha_2 S_{(\text{Riem})^2}+\alpha_3 S_{R^2}.
\end{equation}
After some work, we find a direct analog of the four-dimensional result for the on-shell value of the action:
\begin{equation}
    I_\text{HD}^{5d}=\qty(1+\frac{8\alpha_1-24\alpha_2-205\alpha_3}{2L^2})I_{2\partial}+(\alpha_1-2\alpha_2)I_\text{GB}^\text{susy},
\label{eq:onshellAction}
\end{equation}
where $S_{\text{GB}}^\text{susy}$ is the supersymmetrization of the Gauss-Bonnet action in 5D and $I_{\text{GB}}^\text{susy}$ is its on-shell value. However, this is slightly more complicated than the 4D case, as the Gauss-Bonnet action is no longer topological in 5D, and, as noted in \cite{Bobev:2021qxx}, the two-derivative solutions no longer satisfy the four-derivative equations of motion as was shown in the 4D case \cite{Bobev:2020egg,Bobev:2021oku}.

One important aspect of \eqref{eq:onshellAction} is that, up to field redefinitions and an overall coefficient, the four-derivative action $S_{\mathrm{GB}}^\text{susy}$ is completely fixed.  As shown in \cite{Myers:2009ij}, the five-dimensional Einstein-Maxwell theory admits five independent four-derivative terms up to field redefinitions.  Here we choose a somewhat different but equally valid parametrization from that of \cite{Myers:2009ij}
\begin{equation}
    e^{-1}\cL_{4\partial}=c_1\mathcal I_\text{GB}+c_2 C_{\mu\nu\rho\sigma}F^{\mu\nu}F^{\rho\sigma}+c_3 (F^2)^2+c_4 F^4+c_5\epsilon^{\mu\nu\rho\sigma\lambda}R_{\mu\nu ab}R_{\rho\sigma}{}^{ab}A_{\lambda}.
\label{eq:param4d}
\end{equation}
The fields are normalized according to the two-derivative Lagrangian
\begin{equation}
    e^{-1}\cL_{2\partial}=R+12g^2-\fft14F^2-\fft1{12\sqrt3}\epsilon^{\mu\nu\rho\sigma\lambda}F_{\mu\nu}F_{\rho\sigma}A_\lambda.
\end{equation}
Here $g=1/L$ is the gauge coupling constant, $A_\mu$ is the graviphoton field with $F=\dd{A}$ its corresponding field strength, and $\mathcal I_\text{GB}=R_{\mu\nu\rho\sigma}^2-4R_{\mu\nu}^2+R^2$ is the usual Gauss-Bonnet combination. Our notational convention, here and throughout this chapter, is that $F^2=F_{\mu\nu}F^{\mu\nu}$ and $F^4=F_{\mu\nu}F^{\nu\rho}F_{\rho\sigma}F^{\sigma\mu}$. For the supersymmetric four-derivative invariant, $S_{\mathrm{GB}}$, the above result, \eqref{eq:onshellAction}, fixes these coefficients up to an overall factor
\begin{equation}
    c_1=-2 c_2=8 c_4=2\sqrt{3} c_5,\qquad
    c_3=0.
\label{eq:correctc}
\end{equation}

Although we refer to $S_{\mathrm{GB}}^\text{susy}$ as the supersymmetric Gauss-Bonnet action in 5D, it could equivalently have arisen from integrating out the auxiliary fields of the off-shell Weyl-squared action \cite{Hanaki:2006pj}.  This was done earlier in \cite{Cremonini:2008tw,Myers:2009ij}, which however led to a different set of coefficients \cite{Cremonini:2020smy}
\begin{equation}
    \tilde c_1 = -2\tilde c_2 = -6\tilde c_3 = - \frac{24}{11}\tilde c_4 = 2\sqrt{3}\tilde c_5.
\label{eq:wrongc}
\end{equation}
We resolve this conflict by checking the four-derivative correction to the supersymmetric BMPV solution \cite{Breckenridge:1996mpv}, which corresponds to the ungauged limit ($L\to\infty$) of (\ref{eq:onshellAction}).  In particular, we show that the BPS condition, $M=\fft{\sqrt3}2|Q|$, only remains satisfied for the present choice of coefficients, (\ref{eq:correctc}).  We will comment further on this discrepancy below.

The rest of this chapter is organized as follows. In Section~\ref{sec:sugra}, we show that the three off-shell four-derivative superinvariants reduce to a single Poincar\'e frame invariant, up to field redefinitions and factors of the two-derivative action. We then proceed in Section~\ref{sec:BMPV} to check our results on the BMPV black hole solution. Finally, we conclude in Section \ref{sec:discussion} with some additional comments and open questions. 

\section{Higher-derivative supergravity}\label{sec:sugra}
Minimal $D = 5$ gauged supergravity has a single symplectic Majorana supercharge. The field content is the $\cN=2$ gravity multiplet $(e_\mu^a,\psi_\mu,A_\mu)$. The two-derivative (bosonic) Lagrangian in the Poincar\'e frame is given by
\begin{equation}
    S_{2\partial}=\int\qty[(R+12 g^2)\star1 -\frac{1}{2}F\land \star F -\frac{1}{3\sqrt{3}}F\land F\land A],
\label{eq:2derivAction}
\end{equation}
where $R$ is the Ricci scalar, $F = \dd{A}$ is the field strength of the U(1) graviphoton, and $g=1/L$ is the $U(1)$ gauge coupling that may be identified as the inverse AdS radius. Note that we choose to work in conventions such that $16\pi G_N\equiv 1$, the metric has signature $(-,+,+,+,+)$, and $R_{\mu\nu}=R^{\rho}{}_{\mu\rho\nu}$. The two-derivative equations of motion are
\begin{subequations}
\begin{align}
    0=\mathcal E^\mu&\equiv\nabla_\nu F^{\nu\mu}+\frac{1}{2\sqrt{6}}\epsilon^{\mu\nu\rho\sigma\lambda}F_{\nu\rho}F_{\sigma\lambda},\\
    0=\mathcal E_{\mu\nu}&\equiv R_{\mu\nu}-\left(F_\mu^{\ \,\sigma}F_{\nu\sigma}-\frac{1}{6}g_{\mu\nu}F^2-4g^2 g_{\mu\nu}\right).
\end{align}
\label{eq:eoms}
\end{subequations}

At the four-derivative level, three terms can be added to the action, corresponding to the supersymmetrizations of $(R_{\mu\nu\rho\sigma})^2$, $(R_{\mu\nu})^2$, and $R^2$. However, we choose to parametrize these as $(C_{\mu\nu\rho\sigma})^2+\tfrac{1}{6}R^2$, $(R_{\mu\nu\rho\sigma})^2$, and $R^2$, as the supersymmetrizations of these combinations have been found in \cite{Ozkan:2013nwa, Ozkan:2013uk} via conformal supergravity methods.  We consider these three off-shell invariants below and perform the field redefinitions necessary to transform them into the parametrization of (\ref{eq:param4d}).

\subsection{The action corresponding to $(C_{\mu\nu\rho\sigma})^2+\tfrac{1}{6}R^2$}
\label{sec:fieldredef}

The supersymmetrization of the square of the Weyl tensor was originally considered in \cite{Hanaki:2006pj} using the standard Weyl multiplet, and subsequently in \cite{Ozkan:2013nwa,Ozkan:2013uk} using the dilaton Weyl multiplet.  In the latter case, the supersymmetric completion of $C_{\mu\nu\rho\sigma}^2$ picks up an additional $\fft16R^2$ term, and the Poincar\'e frame action takes the form \cite{Ozkan:2013nwa,Ozkan:2013uk}%
\footnote{Note that our conventions are $16\pi G_N=1$, whereas the conventions in \cite{Ozkan:2013nwa,Ozkan:2013uk} are that $8\pi G_N=1$.}
\begin{align}
    e^{-1}\cL_{C^2+\frac{1}{6}R^2}&=\frac{1}{4} R_{\mu \nu \rho \sigma} R^{\mu \nu \rho \sigma}- R_{\mu \nu} R^{\mu \nu}+\frac{1}{24} R^{2}+\frac{128}{3} D^{2}+\frac{1}{8} \epsilon_{\mu \nu \rho \sigma \lambda} C^{\mu} R^{\nu \rho \tau \delta} R^{\sigma \lambda}_{\ \ \ \tau \delta}\nn \\
    &\quad-\frac{32}{3} R_{\mu \nu \rho \sigma} T^{\mu \nu} T^{\rho \sigma}+4 R_{\mu \nu \rho \sigma} G^{\mu \nu} T^{\rho \sigma}+\frac{2}{3}R T_{\mu \nu} G^{\mu \nu}-\frac{16}{3} R_{\mu \nu} G_{\sigma}^{\ \,\mu} T^{\sigma \nu}\nn \\
    &\quad-\frac{128}{3} R^{\mu \nu} T_{\sigma \mu} T_{\ \,\nu}^{\sigma}+\frac{16}{3} R T^{2}-\frac{64}{3} D T_{\mu \nu} G^{\mu \nu}+\frac{2048}{9} T^{2} D\nn \\
    &\quad-\frac{128}{3} \nabla_{\mu} T_{\nu \rho} \nabla^{\mu} T^{\nu \rho}+\frac{128}{3} \nabla^{\mu} T^{\nu \rho} \nabla_{\nu} T_{\mu \rho}-\frac{256}{3} T_{\mu \nu} \nabla^{\nu} \nabla_{\sigma} T^{\mu \sigma}+2048 T^{4}\nn\\
    &\quad-\frac{5632}{27}\left(T^{2}\right)^{2}-\frac{128}{9} T_{\mu \nu} G^{\mu \nu} T^{2}-\frac{512}{3} T_{\mu \sigma} T^{\sigma \rho} T_{\rho \nu} G^{\nu \mu}\nn\\
    &\quad-\frac{256}{3} \epsilon_{\mu \nu \rho \sigma \lambda} T^{\mu \nu} T^{\rho \sigma} \nabla_{\tau} T^{\lambda \tau}-\frac{64}{3} \epsilon_{\mu \nu \rho \sigma \lambda} G^{\mu \nu} T^{\rho \tau} \nabla_{\tau} T^{\sigma \lambda}\nn\\
    &\quad-32 \epsilon_{\mu \nu \rho \sigma \lambda} G^{\mu \nu} T^{\rho}{ }_{\tau} \nabla^{\sigma} T^{\lambda \tau},\label{eq:Weyl2initial}
\end{align}
where $G=\dd{C}$, and we have relations
\begin{subequations}
\begin{align}
    C_{\mu}&=\sqrt{\frac{2}{3}}A_{\mu},\\
    T_{\mu\nu}&=\frac{3}{16}G_{\mu\nu}=\frac{1}{8}\sqrt{\frac{3}{2}}F_{\mu\nu},\\
    D&=-\frac{1}{32}R-\frac{1}{16}G^2-\frac{26}{3}T^2+2T^{\mu\nu}G_{\mu\nu}=-\frac{1}{32}R+\frac{1}{192}F^2.
\end{align}
\label{eq:identities1}%
\end{subequations}
We note that \cite{Ozkan:2013nwa,Ozkan:2013uk} derived the above action in the context of asymptotically Minkowski space; however, moving to the AdS case (or, equivalently, gauging the supergravity) does not affect \eqref{eq:Weyl2initial}. Rather, this gauging will only affect the two-derivative action by turning on a non-zero gauge parameter $g$, which does affect the field redefinitions used in the simplifications that follow. Making use of \eqref{eq:identities1}, one can simplify \eqref{eq:Weyl2initial} to
\begin{align}
    e^{-1}\cL_{C^2+\frac{1}{6}R^2}&=\frac{1}{4} R_{\mu \nu \rho \sigma} R^{\mu \nu \rho \sigma}- R_{\mu \nu} R^{\mu \nu}+\frac{1}{12} R^{2} +\frac{1}{4}R_{\mu\nu\rho\sigma}F^{\mu\nu}F^{\rho\sigma}-\frac{61}{432}(F^2)^2+\frac{5}{8}F^4\nn \\
    &\quad+\frac{1}{4\sqrt{6}}\epsilon_{\mu\nu\rho\sigma\lambda}A^\mu R^{\nu\rho\tau\delta}R^{\sigma\lambda}{}_{\tau\delta}+\frac{1}{9}RF^2-\frac{5}{3}R^{\mu\nu}F_{\sigma\mu}F^\sigma{}_\nu-(\nabla F)^2\nn\\
    &\quad+3(\nabla_\sigma F^{\mu\sigma})^2-F^{\nu\rho}[\nabla_\mu,\nabla_\nu]F^\mu{}_\rho+\frac{1}{4\sqrt{6}}\epsilon_{\mu\nu\rho\sigma\lambda}F^{\mu\nu}F^{\rho\sigma}\nabla_\tau F^{\tau\lambda}\nn \\
    &\quad-\frac{1}{2}\sqrt{\frac{3}{2}}\epsilon_{\mu\nu\rho\sigma\lambda}F^{\mu\nu}F^{\rho}{}_{\tau}\nabla^\sigma F^{\lambda\tau}.
\label{eq:C2R2}
\end{align}

We now perform a set of field redefinitions to put the action, (\ref{eq:C2R2}), into the canonical form (\ref{eq:param4d}).  Our starting point is a Lagrangian of the form
\begin{equation}
    \cL=\cL_{2\partial}+\alpha\cL_{4\partial},
\label{eq:L24}
\end{equation}
where $\cL_{2\partial}$ is given in (\ref{eq:2derivAction}) and we have introduced a parameter $\alpha$ to keep track of the derivative expansion.  Now consider a field redefinition
\begin{equation}
    g_{\mu\nu}\to g_{\mu\nu}+\alpha\delta g_{\mu\nu},\qquad A_\mu\to A_\mu+\alpha\delta A_\mu.
\label{eq:dgdA}
\end{equation}
Applying this to the full Lagrangian, (\ref{eq:L24}), and allowing for integration by parts results in
\begin{equation}
    e^{-1}\mathcal L\to e^{-1}\mathcal L_{2\partial}+\alpha\left(e^{-1}\mathcal L_{4\partial}+\mathcal E_{\mu\nu}\delta g^{\mu\nu}+\mathcal E^\mu\delta A_\mu\right)+\mathcal O(\alpha^2),
\label{eq:fredef}
\end{equation}
where the two-derivative Einstein and Maxwell equations, $\mathcal E_{\mu\nu}$ and $\mathcal E_\mu$, are given in (\ref{eq:eoms}).  Since we are only interested in the four-derivative effective action, we ignore all terms of $\mathcal O(\alpha^2)$ and higher.  By judicial choice of $\delta g_{\mu\nu}$ and $\delta A_\mu$, we are then able to transform the four-derivative Lagrangian into the form (\ref{eq:param4d}).

As seen in (\ref{eq:fredef}), field redefinitions allow us to shift the four-derivative action by terms proportional to the two-derivative equations of motion.  In practice, this means we can substitute the two-derivative equations of motion, (\ref{eq:eoms}), into the four-derivative action, (\ref{eq:C2R2}) to transform it into canonical form.  However, it is important to note that we are treating the field redefinition, (\ref{eq:dgdA}), perturbatively in the derivative expansion.  In particular, while we are only considering the four-derivative terms, the field redefinition will generate an infinite set of terms beyond four derivatives.  Furthermore, a field redefinition in the path integral will transform the measure and couplings to external sources.  Nevertheless, physical (on-shell) quantities computed before and after the field redefinition will remain unchanged at the four-derivative order.

With the above in mind, we now use integration by parts and the two-derivative equations of motion, \eqref{eq:eoms}, to make the replacements
\begin{subequations}
\begin{align}
    (\nabla F)^2&\to-\frac{1}{3}(F^2)^2-\frac{2}{3}F^4+R_{\mu\nu\rho\sigma}F^{\mu\nu}F^{\rho\sigma}+8g^2 F^2,\\
    (\nabla_\mu F^{\mu\nu})^2 &\to -\frac{1}{3}(F^2)^2+\frac{2}{3}F^4,\\
    \epsilon_{\mu\nu\rho\sigma\lambda}F^{\mu\nu}F^{\rho\sigma}\nabla_\tau F^{\tau\lambda}&\to\frac{4}{\sqrt{6}}\qty[(F^2)^2-2F^4],\\
    F^{\nu\rho}[\nabla_\mu,\nabla_\nu]F^\mu{}_\rho&\to F^4-\frac{1}{6}(F^2)^2-\frac{1}{2}R_{\mu\nu\rho\sigma}F^{\mu\nu}F^{\rho\sigma}-4g^2F^2,
\end{align}
\label{eq:simplifications1}%
\end{subequations}
inside the four-derivative action. These replacement rules are proved in Appendix \ref{app:simplifications}. Using \eqref{eq:eoms} and \eqref{eq:simplifications1}, we find that \eqref{eq:Weyl2initial} can be reduced to the simple form
\begin{align}
    e^{-1}\cL_{C^2+\frac{1}{6}R^2}&=\frac{1}{4}\mathcal I_{\text{GB}}-\frac{1}{4}R_{\mu\nu\rho\sigma}F^{\mu\nu}F^{\rho\sigma}-\frac{1}{16}(F^2)^2+\frac{11}{24}F^4+\frac{1}{4\sqrt{6}}\epsilon^{\mu\nu\rho\sigma\lambda}R_{\mu\nu ab}R_{\rho\sigma}{}^{ab}A_\lambda\nn\\
    &\quad+\frac{2}{3}g^2 F^2+20g^4,
\label{eq:invariant1}
\end{align}
where $\mathcal I_\text{GB}$ is the usual Gauss-Bonnet combination
\begin{equation}
    \mathcal I_\text{GB}=(R_{\mu\nu\rho\sigma})^2-4(R_{\mu\nu})^2+R^2.
\end{equation}

\subsection{The action corresponding to $(R_{\mu\nu\rho\sigma})^2$}

We now turn to the supersymmetrization of $(R_{\mu\nu\rho\sigma})^2$, which was found in \cite{Ozkan:2013nwa,Ozkan:2013uk} to be
\begin{align}
    e^{-1}\cL_{(\text{Riem})^2}&=-\frac{1}{2}\left(R_{\mu \nu a b}\left(\omega_{+}\right)-G_{\mu \nu} G_{a b}\right)\left(R^{\mu \nu a b}\left(\omega_{+}\right)-G^{\mu \nu} G^{a b}\right)\nn \\
    &\quad-\frac{1}{4} \epsilon^{\mu \nu \rho \sigma \lambda}\left(R_{\mu \nu a b}\left(\omega_{+}\right)-G_{\mu \nu} G_{a b}\right)\left(R_{\rho \sigma}^{\ \ \ a b}\left(\omega_{+}\right)-G_{\rho \sigma} G^{a b}\right) C_{\lambda}\nn \\
    &\quad-\epsilon^{\mu \nu \rho \sigma \lambda} B_{\rho \sigma}\left(R_{\mu \nu a b}\left(\omega_{+}\right)-G_{\mu \nu} G_{a b}\right) \nabla_{\lambda}\left(\omega_{+}\right) G^{a b}\nn\\
    &\quad-\nabla_{\mu}\left(\omega_{+}\right) G^{a b} \nabla^{\mu}\left(\omega_{+}\right) G_{a b},
\end{align}
where $H=\dd{B}+\tfrac{1}{2}C\land G$ and we have
\begin{subequations}
\begin{align}
    H_{\mu}^{\ \,ab}&=-\frac{1}{4}e_{f\mu}\epsilon^{fabcd}G_{cd},\\
    \omega_{+\mu}^{\ \ ab}&=\omega_\mu^{ab}+H_\mu^{\ \,ab},
\end{align}
\label{eq:torsionful}%
\end{subequations}
where $\omega^{ab}$ is the torsion-free spin connection. Making use of the standard formula
\begin{equation}
    R^{ab}(\omega_+)=\dd{\omega_+^{ab}}+\omega_+^{ac}\land \omega_{+c}{}^b,
\end{equation}
we may rewrite $R_{\mu\nu ab}$ in a manifestly torsion-free way
\begin{align}
    R_{\mu\nu}{}^{ab}(\omega_+)&=R^{ab}_{\mu\nu}+\frac{1}{2}\epsilon^{fabcd}e_{f[\mu}\nabla_{\nu]}G_{cd}\nn\\
    &\quad+\frac{1}{4}\qty(2G_{[\mu}{}^{a}G_{\nu]}^b+2G^{a\delta}G_{\delta[\mu}e^b_{\nu]}-2G^{b\delta}G_{\delta[\mu}e^a_{\nu]}+G^2e^a_{[\mu}e^b_{\nu]}).
\end{align}
Using \eqref{eq:torsionful}, we also see that
\begin{equation}
    \nabla_\mu(\omega_+)G_{ab}=\nabla_\mu G_{ab}-\frac{1}{2}e^{f}_\mu\epsilon_{f[a|ced}G^{ed}G^c{}_{|b]}.
\end{equation}

It is now straightforward to work out that
\begin{align}
    \left(R_{\mu \nu a b}\bigl(\omega_{+}\right)-G_{\mu \nu} G_{a b}\bigr)^2=&(R_{\mu \nu a b})^2-\frac{3}{2}R^{\mu\nu\rho\sigma}G_{\mu\nu}G_{\rho\sigma}-2R_{\mu\nu}G^{\mu\sigma}G^\nu{}_\sigma\nn\\
    &+\frac{1}{2}G^2R-(\nabla G)^2-(\nabla_\mu G^{\mu\nu})^2+\frac{5}{8}\qty(G^2)^2+\frac{9}{8}G^4\nn\\
    &-\frac{3}{4}\epsilon^{\mu\nu\rho\sigma\lambda}G_{\mu\nu}G_{\rho\sigma}\nabla^\tau G_{\tau\lambda}.
\end{align}
By using some integration by parts to make the gauge invariance manifest, one also finds
\begin{align}
    \epsilon^{\mu \nu \rho \sigma \lambda}\bigl(R_{\mu \nu a b}\left(\omega_{+}\right)-G_{\mu \nu} G_{a b}\bigr)\bigl(R_{\rho \sigma}^{\ \ \ a b}\left(\omega_{+}\right)-G_{\rho \sigma} G^{a b}\bigr) C_{\lambda}\kern-12em&\nn\\
    &\to\epsilon^{\mu\nu\rho\sigma\lambda}R_{\mu\nu ab}R_{\rho\sigma}{}^{ab}C_\lambda+\epsilon^{\mu\rho\sigma\lambda\gamma}G_{\mu\lambda}G_\sigma{}^\delta\nabla_\rho G_{\gamma\delta}\nn\\
    &\quad +2RG^2-8R^{\mu\nu}G_{\mu\sigma}G_\nu{}^\sigma+2R^{\mu\nu\rho\sigma}G_{\mu\nu}G_{\rho\sigma}+G^4\nn\\
    &\quad -2\epsilon^{\mu\nu\rho\sigma\lambda}R_{\mu\nu ab}G_{\rho\sigma}G^{ab}C_\lambda-4\epsilon^{\mu\nu\rho\sigma\lambda}\nabla_\mu H_{\nu ab}G_{\rho\sigma}G^{ab}C_\lambda\nn\\
    &\quad -4\epsilon^{\mu\nu\rho\sigma\lambda}H_{\mu ac}H_{\nu}{}^c{}_{b}G_{\rho\sigma}G^{ab}C_\lambda.
\end{align}
The last three terms in this expression look slightly concerning, but they will be precisely cancelled by those in
\begin{align}
    \epsilon^{\mu \nu \rho \sigma \lambda} B_{\rho \sigma}\bigl(R_{\mu \nu a b}\left(\omega_{+}\right)-G_{\mu \nu} G_{a b}\bigr) \nabla_{\lambda}\left(\omega_{+}\right) G^{a b}\kern-10em&\nn\\
    &\to -R_{\mu\nu\rho\sigma}G^{\mu\nu}G^{\rho\sigma}-\frac{1}{4}\epsilon_{\mu\nu\rho\sigma\lambda}G^{\mu\nu}G^{\rho\sigma}\nabla_{\tau}G^{\tau\lambda}+\frac{1}{2}G^4+\frac{1}{4}\qty(G^2)^2\nn\\
    &\quad +\frac{1}{2}\epsilon^{\mu\nu\rho\sigma\lambda}R_{\mu\nu ab}G_{\rho\sigma}G^{ab}C_\lambda+\epsilon^{\mu\nu\rho\sigma\lambda}\nabla_\mu H_{\nu ab}G_{\rho\sigma}G^{ab}C_\lambda\nn\\
    &\quad +\epsilon^{\mu\nu\rho\sigma\lambda}H_{\mu ac}H_{\nu}{}^c{}_{b}G_{\rho\sigma}G^{ab}C_\lambda.
\end{align}
Finally, we just need
\begin{align}
    \qty(\nabla_{\mu}\left(\omega_{+}\right) G_{a b})^2=&\qty(\nabla G)^2-\epsilon_{\mu\nu\rho\lambda\delta}G^{\lambda\delta}G^\rho{}_\beta\nabla^\mu G^{\nu\beta}-\frac{1}{4}\qty(G^2)^2+\frac{1}{2}G^4.
\end{align}

Using the above expressions along with \eqref{eq:identities1} and making use of appropriate field redefinitions, we get
\begin{align}
    e^{-1}\cL_{(\text{Riem})^2}&=-\frac{1}{2}\mathcal I_{\text{GB}}+\frac{1}{2}R_{\mu\nu\rho\sigma}F^{\mu\nu}F^{\rho\sigma}+\frac{1}{8}(F^2)^2-\frac{11}{12}F^4-\frac{1}{2\sqrt{6}}\epsilon^{\mu\nu\rho\sigma\lambda}R_{\mu\nu ab}R_{\rho\sigma}{}^{ab}A_\lambda\nn\\
    &\quad-\frac{5}{3}g^2 F^2-60g^4.
\label{eq:invariant2}
\end{align}
Thus, we immediately see that
\begin{equation}
    \cL_{(\text{Riem})^2}+2\cL_{C^2+\frac{1}{6}R^2}=-\frac{1}{3}g^2 F^2-20 g^4,
\label{eq:actionRelation}
\end{equation}
which vanishes in the ungauged limit.  We note here also that the supersymmetrized Gauss-Bonnet Lagrangian may be written
\begin{equation}
    \cL_\text{GB}^{\text{susy}}=\cL_{(\text{Riem})^2}+3\cL_{C^2+\tfrac{1}{6}R^2},
\label{eq:GB}
\end{equation}
which we will make use of as an analog to the 4D case \cite{Bobev:2021oku,Bobev:2020egg}.

\subsection{The action corresponding to $R^2$}
Finally, the combination of the $R^2$ Lagrangian with the usual two-derivative action has been found in \cite{Ozkan:2013nwa} (in the language of very special geometry for supergravity coupled to vector multiplets) to be
\begin{align}
    e^{-1}\cL_{R+\alpha R^2}&=\frac{1}{4}(\mathcal{C}+3) R+\frac{2}{3}(104 \mathcal{C}-8) T^{2}+8(\mathcal{C}-1) D+\frac{3}{2} C_{I J K} \rho^{I} F_{a b}^{J} F^{a b K}\nn\\
    &\quad+3 C_{I J K} \rho^{I} \partial_{\mu} \rho^{J} \partial^{\mu} \rho^{K}-24 C_{I J K} \rho^{I} \rho^{J} F_{a b}^{K} T^{a b}+\frac{1}{4} \epsilon^{a b c d e} C_{I J K} A_{a}^{I} F_{b c}^{J} F_{d e}^{K}\nn\\
    &\quad+a_{I} \rho^{I}\left(\frac{9}{64} R^{2}-3 D R-2 R T^{2}+16 D^{2}+\frac{64}{3} D T^{2}+\frac{64}{9}\left(T^{2}\right)^{2}\right),
\end{align}
where $\cC$ is an auxiliary field and $a_I$ parameterizes the $R^2$ corrections. The $D$ equation of motion gives
\begin{equation}
    \cC=1-\frac{1}{8}a_I\rho^I\qty(-3R+32D+\frac{64}{3}T^2).
\end{equation}
Substituting this back in and truncating out the vector multiplets by taking $\rho^I$ to be constant gives
\begin{align}
    e^{-1}\cL_{R+\alpha R^2}&=R+64T^2\frac{3}{2} C_{I J K} \rho^{I} F_{a b}^{J} F^{a b K}\nn\\
    &\quad-24 C_{I J K} \rho^{I} \rho^{J} F_{a b}^{K} T^{a b}+\frac{1}{4} \epsilon^{a b c d e} C_{I J K} A_{a}^{I} F_{b c}^{J} F_{d e}^{K}\nn\\
    &\quad+a_I\rho^I\qty(\frac{15}{64}R^2-DR+\frac{70}{3}RT^2-16D^2-\frac{832}{3}DT^2-\frac{1600}{9}(T^2)^2).
\end{align}
Using the fact that
\begin{subequations}
\begin{align}
    D&=-\frac{1}{32}R+\frac{2}{9}T^2,\\
    R&=\frac{64}{9}T^2-20g^2,
\end{align}
\end{subequations}
we finally get that, after field redefinitions,
\begin{equation}
    e^{-1}\cL_{R^2}=-\frac{205}{24}g^2F^2+100 g^4.
\label{eq:R2os}
\end{equation}

\subsection{The complete four-derivative action}

Given the three invariants, a generic four-derivative action in minimal 5D supergravity can be parameterized by three coefficients
\begin{equation}
    S_{\text{HD}}=S_{2\partial}+\alpha_1S_{C^2+\tfrac{1}{6}R^2}+\alpha_2 S_{(\text{Riem})^2}+\alpha_3 S_{R^2},
\label{eq:gen4ds}
\end{equation}
where the $\alpha_i$ are taken to be small such that the higher-derivative expansion is well-defined. By making use \eqref{eq:actionRelation}, \eqref{eq:GB}, and \eqref{eq:R2os}, this is equivalent (at the four-derivative level) up to field redefinitions to
\begin{align}
    S_{\text{HD}}&=S_{2\partial}+(\alpha_1-2\alpha_2)S_\text{GB}^\text{susy}\nonumber\\
    &\qquad+g^2\int \qty[\frac{8\alpha_1-24\alpha_2-205\alpha_3}{12}F\land\star F+(\alpha_1-2\alpha_2+5\alpha_3)20g^2\star 1].
\end{align}
We would like to make the last portion of this expression manifestly proportional to $S_{2\partial}$. To accomplish this, we perform one additional redefinition
\begin{subequations}
\begin{align}
    A&\to \qty(1+g^2 b_1)A,\\
    g^2&\to\qty(1+g^2b_2)g^2,
\end{align}
\label{eq:Ag2}%
\end{subequations}
where we assume $b_i\sim\cO(\alpha_j)$, so that we may ignore higher-order terms that appear. This field redefinition will then only pick up terms from the two-derivative action
\begin{align}
    S_\text{HD}=&S_{2\partial}+(\alpha_1-2\alpha_2)S_\text{GB}^\text{susy}+g^2\int \left[\qty(\frac{8\alpha_1-24\alpha_2-205\alpha_3}{12}-b_1)F\land\star F\right.\nn\\
    &\left.+\qty(20\alpha_1-40\alpha_2+100\alpha_3+12b_2)g^2\star1-\frac{3b_1}{3\sqrt{3}}F\land F\land A\right].
\end{align}
We may then make use of the two-derivative equations of motion to rewrite this as
\begin{align}
    S_\text{HD}=&S_{2\partial}+(\alpha_1-2\alpha_2)S_\text{GB}^\text{susy}\nn\\
    &+3b_1 g^2\int\left\{\qty[\frac{1}{3b_1}\qty(\frac{8\alpha_1-24\alpha_2-205\alpha_3}{12}-b_1)-\frac{1}{3}]F\land\star F\right.\nn\\
    &\left.+R\star1+\qty(\frac{20\alpha_1-40\alpha_2+100\alpha_3+12b_2}{3b_1}+20)g^2\star1-\frac{1}{3\sqrt{3}}F\land F\land A\right\}.
\end{align}
We must then choose $b_1$ and $b_2$ such that
\begin{subequations}
\begin{align}
    \frac{1}{3b_1}\qty(\frac{8\alpha_1-24\alpha_2-205\alpha_3}{12}-b_1)-\frac{1}{3}&=-\frac{1}{2},\\
    \frac{20\alpha_1-40\alpha_2+100\alpha_3+12b_2}{3b_1}+20&=12.
\end{align}
\end{subequations}
This is solved by
\begin{subequations}
\begin{align}
    b_1&=\frac{1}{6}\qty(8\alpha_1-24\alpha_2-205\alpha_3),\\
    b_2&=\frac{1}{6}\qty(-21\alpha_1+68\alpha_2+360\alpha_3),
\end{align}
\end{subequations}
which finally yields
\begin{equation}
    S_\text{HD}=\qty(1+g^2\frac{8\alpha_1-24\alpha_2-205\alpha_3}{2})S_{2\partial}+(\alpha_1-2\alpha_2)S_\text{GB}^\text{susy}.
\label{eq:HDaction}
\end{equation}
This is in direct analogy to the 4D case \cite{Bobev:2020egg,Bobev:2021oku}.  By use of field redefinitions, we have been able to rewrite the general four-derivative corrected action, (\ref{eq:gen4ds}), in the canonical basis of (\ref{eq:param4d}).  However, it is important to recall that this is a perturbative result valid only to linear order in the $\alpha_i$ coefficients.  From an effective field theory point of view, this is sufficient for most purposes, including computing the on-shell value of the action, as physical observables are invariant under field redefinitions.  However, straightforward use of (\ref{eq:HDaction}) is not valid for off-shell quantities unless the effects of the field redefinitions, (\ref{eq:dgdA}) and (\ref{eq:Ag2}), are fully accounted for.

This result has a strong implication. Just as in the 4D case, at the four-derivative level, the effective supergravity action is completely parametrized by two quantities, $\{S_{2\partial},S_{\mathrm{GB}}\}$, related to the geometry and three independent quantities, $\{\alpha_i,g\}$, related to the particular theory.%
\footnote{While there are three $\alpha_i$ parameters, they only enter in two independent combinations in (\ref{eq:HDaction}).}
Unlike the 4D case, however, $S_{\mathrm{GB}}^\text{susy}$ is not topological, so there can be a potentially richer structure of background geometry dependence in the AdS$_5$/CFT$_4$ setup.

We now give the explicit form of the supersymmetrized Gauss-Bonnet action $S_{\mathrm{GB}}^\text{susy}$ introduced in (\ref{eq:GB}).  Following \cite{Myers:2009ij}, we parameterize the 5D four-derivative Lagrangian as%
\footnote{This parametrization differs from \cite{Myers:2009ij} in $c_1$ and $c_2$, but is chosen for easier comparison with \cite{Cremonini:2020smy}.}
\begin{equation}
    e^{-1}\cL_{4\partial}=c_1\mathcal I_\text{GB}+c_2 C_{\mu\nu\rho\sigma}F^{\mu\nu}F^{\rho\sigma}+c_3 (F^2)^2+c_4 F^4+c_5\epsilon^{\mu\nu\rho\sigma\lambda}R_{\mu\nu ab}R_{\rho\sigma}^{\ \ \ ab}A_{\lambda}.\label{eq:4derivPart}
\end{equation}
Supersymmetry fixes this correction in terms of a single overall coefficient.  Using the definition of the five-dimensional Weyl tensor, we can make the substitution
\begin{equation}
    R_{\mu\nu\rho\sigma}F^{\mu\nu}F^{\rho\sigma}=C_{\mu\nu\rho\sigma}F^{\mu\nu}F^{\rho\sigma}+\frac{4}{3}F^4-\frac{1}{4}\qty(F^2)^2-2g^2F^2,
\end{equation}
obtained in Appendix \ref{app:simplifications} in the expressions (\ref{eq:invariant1}) and (\ref{eq:invariant2}).  Taking the Gauss-Bonnet combination, (\ref{eq:GB}), then gives
\begin{equation}
    S_{\mathrm{GB}}=\int \dd[5]x\sqrt{-g}\qty[\mathcal I_\text{GB}-\frac{1}{2} C_{\mu\nu\rho\sigma}F^{\mu\nu}F^{\rho\sigma}+\frac{1}{8} F^4+\frac{1}{2\sqrt{3}}\epsilon^{\mu\nu\rho\sigma\lambda}R_{\mu\nu ab}R_{\rho\sigma}{}^{ab}A_{\lambda}],\label{eq:GBaction}
\end{equation}
which corresponds to
\begin{equation}
    c_1=-2 c_2=8 c_4=2\sqrt{3} c_5,\qquad c_3=0.
\label{eq:coeffs}
\end{equation}
As mentioned in the introduction, this conflicts with some prior results \cite{Myers:2009ij,Cremonini:2020smy, Cremonini:2021jlmt2}.  However, support for the present result can be obtained from investigating supersymmetric BMPV black holes, which we turn to next.

\section{An application: the BMPV solution}\label{sec:BMPV}

As we have just seen, our main result, \eqref{eq:coeffs}, disagrees with several results in the literature. Thus, we would like to establish some evidence for the present coefficients. We note that the distinction is subtle, as the previously obtained four-derivative action of \cite{Cremonini:2020smy} differs only by a term proportional to $(F^2)^2-2F^4$, which will vanish for purely electric or purely magnetic solutions. This is because a purely electric black hole will have $F_{tr}$ as the only non-vanishing component of the field strength. One then has
\begin{subequations}
\begin{align}
    F^2&=g^{\mu\rho}g^{\nu\sigma}F_{\mu\nu}F_{\rho\sigma}=2g^{tt}g^{rr}(F_{tr})^2,\\
    F^4&=2(g^{tt})^2(g^{rr})^2F_{tr}F_{rt}F_{tr}F_{rt}=2\qty(g^{tt}g^{rr}(F_{tr})^2)^2.
\end{align}
\end{subequations}
So then we see that the combination $(F^2)^2-2F^4$ vanishes.  Another way to see this is to note that this combination can be written as
\begin{equation}
    \epsilon\epsilon F^4\equiv\epsilon_{\alpha\mu_1\mu_2\mu_3\mu_4}\epsilon^{\alpha\nu_1\nu_2\nu_3\nu_4}F^{\mu_1}{}_{\nu_1}F^{\mu_2}{}_{\nu_2}F^{\mu_3}{}_{\nu_3}F^{\mu_4}{}_{\nu_4}=-3((F^2)^2-2F^4),
\label{eq:eeF4}
\end{equation}
where the overall minus sign arises since we are using the signature $(-,+,+,+,+)$.  This also vanishes for purely magnetic objects, as the combination $(F^2)^2-2F^4$ is only sensitive to solutions where both electric and magnetic fields are present. So, to see the distinction, we must either consider a dyonic solution or a charged, rotating solution. Hence we turn to the BMPV solution \cite{Breckenridge:1996mpv}, which is a rotating black hole in five dimensions.

The BMPV solution is an asymptotically Minkowski solution, which corresponds to ungauged supergravity (or, equivalently, the $g\to0$ limit); we consider an asymptotically flat solution here as we can avoid worrying about subtleties having to do with extra factors of the two-derivative action, which simplifies the analysis. The BMPV solution is as follows
\begin{subequations}
\begin{align}
    \dd{s^2}&=-f(r)^{-2}\qty[\dd{t}+\frac{\mu\omega}{r^2}(\sin^2\theta\dd{\phi}-\cos^2\theta\dd{\psi})]^2\nn\\
    &\quad+f(r)\qty[\dd{r}^2+
    r^2\left(\sin^2\theta \dd{\phi}^2 +\cos^2\theta\dd{\psi}^2+\dd{\theta}^2\right)],\\
    A&=\sqrt{3}f(r)^{-1}\qty[\dd{t}+\frac{\mu\omega}{r^2}(\sin^2\theta\dd{\phi}-\cos^2\theta\dd{\psi})],
\end{align}
\end{subequations}
where
\begin{equation}
    f(r)=1+\frac{\mu}{r^2}.
\end{equation}
This solution depends on two parameters, $\mu$, and $\omega$, and describes a charged, spinning black hole with ADM mass
\begin{equation}
    M=\frac{3\pi}{4}\mu,
\end{equation}
two equal magnitude angular momenta in the independent planes defined by $\phi$, $\psi$,
\begin{subequations}
\begin{equation}
    J_{\phi}=\frac{\pi}{4}\mu\omega,
\end{equation}
\begin{equation}
    J_\psi=-\frac{\pi}{4}\mu\omega,
\end{equation}
\end{subequations}
and electric charge
\begin{equation}
    Q=\frac{\sqrt{3}\pi}{2}\mu.
\end{equation}
Being a supersymmetric solution, the BMPV solution satisfies the BPS equation%
\footnote{In AdS, the BPS condition includes the angular momenta, $M=\fft{\sqrt3}2|Q|+(|J_1|+|J_2|)/L$, which is another reason why the asymptotically flat case is simpler to study.}
\begin{equation}
    M=\frac{\sqrt{3}}{2}|Q|.\label{eq:BPS}
\end{equation}
The key point is that we expect \eqref{eq:BPS} to hold even after four-derivative corrections are taken into account, as the system ought to remain supersymmetric.

There is a simple argument that the four-derivative terms \eqref{eq:GBaction} do not modify the charge of the BMPV solution.%
\footnote{The charge may be modified by the factor that appears in front of the two-derivative action after going on-shell, analogous to the four-dimensional case \cite{Bobev:2020egg,Bobev:2021oku}, but we focus only on the four-derivative part here.}
Heuristically, the four-derivative terms look like two-derivative terms squared, so the equations of motion should pick up terms that are more suppressed as $r$ becomes large.\footnote{Note this argument only works for asymptotically Minkowski space, where we expect solutions to fall off at infinity. In AdS, we know that objects such as the Riemann tensor will go to a constant (with respect to $r$) rather than disappear.} Thus, one expects the corrections to $A$ to be subleading in $r$. The charge is computed by integrating over an $S^3$ at $r\to\infty$, so we expect to only pick up the ${1}/{r^3}$ terms in $F$. For example, one might worry about terms of the form $F^3$, Weyl$\cdot F$, or Riem$^2$ contributing, but these must fall off faster than $1/r^3$ since $F$ falls off like $1/r^3$ and Riemann must fall off like $1/r^2$ in an asymptotically flat background.%
\footnote{Any black hole solution with a spatially localized horizon should look more-or-less point-like very far away (near spatial infinity). Hence, the Riemann tensor should fall off no slower than for Schwarzschild, which falls off as $1/r^2$.}
Hence, the subleading corrections from the four-derivative terms should not contribute, and the charge should not change when one introduces higher-derivative corrections. Requiring that \eqref{eq:BPS} hold in the four-derivative case then immediately implies that the mass must not shift when one introduces four-derivative corrections.

The most direct way to access the mass is to find the four-derivative corrected solution and compute the ADM mass. However, this is difficult, so we will use a slightly more indirect approach. The on-shell action is naturally identified with the (classical) Gibbs free energy
\begin{equation}
    I_{\text{HD}}=\beta G,
\end{equation}
where $\beta=T^{-1}$ is the inverse temperature. We have the standard thermodynamic relation
\begin{equation}
    G=U-T\mathcal S-\Phi Q-\Omega J,
\end{equation}
where $U$ is the internal energy, $\cS$ is the entropy, $\Phi$ is the electric potential, and $\Omega$ is the angular velocity. For a black hole, the internal energy should just be the mass $M$.\footnote{Up to a contribution from the cosmological constant, which may be removed by an appropriate boundary counterterm.} Moreover, since the BMPV solution is extremal, it has zero temperature, so we are left with
\begin{equation}
    I_\text{HD}=\beta\qty(M-\sqrt{3}Q),
\end{equation}
where we have substituted in the BPS values of $\Phi$ and $\Omega$. Thus, we see that the change in the action is the change in the BPS equation
\begin{equation}
    \Delta I:=I_\text{HD}-I_{2\partial}=\beta \Delta\qty(M-\sqrt{3}Q)=0.
\end{equation}

In principle, we should evaluate the action on the four-derivative solution; however, it will give the same result as evaluating it on the two-derivative solution. The standard argument is as follows. We write the four-derivative solution as a two-derivative piece plus some perturbing correction (using $\Phi$ schematically for all the fields)
\begin{equation}
    \Phi=\Phi_0+\alpha\delta\Phi+\cO(\alpha^2),
\end{equation}
where $\Phi_0$ is simply the BMPV solution in this case. Then the action is
\begin{align}
    S_\text{HD}[\Phi]&=S_{2\partial}[\Phi_0+\alpha\delta\Phi]+\alpha S_{4\partial}[\Phi_0+\alpha\delta\Phi]\nn\\
    &=S_{2\partial}[\Phi_0]+\alpha\frac{\delta S_{2\partial}}{\delta \Phi}\Big\vert_{\Phi_0}\delta\Phi+\alpha S_{4\partial}[\Phi_0]+\cO(\alpha^2)\nn\\
    &=S_\text{HD}[\Phi_0]+\cO(\alpha^2),
\end{align}
where we have used the fact that ${\delta S_{2\partial}}/{\delta \Phi}\big\vert_{\Phi_0}$ vanishes by the equations of motion.

Thus, we need only evaluate the four-derivative part of the action \eqref{eq:4derivPart} on the two-derivative solution. We will do this for generic coefficients $c_i$, and show that this necessarily leads to \eqref{eq:coeffs}. Note that the BMPV solution has nice asymptotics at infinity, so we do not need to introduce any four-derivative Gibbons-Hawking terms or boundary counterterms to remove divergences.

The Gauss-Bonnet action can be evaluated simply to be
\begin{equation}
    \int\dd[5]{x}e\mathcal I_\text{GB}=\frac{4\pi^2}{5}\frac{5\omega^4+2\omega^2\mu-15\mu^2}{\mu^2}\mathrm{vol}(\bR),
\end{equation}
where $\mathrm{vol}(\bR)$ is the (formally infinite) factor from doing the $t$ integration since we are working with a zero-temperature solution. The Weyl tensor contracted with graviphoton field strengths gives
\begin{equation}
    \int\dd[5]{x}eC_{\mu\nu\rho\sigma}F^{\mu\nu}F^{\rho\sigma}=-\frac{2\pi^2}{5}\frac{-160\omega^4+96\omega^2\mu+15\mu^2}{\mu^2}\mathrm{vol}(\bR).
\end{equation}
The two quartic field strength terms yield
\begin{align}
    \int\dd[5]{x}e(F^2)^2&=\frac{48\pi^2}{5}\frac{40\omega^4-48\omega^2\mu+15\mu^2}{\mu^2}\mathrm{vol}(\bR),\nn\\
    \int\dd[5]{x}eF^4&=\frac{24\pi^2}{5}\frac{20\omega^4-24\omega^2\mu+15\mu^2}{\mu^2}\mathrm{vol}(\bR).
\end{align}
Finally, the mixed Chern-Simons term gives
\begin{equation}
    \int\dd[5]{x}e\epsilon^{\mu\nu\rho\sigma\lambda}R_{\mu\nu ab}R_{\rho\sigma}{}^{ab}A_\lambda=\frac{32\sqrt{3}\pi^2}{5}\frac{5\omega^4-2\omega^2\mu}{\mu^2}\mathrm{vol}(\bR).
\end{equation}
Putting these terms together gives the requirement that
\begin{align}
    0\overset{!}{=}&\frac{2\pi^2\mathrm{vol}(\bR)}{5\mu^2}\left[10(c_1+16c_2+96c_3+24c_4+8\sqrt{3}c_5)\omega^4\right.\nn\\
    &\left.-4(-c_1+24c_2+288c_3+72c_4+8\sqrt{3}c_5)\omega^2\mu-15(2c_1+c_2-24c_3-12c_4)\mu^2\right].
\end{align}
Demanding that the $c_i$ coefficients be independent of the solution parameters $\mu$, $\omega$ then requires that we individually set
\begin{align}
    c_1+16c_2+96c_3+24c_4+8\sqrt{3}c_5&=0,\nn\\
    -c_1+24c_2+288c_3+72c_4+8\sqrt{3}c_5&=0,\nn\\
    2c_1+c_2-24c_3-12c_4&=0.
\end{align}
As we have five coefficients $c_i$ and only two parameters to vary, the solution is not unique.  Solving for the latter $c_i$ in terms of $c_1$ and $c_2$ gives
\begin{align}
    c_3&=-\frac{1}{16}(c_1+2c_2),\nn\\
    c_4&=\frac{1}{24}(7c_1+8c_2),\nn\\
    c_5&=-\frac{\sqrt{3}}{12}(c_1+6c_2).
\end{align}
Since $c_1$ just controls the overall coefficient of the four-derivative action, this is a one-parameter family of solutions. However, there is no ambiguity in the literature as to $c_2$ or $c_5$, so demanding that $c_2=-\frac{1}{2}c_1$ or that $c_5=\frac{1}{2\sqrt{3}}c_1$ immediately gives us
\begin{equation}
    c_2=-\frac{1}{2}c_1,\ c_3=0,\ c_4=\frac{1}{8}c_1, \ c_5=\frac{1}{2\sqrt{3}}c_1,
\end{equation}
in perfect agreement with the present result, \eqref{eq:coeffs}.

\section{Discussion}\label{sec:discussion}

We have shown that the three possible supersymmetric four-derivative terms that we can add to the 5D $\mathcal N=2$ supergravity action reduce after field redefinitions to a single four-derivative superinvariant as well as factors of the original two-derivative action. This, in turn, implied that there is a unique four-derivative piece of the action, up to an overall factor. We checked this explicitly in the case of the BMPV black hole and found excellent agreement.

Of particular note, we found $c_i$ coefficients that disagree with several results in the literature \cite{Cremonini:2008tw,Myers:2009ij,Cremonini:2020smy,Cremonini:2021jlmt2}. However, while we believe the particular $c_i$ determined previously are incorrect, the results that they are used to derive are still generally valid as they are predominantly applied to non-rotating, non-dyonic solutions, for which the discrepancy, in the form of $(F^2)^2-2F^4$, vanishes. As noted in (\ref{eq:eeF4}), this conflict with the previous result is in the $c_3$ and $c_4$ coefficients, and corresponds to a difference in the four-derivative Lagrangians
\begin{equation}
    e^{-1}\mathcal{L}_{4\partial}^{\mathrm{here}}=e^{-1}\mathcal{L}_{4\partial}^{\mathrm{previous}}-\fft1{18}\epsilon_{\alpha\mu_1\mu_2\mu_3\mu_4}\epsilon^{\alpha\nu_1\nu_2\nu_3\nu_4}F^{\mu_1}{}_{\nu_1}F^{\mu_2}{}_{\nu_2}F^{\mu_3}{}_{\nu_3}F^{\mu_4}{}_{\nu_4}.
\end{equation}
This suggests that the previous determination of the $c_i$ had an issue when translating the conventions of \cite{Hanaki:2006pj} to that of \cite{Cremonini:2008tw,Myers:2009ij}.  In particular, $\epsilon\epsilon=+5!$ or $\epsilon\epsilon=-5!$ is signature dependent, and an incorrect sign choice may have arisen when switching conventions. It should also be noted that the same coefficients as in \eqref{eq:coeffs} were also found in \cite{Bobev:2022bjm,Cassani:2022lrk}.

With that being said, the analysis leaves open questions. A natural solution to look at is the Gutowski-Reall solution \cite{Gutowski:2004r}, which is a one-parameter family of charged, spinning, supersymmetric black holes in AdS$_5$. Na\"ively, one expects that the four-derivative correction to the on-shell action should vanish;%
\footnote{The stringy eight-derivative corrections to the Gutowski-Reall solution were recently explored in \cite{Melo:2007} where the shift to the action was indeed shown to vanish in the BPS case.}
however, after appropriate holographic renormalization, we seem to find that the four-derivative correction to the action does not vanish beyond what we expect from the renormalization of the AdS radius (see Appendix \ref{app:GutowskiReall} for some technical details). There is a set of $c_i$ coefficients such that the four-derivative action vanishes, but this requires we have either $c_2\neq-\frac{1}{2}c_1$ or $c_5\neq\frac{1}{2\sqrt{3}}c_1$, which seems to conflict with current results in the literature. The alternative, however, is that there is a non-zero shift in the mass of the black hole. To preserve the BPS relation, this would require a renormalization of either the charge (which seems unlikely given the arguments presented in Section \ref{sec:BMPV}) or of the angular momentum. However, a shift to the mass seems unlikely since it has been shown that there are no corrections when the solution is uplifted to tree-level $\alpha'^3$-corrected type IIB supergravity \cite{Melo:2007}. The resolution seems to be that the field redefinitions we have used alter the boundary values of fields \cite{Cassani:2022lrk}. Indeed, many subtleties are involved when applying the Reall-Santos trick to black holes in AdS \cite{Hu:2023gru}, presumably, this explains the observed discrepancies.

Additional puzzles come from some asymptotically flat solutions like supersymmetric black rings \cite{Elvang:2004emr} and four-dimensional dyonic STU black holes lifted to five dimensions \cite{Cremonini:2021jlmt2}. Of particular note is that the black ring reduces to the BMPV solution in the limit that the ring radius goes to zero. However, both of these solutions again have non-vanishing four-derivative actions. The black ring solution has no set of coefficients $c_i$ such that the four-derivative action vanishes for all choices of solution parameters; however, it does in the BMPV limit. On the other hand, the lift of the dyonic 4D STU black hole has a unique set of $c_i$ coefficients making the four-derivative action vanish with $c_2=-\frac{1}{2}c_1$ and $c_5=\frac{1}{2\sqrt{3}}c_1$, but otherwise unrelated to other coefficients in the literature.

In the future, we would like to resolve these open puzzles regarding the non-vanishing four-derivative actions of BPS solutions. Moreover, we believe it would be fruitful to dimensionally reduce to four dimensions, which would give us $\cN=2$ supergravity coupled to a vector multiplet, which is a truncation of the 4D STU model; we could then compare this with the results of \cite{Bobev:2021oku}. We would also like to extend the results of this note to the 5D STU model, or more generally, $\cN=2$, $D=5$ supergravity coupled to vector multiplets. Finally, although motivated by holography and subleading corrections in supersymmetric partition functions, we have yet to explore this avenue, which we believe will be a worthwhile extension of our results.

%% file: Chapters/chap4.tex
One central organizing principle in the space of QFTs is the \ac{RG} flow. RG flow is often understood as a family of successive quantum field theories starting at some high-energy (UV) CFT and flowing to some low-energy (IR) CFT. As the flow progresses, the effective number of degrees of freedom decreases due to the process of coarse-graining. This reduction can be accurately quantified by ``counting functions,'' which are monotonic along the RG flow and thus render the flows irreversible. Of particular interest are functions that connect quantities in the CFTs, such as $A$-type central charges in even dimensions and sphere free energies in odd dimensions. Both of these quantities will be referred to as central charges in the following discussion. There are well-established theorems regarding such flows, including proofs of the 2d $c$-theorem by Zamolodchikov \cite{Zamolodchikov:1986gt}, the 3d $F$-theorem by Casini and Huerta \cite{Jafferis:2010un,Klebanov:2011gs,Jafferis:2011zi,Casini:2012ei}, and the 4d $a$-theorem by Komargodski and Schwimmer \cite{Cardy:1988cwa,Komargodski:2011vj}; an alternative approach that has been used to great effect involves entanglement entropy and has been quite useful for proving results in $d=2,3,4$ \cite{Casini:2006es,Casini:2017roe,Casini:2017vbe}. There also exist partial results in 5d \cite{Jafferis:2012iv,Chang:2017cdx,Fluder:2020pym} and 6d \cite{Elvang:2012st,Cordova:2015fha,Heckman:2015axa}.

The AdS/CFT correspondence geometrizes many aspects of QFTs and has proven a particularly useful framework for studying the properties of RG flows. Considerable progress on constructing $c$-functions has been made from the holographic perspective:  Various holographic $c$-theorems have been established in this context by making use of the \ac{NEC} \cite{Girardello:1998pd,Freedman:1999gp,Myers:2010tj,Myers:2010xs}, as well as using the entanglement entropy perspective to analyze holographic RG flows \cite{Casini:2011kv,Myers:2012ed}. Holographic methods, for example, permit the construction of certain monotonic $c$-functions in any dimension and at strong coupling, something way beyond the reach of field-theoretic approaches.

Naturally, much work has been done on extending holographic $c$-theorems to include higher-derivative corrections \cite{Anber:2008js,Myers:2010tj,Myers:2010xs,Myers:2012ed,Sinha:2010ai,Gullu:2010pc,Gullu:2010st,Sinha:2010pm,Oliva:2010eb,Myers:2010ru,Oliva:2010zd,Liu:2010xc,Liu:2011iia,Alkac:2018whk,Ghodsi:2019xrx,Anastasiou:2021jcv,Ghodsi:2021xrb,Alkac:2022zda}.  Such extensions allow one to distinguish various central charges \cite{Nojiri:1999mh, Blau:1999vz}. For example, in 4D, we have that $a=c$ at the two-derivative level in gravity or in the large-$N$ limit in field theory. It is well known, however, that $a$ alone has a monotonic flow from the UV to the IR \cite{Komargodski:2011vj}, while $c$ does not. As such, adding higher derivatives allows one to distinguish between the central charges that have monotonic flows and the ones that do not.  Such higher derivatives correspond to sub-leading in $N$ corrections to the central charges. 

In this chapter, we explore the notion of counting functions in RG flows across dimensions, meaning the compactification of a $D$-dimensional CFT, which is the UV fixed point, on a $(D-d)$-dimensional compact space, such that the IR fixed point is a $d$-dimensional CFT.  RG flows across dimensions are particularly amenable to holographic methods; there are many examples of supergravity solutions holographically dual to RG flows interpolating between CFTs of different dimensions \cite{Maldacena:2000mw,Acharya:2000mu,Gauntlett:2000ng,Gauntlett:2001jj,Gauntlett:2001qs,Benini:2013cda,Benini:2015bwz,Bobev:2017uzs}. Some candidate $c$-functions for such flows were studied in \cite{Macpherson:2014eza,Bea:2015fja,Legramandi:2021aqv}, and more recently an explicit $c$-function was constructed in \cite{GonzalezLezcano:2022mcd}. The holographic entanglement entropy picture for such flows was further analyzed in \cite{Deddo:2022wxj}.  In this chapter, we explore the role of higher-derivative corrections in holographic flows across dimensions. As a natural starting point, we generalize some of the results of Myers-Sinha \cite{Myers:2010xs,Myers:2010tj}, who considered the effect of higher-derivative terms in holographic RG flows, to flows across dimensions. 

\subsection*{The holographic setup}

Our starting point is Einstein gravity with a negative cosmological constant.  From an effective field theory point of view, one would expect this to be corrected by a set of higher derivative operators.  The first such terms arise at the four-derivative level and involve a combination of $\hat R_{MNPQ}\hat R^{MNPQ}$, $\hat R_{MN}\hat R^{MN}$, and $\hat R^2$.  However, since the Ricci terms can be shifted by a field redefinition, we may choose the Gauss-Bonnet combination
\begin{equation}
    \chi_4=\hat R_{MNPQ}\hat R^{MNPQ}-4\hat R_{MN}\hat R^{MN}+\hat R^2.
\end{equation}
As a result, we focus on the bulk $(D+1)$-dimensional Lagrangian
\begin{equation}
    e^{-1}\mathcal{L}=\frac{1}{2\kappa^2}\qty[\hat R+\frac{D(D-1)}{L^2}+\alpha\chi_4],
\label{eq:gravLag}
\end{equation}
where $\alpha$ parametrizes the correction.  This choice of the Gauss-Bonnet combination is convenient since in this case, the corrected Einstein equation remains second order in derivatives.  This system admits a maximally symmetric AdS$_{D+1}$ vacuum with an AdS radius $L_{\mathrm{UV}}^2=L^2-\alpha(D-2)(D-3)$, to linear order in $\alpha$.

We are interested in flows from AdS$_{D+1}$ in the UV to AdS$_{d+1}\times M_{D-d}$ in the IR.  Such flows can be induced by coupling the gravitational Lagrangian, (\ref{eq:gravLag}), to a suitable matter sector, \textit{i.e.}, $\mathcal L\to\mathcal L+\mathcal L_{\mathrm{matter}}$.  To parametrize the flow, we split off the would-be internal space $M_{D-d}$ and assume a spacetime metric of the form
\begin{equation}
        \dd s^2=e^{2f(z)}(\eta_{\mu\nu}\dd x^\mu \dd x^\nu+ \dd z^2)+ e^{2g(z)}g_{ij}(y)\dd y^i\dd y^j.
\end{equation}
The flow is along the bulk radial coordinate, $z$, and we take the asymptotics to be such that $e^{2f}\sim e^{2g}\sim 1/z^2$ in the UV ($z\to0$) while $e^{2f}\sim 1/z^2$ with $e^{2g}\sim\mbox{const.}$ in the IR ($z\to\infty$). Note that this metric implicitly assumes flat slicings of $\mathrm{AdS}_{d+1}$, although some authors have considered curved slicings \cite{Ghosh:2017big,Ghosh:2018qtg,Kiritsis:2022oww}.

Given a bulk metric parametrized by the two functions $f(z)$ and $g(z)$, we then explicitly construct a function $c(f,g; z)$ such that $\dd c/\dd z\le 0$ upon imposing the NEC, $T_{MN}\xi^M\xi^N\ge 0$, on the matter sector where $\xi$ is a future-directed null vector. This is the desired monotonicity property. This $c$-function directly generalizes the two-derivative case \cite{GonzalezLezcano:2022mcd} to which it reduces when the Gauss-Bonnet coupling $\alpha$ is set to zero, as well as generalizing the four-derivative case of flows within the same dimension \cite{Myers:2010tj,Myers:2010xs} to which it reduces in the limit that there are no compact internal dimensions. This $c$-function is not unique but has two free parameters characterizing it; despite this mild ambiguity, the IR limit of this central charge is unambiguously the $A$-type central charge, as expected.  In other words, $\lim_{z\to\infty}c(z)= a_\text{IR}$, where $a_{\text{IR}}$ is the four-derivative $A$-type central charge.

As in the two-derivative case \cite{GonzalezLezcano:2022mcd}, this $c$-function diverges in the UV.  However, we show that the divergence of the $c$-function encodes the UV central charge. As we approach the UV, the compact extra dimensions unfurl and our massive KK towers become increasingly light and begin to enter the spectrum, meaning that the number of lower-dimensional degrees of freedom appears to become infinite. Dimensional analysis alone tells us that the central charge must diverge as a pole of order the number of compact dimensions; however, we go further and show that the coefficient of this pole encodes the value of the UV central charge, \textit{i.e.},
\begin{equation}
    c(z)\overset{z\to 0}{\sim}\frac{a_\text{UV}}{z^{D-d}},
\end{equation}
where $a_\text{UV}$ is the (four-derivative) $A$-type central charge in the UV. This is not entirely automatic; it requires an additional constraint on the remaining free parameters of the $c$-function. However, we may always choose the parameters so this is the case.

We also construct $c$-functions from the entanglement entropy. In particular, we consider entangling regions of this CFT which completely wrap the internal space. The entanglement entropy has been shown \cite{Faulkner:2013ana,Hung:2011xb,deBoer:2011wk,Bhattacharyya:2013jma} to then be given by finding the extremal surface that minimizes the Jacobson-Myers functional \cite{Jacobson:1993xs}
    \begin{equation}
    S_\text{JM}=\frac{1}{4G_N}\int_\Sigma \dd[d] x \sqrt{h}\left(1+2\alpha {\cal R}\right)+\frac{1}{2G_N}\int_{\partial\Sigma}\dd[d-1]{x}\sqrt{\tilde h}\,(2\alpha{\cal K}),
\end{equation}
where $\Sigma$ is the extremal surface with boundary $\partial\Sigma$, $h$ is the determinant of the induced metric on $\Sigma$, $\tilde h$ is the induced metric (of the induced metric $h$) on $\partial\Sigma$, ${\cal R}$ is the scalar curvature of $\Sigma$, and $\cal K$ is the trace of the extrinsic curvature of the boundary $\partial\Sigma$.
    For the case of flows from AdS$_{D+1}$ to AdS$_3$, which may be equivalently viewed as flows from CFT$_D$ to CFT$_2$, we explicitly obtain a monotonic $c$-function from the entanglement entropy as
    \begin{equation}
        c_\text{EE}=R\,\partial_R \,\, S_\text{JM},\label{eq:monoC}
    \end{equation}
    where $R$ is the radius of the entangling region. Given a minimal surface whose profile is $r(z)$, $S_\text{JM}$ admits a first integral that can be solved for $r'(z)$. This allows us to explicitly evaluate \eqref{eq:monoC} and subsequently verify its monotonicity due to the NEC. Moreover, it turns out to be the case that this $c$-function which one obtains from the holographic entanglement entropy is indeed related to the local $c$-function obtained directly from the NEC; we show that the monotonicity of one directly implies the monotonicity of the other. Such precise connection of two {\it a priori} differently defined $c$-functions opens the possibility of better understanding the connection between strong subadditivity of the entanglement entropy and the NEC as a condition on the holographic gravity backgrounds. 

    The rest of this chapter is organized as follows. In Section \ref{sec:LH}, we explicitly construct a local $c$-function for the case of Gauss-Bonnet corrected gravity and demonstrate that it flows monotonically from the UV to the IR due to the null energy condition (NEC). In Section \ref{sec:limits}, we show that the IR limit of this $c$-function is the $A$-type central charge and, although the $c$-function diverges in the UV, the coefficient of this divergence encodes the UV central charge. In Section \ref{sec:EE}, we discuss the $c$-function obtained from holographic entanglement entropy and show that it is monotonic, at least when there is no curvature of the internal space, and show that this quantity is related to the NEC-motivated central charge constructed in Section \ref{sec:LH}. A summary and conclusions are given in Section \ref{sec:end}. We relegate some more technical details to Appendix \ref{App:Geom}.

\section{Higher-derivative gravity and  NEC}\label{sec:LH}

We are interested in RG flows from CFT$_D$ to CFT$_d$ triggered by compactification on a $(D-d)$-dimensional manifold, $M_{D-d}$. Holographically, this corresponds to a geometric flow from AdS$_{D+1}$ to AdS$_{d+1}\times M_{D-d}$. The holographic radial coordinate $z$ then naturally functions as the scale for RG flow. We may explicitly realize this setup by choosing a metric
\begin{equation}
\label{Eq:Flowfg}
    \dd s^2 = e^{2f(z)}\qty(\eta_{\mu\nu}\dd x^\mu \dd x^\nu+ \dd z^2)+ e^{2g(z)}g_{ij}(y)\dd y^i\dd y^j,
\end{equation}
such that in the UV region $z\to 0$ the metric is asymptotically AdS$_{D+1}$ and in the IR region $z\to\infty$ the metric asymptotes to AdS$_{d+1}\times M_{D-d}$. To be rigorous, the metric \eqref{Eq:Flowfg} is not the most general metric describing holographic
RG flows across dimensions; for example, there are known holographic RG flows where the
internal space $M_{D-d}$ depends on the holographic radial coordinate $z$ in a non-separable way \cite{Anderson:2011cz,Fluder:2017nww,Bobev:2020jlb}. We restrict our attention to the separable case \eqref{Eq:Flowfg} for simplicity; we leave it as an exercise for future research to extend our analysis of holographic $c$-functions in separable flows to more general non-separable flows.

Furthermore, unless otherwise specified, we will assume that the metric $g_{ij}$ of $M_{D-d}$ is maximally symmetric with Ricci curvature
\begin{equation}
    \tilde R_{ij}=\kappa\frac{D-d-1}{\ell^2}g_{ij},
\end{equation}
where $\kappa=-1$, $0$, or $1$ for negative, flat, or positive curvature, respectively. This is not the most general choice of metric on the internal space, but we make this choice for simplicity; we will generalize this to arbitrary Einstein internal manifolds in Section \ref{sec:generalization}.

As discussed above, we start with a two-derivative theory in the gravitational sector, namely the Einstein-Hilbert Lagrangian with a negative cosmological constant.  At the four-derivative level, we add a Gauss-Bonnet coupling
\begin{equation}
    \chi_4=\hat R_{MNPQ}\hat R^{MNPQ}-4\hat R_{MN}\hat R^{MN}+\hat R^2,
\label{eq:chi4}
\end{equation}
so we end up considering the gravitational Lagrangian
\begin{equation}
    e^{-1}\mathcal{L}=\frac{1}{2\kappa^2}\qty[\hat R+\frac{D(D-1)}{L^2}+\alpha\chi_4],
\label{eq:Lgrav}
\end{equation}
coupled to a matter sector satisfying the null energy condition.

While the NEC is a condition on the matter, namely $T_{MN}\xi^M\xi^N\ge0,$ with $\xi$ a future-directed null vector, the Einstein equation allows this to be recast as a condition on the four-derivative corrected geometry, namely
\begin{equation}
    \qty[\hat R_{MN}+\alpha\qty(\hat R_{MPQR}\hat R_{N}^{\ \ PQR}-2\hat R^{PQ}\hat R_{MPNQ}-2\hat R_{MP}\hat R_N^{\ \ P}+\hat R \hat R_{MN})]\xi^M\xi^N\ge0.
\label{eq:GeneralNEC}
\end{equation}
The main result of this section is to show that the NEC \eqref{eq:GeneralNEC} implies the existence of a monotonic $c$-function from the UV to the IR in the background \eqref{Eq:Flowfg}.


\subsection{Domain wall flows}
Before discussing flows across dimensions, let us first review the case of flows within the same dimension \cite{Myers:2010tj,Myers:2010xs}, \textit{i.e.}, for which we have a metric of the form
\begin{equation}
    \dd{s}^2=e^{2f(z)}\qty(\eta_{\mu\nu}\dd x^\mu\dd x^\nu+\dd z^2).
\end{equation}
Pure AdS corresponds to the solution $f(z)=\log(L/z)$, with $L$ being the AdS radius. Then, in these coordinates, $z=0$ corresponds to the UV, and $z=\infty$ corresponds to the IR. Thus, we have a gravity solution that is a domain wall interpolating between two AdS$_{D+1}$ regions; the corresponding field theory interpretation is that of an RG flow \cite{Girardello:1998pd,Freedman:1999gp}. One can calculate the curvature tensor components
\begin{align}
    \hat R_{\mu\nu\rho\sigma}&=-e^{-2f}(f')^2\qty(\eta_{\mu\rho}\eta_{\nu\sigma}-\eta_{\mu\sigma}\eta_{\nu\rho}),&\hat R_{\mu z\nu z}&=-e^{2f}f''\eta_{\mu\nu},\nonumber\\
    \hat R_{\mu\nu}&=-\qty[f''+(D-1)(f')^2]\eta_{\mu\nu},&\hat R_{zz}&=-Df''.
\end{align}
Choosing a null vector $\xi=\partial_t\pm\partial_z$, the NEC with Gauss-Bonnet corrections is then simply expressed as \cite{Myers:2010tj,Myers:2010xs}
\begin{equation}
    (D-1)\qty(e^{-f})''\qty(1-2\alpha(D-2)(D-3) e^{-2f} (f')^2)\ge0.
\label{eq:dwNEC}
\end{equation}
Note that this will be the only non-trivial NEC due to the planar symmetry of the domain wall.

We now consider flows to the IR.  In the IR, the $A$-type central charge may be computed via the methods of \cite{Henningson:1998gx,Henningson:1998ey} to be \cite{Myers:2010jv}
\begin{equation}
    a_\text{IR}=\frac{L_\text{IR}^{D-1}}{G_N}\qty(1-2(D-1)(D-2)\frac{\alpha}{L_\text{IR}^2}),
\end{equation}
where $G_N$ is the $(D+1)$-dimensional Newton's constant.  In order to obtain a $c$-function, note that in the IR, we expect that $e^f\sim L_\text{IR}/z$, so that $(e^{-f})'\sim 1/L_\text{IR}$.  Replacing $L_{\mathrm{IR}}$ by an effective AdS radius
\begin{equation}
    L_\text{eff}(z)=\frac{1}{(e^{-f})'},
\end{equation}
that interpolates between $L_\text{UV}$ and $L_\text{IR}$ then leads to a natural ansatz for an unnormalized $c$-function
\begin{equation}
    c(z)=\frac{1}{G_N\qty(\qty(e^{-f})')^{D-1}}\qty(1-2\alpha(D-1)(D-2)((e^{-f})')^2).
\label{eq:dwcfnct}
\end{equation}
Taking a derivative with respect to $z$, one gets that
\begin{equation}
    c'(z)=-\frac{(D-1)\qty(e^{-f})''}{G_N\qty(\qty(e^{-f})')^{D}}\qty(1-2\alpha(D-2)(D-3)((e^{-f})')^2)\le 0,
\end{equation}
where the final step makes use of the null energy condition (\ref{eq:dwNEC}). So, there is a monotonically non-increasing flow of $c(z)$ from the UV to the IR. Moreover, one can check that this function $c(z)$ interpolates between the UV and IR central charges in the sense that
\begin{equation}
    c(z=\infty)=a_\text{IR},\qquad c(z=0)=a_\text{UV}.
\end{equation}
Here $a_\text{IR}$ and $a_\text{UV}$ are the $A$-type central charges in the IR and UV, respectively, where \begin{equation}
    a_\text{UV}=\frac{L_\text{UV}^{D-1}}{G_N}\qty(1-2(D-1)(D-2)\frac{\alpha}{L_\text{UV}^2}).
\end{equation}
This expression agrees with \cite{Myers:2010jv}.
\subsection{Two-derivative flows across dimensions}

We now turn to the case at hand, which is flows across dimensions.  Before considering the full case, we review the two-derivative case of flows across dimensions \cite{GonzalezLezcano:2022mcd}, {\it i.e.}, without higher-derivative corrections. For such flows, we use the full metric ansatz \eqref{Eq:Flowfg}, with corresponding Ricci tensor components
\begin{align}
    \hat R_\nu^\mu&=-e^{-2f}\qty[f''+f'\qty((d-1)f'+(D-d)g')]\delta_\nu^\mu,\nonumber\\
    \hat R^i_j&=e^{-2g}\tilde R^i_j-e^{-2f}\qty[g''+g'\qty((d-1)f'+(D-d)g')]\delta^i_j,\nonumber\\
    \hat R^z_z&=-e^{-2f}\qty[df''+(D-d)\qty(g''+g'(g'-f'))].
\end{align}
Note that because we are assuming the AdS$_{d+1}$ in the IR to have flat slicings, the corresponding Ricci tensor $R_{\mu\nu}$ will vanish.

At the two-derivative level, the null energy condition is equivalent to $R_{MN}\xi^M\xi^N\ge0$.  Since the $D$-dimensional isometry is broken by the flow, we end up with two independent inequalities, which correspond to choosing null vectors along $t$-$z$ and $t$-$y$.  These conditions are, respectively,
\begin{subequations}
    \begin{align}
    \text{NEC1: }&-(d-1)\qty(f''-(f')^2)-(D-d)\qty(g''+g'(g'-2f'))&\ge0,\\
    \text{NEC2: }&(f'-g')'+(f'-g')\qty((d-1)f'+(D-d)g')+\kappa\frac{D-d-1}{\ell^2}e^{2f-2g}&\ge0.
\end{align}\label{eq:twoDerivNEC}%
\end{subequations}
NEC1 may be suggestively rewritten as
\begin{equation}
    \qty(e^{-\tilde f})''\ge \frac{(D-1)(D-d)}{(d-1)^2}e^{-\tilde f}(g')^2\ge0,
\label{eq:NEC1rew}
\end{equation}
where $\tilde f$ is an effective warp factor
\begin{equation}
    \tilde f(z)\equiv f(z)+\frac{D-d}{d-1}g(z).
\end{equation}
Likewise, NEC2 can be rearranged into the form
\begin{equation}
    \qty(e^{(d-1)f+(D-d)g}(f'-g'))'\ge -\kappa\frac{D-d-1}{\ell^2}e^{(d+1)f+(D-d-2)g}.
\label{eq:NEC2form}
\end{equation}
Note that the sign of the right-hand side term depends on the sign of the internal curvature, $\kappa$.  For $\kappa=-1$ or $\kappa=0$ the expression on the left-hand side is non-negative.  But for $\kappa=1$ the sign of this term is unconstrained.

As in the domain wall flow, we seek a $c$-function that flows to $a_{\mathrm{IR}}$ in the IR.  Before constructing such a function, we first recall the asymptotics of the flow.  Flowing from AdS$_{D+1}$ in the UV to AdS$_{d+1}$ in the IR, one expects
\begin{align}
    \text{UV }(z=0)&:\ \ (e^{-f})'=(e^{-g})'=\frac{1}{L_\text{UV}},\nonumber\\
    \text{IR }(z=\infty)&:\ \  (e^{-f})'=\frac{1}{L_\text{IR}},\ \ (e^{-g})'=0,
\label{eq:UVIRasy}
\end{align}
For AdS$_{d+1}$ in the IR, we have
\begin{equation}
    a_{\mathrm{IR}}=\frac{L_{\mathrm{IR}}^{d-1}}{G_{d+1}}=\frac{e^{(D-d)g(\infty)}\mathrm{Vol}(M_{D-d})L_{\mathrm{IR}}^{d-1}}{G_N}=\frac{\mathrm{Vol}(M_{D-d})}{G_N}\left(e^{\frac{D-d}{d-1}g(\infty)}L_{\mathrm{IR}}\right)^{d-1},
\end{equation}
where $G_N$ is the $(D+1)$-dimensional Newton's constant, and $G_{d+1}$ is obtained by a standard Kaluza-Klein reduction with internal space metric $\hat g_{ij}=e^{2g(z)}g_{ij}$.  Taking $L_{\mathrm{IR}}\sim1/(e^{-f})'$, it is then natural to write down an unnormalized local holographic $c$-function of the form
\begin{equation}
    c(z)=\frac{1}{((e^{-\tilde f})')^{d-1}}.
\label{eq:twoDerivCfnct}
\end{equation}
In particular, the effective warp factor $\tilde f$ gives the precise combination of internal volume and AdS radius needed to obtain the IR central charge.  As before, one can verify that this is monotonic along flows
\begin{equation}
    c'(z)=-\frac{(d-1)(e^{-\tilde f})''}{((e^{-\tilde f})')^{d}}\le0,
\end{equation}
since $(e^{-\tilde f})''\ge0$ from NEC1, (\ref{eq:NEC1rew}).

As before, we may define an effective AdS radius
\begin{equation}
    L_\text{eff}(z)=\frac{1}{(e^{-\tilde f})'},
\end{equation}
such that $L_\text{eff}'(z)\le 0$. The $c$-function is then simply
\begin{equation}
    c(z)=\frac{L_\text{eff}(z)^{d-1}}{G_N}.
\label{eq:2dcfad}
\end{equation}
Note, however, that $L_{\mathrm{eff}}$ defined here does not correspond directly to the radius of AdS$_{d+1}$; instead, it is the AdS radius modified by the internal volume to account for the dimensionally reduced Newton's constant.  Moreover, unlike the domain wall flow case, this $c(z)$ diverges in the UV. This has a natural explanation: The $D$-dimensional theory appears to have an infinite number of $d$-dimensional degrees of freedom; \textit{i.e.}, as we approach the UV, the compact dimensions become large and we can no longer ignore the infinite KK tower of states. As it turns out, the divergence still encodes the UV central charge; we will return to this point in Section~\ref{sec:UVdiv}.

Note that NEC2, given in the form (\ref{eq:NEC2form}), also leads to a monotonicity of sorts.  In particular, as long as the internal curvature is non-positive, $\kappa\le0$, the quantity
\begin{equation}
    \mathcal{C}(z)=e^{(d-1)\tilde f}(f'-g'),
\end{equation}
satisfies the inequality
\begin{equation}
    \mathcal{C}'(z)\ge 0\qquad(\mbox{provided }\kappa\le0).
\end{equation}
Hence $\mathcal C(z)$ is a monotonically non-decreasing function towards the IR.  Moreover, making use of the IR behavior (\ref{eq:UVIRasy}), we see that
\begin{equation}
    \mathcal{C}\overset{z\to\infty}{\sim}-\frac{e^{(D-d)g_\text{IR}}}{L_\text{IR}}\qty(\frac{L_\text{IR}}{z-z_0})^d<0,
\end{equation}
where $z_0$ is a constant offset. Since this is negative in the IR and the flow is non-decreasing towards the IR, we see that $\mathcal C(z)$ is negative along the entire flow.  Thus it must be the case that $f'<g'$ along the entire flow, so long as $\kappa\le 0$. It would be interesting to explore the implications of this condition as a second constraint on the flow (for $\kappa\le 0$).

\subsection{A concrete example: $\mathrm{AdS}_5\to\mathrm{AdS}_3$}

We now turn to four-derivative flows across dimensions where we include the Gauss-Bonnet coupling.  Since the expressions are somewhat lengthy for arbitrary UV and IR dimensions, $D$ and $d$, we start with a simple example of flowing from AdS$_5$ to AdS$_3\times T^2$ to motivate our procedure.  We thus take a metric of the form
\begin{equation}
    \dd{s}^2=e^{2f(z)}(-\dd{t}^2+\dd{x}^2+\dd{z}^2)+e^{2g(z)}(\dd{y}^2+\dd{w}^2).
\end{equation}
There are various explicit solutions in this class, including supergravity solutions describing flows of ${\cal N}=4$ SYM on $T^2$ \cite{Donos:2011pn,Almuhairi:2011ws,Benini:2015bwz,Uhlemann:2021itz}. The resulting NECs, in the presence of a Gauss-Bonnet term in the action, are obtained by orienting the null vectors along the $t$-$z$ and $t$-$y$ directions, respectively,
\begin{subequations}
    \begin{align}
        \text{NEC1: }&-\left(f''(z)-f'(z)^2\right)-2\left(g''(z)+g'(z)\left(g'(z)-2f'(z)\right)\right)\nonumber\\
        &+4\alpha e^{-2f(z)}g'(z)\Bigl[g'(z)\left(f''(z)-f'(z)^2\right)\nonumber\\
        &\kern3.0cm+2f'(z)\left(g''(z)+g'(z)\left(g'(z)-2f'(z)\right)\right)\Bigr]\ge0,\\
\text{NEC2: }&\left(f'(z)-g'(z)\right)'+\left(f'(z)-g'(z)\right)\left(f'(z)+2g'(z)\right)\nonumber\\
        &+4\alpha e^{-2f(z)}\Bigl[-\left(f'(z)g'(z)\left(f'(z)-g'(z)\right)\right)'\nonumber\\
        &\kern2.2cm+f'(z)g'(z)\left(f'(z)-g'(z)\right)\left(f'(z)-2g'(z)\right)\Bigr]\ge0.
    \end{align}
\label{eq:NEC12AdS53}%
\end{subequations}
These generalize the two-derivative NECs, \eqref{eq:twoDerivNEC}, in the case where $D=4$, $d=2$, and $\kappa=0$.  As a sanity check, note that NEC2 becomes trivial in the domain wall limit, $g=f$,  while NEC1 reduces to 
\begin{equation}
    -3\left(f''(z)-f'(z)^2\right) \left(1-4 \alpha e^{-2f} f'(z)^2\right)\ge0,
\end{equation}
in agreement with the domain wall flow case \eqref{eq:dwNEC}.

In order to obtain a $c$-function, note that, following (\ref{eq:NEC1rew}), the two-derivative NEC1 can be written as
\begin{equation}
    \left(e^{-\tilde f}\right)''\ge 6e^{-\tilde f}(g')^2,
\label{eq:AdS53two}
\end{equation}
where $\tilde f=f+2g$.  Examination of (\ref{eq:NEC12AdS53}) indicates that, in the presence of the Gauss-Bonnet correction, this can be extended to
\begin{equation}
    \qty(\qty(e^{-\tilde f})'+4\alpha e^{-\tilde f-2f}f'g'{}^2)'\ge 6e^{-\tilde f}(g')^2\qty[1+\frac{4}{3}\alpha\, e^{-2f}\qty((f'-g')^2-g'{}^2)].
\label{eq:AdS53four}
\end{equation}
Since $(g')^2$ is non-negative, the right-hand side of the two-derivative expression, (\ref{eq:AdS53two}), is non-negative.  However, the same cannot be said for (\ref{eq:AdS53four}), as the term inside the square brackets can in principle have either sign.  However, as long as we work perturbatively in the higher derivative coupling, $\alpha$, this still leads to a monotonic expression for the left-hand side.

Validity of the perturbative expansion requires that the four-derivative Gauss-Bonnet term be parametrically smaller than the leading-order two derivative term, $\alpha R^2\ll R$, or $\alpha/\ell^2\ll1$ where $\ell$ is some radius of curvature of the background.  For the particular higher derivative flow at hand, (\ref{eq:AdS53four}), this corresponds to the two conditions
\begin{equation}
    \alpha e^{-2f}f'^2\ll1,\qquad \alpha e^{-2f}g'^2\ll1,
\label{eq:pertcond}
\end{equation}
in which case, we can conclude that
\begin{equation}
    \qty(\qty(e^{-\tilde f})'+4\alpha e^{-\tilde f-2f}f'g'{}^2)'\ge0.
\label{eq:Lefficf}
\end{equation}
For a flow interpolating between the asymptotic regions given in (\ref{eq:UVIRasy}), we note that $e^{-f}\sim e^{-g}\sim z/L_{\mathrm{UV}}$ in the UV region, $z\to0$.  Then the perturbative conditions, (\ref{eq:pertcond}), translate into
\begin{equation}
   \frac{\alpha}{L_\text{UV}^2} \ll 1.
\end{equation}
While this changes along the flow, the first condition in (\ref{eq:pertcond}) corresponds to $\alpha/L_{\mathrm{eff}}^2\ll1$ where $L_{\mathrm{eff}}$ is an effective AdS radius interpolating between $L_{\mathrm{UV}}$ and $L_{\mathrm{IR}}$.  For the second condition in (\ref{eq:pertcond}), note that $e^{-g}$ interpolates from $z/L_{\mathrm{UV}}$ to a constant in the IR.  Hence $g'^2$ flows from $1/z^2$ to $0$.  Since $e^{-2f}$ scales as $z^2$ throughout the flow, the combination $e^{-2f}g'^2$ then interpolates between the values
\begin{eqnarray}
      e^{-2f} (g')^2&=&
    \begin{cases}
        0&z\to \infty\text{ (IR)}\\
       \frac{1}{L_\text{UV}^2}& z\to 0\ \text{ (UV)}
    \end{cases}.
\end{eqnarray}
The requirement that we are working perturbatively in $\alpha$ is, therefore,
\begin{equation}
   \left\{\frac{\alpha}{L_\text{UV}^2}, \frac{\alpha}{L_\text{IR}^2}\right\}\ll 1.\,
\end{equation}
at the endpoints of the flow, along with the assumption that the four-derivative corrections remain parametrically small along the flow.  This is equivalent to requiring that our EFT description remains valid.

With this in mind, one may generalize the two-derivative $c$-function defined in (\ref{eq:2dcfad}) by taking
\begin{equation}
    c(z)=\frac{L_{\mathrm{eff}}(z)}{G_N},
\label{eq:exCfnct}
\end{equation}
where now
\begin{equation}
    L_\text{eff}(z)=\frac{1}{\qty(e^{-\tilde{f}})'+4\alpha e^{-\tilde{f}-2f}f'(g')^2},
\label{eq:newLeff}
\end{equation}
is the Gauss-Bonnet corrected effective AdS$_3$ radius, including the internal volume factor.  From (\ref{eq:Lefficf}), we immediately see that $L_{\mathrm{eff}}'(z)\le0$, so that $c'(z)\le0$.  As a result, $c(z)$ is monotonic non-increasing along the flow to the IR, so long as we work perturbatively in $\alpha$. Note that this $c$-function reduces to the two-derivative $c$-function in the IR where $g'=0$; this is a consequence of the fact that the Gauss-Bonnet term is trivial for AdS$_3$  and we might expect otherwise in general dimensions.

Turning our attention to NEC2, we see that it can be written as a total derivative
\begin{equation}
    \qty(e^{\tilde f}(f'-g')\qty(1-4\alpha e^{-2f}f'g'))'\ge0,
\end{equation}
which generalizes (\ref{eq:NEC2form}) for the case $\kappa=0$.  If we commit to being perturbatively small in $\alpha$, (\ref{eq:pertcond}), then the interpretation of NEC2 is almost identical to the two-derivative case \cite{GonzalezLezcano:2022mcd} as summarized above. We can define a function
\begin{equation}
    \mathcal{C}(z)=e^{\tilde f}(f'-g')\qty(1-4\alpha e^{-2f}f'g'),
\label{eq:calCfunc}
\end{equation}
such that $\mathcal{C}'(z)\ge 0$. In the IR, we have that
\begin{equation}
    \mathcal{C}(z)\overset{z\to\infty}{\sim}-\frac{e^{2g_\text{IR}}}{L_\text{IR}}\qty(\frac{L_\text{IR}}{z-z_0})^2<0,
\end{equation}
where $z_0$ is a constant. Since this is a negative in the IR and monotonically non-decreasing with respect to $z$, it must be the case that it is also negative in the UV. Hence, we have that $f'<g'$ along the entire flow.

\subsection{Gauss-Bonnet flows in arbitrary dimensions}
\label{sec:GBad}

Having examined flows from AdS$_5$ to AdS$_3$ we now turn to the general case of Gauss-Bonnet corrected flows in arbitrary dimensions.  Consider a flow from AdS$_{D+1}$ to AdS$_{d+1}$.  As noted above, we consider two conditions arising from the null energy condition, which we denoted NEC1 and NEC2.  Our main interest is in the $c$-function arising from NEC1, although NEC2 will also give rise to a monotonic function from the case $\kappa\le0$.

Making use of the curvature tensor components summarized in Appendix~\ref{app:Riemann}, we find the $t$-$z$ NEC1 to be given by
\begin{align}
    &-(d-1)(f''-(f')^2)-(D-d)(g''+g'(g'-2f'))\nonumber\\
    &+2\alpha e^{-2f}\Bigl[(d-1)(d-2)(f')^2\left((d-3)(f''-f'^2)+(D-d)(g''+g'(g'-2f'))\right)\nn\\
    &+2(D-d)(d-1)f'g'\left((d-2)(f''-f'^2)+(D-d-1)(g''+g'(g'-2f'))\right)\nn\\
    &+(D-d)(D-d-1)g'^2\left((d-1)(f''-f'^2)+(D-d-2)(g''+g'(g'-2f'))\right)\Bigr]\nonumber\\
    &-2\alpha \frac{\kappa}{\ell^2}(D-d)(D-d-1) e^{-2g}\Bigl[(d-1)(f''-f'^2)+(D-d-2)(g''+g'(g'-2f'))\Bigr]\nn\\
    &\ge0.
\label{eq:NEC1}
\end{align}
One can check that upon setting $f=g$ and $\kappa=0$, we get
\begin{align}
    (D-1)\qty((f')^2-f'')\qty(1-2\alpha(D-2)(D-3)e^{-2f} (f')^2)\ge0,
\end{align}
which perfectly agrees with the domain wall flow NEC \eqref{eq:dwNEC}. As a sanity check, one can also see that setting $\alpha=0$ recovers the correct two-derivative result \eqref{eq:twoDerivNEC}.

We now seek a holographic $c$-function which could, {\it a priori}, be any arbitrary function
\begin{equation}
    c(z)=c(f,f',f'',...,g,g',g'',...;z).
\end{equation}
However, inspired by the form of the two-derivative $c$-function \eqref{eq:twoDerivCfnct} and the $\mathrm{AdS}_5\to\mathrm{AdS}_3$ case, namely (\ref{eq:exCfnct}) and (\ref{eq:newLeff}), a natural generalization would to be
\begin{equation}
c(z)=\frac{1}{((e^{-\tilde{f}})')^{d-1}}\to\frac{1+\mathcal O(\alpha)}{((e^{-\tilde{f}})'+\mathcal O(\alpha))^{d-1}},
\end{equation}
where the $\mathcal O(\alpha)$ terms are made from combinations of $f'$, $g'$ and $\kappa$.  Hence, we propose a candidate $c$-function
\begin{align}
    c(z)&=\frac{L_{\mathrm{eff}}(z)^{d-1}}{G_N}\Bigl[1+\alpha \qty(e^{-2f}\qty(a_1 (f')^2+a_2 f'g'+a_3 (g')^2)+b_1 e^{-2g}\frac{\kappa}{\ell^2})\Bigr],\nn\\
    L_{\mathrm{eff}}(z)&=\Bigl[\qty(e^{-\tilde{f}})'+\alpha\, e^{-\tilde{f}}\Bigl(e^{-2f}\qty(a_4 (f')^3+a_5 (f')^2g'+a_6 f'(g')^2+a_7 (g')^3)\nn\\
    &\kern3.5cm+e^{-2g}\frac{\kappa}{\ell^2}\qty(b_2f'+b_3g')\Bigr)\Bigr]^{-1},
\label{eq:LHcfnct}
\end{align}
for some choice of real coefficients $\{a_i, b_j\}$.  The structure of the central charge contains various occurring products of derivatives of the functions $f$ and $g$. Note that we are interested in comparing to NEC1 to obtain monotonicity, and hence have avoided any terms with $f''$ or $g''$  in c(z) as these would lead to $f'''$ and $g'''$ terms in $c'(z)$, as well as $(f'')^2$ and $(g'')^2$ terms.

We now fix the coefficients $\{a_i,b_j\}$ by demanding monotonicity of $c(z)$, namely $c'(z)\le0$ under the assumptions of NEC1 and perturbative control.  To do so, we compute $c'(z)$ and adjust the coefficients to match the $f''$ and $g''$ terms with the structure of NEC1, namely (\ref{eq:NEC1}).  The expression for $c'(z)$ is not particularly illuminating, but it is given in Appendix~\ref{app:c'(z)} for completeness.  Comparing $c'(z)$ to NEC1, we see that for the particular choice of coefficients
\begin{align}
    a_1&=-2(d-1)(d-2),\nonumber\\
    a_2&=-4(D-d)(d-2),\nonumber\\
    a_3&=\mbox{arbitrary},\nn\\
    a_4&=0,\nonumber\\
    a_5&=\frac{4(D-d)(d-2)}{(d-1)},\nonumber\\
    a_6&=2(D-d)\qty(\frac{(2d-3)(D-d)}{(d-1)^2}-1)-\frac{a_3}{d-1},\nonumber\\
    a_7&=\mbox{arbitrary},\nn\\
    b_1&=2(D-d)(D-d-1),\nonumber\\
    b_2&=0,\nonumber\\
    b_3&=\frac{4(D-d)(D-d-1)}{(d-1)},
\label{eq:cfnctCoeffs}
\end{align}
we get monotonicity of the $c$-function, in the sense that
\begin{align}
    c'(z)= - \frac{e^{-\tilde{f}}(L_{\mathrm{eff}})^d}{G_N}\qty[\text{NEC1}+\frac{(D-1)(D-d)}{d-1}(g')^2\qty(1+\mathcal{O}(\alpha))]\le 0,\label{eq:ineq}
\end{align}
where we have made crucial use of the fact that we are working perturbatively in $\alpha$. Here, $\mathcal{O}(\alpha)$ denotes only terms which remain under perturbative control throughout the flow in the sense of (\ref{eq:pertcond}).  Notice also that \eqref{eq:LHcfnct} reduces to \eqref{eq:exCfnct} upon setting $D=4$, $d=2$, and $\kappa=0$, provided we take $a_3=a_7=0$.

Note that the two coefficients $a_3$ and $a_7$ are left undetermined; $a_7$ will be the coefficient of a term proportional to $(g')^2$, and so can never matter within the context of our analysis, and shifting $a_6$ is equivalent to a shift in $a_7$ and a shift in $a_3$ since we are working perturbatively in $\alpha$.  This freedom in choosing $a_3$ and $a_7$ in principle yields a family of $c$-functions that all flow to the same IR central charge as $g'\to0$ in the IR.  However, the UV behavior will be affected, and below we will find a preferred combination of these coefficients.  If one were to relax the above condition \eqref{eq:ineq} by replacing NEC1 with NEC1$\times(1+\mathcal{O}(\alpha))$, then it would become apparent that, due to the perturbative nature of our analysis, there are five free parameters rather than the na\"ively apparent two.  Intuitively, this is equivalent to the freedom of perturbatively combining the numerator of \eqref{eq:LHcfnct} with its denominator.  It is convenient, however, to keep these terms separate when taking the IR limit, as we will see in Section.~\ref{sec:IR}.

We may also consider NEC2, which can be arranged in the form
\begin{align}
    &\Big\{e^{(d-1)\tilde{f}}\Big[(f'-g')+2\alpha\Bigl(e^{-2f}(f'-g')\Big(-(d-1)(d-2)(f')^2\nn\\
    &\kern3cm-2(d-1)(D-d-1)f'g'-(D-d-1)(D-d-2)(g')^2\Big)\nonumber\\
    &\kern3cm+e^{-2g}\frac{\kappa}{\ell^2}\Big((D-4d-1)f'+(-5+(8-3D)D+d(-8+6D))g'\Big)\Bigr)\Big]\Big\}'\nonumber\\
    &\ge-\fft\kappa{\ell^2}e^{(d-1)\tilde{f}+2f-2g}\Bigl[D-d-1+2\alpha\Bigl(e^{-2f}\Bigl(3d(d+1)f''+3d^2g''\nn\\
    &\kern4cm+(D+2d+2)d(d-1)(f')^2-d(D(3-2D)+dD+4d^2+2)f'g'\nonumber\\
    &\kern4cm+(D-d-2)(D^2-2dD-4D-2(d-2)+3)(g')^2\Bigr)\nonumber\\
    &\kern4cm+\fft\kappa{2\ell^2}(D-d-1)(7+2d^2-4d(D-2)+2D(D-4))\Bigr)\Bigr].
\end{align}
One may check that setting $g=f$ and $\kappa=0$ makes the left- and right-hand sides of this inequality identically zero, as it should.  This suggests that we define
\begin{align}
    \mathcal{C}(z)=&e^{(d-1)\tilde{f}}\Big[(f'-g')+2\alpha\Bigl(e^{-2f}(f'-g')\Big(-(d-1)(d-2)(f')^2\nn\\
    &\kern2cm-2(d-1)(D-d-1)f'g'-(D-d-1)(D-d-2)(g')^2\Big)\nonumber\\
    &\kern2cm+e^{-2g}\frac{\kappa}{\ell^2}\Big((D-4d-1)f'+(-5+(8-3D)D+d(-8+6D))g'\Big)\Bigr)\Big],
\end{align}
analogous to (\ref{eq:calCfunc}) for the case of $\mathrm{AdS}_5\to\mathrm{AdS}_3$.  NEC2 is then the statement that
\begin{equation}
    \mathcal C'(z)\ge-\frac\kappa{\ell^2}e^{(d-1)\tilde{f}+2f-2g}(D-d-1)\qty(1+\mathcal{O}(\alpha)).
\end{equation}
Then $\mathcal{C}'(z)>0$ for $\kappa=-1$ and $\mathcal{C}'(z)\ge0$ for $\kappa=0$, so long as the $\mathcal O(\alpha)$ corrections are parametrically small. Then in the IR, we find that
\begin{equation}
    \mathcal{C}(z)\overset{z\to\infty}{\sim}-\frac{e^{(D-d)g_\text{IR}}}{L_\text{IR}}\qty(\frac{L_\text{IR}}{z-z_0})^d\qty(1-\frac{2\alpha(d-1)(d-2)}{L_\text{IR}^2}+\alpha e^{-2g_\text{IR}}\frac{\kappa}{\ell^2}(D-4d-1))<0.
\end{equation}
This should hold so long as $\alpha/L_\text{IR}^2\ll 1$ and $\alpha/\ell^2\ll 1$. Then since $\mathcal{C}(z)$ is negative in the IR and non-decreasing as $z$ increases, we conclude that it must always be negative. This imposes a constraint
\begin{align}
    0>\,&(f'-g')+2\alpha\Bigl(e^{-2f}(f'-g')\Big(-(d-1)(d-2)(f')^2\nn\\
    &\kern2cm-2(d-1)(D-d-1)f'g'-(D-d-1)(D-d-2)(g')^2\Big)\nonumber\\
    &\kern2cm+e^{-2g}\frac{\kappa}{\ell^2}\Big((D-4d-1)f'+(-5+(8-3D)D+d(-8+6D))g'\Big)\Bigr).
\end{align}
Heuristically, this provides an additional constraint to the $c$-function considerations.

\subsubsection{Generic Einstein internal manifolds}\label{sec:generalization}
If we relax the condition that the internal manifold is maximally symmetric and instead only require it to be an Einstein manifold with Ricci curvature
\begin{equation}
    \tilde{R}_{ij}=k g_{ij},
\end{equation}
with $g$ the metric on the internal space, then the null energy condition will be, in general, more complicated. In particular, we no longer know the internal Riemann tensor $\tilde{R}_{ijkl}$; however, the only component of the full Riemann tensor $\hat{R}_{MNPQ}$ that contains the uncontracted internal Riemann tensor is $\hat{R}_{ijkl}$ with all internal indices, which will not affect the $t$-$z$ null energy condition NEC1. Then all the previous arguments hold for the monotonicity of the $c$-function if we replace
\begin{equation}
    \frac{\kappa}{\ell^2}\to\frac{k}{D-d-1}.
\end{equation}
However, the same is not true of NEC2 since it would be dependent on $\tilde{R}_{ijkl}$ in general.

\subsection{Changing coordinates}

While we have parameterized the bulk metric according to (\ref{Eq:Flowfg}), in some situations, it is convenient for one to work in a different gauge,
\begin{equation}
    \dd{s}^2=e^{2A(r)}(-\dd{t}^2+\dd{\vec{x}}^2)+\dd{r}^2+e^{2B(r)}\dd{s}_{M_{D-d}}^2.\label{eq:AltCoords}
\end{equation}
The $c$-functions we have defined do not depend on the choice of coordinates.  Nevertheless, we present NEC1 and the corresponding $c(r)$ function in Appendix~\ref{app:altCoords1} in case such expressions prove useful.

\section{Fixed point limits of the $c$-function}\label{sec:limits}

In Section~\ref{sec:LH}, we have constructed a monotonic $c$-function, (\ref{eq:LHcfnct}) with coefficients given in (\ref{eq:cfnctCoeffs}), for Einstein-Gauss-Bonnet flows across dimensions.  This $c$-function is a natural extension of its two-derivative counterpart, (\ref{eq:twoDerivCfnct}), as well as the higher-derivative $c$-function, (\ref{eq:dwcfnct}), for flows in the same dimension.  To better understand the physics of this NEC-motivated $c$-function, we now consider its UV and IR limits and compare them to the expected central charges at the endpoints of the flow.

\subsection{The IR limit}\label{sec:IR}
One important reason for considering higher derivatives is that they break the degeneracy between the $a$-type and $c$-type central charges. Focusing on $d=4$ for the moment, the Gauss-Bonnet correction splits the two central charges in the IR \cite{Myers:2010xs}
\begin{subequations}
    \begin{align}
        a&=\frac{L_\text{IR}^3}{G_5}\qty(1-\frac{12\alpha}{L_\text{IR}^2}),\\
        c&=\frac{L_\text{IR}^3}{G_5}\qty(1-\frac{4\alpha}{L_\text{IR}^2}).
    \end{align}
\label{eq:4DIRcc}%
\end{subequations}
If we do not include higher derivatives, then these are the same. In particular, a holographic two-derivative flow cannot tell whether $a$ or $c$ is flowing monotonically.  However, for the four-derivative central charge, (\ref{eq:LHcfnct}), we find the IR limit
\begin{equation}
    c(z)\overset{z\to\infty}{\sim}\fft{L_\text{IR}^{d-1}}{G_{d+1}}\qty(1-\frac{2\alpha(d-1)(d-2)}{L_\text{IR}^2}),
\label{eq:IRlimit}
\end{equation}
where, as we show below, the $(d+1)$-dimensional Newton's constant is
\begin{equation}
\fft1{G_{d+1}}=
\frac{e^{(D-d)g_\text{IR}}\mathrm{Vol}(M_{D-d})}{G_N}\left(1+2\alpha(D-d)(D-d-1)\frac{\kappa}{\ell^2}e^{-2g_\text{IR}}\right).
\end{equation}

While the above holds for arbitrary $D$ and $d$, we can compare with the four-dimensional IR central charges, (\ref{eq:4DIRcc}), by setting $d=4$.  In this case, we get that
\begin{equation}
    c(z)\overset{z\to\infty}{\sim}\fft{L_\text{IR}^3}{G_5}\qty(1-\frac{12\alpha}{L_\text{IR}^2}),
\end{equation}
with
\begin{equation}
    \fft1{G_5}=
\frac{e^{(D-4)g_\text{IR}}}{G_N}\left(1+2\alpha(D-4)(D-5)\frac{\kappa}{\ell^2}e^{-2g_\text{IR}}\right).
\label{eq:GNren}
\end{equation}
This then clearly reduces to the $a$ central charge as expected from the $a$-theorem, and notably is not the $c$ central charge.

More generally, we expect the $A$-type central charge in the IR to be \cite{Imbimbo:1999bj,Hung:2011xb}
\begin{equation}
    A=\frac{L_\text{IR}^{d-1}}{G_{d+1}}\qty(1-2(d-1)(d-2)\frac{\alpha}{L_\text{IR}^2}),
\end{equation}
which precisely matches the IR limit, \eqref{eq:IRlimit}. Hence, the $c$-function originating from NEC1 pertains to the monotonicity of what becomes the $A$-type central charge in the IR. Note that we have not imposed this fact; simply solving for the allowed parameters $\{a_i,b_i\}$ in (\ref{eq:LHcfnct}) that give monotonicity from NEC1 has demanded that the IR limit be unambiguously the $A$-type central charge.

We now return to the relation between the $(D+1)$-dimensional and $(d+1)$-dimensional Newton's constant, (\ref{eq:GNren}).  The lower-dimensional Newton's constant is obtained from the compactification of the gravitational part of the Lagrangian (\ref{eq:Lgrav}). In the IR, the spacetime is AdS$_{d+1}\times M_{D-d}$.  Furthermore, in this limit, the Gauss-Bonnet term, (\ref{eq:chi4}), splits as
\begin{equation}
    \chi_4\to\chi_4+\tilde\chi_4+2R\tilde R,
\end{equation}
where $\tilde{R}$ is the internal Ricci scalar and $\tilde{\chi}_4$ is the internal Gauss-Bonnet term
\begin{align}
    \tilde{R}&=\frac{\kappa}{\ell^2}(D-d)(D-d-1),\nn\\
    \tilde{\chi}_4&=\frac{1}{\ell^4}(D-d)(D-d-1)(D-d-2)(D-d-3).
\label{eq:tRtchi}
\end{align}
Then the gravitational action reduces as
\begin{align}
    S&=\fft1{16\pi G_N}\int d^{D+1}x\sqrt{-g}\qty[R+\fft{D(D-1)}{L^2}+\alpha\chi_4]\nn\\
    &=\fft1{16\pi G_N}\int d^{d+1}x\sqrt{-g_{d+1}}\int d^{D-d}ye^{(D-d)g_{IR}}\sqrt{g_{D-d}}\nn\\
    &\kern3cm\times\biggl[\left(1+2\alpha\tilde{R}\right)R+\fft{D(D-1)}{L^2}+\alpha\chi_4+\tilde R+\alpha\tilde{\chi}_4\biggr].
\end{align}
Integrating out the internal coordinates gives the $(d+1)$-dimensional Newton's constant
\begin{equation}
    \fft1{G_{d+1}}=\fft{e^{(D-d)g_{IR}}\mathrm{Vol}(M_{D-d})}{G_N}\left(1+2\alpha\tilde{R}\right),
\end{equation}
where $\tilde R$ is given in (\ref{eq:tRtchi}).  Making this substitution for $\tilde R$ yields the above expression in (\ref{eq:GNren}).

\subsection{The UV divergence}\label{sec:UVdiv}

We now turn to the UV behavior of the $c$-function, (\ref{eq:LHcfnct}).  As is often the case when defining $c$-functions in flows across dimensions, this function diverges in the UV.  This, of course, is not surprising since we will see an infinite number of lower-dimensional degrees of freedom in the UV. While (\ref{eq:LHcfnct}) does not interpolate between the UV and IR central charges, we can still ask whether its UV divergence can be related to the UV central charge.  To answer this question, we first look at the two-derivative case.

\subsubsection{Two-derivative case}
Ignoring higher-derivative corrections for the moment, for general $D$ to $d$ dimensional flows, the $c$-function is given by (\ref{eq:twoDerivCfnct}), which we write out as
\begin{equation}
    c(z) = e^{(d-1)f+(D-d)g}\left(-f' - \frac{D-d}{d-1}g'\right)^{-(d-1)}.
\end{equation}
In the UV we have $e^f\sim e^g\sim L_\text{UV}/z$, so
\begin{equation}
    c(z) \overset{z\to 0}{\sim} \left(\frac{L_\text{UV}}{z}\right)^{D-1}\left(\frac{d-1}{D-1}z\right)^{d-1} ~\propto \quad \frac{(L_\text{UV})^{D-1}}{z^{D-d}}.
\end{equation}
The numerator gives the unnormalized UV central charge. The denominator diverges with increasing energy scale, and the power is the number of compact dimensions.

\subsubsection{Gauss-Bonnet}
Now we consider what happens when we reintroduce the Gauss-Bonnet term. For the case of no internal curvature, $\kappa=0$, the $c$-function in (\ref{eq:LHcfnct}) reduces to
\begin{equation} \label{k=0 c func}
    c(z)=e^{(d-1)f+(D-d)g}\frac{1+\alpha e^{-2f}\qty(a_1 (f')^2+a_2f'g'+a_3(g')^2)}{\qty(-f'-\frac{D-d}{d-1}g'+\alpha e^{-2f}\qty(a_5(f')^2g'+a_6f'(g')^2+a_7(g')^3))^{d-1}}\;,
\end{equation}
which, in the UV limit, behaves as 
\begin{align}
    c(z)&\overset{z\to 0}{\sim}\left(\frac{L_\text{UV}}{z}\right)^{D-1}\frac{1+\frac{\alpha}{{L_\text{UV}^2}}\qty(a_1 +a_2+a_3)}{\qty(\frac{D-1}{d-1}\frac{1}{z} - \frac{\alpha}{L_\text{UV}^2}\qty(a_5+a_6+a_7)\frac{1}{z})^{d-1}} \nonumber\\
    &= \qty(\frac{d-1}{D-1})^{d-1}\frac{(L_\text{UV})^{D-1}}{z^{D-d}} \frac{1+\frac{\alpha}{{L_\text{UV}^2}}\qty(a_1 +a_2+a_3)}{\qty(1-\frac{\alpha}{{L_\text{UV}^2}}\frac{d-1}{D-1}\qty(a_5+a_6+a_7))^{d-1}} \label{4 deriv UV}\;.
\end{align}
Note that the curvature terms proportional to $\kappa/\ell^2$ do not affect the UV limit \eqref{4 deriv UV} since $e^{-2g}f'\sim z$ and $e^{-2g}\sim z^2$ in the UV, which is to be expected since intuitively the ``compact'' dimensions will appear large at very high energies. If we demand that $c(z) \propto a_\text{UV}/z^{D-d}$ in the UV limit, (\ref{4 deriv UV}) places constraints on sums of the $a$ coefficients. In particular, compared to the known result,
\begin{equation}
    a_\text{UV}=\frac{L_\text{UV}^{D-1}}{G_N}\qty(1-2(D-1)(D-2)\frac{\alpha}{L_\text{UV}^2}),
\end{equation}
we must satisfy
\begin{equation}
    a_1+a_2+a_3+\frac{(d-1)^2}{D-1}(a_5+a_6+a_7)=-2(D-1)(D-2),\label{eqn:UVacons}
\end{equation}
which corresponds to the requirement that
{\footnotesize
\begin{align}
    a_7=&\frac{D-1}{(d-1)^2}\Bigg[2\frac{(D-1)((D+1)(D-4)-3d(d-3))+(d-3)(d-1)(D-d)+(2d-3)(D-d)^2}{(D-1)}\nn\\
    &\qquad\qquad-\frac{D-d}{d-1}a_3\Bigg].
\label{eqn:a7req}
\end{align}}%
This provides an additional constraint on the coefficients \eqref{eq:cfnctCoeffs}, reducing the number of free coefficients from two to one. Given the discussion hitherto, we may always impose this additional requirement.

\section{Higher-derivative gravity and holographic entanglement entropy}\label{sec:EE}

In this section, we discuss the construction of monotonic $c$-functions from the perspective of holographic entanglement entropy. It is well-known that finding the entanglement entropy of a region in a holographic CFT is equivalent to finding a bulk surface minimizing some choice of functional; at the two-derivative level, this is just the Ryu-Takayanagi (RT) area functional \cite{Ryu:2006bv,Ryu:2006ef}. However, minimizing the area of the extremal surface is insufficient when higher derivatives are present; in particular, it has been argued  \cite{Faulkner:2013ana,Hung:2011xb,deBoer:2011wk,Bhattacharyya:2013jma} that, given a theory described by the Einstein-Gauss-Bonnet action
\begin{eqnarray}
    I_{\rm total}&=& I_{\rm bulk} + I_{\rm GH} + I_{\rm ct}, \nonumber \\
    I_{\rm bulk}&=& \int\dd[D+1]{x}\qty[R+\frac{D(D-1)}{L^2}+\alpha\chi_4], \nonumber \\
    I_{\rm GH}&=&\int\dd[D]{x}\qty[ K -2\alpha \qty(G_{ab}K^{ab}+\frac{1}{3}\qty(K^3-3KK_2 +2K_3))],
\end{eqnarray}
where $K_{ab}$ is the extrinsic curvature with trace $K$, $K_2=(K_{ab})^2$, and $K_3=K_{ab}K^{bc}K_{c}^{\ \, a}$, the RT functional must be replaced with the \ac{JM} functional\footnote{Note that for black holes, the Jacobson-Myers functional leads to the same result as Wald's entropy \cite{Wald:1993nt,Iyer:1994ys,Jacobson:1993vj}, but it is generically different.} \cite{Jacobson:1993xs}
\begin{equation}
    S_\text{JM}=\frac{1}{4G_N}\int_\Sigma \dd[d] x \sqrt{h}\left(1+2\alpha {\cal R}\right)+\frac{1}{2G_N}\int_{\partial\Sigma}\dd[d-1]{x}\sqrt{\tilde h}2\alpha{\cal K},
\end{equation}
where $\Sigma$ is the surface over which the functional is being minimized with boundary $\partial\Sigma$, $h$ is the determinant of the induced metric on $\Sigma$, $\tilde h$ is the induced metric (of the induced metric $h$) on $\partial\Sigma$, ${\cal R}$ is the scalar curvature of $\Sigma$, and $\cal K$ is the trace of the extrinsic curvature of the boundary $\partial\Sigma$. The term containing $\cal{K}$ may be viewed as a Gibbons-Hawking term that renders the variational principle well-defined. The equation  of motion that follows from the JM functional is
\begin{equation}
    {\cal K}+2\alpha ({\cal R}{\cal K}-2{\cal R}_{ij}{\cal K}^{ij})=0.
\end{equation}
We may then compute the holographic entanglement entropy of a region $A$ by minimizing this functional over all surfaces homologous to $A$
\begin{equation}
    S_\text{EE}=\min_{\Sigma\sim A}S_\text{JM}(\Sigma).
\end{equation}

The goal of this section is to construct monotonic $c$-functions from the entanglement entropy. For flows down to AdS$_3$, it is natural to obtain a monotonic $c$-function as the coefficient of the logarithmic term \cite{Casini:2006es}
\begin{equation} \label{eqn:c from S}
    c_\text{EE}=R\partial_R S_\text{EE},
\end{equation}
where $R$ is the radius of the entangling region. An analogous quantity that interpolates between free energies in AdS$_4$ flows is
\begin{equation}\label{eqn:F from S}
    c_\text{EE} = R\partial_R S_\text{EE} - S_\text{EE}\;,
\end{equation}
and its monotonicity can be proven using strong subadditivity on field-theoretic grounds \cite{Casini:2017vbe}. However, how to define similar quantities for AdS$_5$ and above is unclear. Strong subadditivity may be used to construct monotonic functions in higher dimensions, but they no longer interpolate between central charges at the fixed points.

\subsection{AdS$_{D+1}\to$ AdS$_3$}\label{sec:EEcfnct}
Looking at flows from AdS$_{D+1}$ down to AdS$_3$ is the most tractable. Equivalently, this may be viewed as a flow from CFT$_D$ to CFT$_2$. Generically, we have a metric of the form \eqref{Eq:Flowfg}, but we will further specify the metric to be
\begin{equation}
    \dd s^2=e^{2f(z)}\qty(-\dd t^2+\dd z^2+\dd r^2)+e^{2g(z)}\dd s_{M_{D-2}}^2,
\end{equation}
with asymptotic behavior
\begin{align}
    &z\to0:\ &f(z)\to\log\qty(L_\text{UV}/z),&\qquad g(z)\to\log\qty(L_\text{UV}/z),\nonumber\\
    &z\to\infty:\ &f(z)\to\log\qty(L_\text{IR}/z),&\qquad g(z)\to g_\text{IR}.
\end{align}
Our CFT$_D$ lives on $\mathbb{R}^{1,1}\times M_{D-2}$. We will consider entangling regions%
\footnote{For a more detailed discussion of choices of entangling regions in flows across dimensions, see \cite{GonzalezLezcano:2022mcd}.}
which wrap the internal $M_{D-2}$.  The induced metric on a constant time slice parameterized by a profile $r(z)$ is
\begin{equation}
    \dd{\sigma}^2=e^{2f}(1+r'(z)^2)\dd{z}^2+e^{2g}\dd{s}^2_{M_{D-2}}.
\end{equation}
We will assume boundary conditions
\begin{equation}
    r(0)=R,\qquad r(z_0)=0,\qquad r'(z_0)=-\infty, 
\end{equation}
where $R$ is the radius of the entangling region, and $z_0$ is the deepest point in the bulk that the minimal surface probes along the holographic radial coordinate, that is, the turning point of the surface in the mechanical analogy. In terms of this profile, the induced Ricci scalar is
\begin{align}
    \mathcal{R}=&(D-2)(D-3)\frac{\kappa}{\ell^2}e^{-2g}\nonumber\\
    &+(D-2)\frac{e^{-2f}}{\left(1+(r')^2\right)^2}\Bigg[\qty(1+(r')^2)\qty(2f'g'-(D-1)(g')^2-2g'')+2g'r'r''\Bigg],\label{eq:Rind}
\end{align}
which, after some integration by parts, leads to a JM functional 
\begin{align} \label{eqn:AdS3JM}
    S_\text{JM}&=\frac{2\text{Vol}\qty(M_{D-2})}{4G_N}\int\dd{z}e^{\tilde f}\qty[\sqrt{1+(r')^2}\qty(1+2\tilde\alpha\frac{\kappa}{\ell^2}e^{-2g})+2\tilde\alpha \frac{e^{-2f}(g')^2}{\sqrt{1+(r')^2}}],
\end{align}
where the rescaled Gauss-Bonnet coupling
\begin{equation}
    \tilde\alpha\equiv\alpha(D-2)(D-3),
\end{equation}
is introduced for convenience.  Here, we have ignored the boundary term from integrating by parts since it will automatically cancel with the Gibbons-Hawking term $\cal K$. Since this functional is independent of $r(z)$, $S_\text{JM}$ admits a first integral
\begin{equation}
    C=\frac{r' e^{\tilde f} \left(\left((r')^2+1\right) \left(1+2 \tilde\alpha \frac{\kappa}{\ell^2}e^{-2g} \right)-2\tilde\alpha e^{-2f} (g')^2\right)}{\left((r')^2+1\right)^{3/2}},
\end{equation}
which can be solved to give
\begin{equation}
    r'(z)=-\frac{\mathcal{F}}{\sqrt{1-\mathcal{F}^2+4\tilde\alpha\qty(\frac{\kappa}{\ell^2}e^{-2g}-e^{-2f}(g')^2(1-\mathcal{F}^2))}},\qquad \mathcal{F}(r)\equiv C e^{-\tilde f},\label{eq:rprime}
\end{equation}
or, equivalently,
\begin{equation}
    z'(r)=-\frac{\sqrt{1-\mathcal{F}^2+4\tilde\alpha\qty(\frac{\kappa}{\ell^2}e^{-2g}-e^{-2f}(g')^2(1-\mathcal{F}^2))}}{\mathcal{F}}.\label{eq:zprime}
\end{equation}
To fix the value of $C$, we note that we should have $r'(z)\to-\infty$ as $z\to z_0$; this then requires that
\begin{equation}
    C=e^{\tilde f_0}\qty(1+2\tilde\alpha\frac{\kappa}{\ell^2}e^{-2g_0})\quad\text{where}\quad \tilde f_0=\tilde f(z_0),\ g_0=g(z_0).
\end{equation}

Recall that we are interested in obtaining a monotonic $c$-function from the entanglement entropy following \eqref{eqn:c from S}, where $R$ is given by
\begin{align}
    R=&-\int_0^{z_0}\dd{z}r'(z).
\end{align}
The negative sign is because $r'(z)$ is negative in this parameterization. We know $r'(z)$ from the integral of motion, \eqref{eq:rprime}, and so we may write
\begin{align}
    R=&\int_0^{z_0}\dd{z}\frac{\mathcal{F}}{\sqrt{1-\mathcal{F}^2+4\tilde\alpha\qty(\frac{\kappa}{\ell^2}e^{-2g}-e^{-2f}(g')^2(1-\mathcal{F}^2))}}\nonumber\\
    =&\int_0^{z_0}\dd{z}\qty[\frac{\mathcal{F}}{\sqrt{1-\mathcal{F}^2+4\tilde\alpha\frac{\kappa}{\ell^2}e^{-2g}}}+2\tilde\alpha \frac{e^{-2f}(g')^2\mathcal{F}}{\sqrt{1-\mathcal{F}^2}}]+\mathcal{O}(\tilde\alpha^2).
\end{align}
Note that in the second line, we have partially expanded the denominator; this will be important to avoid triple derivatives from integrating by parts. As in the two-derivative case, the integrand is divergent at the cap-off point $z_0$, so it must be integrated by parts to give
\begin{align}
    R=&\lim_{\epsilon\to0}\int_{r_0}^{r_c}\dd{r}\left[\sqrt{1-\mathcal{F}^2+4\tilde\alpha\frac{\kappa}{\ell^2}e^{-2g}}\dv{}{r}\frac{1}{\mathcal{F}'+\frac{4\tilde\alpha}{\mathcal{F}}\frac{\kappa}{\ell^2}e^{-2g}g'}\right.\nn\\
    &\left.\qquad\qquad\qquad+2\tilde\alpha \sqrt{1-\mathcal{F}^2}\dv{}{r}\qty(\frac{e^{-2f}(g')^2}{\mathcal{F}'})\right]+2\tilde\alpha\lim_{\epsilon\to0}\frac{e^{-2f}(g')^2}{\mathcal{F}'}\Bigg\vert_{z=\epsilon}+\mathcal{O}(\tilde\alpha^2).\label{eq:R}
\end{align}
The profile $r(z)$ has been useful for obtaining an expression for $R$, but it will now be useful to phrase matters in terms of a profile $z(r)$ with boundary conditions
\begin{equation}
    z(0)=z_0,\ \ z'(0)=0,\ \ z(R)=0.
\end{equation}
The induced Ricci scalar for this profile is
\begin{align}
    \mathcal{R}=&(D-2)(D-3)\frac{\kappa}{\ell^2}e^{-2g}\nonumber\\
    &+(D-2)\frac{e^{-2f}(z')^2}{\left(1+(z')^2\right)^2}\Bigg[\qty(1+(z')^2)\qty(2f'g'-(D-1)(g')^2-2g'')+2g'z'z''\Bigg],
\end{align}
which leads to a Jacobson-Myers functional of the form
\begin{align}
    S_\text{JM}=&\frac{2\text{Vol}(M_{D-2})}{4G_N}\int_0^{R_c}\dd{r}e^{\tilde f(z(r))}\Bigg[\sqrt{1+(z')^2}\qty(1+2\tilde\alpha\frac{\kappa}{\ell^2}e^{-2g(z(r))})\nonumber\\
    &\kern5cm+2\tilde\alpha\frac{e^{-2f(z(r))}(g')^2(z')^2}{\sqrt{1+(z')^2}}\Bigg]-2\tilde\alpha e^{\tilde f-2f}g'\Big\vert_{r=R_c},\label{eq:SJM}
\end{align}
where $R_c$ is the cutoff value of $R$ such that $z(R_c)=\epsilon$. The boundary term, while divergent, is independent of $R$ and so will not cause us any issues. Since the UV boundary condition has the form $z_R(r=R_c)=\epsilon$, varying this boundary condition with respect to $R$ gives the relation
\begin{equation}
    z'\dv{R_c}{R}+\dv{z}{R}=0.\label{id1}
\end{equation}
Moreover, as $\epsilon\to 0$, $\dd R_c/\dd R\to 1$ at the boundary. One may now apply $R\partial_R$ to \eqref{eq:SJM} and impose the equations of motion. Using the relation \eqref{id1}, the monotonic central charge is then given by
\begin{equation}
    c_\text{EE}=\frac{2\text{Vol}(M_{D-2})}{4G_N}e^{\tilde f_0}\qty(1+2\tilde\alpha\frac{\kappa}{\ell^2}e^{-2g_0})R.\label{eq:EEcfnct}
\end{equation}
This generalizes the two-derivative case \cite{GonzalezLezcano:2022mcd} by simply using the four-derivative first integral $C$ rather than the two-derivative one $e^{\tilde{f}_0}$. Using the identity
\begin{equation}
    \pdv{\mathcal{F}}{z_0}=\qty(\tilde f'(z_0)-4\tilde\alpha\frac{\kappa}{\ell^2}e^{-2g_0}g'(z_0))\mathcal{F},
\end{equation}
substituting our expression for $R$ \eqref{eq:R} into $c_\text{EE}$, and differentiating with respect to $z_0$, we can show that
\begin{align}
    \dv{c_\text{EE}}{z_0}=&\frac{2\text{Vol}\qty(M_{D-2})}{4G_N}\int_0^{R}\dd r\frac{e^{\tilde f}\mathcal{F}^2\tilde f'_0}{\sqrt{1-\mathcal{F}^2}(\tilde f')^2}\Bigg\{(\tilde f')^2-\tilde f''+\tilde\alpha\frac{\kappa}{\ell^2}\frac{e^{-2g_0}g_0'}{\tilde f_0'}\qty((\tilde f')^2-\tilde f'')\nonumber\\
    &+\tilde\alpha e^{-2f}\qty[-2f''(g')^2+4g''f'g'+2(D-2)g''(g')^2-2(f')^2(g')^2+2(D-2)(g')^4]\nonumber\\
    &+2\tilde\alpha\frac{\kappa}{\ell^2}\frac{e^{2(\tilde f-2\tilde f_0-g)}}{\qty(e^{2\tilde f_0}-e^{2\tilde f})\tilde f'}\Bigg[-4e^{4\tilde f_0}(f')^3+3\qty(4e^{\tilde f}-(D-4)e^{4\tilde f_0}-6e^{2(\tilde f+\tilde f_0)})(f')^2g'\nonumber\\
    &-(D-2)\qty(-4(3D-8)e^{\tilde f}+(D(D+10)+20)e^{4\tilde f_0}+6(3D-8)e^{2(\tilde f+\tilde f_0)})(g')^3\nonumber\\
    &+\qty(-8e^{\tilde f}+(D-6)e^{4\tilde f_0}+12e^{2(\tilde f+\tilde f_0)})g'f''\nn\\
    &+(D-2)\qty(-4e^{\tilde f}+(D-4)e^{4\tilde f_0}+6e^{2(\tilde f+\tilde f_0)})g'g''\nonumber\\
    &+\qty(2(3D-7)(g')^2+g'')f'\qty(4e^{\tilde f}-6e^{2(\tilde f+\tilde f_0)})\nonumber\\
    &+4e^{4\tilde f_0}f'\qty(-(3D(D-8)+40)(g')^2+f''+Dg'')\Bigg]\Bigg\},\label{eq:dcmono/dz0}
\end{align}
where, for notational simplicity, we have denoted
\begin{equation}
    \tilde f_0'=\tilde f'(z_0),\qquad g_0'=g'(z_0).
\end{equation}
The above formula \eqref{eq:dcmono/dz0} of course assumes the use of the integral of motion \eqref{eq:zprime}. Note that this agrees with \cite{GonzalezLezcano:2022mcd} for $\tilde\alpha=0$.

If one sets $\kappa=0$, then we see that, schematically,
\begin{align}
    \dv{c_\text{EE}}{z_0}&=\frac{2\text{Vol}\qty(M_{D-2})}{4G_N}\int_0^{R}\dd r\frac{e^{\tilde f}\mathcal{F}^2\tilde f'(z_0)}{\sqrt{1-\mathcal{F}^2}(\tilde f')^2}\qty[\text{NEC1}+(D-1)(D-2)(g')^2\qty(1+\mathcal{O}\qty(\tilde\alpha))]\nn\\
    &\le 0,
\end{align}
where the sign of $\tilde f'(z_0)$ is expected to be negative from NEC1. Thus, for $\kappa=0$, we recover a notion of monotonicity along flows from the UV to the IR. Unfortunately, for $\kappa\ne0$, it is unclear what to make of the resulting expression. 

Note that upon setting $D=2$, $g=f$, and $\kappa=0$, one recovers the result for the strip in flows within the same dimension \cite{Myers:2012ed}. The comparison is more direct in the coordinates \eqref{eq:AltCoords}; the expression \eqref{eq:dcmono/dz0} is reexpressed in said coordinates in Appendix \ref{app:altCoords2}.

\subsection{Relation to the NEC-motivated $c$-function}
It is interesting to note that the monotonic $c$-function \eqref{eq:EEcfnct} constructed from the entanglement entropy is related to the NEC-motivated $c$-function \eqref{eq:LHcfnct}, at least for $\kappa=0$. For arbitrary $D$ with $\kappa=0$, we have 
\begin{equation}
    R = \int_0^{z_0}\dd z\frac{\mathcal{F}}{\sqrt{1-\mathcal{F}^2}}\left(1+2\tilde\alpha e^{-2f}(g')^2\right),
\end{equation}
and the entropic $c$-function is
\begin{equation}
    c_{\text{EE}}(z_0) \propto e^{f_0+(D-2)g_0}R
    = \int_0^{z_0}\dd z\frac{e^{f + (D-2) g}\mathcal{F}^2}{\sqrt{1-\mathcal{F}^2}}\left(1+2\tilde\alpha e^{-2f}(g')^2\right) .
\end{equation}
We may split up the integrand as
\begin{equation}
    \left(\frac{-e^{f + (D-2) g}}{\left(f'+(D-2) g'\right)}\left(1+2\tilde\alpha e^{-2f}(g')^2\right)\right)\left(\frac{-\left(f'+(D-2) g'\right)\mathcal{F}^2}{\sqrt{1-\mathcal{F}^2}}\right),
\end{equation}
such that the right term is a total derivative. Conveniently, the left term can be identified as
\begin{equation}
    c_{\text{NEC}}(z) = \frac{-e^{f + (D-2) g}}{\left(f'+(D-2) g'\right)}\left(1+2\tilde\alpha e^{-2f}(g')^2\right), \label{eqn:cnecfromEE}
\end{equation}
the NEC-motivated $c$-function \eqref{k=0 c func}. This $c$-function follows the coefficient constraints presented in \eqref{eq:cfnctCoeffs}, and further constrains $a_3=2$, where $a_3$ was previously free. However, \eqref{eqn:cnecfromEE} does not give $a_\text{UV}$ as its residue since it does not follow \eqref{eqn:UVacons}. The expression for $c_\text{EE}$ can then be integrated by parts:
\begin{equation} \label{c_mono_arbitrary_D}
     c_{\text{EE}}(z_0) \propto -\sqrt{1-\mathcal{F}^2} \;c_{\text{NEC}}(z)\Bigg |_0^{z_0} + \int_0^{z_0}\dd z\sqrt{1-\mathcal{F}^2}\left(\dv{c_{\text{NEC}}}{z}\right).
\end{equation}
After differentiating with respect to $z_0$, the surface term disappears since $\mathcal{F}(z_0)=1$. Similarly, the derivative hitting the upper integration bound gives no contribution. When computing $\dd c_{\text{EE}}/\dd z_0$ the $z_0$ derivative does not modify $\dd c_{\text{NEC}}/\dd z$, so the monotonicity of $c_{\text{NEC}}$ directly translates to monotonicity of $c_{\text{EE}}$.

\subsection{AdS$_{D+1}\to$ AdS$_{d+1}$ for general $d$}

One might also consider the more general case of flows down to AdS$_{d+1}$ with $d>2$. We will specialize our metric to be
\begin{equation}
    \dd s^2=e^{2f(z)}\qty(-\dd t^2+\dd z^2+\dd r^2+r^2 \dd\Omega_{d-2}^2)+e^{2g(z)}\dd s_{M_{D-d}}^2,
\end{equation}
and we will specify that our entangling region wraps $M_{D-d}$ and has a spherical cross-section of radius $R$. Given a profile $r(z)$, this then results in an induced metric
\begin{equation}
    \dd \sigma^2=e^{2f(z)}\qty(1+r'(z)^2)\dd z^2+e^{2f(z)}r(z)^2\dd\Omega_{d-2}^2+e^{2g(z)}\dd s_{M_{D-d}}^2,
\end{equation}
with induced Ricci scalar
\begin{align}
    \mathcal{R}=&(d-2)(d-3)\frac{e^{-2f}}{r^2}+(D-d)(D-d-1)\frac{\kappa}{\ell^2}e^{-2g}\nonumber\\
    &+\frac{e^{-2f}}{r^2(1+(r')^2)^2}\Big[-\qty(1+(r')^2)\Big(2(d-2)\qty((d-2)f'+(D-d)g')rr'\nonumber\\
    &+(d-2)(d-3)(r')^2+\big((d-2)(d-3)(f')^2+2(d-3)(D-d)f'g'\nonumber\\
    &+(D-d)(D-d-1)(g')^2+2(d-2)f''+2(D-d)g'')\big)r^2\nn\\
    &+2\qty(-(d-2)+\qty((d-2)f'+(D-d)g')rr')rr''\Big].
\end{align}
For more details, see Appendix \ref{app:inducedR}. It is straightforward to check that for $d=2$, $\mathcal{R}$ reduces to \eqref{eq:Rind}. Similar to (\ref{eqn:AdS3JM}), the JM functional is then
\begin{align}
    S_\text{JM}=&\frac{\text{Vol}\qty(S^{d-2})\text{Vol}\qty(M_{D-d})}{4G_N}\nn\\
    &\times\int\dd{z}\Bigg\{r^{d-2} e^{(d-1)\tilde f}\sqrt{1+(r')^2}\qty(1+2\alpha(D-d)(D-d-1)\frac{\kappa}{\ell^2}e^{-2g})\nonumber\\
    &\qquad\qquad+2\alpha\, r^{d-4}\frac{e^{(d-1)\tilde f-2f}}{\sqrt{1+(r')^2}}\Big[r^2\big((d-2)(d-3)(f')^2+2(d-2)(D-d)f'g'\nonumber\\
    &\qquad\qquad+(D-d)(D-d-1)(g')^2\big)+2(d-2)rr'\qty((d-3)f'+(D-d)g')\nonumber\\
    &\qquad\qquad+(d-2)(d-3)\qty(1+2(r')^2)\Big]\Bigg\},\label{eqn:genJM}
\end{align}
where we have we have again integrated by parts and used the boundary term to cancel the Gibbons-Hawking term. If we set $\alpha=0$, this agrees with the two-derivative case \cite{GonzalezLezcano:2022mcd,Deddo:2022wxj}. Moreover, setting $D=d$, we recover
\begin{align}
    S_\text{JM}=&\frac{\text{Vol}\qty(S^{d-2})}{4G_N}\int\dd{z}r^{D-2} e^{(D-1)f}\sqrt{1+(r')^2}\left\{1+2\tilde\alpha\qty[\qty(f'+\frac{r'}{r})^2+\frac{1+(r')^2}{r^2}]\right\},\label{eqn:a=0JM}
\end{align}
which corresponds to the entanglement entropy of a spherical entangling region in flows within the same dimension, as studied in \cite{Hung:2011xb,deBoer:2011wk}.

However, the method applied in the $d=2$ case relied heavily on the fact that the integrand of $S_\text{JM}$ admitted a first integral. Since \eqref{eqn:genJM} contains an explicit factor of $r(z)$, one cannot use the same technique. Without a first integral, we cannot rewrite $r'(z)$ in terms of the turning point $z_0$ to produce an expression like (\ref{eq:EEcfnct}). Recall that monotonicity for the $d=2$ was demonstrated with respect to $z_0$, and it is not clear how one would proceed when $c_\text{EE}$ is not expressed as a function of $z_0$.


\section{Discussion}\label{sec:end}

In this chapter, we have explored higher-derivative renormalization group flows across dimensions. Our first look at holographic flows across dimensions involved explicitly constructing a $c$-function that monotonically decreases along flows from the UV to the IR due to the NEC. This $c$-function, just as the one constructed in the two-derivative case \cite{GonzalezLezcano:2022mcd}, is divergent; we have, however, shown that this divergence can be made to encode the UV central charge. Our second approach was to construct a monotonic $c$-function from the holographic entanglement entropy, given by a minimal surface prescription minimizing the Jacobson-Myers functional. We looked specifically at flows from AdS$_{D+1}$ to AdS$_3$ and explicitly constructed a monotonic $c$-function. More surprising is that this $c$-function is related to the NEC-motivated  $c$-function.

Of course, one could ask: Given that the higher curvature corrections must be treated perturbatively, how could our story have failed? Considering that we are working perturbatively in $\alpha$, we can move terms from the numerator of our $c$-function \eqref{eq:LHcfnct} into the denominator, and so there are only 5 free parameters to consider. On the other hand, the four-derivative part of NEC1 \eqref{eq:NEC1} has, up to our perturbative omission of terms proportional to $(g')^2$, 10 terms that must be matched in $c'(z)$. So, the fact that the NEC-motivated $c$-function evolves monotonically is a non-trivial statement and could have easily not been the case.

We note that we could have additionally included in the action the quasi-topological term $\mathcal{Z}_{D+1}$ given by
\begin{align}
    \mathcal{Z}_{D+1}=&\hat R_{M\ \ N}^{\ \ \,P\ \ \, Q}\hat R_{P\ \,Q}^{\ \ R\ \ S}\hat R_{R\ \ S}^{\ \ M\ \ N}+\frac{1}{(2D-1)(D-3)}\bigg[\frac{3(3D-5)}{8}\hat R_{MNPQ}\hat R^{MNPQ}\hat R\nonumber\\
    &-3(D-1)\hat R_{MNPQ}\hat R^{MNP}_{\ \ \ \ \ \ R}\hat R^{QR}+3(D-1)\hat R_{MNPQ}\hat R^{MP}\hat R^{NQ}
    \nonumber\\
    &+6(D-1)\hat R_{MN}\hat R^{NP}\hat R_{P}^{\ \ \,M}-\frac{3(3D-1)}{2}\hat R_{MN}\hat R^{MN} \hat R+\frac{3(D-1)}{8}\hat R^3\bigg],
\end{align}
which was constructed in \cite{Myers:2010ru,Oliva:2010eb}; this term played a prominent role in \cite{Myers:2010xs,Myers:2010tj}. For our purposes, however, this term presents some difficulties. In contrast to the Gauss-Bonnet term, or even more generally Lovelock terms, the coefficients of $\mathcal{Z}_{D+1}$ are dimension-dependent. This presents us with a problem: We must either choose $\mathcal{Z}_{D+1}$, which is quasi-topological in the UV but which yields unsavory terms in the IR, or we could choose $\mathcal{Z}_{d+1}$ which is quasi-topological in the IR but not the UV.  The conundrum originates from the fact that $\mathcal{Z}_{D+1}$ was engineered to be quasi-topological for maximally symmetric backgrounds, and our background \eqref{Eq:Flowfg} does not satisfy this criterion. Hence, we would generically have to deal with fourth-order derivatives in the NEC.

As mentioned at the beginning of the chapter, the field theory techniques required for proving monotonicity theorems are very dimension-dependent. Recall that Zamolodchikov's proof of the $c$-theorem in 2d relied on properties of the correlator of two stress-energy tensors \cite{Zamolodchikov:1986gt} while in 4d Schwimmer and Komargodski relied on properties of certain four-point amplitudes to prove the $a$-theorem \cite{Komargodski:2011vj}. The entropic approach, due largely to Casini and collaborators, relied almost exclusively on strong subadditivity of the relative entropy \cite{Casini:2017vbe}. It is an outstanding problem to connect these different approaches. Holography has furnished two sets of proofs, one following from NEC and another related to the entropy via the Ryu-Takayanagi prescription. We have found, at least in a particular case, that the proofs are connected. We hope to explore this connection in more detail in the future and hope to draw lessons that might translate to field theoretic approaches. Another question that seems particularly suitable for holographic attacks is the nature of supersymmetric flows; in this case, Einstein's equations can be replaced by a set of linear differential equations.  

%% file: Chapters/chap5.tex
Consistent truncations have played a pivotal role in theoretical physics, ranging from string theory and supergravity to brane-world scenarios. As we saw in Chapter \ref{chap:Intro}, the general principle is that, given a Kaluza-Klein reduction on some compact manifold, one is interested in removing all but a finite number of modes from the infinite Kaluza-Klein tower in such a way as to maintain consistency of the theory, \textit{i.e.}, such that the solutions to the equations of motion of the truncated theory are also solutions of the original theory. The classic example is the Scherk-Schwarz reduction \cite{Scherk:1979zr} wherein the internal space is taken to be a group manifold which becomes the gauge group of the effective lower-dimensional theory. In this case, one can obtain a consistent truncation by restricting to the singlet sector, which enforces consistency via a symmetry principle. More generally, however, in the absence of a manifest symmetry principle, consistent truncations have traditionally been rare and difficult to construct; see for example \cite{Duff:1984hn,Cvetic:2000dm}. In particular, there is no such simple rule for general reductions, and the consistency of a truncation imposes stringent requirements on the field content and couplings of both the higher and lower dimensional theories.

Naturally, there has been much work on non-trivial consistent truncations. In particular, there are many examples of coset reductions, including sphere truncations \cite{deWit:1986oxb,Nastase:1999kf,Lu:1999bc,Cvetic:1999un,Lu:1999bw,Cvetic:1999au,Cvetic:2000dm,Cvetic:2000nc,Lee:2014mla,Nicolai:2011cy,Samtleben:2019zrh,Bonetti:2022gsl,Ciceri:2023bul} and more general coset reductions \cite{Cvetic:2003jy,House:2005yc,Cassani:2009ck,Cassani:2010na,Bena:2010pr}; in such cases, the massless sector contains charged (non-singlet) fields that one is interested in keeping and care must be taken that these do not source the fields that one wishes to truncate away. There are also examples of reductions wherein one is interested in keeping a finite number of massive modes, such as those on Sasaki-Einstein spaces \cite{Buchel:2006gb,Gauntlett:2009zw,Cassani:2010uw, Liu:2010sa,Gauntlett:2010vu,Skenderis:2010vz,Bah:2010cu,Liu:2010pq,Bah:2010yt} and $T^{1,1}$ \cite{Liu:2011dw}, where one is often interested in keeping massive breathing and squashing modes. Despite the difficulty of finding consistent truncations, there are powerful results. Indeed, one has the conjecture that any warped product AdS$_D\times M_d$ supergravity solution in ten or eleven dimensions has a consistent truncation to a solution of pure gauged supergravity in $D$ dimensions with the same amount of supersymmetry as the original solution \cite{Gauntlett:2007ma}, with additional evidence having been constructed in \cite{Gauntlett:2007sm,OColgain:2011ng,Jeong:2013jfc,Passias:2015gya,Hong:2018amk,MatthewCheung:2019ehr,Larios:2019kbw,Cheung:2022ilc,Couzens:2022aki}. It is also generally believed that truncating to just the massless graviton multiplet is consistent \cite{Pope:1987ad}. Note that all of these results are at the two-derivative level.

An important, more recent development has been the use of exceptional field theory \cite{Hohm:2013pua,Hohm:2013vpa,Hohm:2013uia,Hohm:2014fxa,Godazgar:2014nqa}
as a means to construct consistent truncations \cite{Hohm:2014qga}.  The power of exceptional field theory is that consistency is guaranteed by using a generalized Scherk-Schwarz reduction.  This has led to many new examples of consistent truncations \cite{Cassani:2016ncu,Malek:2016bpu,Malek:2017njj,Malek:2017cle,Malek:2018zcz,Malek:2019ucd,Cassani:2019vcl,Malek:2020jsa} as well as analysis of the Kaluza-Klein spectra around such truncations 
\cite{Malek:2019eaz,Malek:2020yue,Eloy:2020uix,Varela:2020wty,Cesaro:2020soq,Bobev:2020lsk,Cesaro:2021haf}.  Such developments have put the Gauntlett-Varela conjecture, \cite{Gauntlett:2007ma}, on firm ground.  Nevertheless, despite such enormous progress in the construction of non-trivial consistent truncations, many of the results are currently limited to the leading-order two-derivative theory.

While it seems reasonable that consistency of a truncation at the two-derivative level would imply consistency at higher-derivative order, it is unclear that this necessarily holds.  After all, one possible obstruction could be a higher-derivative coupling between the retained modes and states in the Kaluza-Klein tower.  In the supergravity context, this could appear as additional couplings between the supergravity multiplet and matter multiplets in the spectrum.  To examine this possibility,  we will work specifically in the context of four-derivative heterotic supergravity reduced on a torus.  This is a very standard Kaluza-Klein reduction, and by restricting to zero modes on the torus (\textit {i.e.}, the singlet sector), one is ensured to obtain a consistent truncation.  The bosonic reduction of the four-derivative theory was obtained in \cite{Eloy:2020dko}.

It is important to note that reducing heterotic supergravity on $T^n$ leads to a half-maximal supergravity theory in $10-n$ dimensions coupled to $n$ vector multiplets.  The question then arises whether it is consistent to truncate out the vector multiplets, as they naturally arise at the same massless Kaluza-Klein level from the same ten-dimensional fields that give rise to the lower-dimensional supergravity multiplet.  We answer this in the affirmative by explicitly truncating out the bosonic fields in the vector multiplets at the level of their equations of motion as well as their superpartners at the level of the supersymmetry variations.

While we work in general dimensions, the reduction on $T^4$ to six dimensions was considered in \cite{Chang:2021tsj}, which performed a truncation to $\mathcal N=(1,0)$ supergravity coupled to one tensor multiplet and four hypermultiplets.  This truncation further reduces the supersymmetry and was indeed shown to be consistent.  On the other hand, our truncation gives $\mathcal N=(1,1)$ supergravity which, in the $\mathcal N=(1,0)$ language corresponds to supergravity coupled to one tensor and two gravitino multiplets.  We show that, while the gravitino multiplet can be consistently truncated, the tensor multiplet cannot be removed at the four-derivative level, even though it can be decoupled from the two-derivative theory.  This is a concrete example of a higher-derivative obstruction to a consistent truncation, even in the relatively simple example of a torus reduction.

\subsection*{The torus reduction}

We work with the fields of ten-dimensional heterotic supergravity, $(g_{MN},\psi_M,B_{MN},\lambda,\phi)$, disregarding the heterotic vector multiplets.  Our starting point is the torus reduction of the metric
\begin{equation}
    \dd s_{10}^2=g_{\mu\nu}\dd x^\mu \dd x^\nu+g_{ij}\eta^i\eta^j,\qquad\eta^i=\dd y^i+A_\mu^i\dd x^\mu,
\end{equation}
where $x^\mu$ are coordinates on the base space and $y^i$ are coordinates on the internal space. The two-form $B$ is reduced as
\begin{equation}
    B=\ft12b_{\mu\nu}\dd x^\mu\wedge \dd x^\nu+B_{\mu i}\dd x^\mu\wedge\eta^i+\ft12b_{ij}\eta^i\wedge\eta^j.
\end{equation}
Naturally, reducing the 10D gravity multiplet on an $n$-dimensional torus leads to a half-maximal gravity multiplet coupled to $n$ vector multiplets. By analyzing the bosonic equations of motion, we show that it is consistent to truncate out the vector multiplet, and we write down the reduced Lagrangian.  The resulting bosonic reduction ansatz, $(g_{MN},B_{MN},\phi)\to(g_{\mu\nu},b_{\mu\nu},A_\mu^{(-)\,i},\varphi)$, takes the form
\begin{align}
    &g_{\mu\nu}=g_{\mu\nu},\qquad A_\mu^i=\fft12A_\mu^{(-)\,i},\kern2.9em g_{ij}=\delta_{ij}+\frac{\alpha'}{16}F^{(-)\,i}_{\mu\nu}F^{(-)\,i}_{\mu\nu},\nn\\
    &b_{\mu\nu}=b_{\mu\nu},\qquad B_{\mu i}=-\fft12A_\mu^{(-)\,i},\qquad b_{ij}=0,\nn\\
    &\phi=\varphi.
\label{eq:truncanz1}
\end{align}
Here we have introduced the notation $A^{(\pm)\,i}=A_\mu^i\pm \delta^{ij}B_{\mu i}$, or equivalently $F^{(\pm)\,i}=F^{i}\pm \delta^{ij}G_j$, where $F^i=\dd A^i$ and $G_i=\dd B_i$.  The $A^{(+)\,i}$ are in the vector multiplet and are truncated out along with the scalars $g_{ij}$ and $b_{ij}$.  Note, in particular, the $\mathcal O(\alpha')$ addition to $g_{ij}$ that is required for the truncation to be consistent.

For the fermions, the gravitino $\psi_M$ splits into a lower-dimensional gravitino $\psi_\mu$ along the uncompactified directions and gaugini $\psi_i$ along the compact directions. After an appropriate shift, we show that the truncation of the bosonic sector is consistent with supersymmetry, in the sense that $\delta\tilde\psi_i=0$ where $\tilde\psi_i$ are the $\mathcal O(\alpha')$ corrected gaugini. In particular, if we write the gravitino variation as $\delta\psi_\mu=\mathcal D_\mu\epsilon$, then this redefinition takes the elegant form
\begin{equation}
    \tilde\psi_i=\psi_i-\frac{\alpha'}{4}F^{(-)\,i}_{\mu\nu}\mathcal D_\mu\psi_\nu,
\end{equation}
where $F^{(-)\,i}$ is the combination of field strengths that remains after our truncation, and this is specifically selected out by the gaugino variation. 

\subsection*{An \texorpdfstring{$
\mathcal O(\alpha')$}{O(alpha')} corrected black string}

Finally, for illustrative purposes, we look at the four-derivative corrected BPS black string in ten dimensions which reduces to a nine-dimensional black hole.  The leading order black hole solution takes the well-known form \cite{Lu:1995cs}
\begin{align}
    \dd s_9^2&=-\qty(1+\frac{k}{r^6})^{-2}\dd t^2+\dd r^2+r^2\dd\Omega_{7}^2,\nn\\
    A&=\frac{1}{1+\frac{k}{r^6}}\dd t,\nn\\
    e^{\varphi}&=\qty(1+\frac{k}{r^6})^{-1/2}.
\end{align}
We find that the four-derivative corrections to the 10D uplifted metric are then
\begin{align}
    \dd s_{10}^2=&-\qty(1+\frac{k}{r^6})^{-2}\qty(1+\frac{18\alpha' k^2}{r^2(k+r^6)^2})\dd t^2+\dd r^2+r^2\dd\Omega_7^2\nn\\
    &+\qty(1-\frac{18\alpha' k^2}{r^2(k+r^6)^2})\qty(\dd z-\frac{1}{1+\frac{k}{r^6}}\qty(1+\frac{18\alpha' k^2}{r^2(k+r^6)^2})\dd t)^2+\mathcal{O}(\alpha'^2),
\end{align}
while the $B$-field remains unchanged. Similar $\alpha'$-corrected heterotic black holes in lower dimensions were considered in \cite{Natsuume:1994hd,Kats:2006xp,Castro:2007hc,Cvitan:2007pk,Cano:2018brq,Cano:2019ycn,Cano:2021nzo,Ortin:2021win,Cano:2022tmn}. In particular, the truncation places requirements on the components of the metric in the compactified direction, and we find that these are precisely in agreement with the four-derivative corrected black hole solution.

The rest of this chapter is organized as follows. In Section \ref{sec:torus}, we review four-derivative heterotic supergravity and discuss the torus reduction. In Section \ref{sec:bosonic}, we verify the consistency of truncating out the vector multiplets by analyzing the bosonic equations of motion, and in Section \ref{sec:fermionic}, we verify the consistency by analysis of the gaugini variations. In Section \ref{sec:example}, we derive the four-derivative corrections to the ten-dimensional BPS black string geometry and compare it with the field redefinitions required in Section \ref{sec:bosonic}. Finally, we conclude in Section \ref{sec:conclusion} and discuss some further truncations.

\section{Heterotic torus reduction}\label{sec:torus}

In this section, we reduce the bosonic fields of four-derivative heterotic supergravity on a torus. Our notation is such that we use $M, N,\ldots$ for curved indices in 10D and $A, B,\ldots$ for rigid indices in 10D, as well as $\mu,\nu,\ldots$ for curved indices along the base space, $\alpha,\beta,\ldots$ for rigid indices along the base space, $i,j,\ldots$ for curved indices along the internal torus, and $a,b,\ldots$ for rigid indices along the internal torus. That is, we split our curved indices as $M\to\{\mu, i\}$ and our rigid indices as $A\to\{\alpha, a\}$. We use $\hat\nabla$ to mean the Levi-Civita connection in 10D, while we use $\nabla$ for the Levi-Civita connection on the base space.

\subsection{Four-derivative heterotic supergravity}
Heterotic supergravity is a ten-dimensional, $\mathcal N=(1,0)$ theory with a single Majorana-Weyl supercharge. The field content is simply the half-maximal gravity multiplet, consisting of the metric $g_{MN}$, the Majorana-Weyl gravitino $\psi_M$, the two-form $B_{MN}$, the Majorana-Weyl dilatino $\lambda$, and the dilaton $\phi$. In the string frame, the ten-dimensional bosonic Lagrangian up to four-derivative corrections takes the form \cite{Bergshoeff:1988nn,Bergshoeff:1989de,Metsaev:1987zx,Chemissany:2007he}
\begin{equation}
    e^{-1}\mathcal L=e^{-2\phi}\left[R+4(\partial_M\phi)^2-\fft1{12}\tilde H_{MNP}^2+\frac{\alpha'}{8}\big(R_{MNAB}(\Omega_+)\big)^2\right]+\mathcal{O}(\alpha'^3),
\label{eq:Lhet}
\end{equation}
where $R$ is the Ricci scalar and we have defined
\begin{equation}
    \tilde H= H-\frac{\alpha'}{4}\omega_{3L}(\Omega_+),
\label{eq:Htilde}
\end{equation}
where $H=\dd B$ is the three-form flux. We have implicitly truncated the heterotic gauge fields, as they will not play an important role in our discussion. Here we have introduced the torsionful connection
\footnote{Note that the choice of $\Omega_+$ versus $\Omega_-$ is equivalent to a choice of the sign of $H$ in the gravitino variation, and we may always switch conventions by doing a sign flip $B\to -B$. This is just a choice of worldsheet parity.  Our convention is opposite that used in \cite{Bergshoeff:1989de}.}
\begin{equation}
    \Omega_+=\Omega+\fft12\mathcal H,\qquad\mathcal H^{AB}\equiv \tilde H_M{}^{AB}\dd x^M,
\end{equation}
where $\Omega$ is the spin connection, and the corresponding curvature is
\begin{equation}
    R(\Omega_+)=\dd\Omega_++\Omega_+\wedge\Omega_+.
\end{equation}
Such choice of connection is required so that $(\Omega_+,\psi_{MN})$ transforms as an $SO(9,1)$ gauge multiplet \cite{Bergshoeff:1989de}, where $\psi_{MN}=2\nabla_{[M}(\Omega_-)\psi_{N]}$ is the supercovariant gravitino curvature. The Lorentz Chern-Simons form is
\begin{equation}
    \omega_{3L}(\Omega_+)=\Tr\left(\Omega_+\wedge \dd\Omega_++\fft23\Omega_+\wedge\Omega_+\wedge\Omega_+\right),
\label{eq:LCS+}
\end{equation}
and is required by anomaly cancellation. This immediately leads to the Bianchi identity
\begin{equation}
    \dd\tilde H=-\frac{\alpha'}{4}\Tr \qty[R(\Omega_+)\land R(\Omega_+)].
\end{equation}
This is characteristic of the two-group structure.

Note that we can break up the Lagrangian (\ref{eq:Lhet}), into two- and four-derivative parts
\begin{align}
    e^{-1}\mathcal L_{2\partial}&=e^{-2\phi}\left[R+4(\partial_M\phi)^2-\fft1{12}H_{MNP}^2\right],\nn\\
    e^{-1}\mathcal L_{4\partial}&=\frac{\alpha'}{8}e^{-2\phi}\left[\qty(R_{MNAB}(\Omega_+))^2+\fft13H^{MNP}\omega_{3L\,MNP}(\Omega_+)\right].
\label{eq:Lags}
\end{align}
The bosonic equations of motion are
\begin{align}
    0&=\mathcal E_\phi\equiv R-4\qty(\partial_M\phi)^2+4\hat\Box\phi-\fft1{12}\tilde H_{MNP}^2+\frac{\alpha'}{8}\qty(R_{MNAB}(\Omega_+))^2,\nn\\
    0&=\mathcal E_{g,MN}\equiv R_{MN}+2\hat\nabla_M\hat\nabla_N\phi-\fft14\tilde H_{MAB}\tilde H_N{}^{AB}+\frac{\alpha'}{4}R_{MPAB}(\Omega_+)R_N{}^{PAB}(\Omega_+),\nn\\    0&=\mathcal E_{H,NP}\equiv\hat\nabla^M\qty(e^{-2\phi}\tilde H_{MNP}),
\end{align}
where we have used the dilaton equation $\mathcal E_\phi$ to simplify the Einstein equation $\mathcal E_{g,MN}$. Equivalently, one may use the fact that the variation of the action with respect to $\Omega_+$ is proportional to the two-derivative equations of motion \cite{Bergshoeff:1988nn}. These can also be broken up into two- and four-derivative parts, and we write $\mathcal E=\mathcal E^{(0)}+\alpha'\mathcal E^{(1)}$.  Then
\begin{align}
    \mathcal E_\phi^{(0)}&=R-4(\partial_M\phi)^2+4\hat\Box\phi-\fft1{12} H_{MNP}^2,\nn\\
    \mathcal E_{g,MN}^{(0)}&=R_{MN}+2\hat\nabla_M\hat\nabla_N\phi-\fft14 H_{MAB} H_N{}^{AB},\nn\\
    \mathcal E_{H,NP}^{(0)}&=e^{2\phi}\hat\nabla^M\qty(e^{-2\phi} H_{MNP}),
\label{eq:eom2d}
\end{align}
and
\begin{align}
    \mathcal E_\phi^{(1)}&=\fft1{24}H_{MNP}\omega_{3L}^{MNP}(\Omega_+)+\fft18\qty(R_{MNAB}(\Omega_+))^2,\nn\\
    \mathcal E_{g,MN}^{(1)}&=\fft18 H_{MAB} \omega_{3L\,N}{}^{AB}(\Omega_+)+\fft14R_{MPAB}(\Omega_+)R_N{}^{PAB}(\Omega_+),\nn\\
    \mathcal E_{H,NP}^{(1)}&=-\fft14e^{2\phi}\hat\nabla^M\qty(e^{-2\phi} \omega_{3L,MNP}(\Omega_+)).
\label{eq:eom4d}
\end{align}

\subsubsection{Supersymmetry variations}

Although we primarily focus on the reduction of the bosonic fields, the supersymmetry variations of the fermionic fields also need to be considered to ensure a consistent truncation.  Up to $\mathcal O(\alpha')$, the supersymmetry transformations of the gravitino and dilatino are
\cite{Strominger:1986uh, Bergshoeff:1989de}%
\footnote{To avoid confusion with $\delta$ denoting $\mathcal O(\alpha')$ corrections, we use $\delta_\epsilon$ for supersymmetry transformations parameterized by a spinor $\epsilon$.}
\begin{align}
    \delta_\epsilon\psi_M&=\nabla_M(\Omega_-)\epsilon=\left(\partial_\mu+\fft14\Omega_{-\,M}{}^{AB}\Gamma_{AB}\right)\epsilon=\left(\nabla_M-\fft18\tilde H_{MNP}\Gamma^{NP}\right)\epsilon,\nn\\
    \delta_\epsilon\lambda&=\left(\Gamma^M\partial_M\phi-\fft1{12}\tilde H_{MNP}\Gamma^{MNP}\right)\epsilon.
\end{align}
The structure of these variations is such that the $\mathcal O(\alpha')$ corrections are entirely contained in the definition of $\tilde H$ given in (\ref{eq:Htilde}).  As above, we can write
\begin{equation}
    \delta_\epsilon \psi_M=\delta_\epsilon\psi_M^{(0)}+\alpha'\delta_\epsilon\psi_M^{(1)},\qquad\delta_\epsilon\lambda=\delta_\epsilon\lambda^{(0)}+\alpha'\delta_\epsilon\lambda^{(1)},
\end{equation}
where
\begin{align}
    \delta_\epsilon\psi_M^{(0)}&=\left(\nabla_M-\fft18H_{MNP}\Gamma^{NP}\right)\epsilon,&\delta_\epsilon\psi_M^{(1)}&=\fft1{32}\omega_{3L,MNP}\Gamma^{NP}\epsilon,\nn\\
    \delta_\epsilon\lambda^{(0)}&=\left(\Gamma^M\partial_M\phi-\fft1{12} H_{MNP}\Gamma^{MNP}\right)\epsilon,&\delta_\epsilon\lambda^{(1)}&=\fft1{48}\omega_{3L,MNP}\Gamma^{MNP}\epsilon.
\label{eq:deltas}
\end{align}

\subsection{Torus reduction}
We perform a standard Kaluza-Klein reduction on an $n$-dimensional torus $T^n$ by taking our metric to be
\begin{equation}
    \dd s_{10}^2=g_{\mu\nu}\dd x^\mu \dd x^\nu+g_{ij}\eta^i\eta^j,\qquad\eta^i=\dd y^i+A_\mu^i\dd x^\mu,
\label{eq:redMet}
\end{equation}
where $x^\mu$ are coordinates on the base space and $y^i$ are coordinates on the internal space. We can introduce a natural zehnbein basis
\begin{equation}
    E^\alpha=e_\mu^\alpha \dd x^\mu,\qquad E^a=e_i^a\eta^i,
\end{equation}
where $e^\alpha$ is a vielbein for $g_{\mu\nu}$ and $e^a$ is a vielbein for $g_{ij}$, so that $\dd s_{10}^2=\eta_{\alpha\beta}E^\alpha E^\beta+\delta_{ab}E^aE^b$.  Then
\begin{equation}
    E=\begin{pmatrix}e^\alpha\\e_i^a\eta^i\end{pmatrix},\qquad \dd E=\begin{pmatrix}\dd e^\alpha\\ \dd e_i^a\wedge\eta^i+e_i^aF^i\end{pmatrix},
\end{equation}
where the abelian field strength is given locally by $F^i=\dd A^i$.  In components, we have
\begin{equation}
    E_M{}^A=\begin{pmatrix}e_\mu^\alpha~&e_i^aA_\mu^i\\0&e_i^a\end{pmatrix},\qquad E_A{}^M=\begin{pmatrix}e_\alpha^\mu~&-e_\alpha^\mu A_\mu^i\\0&e_a^i\end{pmatrix}.
\end{equation}

The torsion-free spin connection can be computed to be
\begin{equation}
    \Omega=\begin{pmatrix}\omega^{\alpha\beta}-\fft12g_{ij}F_{\alpha\beta}^i\eta^j&\fft12e_i^bF_{\mu\alpha}^i\dd x^\mu-\fft12e^i_b\partial_\alpha g_{ij}\eta^j\\-\fft12e_i^aF_{\mu\beta}^i\dd x^\mu+\fft12e^i_a\partial_\beta g_{ij}\eta^j&\fft12\qty(e^{ia}\dd e_i^b-e^{ib}\dd e_i^a)\end{pmatrix},
\label{eq:tfc}
\end{equation}
where $\omega$ is the torsion-free spin connection on the base manifold.

\subsubsection{Inclusion of torsion}

In addition to the metric, the $B$-field is reduced according to
\begin{equation}
    B=\ft12b_{\mu\nu}\dd x^\mu\wedge \dd x^\nu+B_{\mu i}\dd x^\mu\wedge\eta^i+\ft12b_{ij}\eta^i\wedge\eta^j.
\label{eq:redB}
\end{equation}
Computing $H=\dd B$ then gives
\begin{equation}
    H=h+\tilde G_i\wedge\eta^i+\ft12\dd b_{ij}\wedge\eta^i\wedge\eta^j,
\end{equation}
where
\begin{equation}
    h=\dd b-F^i\wedge B_i,\qquad\tilde G_i=G_i-b_{ij}F^j,\qquad G_i=\dd B_i.
\end{equation}
The one-form $\mathcal H^{AB}$ is then
\begin{equation}
    \mathcal H=\begin{pmatrix}h_\mu{}^{\alpha\beta}\dd x^\mu+\tilde G^{\alpha\beta}{}_i\eta^i&e^{ib}\qty(\tilde G_{\mu\alpha i}\dd x^\mu+\partial_\alpha b_{ij}\eta^j)\\-e^{ia}\qty(\tilde G_{\mu\beta i}\dd x^\mu+\partial_\beta b_{ij}\eta^j)&e^{ia}e^{jb}\dd b_{ij}\end{pmatrix}.
\end{equation}
Combining $\mathcal H$ with the torsion-free connection $\Omega$ in (\ref{eq:tfc}) then gives the torsional connection
\begin{align}
    \Omega_+=&\left(\begin{matrix}
        \omega_+^{\alpha\beta}-\fft12\qty(g_{ij}F_{\alpha\beta}^j-\tilde G_{\alpha\beta i})\eta^i\\
        -\fft12e^{ia}\qty(\qty(g_{ij}F_{\mu\beta}^j+\tilde G_{\mu\beta i})\dd x^\mu-\partial_\beta(g_{ij}-b_{ij})\eta^j)
    \end{matrix}\right.\nn\\
    &\kern12em\left.\begin{matrix}
        \fft12e^{ib}\qty(\qty(g_{ij}F_{\mu\alpha}^j+\tilde G_{\mu\alpha i})\dd x^\mu-\partial_\alpha\qty(g_{ij}-b_{ij})\eta^j)\\
        \fft12e^{ia}e^{jb}\qty(e_j^c\dd e_i^c-e_i^c\dd e_j^c+\dd b_{ij})
    \end{matrix}\right).
\label{eq:Omega+}
\end{align}
The connection $\Omega_-$ may be obtained by taking $H\to-H$.  The torsionful Riemann tensor can be calculated from $R(\Omega_+)=\dd \Omega_++\Omega_+\wedge\Omega_+$.  The frame components are given in Appendix \ref{app:riemann}.

\subsection{The bosonic reduction at leading order}

Before proceeding with the truncation of the reduced vector multiplets, it is instructive to review the standard Kaluza-Klein reduction of the two-derivative action and equations of motion.  Since the truncation to the zero modes on the torus, (\ref{eq:redMet}) and (\ref{eq:redB}), is guaranteed to be consistent, we can directly reduce the two-derivative Lagrangian, (\ref{eq:Lags}).  This yields the standard Kaluza-Klein result \cite{Maharana:1992my}
\begin{align}
    e^{-1}\mathcal L^{(0)}&=e^{-2\varphi}\Bigl[R(\omega)+4\partial_\mu\varphi^2-\fft1{12}h_{\mu\nu\rho}^2-\fft14\left(g_{ij}F_{\mu\nu}^iF^{\mu\nu\,j}+g^{ij}\tilde G_{\mu\nu\,i}\tilde G^{\mu\nu}_j\right)\nn\\
    &\kern4em-\fft14g^{ij}g^{kl}(\partial_\mu g_{ik}\partial^\mu g_{jl}+\partial_\mu b_{ik}\partial^\mu b_{jl})\Bigr],
\label{eq:redlag2d}
\end{align}
where the reduced dilaton $\varphi$ is given by
\begin{equation}
    \varphi=\phi-\fft14\log\det g_{ij}.
\label{eq:dilshift}
\end{equation}

It is also straightforward to directly reduce the leading order ten-dimensional equations of motion (\ref{eq:eom2d}).  Making use of some of the reduction expressions in the Appendix, we obtain the reduced two-derivative Einstein equations
\begin{align}
    \mathcal E_{g,\alpha\beta}^{(0)}=&R(\omega)_{\alpha\beta}-\fft12\qty(g_{ij}F_{\alpha\gamma}^iF_{\beta\gamma}^j+g^{ij}\tilde G_{\alpha\gamma\,i}\tilde G_{\beta\gamma\,j})-\fft14h_{\alpha\gamma\delta}h_{\beta\gamma\delta}+2\nabla_\alpha\nabla_\beta\varphi\nonumber\\
    &-\fft14g^{ij}g^{kl}\qty(\partial_\alpha g_{ik}\partial_\beta g_{jl}+\partial_\alpha b_{ik}\partial_\beta b_{jl}),\nn\\
    \mathcal E_{g,\alpha b}^{(0)}=&\fft12e^i_b\left(e^{2\varphi}\nabla_\gamma(e^{-2\varphi}g_{ij}F_{\alpha\gamma}^j)-\fft12h_{\alpha\gamma\delta}\tilde G_{\gamma\delta\,i}-g^{jk}\tilde G_{\alpha\gamma\,j}\partial_\gamma b_{ki}\right),\nn\\
    \mathcal E_{g,ab}^{(0)}=&-\fft12e^i_ae^j_b\biggl(e^{2\varphi}\nabla^\gamma(e^{-2\varphi}\nabla_\gamma g_{ij})-\fft12(g_{ik}g_{jl}F_{\alpha\beta}^kF_{\alpha\beta}^l-\tilde G_{\alpha\beta\,i}\tilde G_{\alpha\beta\,j})\nn\\
    &\qquad-g^{kl}\qty(\partial_\gamma g_{ik}\partial_\gamma g_{jl}-\partial_\gamma b_{ik}\partial_\gamma b_{jl})\biggr),
\label{eq:eom2dred}
\end{align}
and the reduced $H$-field equations of motion
\begin{align}
    \mathcal E_{H,\alpha\beta}^{(0)}&=e^{2\varphi}\nabla^\gamma\qty(e^{-2\varphi}h_{\alpha\beta\gamma}),\nn\\
    \mathcal E_{H,\alpha b}^{(0)}&=e_i^b\left(e^{2\varphi}\nabla^\gamma(e^{-2\varphi}g^{ij}\tilde G_{\gamma\alpha\,j})+\fft12h_{\alpha\gamma\delta}F_{\gamma\delta}^i\right),\nn\\
    \mathcal E_{H,ab}^{(0)}&=e^{[i}_ae^{j]}_b\left(e^{2\varphi}\nabla^\gamma(e^{-2\varphi}\nabla_\gamma b_{ij})-g_{ik}F_{\gamma\delta}^k\tilde G_{\gamma\delta\,j}+2g^{kl}\partial_\gamma g_{ik}\partial_\gamma b_{jl}\right).
\label{eq:eom2dredH}
\end{align}
Finally, the reduced dilaton equation is
\begin{align}
    \mathcal E_\phi^{(0)}=&R(\omega)-\fft14\qty(g_{ij}F_{\alpha\beta}^iF_{\alpha\beta}^j+g^{ij}\tilde G_{\alpha\beta\,i}\tilde G_{\alpha\beta\,j})-\fft1{12}h_{\alpha\beta\gamma}^2+4\Box\varphi-4\qty(\partial_\alpha\varphi)^2\nn\\
    &-\fft14g^{ij}g^{kl}\qty(\partial_\alpha g_{ik}\partial_\alpha g_{jl}+\partial_\alpha b_{ik}\partial_\alpha b_{jl}).
\label{eq:eom2dredphi}
\end{align}
Since the torus reduction is consistent, these equations can also be directly obtained from the reduced Lagrangian (\ref{eq:redlag2d}).

\subsection{Supersymmetry variations at leading order}

Along with the leading order bosonic reduction, we can consider the supersymmetry variations of the gravitino and dilatino.  When dimensionally reduced, we have $\{\psi_M,\lambda\}\longrightarrow\{\psi_\mu, \psi_i, \lambda\}$.  As in the case of the lower-dimensional dilaton shift (\ref{eq:dilshift}), the dilatino also requires a shift of the form
\begin{equation}
    \tilde\lambda=\lambda-\Gamma^i\psi_i.
\end{equation}
With this in mind, the reduction of the lowest-order transformations, (\ref{eq:deltas}), gives
\begin{align}
    \delta_\epsilon\psi_\mu^{(0)}&=\left(\nabla_\mu(\omega_-)+\fft14\qty(g_{ij}F_{\mu\nu}^j-\tilde G_{\mu\nu\,i})\gamma^\nu\Gamma^i-\fft18\qty(2e_i^c\partial_\mu e_j^c+\partial_\mu b_{ij})\Gamma^{ij}\right)\epsilon,\nn\\
    \delta_\epsilon\psi_i^{(0)}&=\left(-\fft18\qty(g_{ij}F_{\mu\nu}^j+\tilde G_{\mu\nu\,i})\gamma^{\mu\nu}-\fft14\partial_\mu\qty(g_{ij}-b_{ij})\gamma^\mu\Gamma^j\right)\epsilon,\nn\\
    \delta_\epsilon\tilde\lambda^{(0)}&=\left(\gamma^\mu\partial_\mu\varphi-\fft1{12}h_{\mu\nu\lambda}\gamma^{\mu\nu\lambda}+\fft18(g_{ij}F_{\mu\nu}^j-\tilde G_{\mu\nu\,i})\gamma^{\mu\nu}\Gamma^i\right)\epsilon.
\label{eq:deltalo}
\end{align}
At this order, the gravitino $\psi_\mu^{(0)}$ and dilatino $\tilde\lambda^{(0)}$, belong in the supergravity multiplet, while the internal components $\psi_i^{(0)}$ fall into vector multiplets.  This allows us to identify the graviphoton and vector multiplet gauge field combinations as
\begin{align}
    F_{\mu\nu}^{a\,(-)}&=e^a_iF_{\mu\nu}^i-e_a^i\tilde G_{\mu\nu\,i},&&(\hbox{graviphoton})\nn\\
    F_{\mu\nu}^{a\,(+)}&=e^a_iF_{\mu\nu}^i+e_a^i\tilde G_{\mu\nu\,i}.&&(\hbox{vector})
\label{eq:gpv}
\end{align}
This will serve as a guide for truncating out the vector multiplets below.

\section{Truncating out the vector multiplets}\label{sec:bosonic}

Reducing the ten-dimensional heterotic action on $T^n$ gives rise to a lower-dimensional half-maximal supergravity coupled to $n$ vector multiplets.  Here we proceed to truncate out the vector multiplets, leading to a pure half-maximal supergravity in lower dimensions.  The truncation of the two-derivative theory is straightforward, and our main intent is to highlight that the truncation remains consistent at the four-derivative level.  We start by considering the two-derivative truncation.

\subsection{The supergravity truncation at leading order}

As indicated in (\ref{eq:deltalo}) and (\ref{eq:gpv}), the bosonic fields in the vector multiplet consist of the vectors $F_{\mu\nu}^{a\,(+)}$ along with their scalar superpartners $g_{ij}-b_{ij}$.  This suggests that, at least at leading order, we can truncate out the vector multiplets by taking
\begin{equation}
    g_{ij}=\delta_{ij},\qquad b_{ij}=0,\qquad G_{\mu\nu\,i}=-F_{\mu\nu}^i.
\end{equation}
(Note that, with $g_{ij}=\delta_{ij}$, the internal indices $i,j,\ldots$ are raised and lowered using $\delta_{ij}$.)  However, as an intermediate step, it is instructive to truncate the scalars before considering the gauge fields.  Thus we let
\begin{equation}
    g_{ij}=\delta_{ij},\qquad b_{ij}=0,\qquad F_{\mu\nu}^{(\pm)\,i}=F_{\mu\nu}^i\pm G_{\mu\nu\,i}.
\end{equation}
In this case, the two-derivative equations of motion, (\ref{eq:eom2dred}), (\ref{eq:eom2dredH}), and (\ref{eq:eom2dredphi}), take the form
\begin{align}
    \mathcal E_{g,\alpha\beta}^{(0)}&=R(\omega)_{\alpha\beta}-\fft14\qty(F_{\alpha\gamma}^{(+)\,i}F_{\beta\gamma}^{(+)\,i}+F_{\alpha\gamma}^{(-)\,i}F_{\beta\gamma}^{(-)\,i})-\fft14h_{\alpha\gamma\delta}h_{\beta\gamma\delta}+2\nabla_\alpha\nabla_\beta\varphi,\nn\\
    \mathcal E_{g,\alpha i}^{(0)}&=-\fft14\left(e^{2\varphi}\nabla_\gamma(e^{-2\varphi}F_{\gamma\alpha}^{(+)\,i})+\fft12h_{\alpha\gamma\delta} F_{\gamma\delta}^{(+)\,i}\right)\nn\\
    &\qquad-\fft14\left(e^{2\varphi}\nabla_\gamma(e^{-2\varphi}F_{\gamma\alpha}^{(-)\,i})-\fft12h_{\alpha\gamma\delta} F_{\gamma\delta}^{(-)\,i}\right),\nn\\
    \mathcal E_{g,ij}^{(0)}&=\fft18\qty(F_{\alpha\beta}^{(+)\,i}F_{\alpha\beta}^{(-)\,j}+F_{\alpha\beta}^{(-)\,i}F_{\alpha\beta}^{(+)\,j}),\nn\\
    \mathcal E_\phi^{(0)}&=R(\omega)-\fft18\qty(F_{\alpha\beta}^{(+)\,i}F_{\alpha\beta}^{(+)\,i}+F_{\alpha\beta}^{(-)\,i}F_{\alpha\beta}^{(-)\,i})-\fft1{12}h_{\alpha\beta\gamma}^2+4\Box\varphi-4(\partial_\alpha\varphi)^2,\nn\\
    \mathcal E_{H,\alpha\beta}^{(0)}&=e^{2\varphi}\nabla^\gamma\qty(e^{-2\varphi}h_{\alpha\beta\gamma}),\nn\\
    \mathcal E_{H,\alpha i}^{(0)}&=\fft12\left(e^{2\varphi}\nabla^\gamma(e^{-2\varphi}F_{\gamma\alpha}^{(+)\,i})+\fft12h_{\alpha\gamma\delta}F_{\gamma\delta}^{(+)\,i}\right)\nn\\
    &\qquad-\fft12\left(e^{2\varphi}\nabla^\gamma(e^{-2\varphi}F_{\gamma\alpha}^{(-)\,i})-\fft12h_{\alpha\gamma\delta}F_{\gamma\delta}^{(-)\,i}\right),\nn\\
    \mathcal E_{H,ij}^{(0)}&=\fft14\qty(F_{\alpha\beta}^{(+)\,i}F_{\alpha\beta}^{(-)\,j}-F_{\alpha\beta}^{(-)\,i}F_{\alpha\beta}^{(+)\,j}).
\end{align}

At the bosonic level, we can proceed in two ways, by either setting $F^{(+)}=0$ or $F^{(-)}=0$.  The former case will truncate the gauge fields in the vector multiplet, while the latter will remove the graviphotons, leading to a consistent but non-supersymmetric truncation.  Note, in particular, that the two-derivative bosonic Lagrangian, (\ref{eq:Lags}), is invariant under $B\to-B$.  This is what underlies the symmetry between $F^{(+)}$ and $F^{(-)}$ at the leading order.

We are, of course, mainly interested in a supersymmetric consistent truncation.  Thus we proceed by setting $F^{(+)}=0$.  Specifically, we take
\begin{equation}
    g_{ij}=\delta_{ij},\qquad b_{ij}=0,\qquad A_\mu^i=\fft12A_\mu^{(-)\,i},\qquad B_{\mu\,i}=-\fft12A_\mu^{(-)\,i}.
\label{eq:red2d}
\end{equation}
Doing so then yields the two-derivative equations of motion
\begin{align}
    \mathcal E_{g,\alpha\beta}^{(0)}&=R(\omega)_{\alpha\beta}-\fft14F_{\alpha\gamma}^{(-)\,i}F_{\beta\gamma}^{(-)\,i}-\fft14h_{\alpha\gamma\delta}h_{\beta\gamma\delta}+2\nabla_\alpha\nabla_\beta\varphi,\nn\\
    \mathcal E_{g,\alpha i}^{(0)}&=-\fft14\left(e^{2\varphi}\nabla_\gamma\qty(e^{-2\varphi}F_{\gamma\alpha}^{(-)\,i})-\fft12h_{\alpha\gamma\delta} F_{\gamma\delta}^{(-)\,i}\right),\nn\\
    \mathcal E_{g,ij}^{(0)}&=0,\nn\\
    \mathcal E_\phi^{(0)}&=R(\omega)-\fft18F_{\alpha\beta}^{(-)\,i}F_{\alpha\beta}^{(-)\,i}-\fft1{12}h_{\alpha\beta\gamma}^2+4\Box\varphi-4(\partial_\alpha\varphi)^2,\nn\\
    \mathcal E_{H,\alpha\beta}^{(0)}&=e^{2\varphi}\nabla^\gamma\qty(e^{-2\varphi}h_{\alpha\beta\gamma}),\nn\\
    \mathcal E_{H,\alpha i}^{(0)}&=-\fft12\left(e^{2\varphi}\nabla^\gamma\qty(e^{-2\varphi}F_{\gamma\alpha}^{(-)\,i})-\fft12h_{\alpha\gamma\delta}F_{\gamma\delta}^{(-)\,i}\right),\nn\\
    \mathcal E_{H,ij}^{(0)}&=0.
\label{eq:2dtrunc}
\end{align}
Note, in particular, that the internal Einstein and $H$ equations are trivial and that the mixed Einstein and $H$ equations are consistent with each other.  This set of equations can be derived from the reduced Lagrangian
\begin{equation}
    e^{-1}\mathcal L=e^{-2\varphi}\left(R+4\qty(\partial\varphi)^2-\fft1{12}h_{\mu\nu\rho}^2-\fft18\qty(F_{\mu\nu}^{(-)\,i})^2\right),
\end{equation}
where the $h$ Bianchi identity is given by
\begin{equation}
    h=\dd b+\fft14F^{(-)\,i}\wedge A^{(-)\,i}\qquad\Rightarrow\qquad \dd h=\fft14F^{(-)\,i}\wedge F^{(-)\,i}.
\label{eq:hbian}
\end{equation}
This can equally well be obtained by directly substituting the truncation ansatz, (\ref{eq:red2d}), into the two-derivative Lagrangian (\ref{eq:Lags}).

\subsection{The supergravity truncation at \texorpdfstring{$\mathcal O(\alpha')$}{O(alpha prime)}}

We now wish to extend the truncation of the vector multiplets to the four-derivative level.  Working to $\mathcal O(\alpha')$, the supergravity truncation, (\ref{eq:red2d}), is expected to receive corrections.  With a slight abuse of notation, we thus let
\begin{align}
    &g_{\mu\nu}=g_{\mu\nu}+\alpha'\delta g_{\mu\nu},\qquad b_{\mu\nu}=b_{\mu\nu}+\alpha'\delta b_{\mu\nu},\qquad \varphi=\varphi+\alpha'\delta\varphi,\nn\\
    &A_\mu^i=\fft12A_\mu^{(-)\,i}+\alpha'\delta A_\mu^i,\qquad B_{\mu i}=-\fft12A_\mu^{(-)\,i}+\alpha'\delta B_{\mu i},\nn\\
    &g_{ij}=\delta_{ij}+\alpha'\delta g_{ij},\qquad b_{ij}=0+\alpha'\delta b_{ij}.
\label{eq:corr1}
\end{align}
The equations of motion to $\mathcal O(\alpha')$ then take the form
\begin{equation}
    \mathcal E=\mathcal E^{(0)}+\alpha'\qty(\delta\mathcal E^{(0)}+\mathcal E^{(1)}).
\end{equation}
Here $\delta\mathcal E^{(0)}$ arises from inserting the corrected fields into the two-derivative equations, and $\mathcal E^{(1)}$ can be obtained from inserting the leading order fields into the four-derivative equations.

Extending the leading order equations of motion (\ref{eq:2dtrunc}) to the next order, we see that the necessary conditions for maintaining a consistent truncation are
\begin{equation}
    \delta\mathcal E_{g,ij}^{(0)}+\mathcal E_{g,ij}^{(1)}=0,\qquad \delta\mathcal E_{H,ij}^{(0)}+\mathcal E_{H,ij}^{(1)}=0,
\label{eq:seoms}
\end{equation}
to ensure truncation of the scalars, and
\begin{equation}
    \delta\mathcal E_{g,\alpha i}^{(0)}+\mathcal E_{g,\alpha i}^{(1)}=\fft12\qty(\delta\mathcal E_{H,\alpha i}^{(0)}+\mathcal E_{H,\alpha i}^{(1)}).
\label{eq:max2}
\end{equation}
to ensure truncation of the vector multiplet gauge fields.  Solving these conditions will provide constraints on the correction terms in (\ref{eq:corr1}).

To calculate $\mathcal E^{(1)}$, we only need to work with the leading order truncation.  This simplifies various objects needed in the calculation.  In particular, the torsionful spin connection reduces to
\begin{equation}
    \Omega_+=\begin{pmatrix}\omega_+^{\alpha\beta}-\fft12F_{\alpha\beta}^{(-)\,i}\eta^i&0\\0&0\end{pmatrix}.
\end{equation}
This gives the torsionful Riemann tensor
\begin{align}
    R_{\gamma\delta}{}^{\alpha\beta}(\Omega_+)&=R_{\gamma\delta}{}^{\alpha\beta}(\omega_+)-\fft14F_{\gamma\delta}^{(-)\,i}F_{\alpha\beta}^{(-)\,i},\nn\\
    R_{\gamma d}{}^{\alpha\beta}(\Omega_+)&=-\fft12\delta^i_d\nabla_\gamma^{(+)}F_{\alpha\beta}^{(-)\,i},\nn\\
    R_{cd}{}^{\alpha\beta}(\Omega_+)&=\fft12\delta_c^{[i}\delta_d^{j]}F_{\alpha\gamma}^{(-)\,i}F_{\gamma\beta}^{(-)\,j},
\end{align}
and Lorentz Chern-Simons form
\begin{align}
    \omega_{3L,\alpha\beta\gamma}(\Omega_+)&=\omega_{3L,\alpha\beta\gamma}(\omega_+),\nn\\
    \omega_{3L,\alpha\beta c}(\Omega_+)&=\delta^i_c\qty(R_{\alpha\beta}{}^{\gamma\delta}(\omega_+)F_{\gamma\delta}^{(-)\,i}-\fft18F_{\alpha\beta}^{(-)\,j}F_{\gamma\delta}^{(-)\,j}F_{\gamma\delta}^{(-)\,i}),\nn\\
    \omega_{3L,\alpha bc}(\Omega_+)&=\delta^{[i}_b\delta^{j]}_c\qty(\fft12F_{\gamma\delta}^{(-)\,i}\nabla_\alpha^{(+)}F_{\gamma\delta}^{(-)\,j}),\nn\\
    \omega_{3L,abc}(\Omega_+)&=\delta_a^{[i}\delta_b^j\delta_c^{k]}\qty(-\fft12F_{\alpha\beta}^{(-)\,i}F_{\beta\gamma}^{(-)\,j}F_{\gamma\alpha}^{(-)\,k}).
\label{eq:LCSf}
\end{align}
Note that we have dropped an exact term from $\omega_{3L}(\Omega_+)$, which is implicitly absorbed into a field redefinition of $B$.  For details, see Appendix~\ref{sec:CSterm}

\subsubsection{Truncating the internal Einstein equation}
We first check the scalar equations of motion (\ref{eq:seoms}), corresponding to the internal Einstein equation.  Starting with the $\mathcal E_{g,MN}^{(1)}$ from (\ref{eq:eom4d}), we find
\begin{align}
    \mathcal E_{g,ij}^{(1)}&=\fft1{16}\Bigl[-R_{\alpha\beta\gamma\delta}(\omega_+)F_{\alpha\beta}^{(-)\,(i}F_{\gamma\delta}^{(-)\,j)}+\nabla_\gamma^{(+)}F_{\alpha\beta}^{(-)\,i}\nabla_\gamma^{(+)}F_{\alpha\beta}^{(-)\,j}+\fft18F_{\alpha\beta}^{(-)\,i}F_{\alpha\beta}^{(-)\,k}F_{\gamma\delta}^{(-)\,j}F_{\gamma\delta}^{(-)\,k}\nn\\
    &\qquad+\fft12F_{\alpha\beta}^{(-)\,i}F_{\beta\gamma}^{(-)\,j}F_{\gamma\delta}^{(-)\,k}F_{\delta\alpha}^{(-)\,k}-\fft12F_{\alpha\beta}^{(-)\,i}F_{\beta\gamma}^{(-)\,k}F_{\gamma\delta}^{(-)\,j}F_{\delta\alpha}^{(-)\,k}\Bigr].
\end{align}
Since this is non-zero, it would have to cancel against a similar expression in $\delta\mathcal E_{g,ij}^{(0)}$.  To find this correction, we need to start from the full expression for $\mathcal E_{g,ij}^{(0)}$ in (\ref{eq:eom2dred}).  To first order, we find
\begin{align}
    \delta\mathcal E_{g,ij}^{(0)}&=-\fft12e^{2\varphi}\partial_\gamma\qty(e^{-2\varphi}\partial_\gamma\delta g_{ij})+\fft18F_{\alpha\beta}^{(-)\,k}F_{\alpha\beta}^{(-)\,j}\delta g_{ik}\nn\\
    &\qquad+\fft14F_{\alpha\beta}^{(-)\,i}\qty(\delta F_{\alpha\beta}^j+\delta G_{\alpha\beta\,j}-\fft12F_{\alpha\beta}^{(-)\,k}\delta b_{jk}),
\label{eq:dEgij}
\end{align}
where symmetry of $(ij)$ is implied.  Our main focus is on $\delta g_{ij}$ and $\delta b_{ij}$.  Since these carry $i$ and $j$ indices, and since we want them to be two-derivative terms, we expect them to be built out of bilinears in the field strengths, $F_{\alpha\beta}^{(-)\,i}$.  We will confirm below that an appropriate choice is to take
\begin{equation}
    \delta g_{ij}=\fft1{16}F_{\alpha\beta}^{(-)\,i}F_{\alpha\beta}^{(-)\,j},\qquad\delta b_{ij}=0.
\label{eq:deltagb}
\end{equation}
Note that there is no obvious antisymmetric choice for $\delta b_{ij}$, so the only natural result is to set it to zero.

After some manipulation, we find
\begin{align}
    -\fft12e^{2\varphi}\partial_\gamma\qty(e^{-2\varphi}\partial_\gamma\delta g_{ij})&=-\fft1{16}\Bigl[-R_{\alpha\beta\gamma\delta}F_{\alpha\beta}^{(-)\,i}F_{\gamma\delta}^{(-)\,j}+\nabla_\gamma F_{\alpha\beta}^{(-)\,i}\nabla_\gamma F_{\alpha\beta}^{(-)\,j}\nn\\
    &\qquad\qquad+2F_{\alpha\gamma}^{(-)\,i}F_{\beta\gamma}^{(-)\,j}\qty(R_{\alpha\beta}+2\nabla_\alpha\nabla_\beta\varphi)\nn\\
    &\qquad\qquad+2F_{\alpha\beta}^{(-)\,i}\nabla_\alpha\qty(e^{2\varphi}\nabla_\gamma(e^{-2\varphi}F_{\gamma\beta}^{(-)\,j}))\Bigl].
\end{align}
The terms in parentheses in the second line are almost the leading order equations of motion (\ref{eq:2dtrunc}), but are missing a few terms.  By adding and subtracting, we can arrive at
\begin{align}
    -\fft12e^{2\varphi}\partial_\gamma\qty(e^{-2\varphi}\partial_\gamma\delta g_{ij})&=-\fft1{16}\Bigl[-R_{\alpha\beta\gamma\delta}F_{\alpha\beta}^{(-)\,i}F_{\gamma\delta}^{(-)\,j}+\nabla_\gamma F_{\alpha\beta}^{(-)\,i}\nabla_\gamma F_{\alpha\beta}^{(-)\,j}\nn\\
    &\qquad\qquad+\fft12F_{\alpha\beta}^{(-)\,i}F_{\beta\gamma}^{(-)\,j}F_{\gamma\delta}^{(-)\,k}F_{\delta\alpha}^{(-)\,k}+\fft12h_{\alpha\delta\epsilon}h_{\beta\delta\epsilon}F_{\alpha\gamma}^{(-)\,i}F_{\beta\gamma}^{(-)\,j}\nn\\
    &\qquad\qquad+F_{\alpha\beta}^{(-)\,i}F_{\gamma\delta}^{(-)\,j}\nabla_\alpha h_{\beta\gamma\delta}+h_{\beta\gamma\delta}F_{\alpha\beta}^{(-)\,i}\nabla_\alpha F_{\gamma\delta}^{(-)\,j}\nn\\
    &\qquad\qquad+2F_{\alpha\gamma}^{(-)\,i}F_{\beta\gamma}^{(-)\,j}\mathcal E_{\alpha\beta}^{(0)}+2F_{\alpha\beta}^{(-)\,i}\nabla_\alpha\mathcal E_\beta^{(0)\,j}\Bigr],
\end{align}
where we have normalized the graviphoton equation of motion according to $\mathcal E_\alpha^{(0)\,i}=-4\mathcal E_{g,\alpha i}^{(0)}=-2\mathcal E_{H,\alpha i}^{(0)}$.  We can rewrite the torsion-free Riemann and covariant derivatives in terms of their torsionful versions.  The result is
\begin{align}
    -\fft12e^{2\varphi}\partial_\gamma\qty(e^{-2\varphi}\partial_\gamma\delta g_{ij})&=-\fft1{16}\Bigl[-R_{\alpha\beta\gamma\delta}(\omega_+)F_{\alpha\beta}^{(-)\,i}F_{\gamma\delta}^{(-)\,j}+\nabla_\gamma^{(+)}F_{\alpha\beta}^{(-)\,i}\nabla_\gamma^{(+)}F_{\alpha\beta}^{(-)\,j}\nn\\
    &\qquad\qquad+\fft12F_{\alpha\beta}^{(-)\,i}F_{\beta\gamma}^{(-)\,j}F_{\gamma\delta}^{(-)\,k}F_{\delta\alpha}^{(-)\,k}+2F_{\alpha\beta}^{(-)\,i}F_{\gamma\delta}^{(-)\,j}\nabla_\alpha h_{\beta\gamma\delta}\nn\\
    &\qquad\qquad+2F_{\alpha\gamma}^{(-)\,i}F_{\beta\gamma}^{(-)\,j}\mathcal E_{\alpha\beta}^{(0)}+2F_{\alpha\beta}^{(-)\,i}\nabla_\alpha\mathcal E_\beta^{(0)\,j}\Bigr],
\end{align}
Taking into account the implicit symmetrization of $(ij)$, the term involving $\nabla_\alpha h_{\beta\gamma\delta}$ can be simplified using the Bianchi identity (\ref{eq:hbian}).  The result is then
\begin{align}
    -\fft12e^{2\varphi}\partial_\gamma\qty(e^{-2\varphi}\partial_\gamma\delta g_{ij})=&-\fft1{16}\Bigl[-R_{\alpha\beta\gamma\delta}(\omega_+)F_{\alpha\beta}^{(-)\,i}F_{\gamma\delta}^{(-)\,j}+\nabla_\gamma^{(+)}F_{\alpha\beta}^{(-)\,i}\nabla_\gamma^{(+)}F_{\alpha\beta}^{(-)\,j}\nn\\
    &\qquad\quad+\fft14F_{\alpha\beta}^{(-)\,i}F_{\alpha\beta}^{(-)\,k}F_{\gamma\delta}^{(-)\,j}F_{\gamma\delta}^{(-)\,k}+\fft12F_{\alpha\beta}^{(-)\,i}F_{\beta\gamma}^{(-)\,j}F_{\gamma\delta}^{(-)\,k}F_{\delta\alpha}^{(-)\,k}\nn\\
    &\qquad\quad-\fft12F_{\alpha\beta}^{(-)\,i}F_{\beta\gamma}^{(-)\,k}F_{\gamma\delta}^{(-)\,j}F_{\delta\alpha}^{(-)\,k}+2F_{\alpha\gamma}^{(-)\,i}F_{\beta\gamma}^{(-)\,j}\mathcal E_{\alpha\beta}^{(0)}\nn\\
    &\qquad\quad+2F_{\alpha\beta}^{(-)\,i}\nabla_\alpha\mathcal E_\beta^{(0)\,j}\Bigl],
\end{align}
Inserting this into (\ref{eq:dEgij}) and taking into account the second term in (\ref{eq:dEgij}) as well, we find
\begin{align}
    \delta\mathcal E_{g,ij}^{(0)}&=-\fft1{16}\Bigl[-R_{\alpha\beta\gamma\delta}(\omega_+)F_{\alpha\beta}^{(-)\,i}F_{\gamma\delta}^{(-)\,j}+\nabla_\gamma^{(+)}F_{\alpha\beta}^{(-)\,i}\nabla_\gamma^{(+)}F_{\alpha\beta}^{(-)\,j}+\fft18F_{\alpha\beta}^{(-)\,i}F_{\alpha\beta}^{(-)\,k}F_{\gamma\delta}^{(-)\,j}F_{\gamma\delta}^{(-)\,k}\nn\\
    &\qquad+\fft12F_{\alpha\beta}^{(-)\,i}F_{\beta\gamma}^{(-)\,j}F_{\gamma\delta}^{(-)\,k}F_{\delta\alpha}^{(-)\,k}-\fft12F_{\alpha\beta}^{(-)\,i}F_{\beta\gamma}^{(-)\,k}F_{\gamma\delta}^{(-)\,j}F_{\delta\alpha}^{(-)\,k}\nn\\
    &\qquad+2F_{\alpha\gamma}^{(-)\,i}F_{\beta\gamma}^{(-)\,j}\mathcal E_{\alpha\beta}^{(0)}+2F_{\alpha\beta}^{(-)\,i}\nabla_\alpha\mathcal E_\beta^{(0)\,j}-4F_{\alpha\beta}^{(-)\,i}\qty(\delta F_{\alpha\beta}^j+\delta G_{\alpha\beta\,j})\Bigr],
\end{align}
As a result, we are left with
\begin{equation}
     \delta\mathcal E_{g,ij}^{(0)}+\mathcal E_{g,ij}^{(1)}=-\fft18\left[F_{\alpha\gamma}^{(-)\,i}F_{\beta\gamma}^{(-)\,j}\mathcal E_{\alpha\beta}^{(0)}+F_{\alpha\beta}^{(-)\,i}\nabla_\alpha\mathcal E_\beta^{(0)\,j}-2F_{\alpha\beta}^{(-)\,i}\qty(\delta F_{\alpha\beta}^j+\delta G_{\alpha\beta\,j})\right],
\end{equation}
which vanishes by the leading order equations of motion provided
\begin{equation}
    \delta F_{\alpha\beta}^j+\delta G_{\alpha\beta\,j}=0.
\label{eq:deltaFG}
\end{equation}

\subsubsection{Truncating the internal \texorpdfstring{$H$}{H} equation}

We now turn to the internal components of the $H$ equation of motion (\ref{eq:seoms}).  For $\mathcal E_{H,ij}^{(1)}$, we find
\begin{align}
    \mathcal E_{H,ij}^{(1)}&=-\fft14\Bigg[\fft12e^{2\phi}\nabla_\gamma\qty(e^{-2\phi}F_{\alpha\beta}^{(-)\,i}\nabla_\gamma F_{\alpha\beta}^{(-)\,j})+\fft12e^{2\phi}\nabla_\gamma\qty(e^{-2\phi}h_{\gamma\alpha\beta}F_{\alpha\delta}^{(-)\,i}F_{\beta\delta}^{(-)\,j})\nn\\
    &\qquad\qquad-\fft12\nabla_\gamma h_{\delta\alpha\beta}F_{\gamma\delta}^{(-)\,i}F_{\alpha\beta}^{(-)\,j}\Bigg].
\label{eq:cEHij1}
\end{align}
Note that here we implicitly assume antisymmetry on $[ij]$.  This antisymmetry will be very useful in making many terms disappear.  Along with $\mathcal E_{H,ij}^{(1)}$, we also have
\begin{align}
    \delta\mathcal E_{H,ij}^{(0)}&=e^{2\varphi}\nabla^\gamma\qty(e^{-2\varphi}\nabla_\gamma\delta b_{ij})+\fft14F_{\alpha\beta}^{(-)\,j}F_{\alpha\beta}^{(-)\,k}\delta g_{ik}\nn\\
    &\qquad+\fft12F_{\alpha\beta}^{(-)\,j}\qty(\delta F_{\alpha\beta}^i+\delta G_{\alpha\beta\,i}-\fft12F_{\alpha\beta}^{(-)\,k}\delta b_{ik}).
\end{align}
If we take (\ref{eq:deltagb}) for the corrections $\delta g_{ij}$ and $\delta b_{ij}$, along with (\ref{eq:deltaFG}), we see that this actually vanishes, namely $\delta\mathcal E_{H,ij}^{(0)}=0$.  Thus, to be consistent, we then need to have $\mathcal E_{H,ij}^{(1)}$ vanishing as well.  To see that this is indeed the case, we can manipulate (\ref{eq:cEHij1}) by expanding out the $\nabla_\gamma$ derivative in the first two terms while making use of antisymmetry on $[ij]$
\begin{align}
    \mathcal E_{H,ij}^{(1)}&=-\fft14\Bigg[\fft12e^{2\phi}F_{\alpha\beta}^{(-)\,i}\nabla_\gamma\qty(e^{-2\phi}\nabla_\gamma F_{\alpha\beta}^{(-)\,j})+\fft12e^{2\phi}\nabla_\gamma\qty(e^{-2\phi}h_{\gamma\alpha\beta})F_{\alpha\delta}^{(-)\,i}F_{\beta\delta}^{(-)\,j}\nn\\
    &\qquad+h_{\gamma\alpha\beta}\nabla_\gamma\qty(F_{\alpha\delta}^{(-)\,i}F_{\beta\delta}^{(-)\,j})-\fft12\nabla_\gamma h_{\delta\alpha\beta}F_{\gamma\delta}^{(-)\,i}F_{\alpha\beta}^{(-)\,j}\Bigg].
\end{align}
We can rewrite the first term using the Bianchi identity $\dd F^{(-)\,j}=0$
\begin{equation}
    \fft12e^{2\phi}F_{\alpha\beta}^{(-)\,i}\nabla_\gamma\qty(e^{-2\phi}\nabla_\gamma F_{\alpha\beta}^{(-)\,j})=e^{2\phi}F_{\alpha\beta}^{(-)\,i}\nabla_\gamma\qty(e^{-2\phi}\nabla_\alpha F_{\gamma\beta}^{(-)\,j}).
\end{equation}
After moving $\nabla_\alpha$ past the dilaton factor and commuting it past the $\nabla_\gamma$, we end up with part of the graviphoton equation of motion, $\mathcal E_\beta^{(0)\,j}$.  Collecting terms and simplifying then gives
\begin{equation}
    \mathcal E_{H,ij}^{(1)}=-\fft14\left[F_{\alpha\beta}^{(-)\,i}\nabla_\alpha\mathcal E_\beta^{(0)\,j}+\fft12\mathcal E_{H,\alpha\beta}^{(0)}F_{\alpha\delta}^{(-)\,i}F_{\beta\delta}^{(-)\,j}\right].
\end{equation}
This now vanishes by the lowest order equations of motion.

\subsubsection{Compatibility of the Maxwell equations}

The final expression to verify is (\ref{eq:max2}), namely the consistency of the two Maxwell equations. The shifts of the two-derivative equations of motion are straightforwardly found to be
\begin{align}
    \delta\mathcal E^{(0)}_{g,\alpha a}&=\delta e^i_a\qty[\frac{1}{2}e^{2\varphi}\nabla^\gamma\qty(e^{-2\varphi}g_{ij}F^{j}_{\alpha\gamma})-\frac{1}{4}h_{\alpha\gamma\delta}G^{\gamma\delta}_i]+\frac{1}{2}e^i_ae^{2\varphi}\nabla^\gamma\qty(e^{-2\varphi}\delta g_{ij} F^j_{\alpha\gamma}),\nonumber\\
    \delta\mathcal E^{(0)}_{H,\alpha a}&=\delta e^a_i\qty[e^{2\varphi}\nabla^\gamma\qty(e^{-2\varphi}g^{ij}G_{j\alpha\gamma})+\frac{1}{2}h_{\alpha\gamma\delta}F^{i\gamma\delta}]+e^a_ie^{2\varphi}\nabla^\gamma\qty(e^{-2\varphi}\delta g^{ij}G_{j\alpha\gamma}),
\end{align}
so that the difference, after imposing our truncation, is simply
\begin{align}
    \delta\mathcal E^{(0)}_{H,\alpha a}-2\delta\mathcal E^{(0)}_{g,\alpha a}=&\delta e^a_i \delta^i_a\mathcal E^{(0)}_{H,\alpha a}-2\delta e^i_a\delta^a_i\mathcal E^{(0)}_{g,\alpha a}+\frac{1}{32}e_i^a h_{\alpha\gamma\epsilon}F^{(-)\,j}_{\gamma\epsilon}F^{(-)\,i}_{\beta\delta}F^{(-)\,j}_{\beta\delta}\nn\\
    &+\frac{1}{8}\delta_i^a F^{(-)\,j}_{\gamma\alpha}F^{(-)\,(i}_{\beta\delta}\nabla_\gamma F^{(-)\,j)}_{\beta\delta}.
\end{align}
The four-derivative parts of the equations of motion are simply
\begin{align}
    \mathcal E^{(1)}_{g,\alpha a}=&\delta_i^a\bigg[\frac{1}{16} R_{\beta\gamma} ^{\ \ \ \delta\epsilon}(\omega_+)h_{\alpha\beta\gamma}F_{\delta\epsilon}^{(-)\,i}+\frac{1}{8}R_{\alpha\beta}^{\ \ \gamma\delta}(\omega_+)\nabla_\beta^{(+)}F^{(-)\,i}_{\gamma\delta}-\frac{1}{32}F^{(-)\,i}_{\beta\gamma}\omega_{3L,\alpha\beta\gamma}(\omega_+)\nn\\
    &\ \ \ -\frac{1}{128} h_{\alpha\beta\gamma}F^{(-)\,j}_{\beta\gamma}F^{(-)\,j}_{\delta\epsilon}F^{(-)\,i}_{\delta\epsilon}-\frac{1}{16}F^{(-)\,[i}_{\gamma\epsilon}F^{(-)\,j]}_{\epsilon\delta}\nabla_\alpha^{(+)}F_{\gamma\delta}^{(-)\,j}\bigg],
\end{align}
and
\begin{align}
    \mathcal E^{(1)}_{H,\alpha a}&=-\frac{1}{4}\delta_i^a e^{2\varphi}\nabla^\beta\qty[e^{-2\varphi}\qty(R_{\beta\alpha}^{\ \ \ \gamma\delta}(\omega_+)F_{\gamma\delta}^{(-)\,i}-\frac{1}{8}F^{(-)\,j}_{\beta\alpha}F^{(-)\,j}_{\gamma\delta}F^{(-)\,i}_{\gamma\delta})]\nn\\
    &\qquad-\frac{1}{16}\delta_i^aF^{(-)\,i}_{\beta\gamma}\omega_{3L,\alpha\beta\gamma}(\omega_+).
\end{align}
By use of the torsion-free differential Bianchi identity $\nabla_{[\alpha} R_{\beta\gamma]\delta\epsilon}=0$, we have that \begin{equation}
    \nabla^\beta R_{\alpha\beta\gamma\delta}=-2\nabla_{[\gamma}R_{\delta]\alpha},
\end{equation} 
and so, after appropriate substitution of equations of motion and use of the $h$ Bianchi identity \eqref{eq:hbian}, we find that
\begin{align}
&e^{2\varphi}\nabla^\beta\qty(e^{-2\varphi}R_{\beta\alpha}^{\ \ \ \gamma\delta}(\omega_+)F_{\gamma\delta}^{(-)\,i})=\nn\\
&\qquad F^{(-)\,i}_{\gamma\delta}\Bigg[-\frac{1}{4}\nabla_\gamma\qty(F^{(-)\,j}_{\delta\epsilon}F^{(-)\,j}_{\epsilon\alpha})-\frac{1}{4}\nabla_\gamma\qty(h_{\delta\beta\epsilon}h_{\alpha\beta\epsilon})+\frac{1}{2}R_{\gamma[\delta|\beta\epsilon}h_{|\alpha]\beta\epsilon}-\frac{1}{8}h_{\delta\alpha\epsilon}F^{(-)j}_{\gamma\beta}F^{(-)j}_{\epsilon\beta}\nn\\
&\qquad\qquad\quad-\frac{1}{8}h_{\delta\alpha\epsilon}h_{\gamma\beta\omega}h_{\epsilon\beta\omega}-\frac{1}{16}h_{\alpha\beta\epsilon}F^{(-)\,j}_{\beta\epsilon}F^{(-)\,j}_{\gamma\delta}+\frac{1}{8}F^{(-)\,j}_{\alpha\beta}\nabla^\beta F^{(-)\,j}_{\gamma\delta}\nn\\
&\qquad\qquad\quad-\frac{1}{8}h_{\delta\beta\epsilon}F^{(-)\,j}_{\beta\epsilon}F^{(-)\,j}_{\alpha\gamma}-\frac{1}{4}F^{(-)\,j}_{\beta\delta}\nabla^\beta F^{(-)\,j}_{\alpha\gamma}+\frac{1}{4}h_{\beta\epsilon\delta}\nabla^\beta h_{\alpha\gamma\epsilon}-\nabla_{[\gamma}\mathcal E^{(0)}_{g,\delta]\alpha}\nn\\
&\qquad\qquad\quad-\frac{1}{2}\nabla_\gamma\mathcal E^{(0)}_{H,\delta\alpha}-\partial_\gamma\varphi\mathcal E^{(0)}_{H,\delta\alpha}-\frac{1}{2}h_{\delta\alpha\epsilon}\mathcal E^{(0)}_{H,\gamma\epsilon}+\frac{1}{4}h_{\alpha\gamma\epsilon}\mathcal E^{(0)}_{H,\epsilon\delta}\Bigg]\nn\\
&\qquad+R_{\alpha\beta}^{\ \ \ \gamma\delta}(\omega_+)\nabla^\beta F^{(-)\,i}_{\gamma\delta}.
\end{align}
We then substitute this back into $\mathcal E^{(0)}_{H,\alpha a}$, and take the difference $\qty(\delta \mathcal E^{(0)}_{H,\alpha a}+\mathcal E^{(1)}_{H,\alpha a})-2\qty(\delta \mathcal E^{(0)}_{g,\alpha a}+\mathcal E^{(1)}_{g,\alpha a})$. We then expand out the torsionful Riemann tensors as
\begin{equation}
    R_{\alpha\beta}^{\ \ \ \gamma\delta}(\omega_+)=R_{\alpha\beta}^{\ \ \ \gamma\delta}+\nabla_{[\alpha}h_{\beta]}^{\ \ \gamma\delta}+\frac{1}{2}h_{[\alpha}^{\ \ \gamma\epsilon}h_{\beta]}^{\ \ \epsilon\delta},
\end{equation}
and use the Riemann algebraic Bianchi identity $R_{[\alpha\beta\gamma]\delta}=0$, as well as the $h$ Bianchi identity \eqref{eq:hbian} and the $F$ Bianchi identity $\dd F^{(-)\,i}=0$, to get that
\begin{align}
    &\qty(\delta \mathcal E^{(0)}_{H,\alpha a}+\mathcal E^{(1)}_{H,\alpha a})-2\qty(\delta \mathcal E^{(0)}_{g,\alpha a}+\mathcal E^{(1)}_{g,\alpha a})=\nn\\
    &\qquad-\frac{1}{4}\delta_i^aF^{(-)\,i}_{\gamma\delta}\Bigg[-\nabla_{[\gamma}\mathcal E^{(0)}_{g,\delta]\alpha}-\frac{1}{2}\nabla_\gamma\mathcal E^{(0)}_{H,\delta\alpha}-\partial_\gamma\varphi\mathcal E^{(0)}_{H,\delta\alpha}-\frac{1}{2}h_{\delta\alpha\epsilon}\mathcal E^{(0)}_{H,\gamma\epsilon}+\frac{1}{4}h_{\alpha\gamma\epsilon}\mathcal E^{(0)}_{H,\epsilon\delta}\Bigg]\nn\\
    &\qquad+\delta e^a_i \delta^i_a\mathcal E^{(0)}_{H,\alpha a}-2\delta e^i_a\delta^a_i\mathcal E^{(0)}_{g,\alpha a},
\end{align}
which vanishes after the application of the two-derivative equations of motion. This verifies \eqref{eq:max2}. Hence, the truncation of the vector multiplet is consistent with the bosonic equations of motion.

\subsubsection{The surviving equations of motion}

The above shows that it is consistent to truncate away the bosonic equations of motion related to the reduced vector multiplet fields, namely $\mathcal E_{g,ij}$, $\mathcal E_{H,ij}$ and $\mathcal E_{H,\alpha i}-2\mathcal E_{g,\alpha i}$, corresponding to the equations of motion for $g_{ij}$, $b_{ij}$ and $F_{\mu\nu}^{(+)\,i}$, respectively.  The remaining untruncated equations of motion are those of the reduced supergravity multiplet fields.  These are the Einstein equation, $\mathcal E_{g,\alpha\beta}$, dilaton equation, $\mathcal E_\phi$, $h$-field equation, $\mathcal E_{H,\alpha\beta}$ and graviphoton equation, $-\mathcal E_{H,\alpha i}-2\mathcal E_{g,\alpha i}$.

Combined with the two-derivative equations of motion in (\ref{eq:2dtrunc}), the reduced Einstein equation is
\begin{align}
    \mathcal E_{g,\alpha\beta}&=R(\omega)_{\alpha\beta}-\fft14F_{\alpha\gamma}^{(-)\,i}F_{\beta\gamma}^{(-)\,i}-\fft14\tilde h_{\alpha\gamma\delta}\tilde h_{\beta\gamma\delta}+2\nabla_\alpha\nabla_\beta\varphi\nn\\
    &\quad+\frac{\alpha'}{4}\biggl(R_{\alpha\gamma\delta\epsilon}(\omega_+)R_\beta{}^{\gamma\delta\epsilon}(\omega_+)-R_{\alpha}{}^{\gamma\delta\epsilon}(\omega_+)F^{(-)\,i}_{\beta\gamma}F^{(-)\,i}_{\delta\epsilon}\nn\\
    &\kern4em+\frac{1}{8}F^{(-)\,i}_{\alpha\gamma}F_{\beta}{}^{\gamma(-)\,j}F^{(-)\,i}_{\delta\epsilon}F^{\delta\epsilon(-)\,j}+\frac{1}{4}\nabla^{(+)}_\alpha F^{(-)\,i}_{\gamma\delta}\nabla^{(+)}_
    \beta F^{\gamma\delta(-)\,i}\biggr),
\label{eq:redeins}
\end{align}
where symmetrization on $(\alpha\beta)$ is implicitly assumed.  Here we have introduced
\begin{equation}
    \tilde h=h-\fft{\alpha'}4\omega_{3L}(\omega_+)=\dd b+\fft14A^{(-)\,i}\wedge F^{(-)\,i}-\fft{\alpha'}4\omega_{3L}(\omega_+),
\label{eq:tildeH}
\end{equation}
such that
\begin{equation}
    \dd\tilde h=\fft14F^{(-)\,i}\wedge F^{(-)\,i}-\fft{\alpha'}4\Tr R(\omega_+)\wedge R(\omega_+).
\end{equation}
This reduced $\tilde h$ has the simple equation of motion
\begin{equation}
    \mathcal E_{H,\alpha\beta}=e^{2\varphi}\nabla^\gamma\qty(e^{-2\varphi}\tilde h_{\alpha\beta\gamma}).
\label{eq:redH1}
\end{equation}

The remaining equations of motion are the dilaton equation
\begin{align}
    \mathcal E_\phi&=R(\omega)-\fft18F_{\alpha\beta}^{(-)\,i}F_{\alpha\beta}^{(-)\,i}-\fft1{12}\tilde h_{\alpha\beta\gamma}^2+4\Box\varphi-4(\partial_\alpha\varphi)^2,\nn\\
    &\quad+\fft{\alpha'}8\biggl((R_{\alpha\beta\gamma\delta}(\omega_+))^2-R_{\alpha\beta\gamma\delta}(\omega_+)F_{\alpha\beta}^{(-)\,i}F_{\gamma\delta}^{(-)\,i}+\fft12\qty(\nabla_\alpha^{(+)}F_{\beta\gamma}^{(-)\,i})^2\nn\\
    &\kern3em+\fft18F_{\alpha\beta}^{(-)\,i}F_{\beta\gamma}^{(-)\,i}F_{\gamma\delta}^{(-)\,j}F_{\delta\alpha}^{(-)\,j}-\fft18F_{\alpha\beta}^{(-)\,i}F_{\beta\gamma}^{(-)\,j}F_{\gamma\delta}^{(-)\,i}F_{\delta\alpha}^{(-)\,j}\nn\\
    &\kern3em+\fft18F_{\alpha\beta}^{(-)\,i}F_{\alpha\beta}^{(-)\,j}F_{\gamma\delta}^{(-)\,i}F_{\gamma\delta}^{(-)\,j}
    \biggr),
\label{eq:reddil1}
\end{align}
and the graviphoton equation
\begin{align}
    \mathcal E_{A,\alpha i}&=e^{2\varphi}\nabla^\gamma\qty(e^{-2\varphi}F_{\gamma\alpha}^{(-)\,i})-\fft12\tilde h_{\alpha\beta\gamma}F_{\beta\gamma}^{(-)\,i}\nn\\
    &\quad+\fft{\alpha'}4\biggl(-\tilde h_{\alpha\beta\gamma}R_{\beta\gamma\delta\epsilon}(\omega_+)F_{\delta\epsilon}^{(-)\,i}+\fft14\tilde h_{\alpha\beta\gamma}F_{\beta\gamma}^{(-)\,j}F_{\delta\epsilon}^{(-)\,i}F_{\delta\epsilon}^{(-)\,j}-2R_{\alpha\beta\gamma\delta}(\omega_+)\nabla_\beta^{(+)}F_{\gamma\delta}^{(-)\,i}\nn\\
    &\kern4em+\fft12F_{\alpha\gamma}^{(-)\,j}F_{\delta\epsilon}^{(-)\,j}\nabla_\gamma^{(+)}F_{\delta\epsilon}^{(-)\,i}-\fft12F_{\alpha\gamma}^{(-)\,j}F_{\delta\epsilon}^{(-)\,i}\nabla_\gamma^{(+)}F_{\delta\epsilon}^{(-)\,j}\nn\\
    &\kern4em+F_{\beta\gamma}^{(-)\,i}F_{\gamma\delta}^{(-)\,j}\nabla_\alpha^{(+)}F_{\beta\delta}^{(-)\,j}\biggr).
\label{eq:redA1}
\end{align}
Note that, at the order we are working at, we cannot distinguish between $h$ and $\tilde h$ in the $\mathcal O(\alpha')$ terms.  Nevertheless, the use of $\tilde h$ is expected to be natural when extending to additional higher orders in $\alpha'$.

\subsection{The reduced Lagrangian}

Since a torus reduction is known to be consistent, the reduced equations of motion will necessarily be consistent with the Kaluza-Klein reduced Lagrangian.  With the further consistent truncation of the vector multiplets, the supergravity multiplet equations of motion, (\ref{eq:redeins}), (\ref{eq:redH1}), (\ref{eq:reddil1}), and (\ref{eq:redA1}), may be derived from the effective Lagrangian given by substituting the truncation into our Lagrangian \eqref{eq:Lhet}, which gives
\begin{align}
    e^{-1}\mathcal L=&e^{-2\varphi}\biggl[R+4\qty(\partial\varphi)^2-\fft1{12}\tilde h_{\mu\nu\rho}^2-\fft18\qty(F_{\mu\nu}^{(-)\,i})^2\nn\\
    &\qquad+\frac{\alpha'}{8}\biggl(\qty(R_{\mu\nu\rho\sigma}(\omega_+))^2-R^{\mu\nu\rho\sigma}(\omega_+)F_{\mu\nu}^{(-)\,i}F_{\rho\sigma}^{(-)\,i} +\frac{1}{2}\qty(\nabla_\rho^{(+)}F_{\mu\nu}^{(-)\,i})^2\nn\\
    &\kern5em+\frac{1}{8}F_\mu{}^{\nu\,(-)\,i}F_\nu{}^{\rho(-)\,i}F_\rho{}^{\sigma\,(-)\,j}F_\sigma{}^{\mu\,(-)\,j}-\frac{1}{8}F_\mu{}^{\nu\,(-)\,i}F_\nu{}^{\rho(-)\,j}F_\rho{}^{\sigma\,(-)\,i}F_\sigma{}^{\mu\,(-)\,j}\nn\\
    &\kern5em+\frac{1}{8}F_{\mu\nu}^{(-)\,i}F^{\mu\nu\,(-)\,j}F^{(-)\,i}_{\rho\sigma}F^{\rho\sigma\,(-)\,j}\biggr)\biggr].
\end{align}
While higher-derivative Lagrangians can be transformed by field redefinitions, it is important to note that such field redefinitions will also transform the equations of motion.  This form of the Lagrangian is what matches the equations of motion given above.

Using our freedom to perform field redefinitions, we may transform this Lagrangian into a more standard form.  In particular, we can rewrite the $(\nabla F)^2$ term using the identity
\begin{align}
    \qty(\nabla_\rho^{(+)}F_{\mu\nu}^{(-)\,i})^2&=R^{\mu\nu\rho\sigma}(\omega_+)F_{\mu\nu}^{(-)\,i}F_{\rho\sigma}^{(-)\,i}-\fft14F_{\mu\nu}^{(-)\,i}F^{\mu\nu\,(-)\,j}F^{(-)\,i}_{\rho\sigma}F^{\rho\sigma\,(-)\,j}\nn\\
    &\quad-\fft12F_\mu{}^{\nu\,(-)\,i}F_\nu{}^{\rho(-)\,i}F_\rho{}^{\sigma\,(-)\,j}F_\sigma{}^{\mu\,(-)\,j}+\fft12F_\mu{}^{\nu\,(-)\,i}F_\nu{}^{\rho(-)\,j}F_\rho{}^{\sigma\,(-)\,i}F_\sigma{}^{\mu\,(-)\,j}\nn\\
    &\quad+2e^{2\varphi}\nabla_\rho(e^{-2\varphi}F_{\mu\nu}^{(-)\,i}\nabla^\mu F^{\rho\nu\,(-)\,i})-2F^{\mu\nu\,(-)\,i}\nabla_\mu\mathcal E_\nu^{(0)\,i}\nn\\
    &\quad-2F_{\mu\rho}^{(-)\,i}F_\nu{}^{\rho\,(-)\,i}\mathcal E_{g,\mu\nu}^{(0)}.
\end{align}
Here, the last line includes a total derivative and terms proportional to the leading order equations of motion.  The equation of motion terms can be removed by a suitable field redefinition, in which case we end up with the concise expression
\begin{align}
    e^{-1}\mathcal L=&e^{-2\varphi}\biggl[R+4\qty(\partial\varphi)^2-\fft1{12}\tilde h_{\mu\nu\rho}^2-\fft18\qty(F_{\mu\nu}^{(-)\,i})^2\nn\\
    &\qquad+\frac{\alpha'}{8}\biggl(\qty(R_{\mu\nu\rho\sigma}(\omega_+))^2-\fft12R^{\mu\nu\rho\sigma}(\omega_+)F_{\mu\nu}^{(-)\,i}F_{\rho\sigma}^{(-)\,i} \nn\\
    &\kern2em-\frac{1}{8}F_\mu{}^{\nu\,(-)\,i}F_\nu{}^{\rho(-)\,i}F_\rho{}^{\sigma\,(-)\,j}F_\sigma{}^{\mu\,(-)\,j}+\frac{1}{8}F_\mu{}^{\nu\,(-)\,i}F_\nu{}^{\rho(-)\,j}F_\rho{}^{\sigma\,(-)\,i}F_\sigma{}^{\mu\,(-)\,j}\biggr)\biggr].
\label{eq:Ltruncred}
\end{align}

This reduced Lagrangian can be compared with that of Ref.~\cite{Eloy:2020dko}.  Note, however, that Ref.~\cite{Eloy:2020dko} does not use a torsionful connection.  To make the comparison, we may use the identities
\begin{align}
    R_{\mu\nu\rho\sigma}(\omega_+)^2&=R_{\mu\nu\rho\sigma}^2+\fft12R^{\mu\nu\rho\sigma}h_{\mu\nu\lambda}h_{\rho\sigma}{}^\lambda-\fft18(h^2_{\mu\nu})^2-\fft18h^4\nn\\
    &\quad+\fft3{32}F_{\mu\nu}^{(-)\,i}F^{\mu\nu\,(-)\,j}F^{(-)\,i}_{\rho\sigma}F^{\rho\sigma\,(-)\,j}-\fft3{16}F_\mu{}^{\nu\,(-)\,i}F_\nu{}^{\rho(-)\,j}F_\rho{}^{\sigma\,(-)\,i}F_\sigma{}^{\mu\,(-)\,j}\nn\\
    &\quad-\fft14h^{2\,\mu\nu}F_{\mu\rho}^{(-)\,i}F_\nu{}^{\rho\,(-)\,i}+\fft18h^{\mu\nu\lambda}h^{\rho\sigma}{}_\lambda F_{\mu\nu}^{(-)\,i}F_{\rho\sigma}^{(-)\,i}-\fft14h^{\mu\nu\lambda}h^{\rho\sigma}{}_\lambda F_{\mu\rho}^{(-)\,i}F_{\nu\sigma}^{(-)\,i}\nn\\
    &\quad+e^{2\varphi}\nabla_\mu(e^{-2\varphi}h_{\nu\rho\sigma}\nabla^\nu h^{\mu\rho\sigma})-h^{\nu\rho\sigma}\nabla_\nu\mathcal E_{H,\rho\sigma}^{(0)}-h^{2\,\mu\nu}\mathcal E_{g,\mu\nu}^{(0)},
\end{align}
and
\begin{align}
    R^{\mu\nu\rho\sigma}(\omega_+)F_{\mu\nu}^{(-)\,i}F_{\rho\sigma}^{(-)\,i}&=R^{\mu\nu\rho\sigma}F_{\mu\nu}^{(-)\,i}F_{\rho\sigma}^{(-)\,i}+\fft18F_{\mu\nu}^{(-)\,i}F^{\mu\nu\,(-)\,j}F^{(-)\,i}_{\rho\sigma}F^{\rho\sigma\,(-)\,j}\nn\\
    &\quad-\fft14F_\mu{}^{\nu\,(-)\,i}F_\nu{}^{\rho(-)\,j}F_\rho{}^{\sigma\,(-)\,i}F_\sigma{}^{\mu\,(-)\,j}-\fft12h^{\mu\nu\lambda}h^{\rho\sigma}{}_\lambda F_{\mu\rho}^{(-)\,i}F_{\nu\sigma}^{(-)\,i},
\end{align}
where we have defined $h^4=h^{\mu\nu\rho}h_{\mu\sigma\lambda}h_{\nu\sigma\epsilon}h_{\rho\lambda\epsilon}$. Note that $\tilde h$ defined in (\ref{eq:tildeH}) also contributes to the four derivative action through $\tilde h_{\mu\nu\rho}^2$.  At $\mathcal O(\alpha')$ we only need to worry about the cross-term
\begin{align}
    h^{\mu\nu\rho}\omega_{3L\,\mu\nu\rho}(\omega_+)&=h^{\mu\nu\rho}\omega_{3L\,\mu\nu\rho}-3R^{\mu\nu\rho\sigma}h_{\mu\nu\lambda}h_{\rho\sigma}{}^\lambda-\fft3{16}h^{\mu\nu\lambda}h^{\rho\sigma}{}_\lambda F_{\mu\nu}^{(-)\,i}F_{\rho\sigma}^{(-)\,i}\nn\\
    &\quad+\fft38h^{\mu\nu\lambda}h^{\rho\sigma}{}_\lambda F_{\mu\rho}^{(-)\,i}F_{\nu\sigma}^{(-)\,i}+\fft12h^4.
\end{align}
Making the above substitutions in (\ref{eq:Ltruncred}) then gives us
\begin{align}
    e^{-1}\mathcal L=&e^{-2\varphi}\biggl[R+4\qty(\partial\varphi)^2-\fft1{12} h_{\mu\nu\rho}^2-\fft18\qty(F_{\mu\nu}^{(-)\,i})^2\nn\\
    &+\frac{\alpha'}{8}\biggl(\qty(R_{\mu\nu\rho\sigma})^2-\frac{1}{2}R_{\mu\nu\rho\sigma}h^{\mu\nu\lambda}h^{\rho\sigma}{}_{\lambda}-\frac{1}{2}R^{\mu\nu\rho\sigma}F^{(-)\,i}_{\mu\nu}F^{(-)\,i}_{\rho\sigma}-\frac{1}{8}(h^2_{\mu\nu})^2+\frac{1}{24}h^4\nn\\
    &-\frac{1}{4}h^2_{\mu\nu} F^{(-)\,i}_{\mu\rho}F^{(-)\,i}_{\nu\rho}+\frac{1}{16}h^{\mu\nu\lambda}h^{\rho\sigma}{}_{\lambda}F^{(-)\,i}_{\mu\nu}F^{(-)\,i}_{\rho\sigma}+\frac{1}{8}h^{\mu\nu\lambda}h^{\rho\sigma}{}_\lambda F^{(-)\,i}_{\mu\rho}F^{(-)\,i}_{\nu\sigma}\nn\\
    &+\frac{1}{32}F^{(-)\,i}_{\mu\nu}F^{(-)\,j}_{\mu\nu}F^{(-)\,i}_{\rho\sigma}F^{(-)\,j}_{\rho\sigma}-\frac{1}{8}F^{(-)\,i}_{\mu\nu}F^{(-)\,i}_{\nu\sigma}F^{(-)\,j}_{\sigma\rho}F^{(-)\,j}_{\rho\mu}\nn\\
    &+\frac{1}{16}F^{(-)\,i}_{\mu\nu}F^{(-)\,j}_{\nu\sigma}F^{(-)\,i}_{\sigma\rho}F^{(-)\,j}_{\rho\mu}+\fft13h^{\mu\nu\rho}\omega_{3L\,\mu\nu\rho}\biggr)\biggr].
\end{align}
In particular, this agrees with the result of \cite{Eloy:2020dko} after appropriate truncation.\footnote{It is important to note that our conventions differ from those of \cite{Eloy:2020dko}. In particular, one must send $B\to-B$ to compare, so our truncation $F^{(+)}=0$ is equivalent to $F^{(-)}=0$ in the notation of \cite{Eloy:2020dko}.}

\section{Truncating the fermionic sector}\label{sec:fermionic}

Hitherto, we have only looked at the bosonic sector of the theory; it is a non-trivial test to additionally check that the truncation extends to the fermion sector.  We start with the $\mathcal O(\alpha')$ truncation
\begin{equation}
    g_{ij}=\delta_{ij}+\alpha'\delta g_{ij},\qquad b_{ij}=0,\qquad F_{\mu\nu}^i=\fft12F_{\mu\nu}^{(-)\,i},\qquad\tilde G_{\mu\nu\,i}=-\fft12F_{\mu\nu}^{(-)\,i},
\label{eq:truncanz}
\end{equation}
and consider the leading order supersymmetry variations (\ref{eq:deltalo}).  We find, at the lowest order
\begin{align}
    \delta_\epsilon\psi_\mu^{(0)}&=\left(\nabla_\mu(\omega_-)+\fft14F_{\mu\nu}^{(-)\,i}\gamma^\nu\Gamma^{\underline i}\right)\epsilon,\nn\\
    \delta_\epsilon\psi_i^{(0)}&=0,\nn\\
    \delta_\epsilon\tilde\lambda^{(0)}&=\left(\gamma^\mu\partial_\mu\varphi-\fft1{12}h_{\mu\nu\lambda}\gamma^{\mu\nu\lambda}+\fft18F_{\mu\nu}^{(-)\,i}\gamma^{\mu\nu}\Gamma^{\underline i}\right)\epsilon,
\label{eq:dello}
\end{align}
where the internal Dirac matrices $\Gamma^{\underline i}$ have flat-space indices.  We see that the gaugini are consistently truncated out at this order.  However, with the $\delta g_{ij}$ shift in (\ref{eq:truncanz}), the lowest order transformations, (\ref{eq:deltalo}), also give rise to the $\mathcal O(\alpha')$ terms
\begin{align}
    \delta(\delta_\epsilon\psi_\mu^{(0)})&=0,\nn\\
    \delta(\delta_\epsilon\psi_i^{(0)})&=\left(-\fft1{16}\delta g_{ij}F_{\mu\nu}^{(-)\,j}\gamma^{\mu\nu}-\fft14\partial_\mu\delta g_{ij}\gamma^\mu\Gamma^{\underline{j}}\right)\epsilon,\nn\\
    \delta(\delta_\epsilon\tilde\lambda^{(0)})&=0.
\label{eq:3dfromlo}
\end{align}

\subsection{The variations at \texorpdfstring{$\mathcal O(\alpha')$}{O(alpha prime)}}

The shift in the lowest order internal gravitino variation, $\delta(\delta_\epsilon\psi_i^{(0)})$,  will combine with the higher order term, $\delta_\epsilon\psi_i^{(1)}$, to yield the complete gaugino variation.  As we are aiming to truncate away the vector multiplets, this combined variation ought to vanish.

Reduction of the first order internal gravitino variation, $\delta\psi_i^{(1)}$, in (\ref{eq:deltas}) yields
\begin{align}
    \delta_\epsilon\psi_i^{(1)}&=\fft1{32}\Bigl(R_{\alpha\beta}{}^{\gamma\delta}(\omega_+)F_{\gamma\delta}^{(-)\,i}\gamma^{\alpha\beta}-\fft18F_{\alpha\beta}^{(-)\,j}F_{\gamma\delta}^{(-)\,j}F_{\gamma\delta}^{(-)\,i}\gamma^{\alpha\beta}+F_{\gamma\delta}^{(-)\,[j}\nabla_\alpha^{(+)}F_{\gamma\delta}^{(-)\,i]}\gamma^\alpha\Gamma^{\underline j}\nn\\
    &\qquad\qquad-\fft12F_{\alpha\beta}^{(-)\,i}F_{\beta\gamma}^{(-)\,j}F_{\gamma\alpha}^{(-)\,k}\Gamma^{\underline{jk}}\Bigr)\epsilon.
\end{align}
Combining this with (\ref{eq:3dfromlo}) gives
\begin{align}
    \delta(\delta_\epsilon\psi_i^{(0)})+\delta_\epsilon\psi_i^{(1)}&=\fft1{32}F_{\alpha\beta}^{(-)\,i}\bigg(R_{\gamma\delta}{}^{\alpha\beta}(\omega_+)\gamma^{\gamma\delta}-\fft14F_{\alpha\beta}^{(-)\,j}F_{\gamma\delta}^{(-)\,j}\gamma^{\gamma\delta}-\nabla_\gamma^{(+)} F_{\alpha\beta}^{(-)\,j}\gamma^\gamma\Gamma^{\underline j}\nn\\
    &\qquad\qquad\qquad-\fft12F_{\beta\gamma}^{(-)\,j}F_{\gamma\alpha}^{(-)\,k}\Gamma^{\underline{jk}}\bigg)\epsilon.
\label{eq:delgau}
\end{align}
As this is non-vanishing, the gaugino must be shifted if we are to truncate it away.

The Riemann tensor term in the variation suggests that we use the commutator of two covariant derivatives.  This can arise from the variation of a covariant derivative of the gravitino.  Since we work only to $\mathcal O(\alpha')$, we can take something like $\nabla_{[\mu}\psi_{\nu]}$ whose variation will gives $\nabla_{[\mu}\nabla_{\nu]}\epsilon$.  To be more precise, consider the variation $\delta\psi_\mu^{(0)}$ in (\ref{eq:dello}), which we write as $\delta\psi_\mu^{(0)}=\mathcal D_\mu\epsilon$ where
\begin{equation}
    \mathcal D_\mu\equiv\nabla_\mu(\omega_-)+\fft14F_{\mu\nu}^{(-)\,i}\gamma^\nu\Gamma^{\underline i}=\nabla_\mu-\fft18h_{\mu\nu\lambda}\gamma^{\nu\lambda}+\fft14F_{\mu\nu}^{(-)\,i}\gamma^\nu\Gamma^{\underline i}.
\end{equation}
The commutator we want is then
\begin{align}
    [\mathcal D_\mu,\mathcal D_\nu]=&\fft14\bigg(R_{\mu\nu}{}^{\alpha\beta}(\omega_-)\gamma^{\alpha\beta}+\left(2\nabla_{[\mu}F_{\nu]\alpha}^{(-)\,i}-h_{[\mu}{}^{\alpha\beta}F_{\nu]\beta}^{(-)\,i}\right)\gamma^\alpha\Gamma^{\underline i}-\fft12F_{\mu\alpha}^{(-)\,i}F_{\nu\beta}^{(-)\,i}\gamma^{\alpha\beta}\nn\\
    &\qquad-\fft12F_{\mu\alpha}^{(-)\,i}F_{\nu\alpha}^{(-)\,j}\Gamma^{\underline{ij}}\bigg).
\end{align}
Comparing this with (\ref{eq:delgau}) indicates that we need to convert between $R(\omega_+)$ and $R(\omega_-)$.  Using the $h$ Bianchi identity, $\dd h=\fft14F^{(-)\,i}\wedge F^{(-)\,i}$, we have
\begin{equation}
    R_{\alpha\beta\gamma\delta}(\omega_-)=R_{\gamma\delta\alpha\beta}(\omega_+)-\fft34F_{[\alpha\beta}^{(-)\,i}F_{\gamma\delta]}^{(-)\,i}.
\end{equation}
A bit of manipulation, including use of the $F$ Bianchi $\dd F^{(-)\,i}=0$, then shows that $\delta_\epsilon\tilde\psi_i=0$ where we have defined the shifted gaugino
\begin{equation}
    \tilde\psi_i=\psi_i-\frac{\alpha'}{4}F_{\mu\nu}^{(-)\,i}\mathcal D_\mu\psi_\nu.
\end{equation}
This demonstrates that we can consistently truncate $\tilde\psi_i$ out of the fermion sector while preserving the supersymmetry of the solution.

For completeness, we also note that the gravitino and dilatino variations have four-derivative corrections
\begin{align}
    \delta_\epsilon\psi_\mu^{(1)}=&\frac{1}{32}\Bigg[\omega_{3L,\mu\nu\rho}(\omega_+)\gamma^{\nu\rho}+2\qty(R_{\mu\nu}{}^{\alpha\beta}(\omega_+)-\frac{1}{8}F^{(-)\,j}_{\mu\nu}F^{(-)\,j}_{\alpha\beta})F_{\alpha\beta}^{(-)\,i}\gamma^\nu\Gamma^{\underline i}\nn\\
    &\qquad+\frac{1}{2}F^{(-)\,i}_{\alpha\beta}\nabla^{(+)}_\mu F^{(-)\,j}_{\alpha\beta}\Gamma^{\underline{ij}}\Bigg]\epsilon,
\end{align}
and
\begin{align}
    \delta_\epsilon\tilde\lambda^{(1)}&=\frac{1}{48}\biggl[\omega_{3L,\mu\nu\rho}(\omega_+)\gamma^{\mu\nu\rho}+\frac{3}{2}\left(R_{\mu\nu}{}^{\alpha\beta}(\omega_+)-\fft18F_{\mu\nu}^{(-)\,j}F_{\alpha\beta}^{(-)\,j}\right)F_{\alpha\beta}^{(-)\,i}\gamma^{\mu\nu}\Gamma^{\underline{i}}\nn\\
    &\kern4em+\fft14F^{(-)\,i}_{\alpha\beta}F^{(-)\,j}_{\beta\gamma}F^{(-)\,k}_{\gamma\alpha}\Gamma^{\underline{ijk}}\biggr]\epsilon.
\end{align}
%
%

\section{Comparing with the four-derivative corrected BPS black string}\label{sec:example}

A natural test for our truncation is considering $\mathcal N=1$, $D=9$ supergravity \cite{Gates:1984kr}. The field content is the 9D metric $g_{\mu\nu}$, the gravitino $\psi_\mu$, a vector $A_\mu$, a two-form $B_{\mu\nu}$, a dilatino $\lambda$, and a dilaton $\varphi$.  Black hole solutions to this minimal supergravity lift to ten-dimensional strings, so we may compare these two cases.

\subsection{Two-derivative solution}

Balancing the nine-dimensional graviphoton charge with the mass gives rise to a supersymmetric black hole solution given by \cite{Lu:1995cs}
\begin{align}
    \dd s_9^2&=-\qty(1+\frac{k}{r^6})^{-2}\dd t^2+\dd r^2+r^2\dd\Omega_{7}^2,\nn\\
    A&=\frac{1}{1+\frac{k}{r^6}}\dd t,\nn\\
    e^{\varphi}&=\qty(1+\frac{k}{r^6})^{-1/2},
\end{align}
with all other fields vanishing. Here we have denoted the metric of the round $S^7$ by $\dd\Omega_7^2$. This two-derivative solution uplifts to a 10D black string solution via \eqref{eq:redMet}
\begin{align}
    \dd s_{10}^2&=-\qty(1+\frac{k}{r^6})^{-2}\dd t^2+\dd r^2+r^2\dd\Omega_{7}^2+\qty(\dd z+\frac{\dd t}{1+\frac{k}{r^6}})^2,\nn\\
    B&=-\frac{1}{1+\frac{k}{r^6}}\dd t\land \dd z,\nn\\
    e^{\phi}&=\qty(1+\frac{k}{r^6})^{-1/2}.
\end{align}
It is straightforward to check that this satisfies the 10D two-derivative equations of motion \eqref{eq:eom2dred}. Moreover, this is a BPS solution, in the sense that $\delta\psi_M^{(0)}=0$ and $\delta\lambda^{(0)}=0$ with Killing spinor
\begin{equation}
    \epsilon^{(0)}=\qty(1+\frac{k}{r^6})^{-1/2}\qty(1-\gamma^{0}\Gamma^{\underline z})\epsilon_0(\Omega_7),
\end{equation}
where $\epsilon_0$ is a covariantly constant spinor\footnote{Concretely, if we write the seven angles of $S^7$ as $\theta_n$, $n=1,...,7$, such that the angles are defined recursively as $\dd \Omega_{n}^2=\dd\theta_{8-n}^2+\sin^2\theta_{8-n}\dd \Omega_{{n-1}}^2$, then $\epsilon_0(\Omega_7)$ is given by $\epsilon_0=\prod_{n=1}^7\exp[\frac{\theta_n}{2}\gamma^{n+1,n+2}]\eta_0$, where $\eta_0$ is a constant spinor, \textit{i.e.}, with no coordinate dependence. This can be seen via an identical analysis to that in \cite{Lu:1998nu}.} on $S^7$. Thus, the solution is $\frac{1}{2}$-BPS. 

\subsection{Four-derivative correction}

We now solve the ten-dimensional equations of motion, including $\mathcal O(\alpha')$ corrections.  We do so by making an ansatz for the four-derivative corrected black string as
\begin{align}
    \dd s_{10}^2=&-\qty(1+\frac{k}{r^6})^{-2}\qty(1+\alpha' f(r))\dd t^2+\qty(1+\alpha' g(r))\dd r^2+r^2\dd\Omega_{7}^2\nn\\
    &+\qty(1+\alpha' h(r))\qty(\dd z+\qty(1+\alpha' j(r))\frac{\dd t}{1+\frac{k}{r^6}})+\mathcal{O}(\alpha'^2),\nn\\
    B=&-\frac{1+\alpha' \kay(r)}{1+\frac{k}{r^6}}\dd t\land \dd z+\mathcal{O}(\alpha'^2),\nn\\
    e^{\phi}=&\qty(1+\alpha' \ell(r))\qty(1+\frac{k}{r^6})^{-1/2}+\mathcal{O}(\alpha'^2),
\end{align}
where we have explicitly assumed that the $SO(8)$ symmetry is preserved and that the solution continues to have $\partial_t$ and $\partial_z$ as Killing vectors. Demanding that our fermion variations vanish implies the conditions
\begin{align}
    0&=g(r),\nn\\
    0&=\dv{}{r}\qty(\frac{f(r)-h(r)-2j(r)}{1+\frac{k}{r^6}}),\nn\\
    0&=\dv{}{r}\qty(\frac{h(r)+j(r)-\kay(r)}{1+\frac{k}{r^6}}),\nn\\
    0&=6kf(r)-\qty(1+\frac{k}{r^6})^2\dv{}{r}\qty(\frac{j(r)+\kay(r)}{1+\frac{k}{r^6}})+4r(k+r^6)m'(r),\nn\\
    0&=3k\qty(2\kay(r)-f(r)-h(r))+r(k+r^6)\qty(\kay'(r)-2\ell'(r)),\label{eq:BPScond1}
\end{align}
where we have written our BPS spinor as $\epsilon=(1+\alpha'm(r))\epsilon^{(0)}$. Note that since we are demanding $\delta\psi_\mu=0$ and $\delta\lambda=0$ for a BPS solution, it is fine to require that $\delta\psi_z=0$ without any field redefinitions. In particular, \eqref{eq:BPScond1} allows us to write
\begin{align}
    f(r)-h(r)-2j(r)=&c_1\qty(1+\frac{k}{r^6}),\nn\\
    h(r)+j(r)-\kay(r)=&-c_2\qty(1+\frac{k}{r^6}),\nn\\
    \kay'(r)-2\ell'(r)=&\frac{3k(c_1-2c_2)}{r(k+r^6)}\qty(1+\frac{k}{r^6}),\label{eq:BPScond2}
\end{align}
where $c_1$ and $c_2$ are undetermined constants of integration. Substituting these conditions \eqref{eq:BPScond2} into our equations of motion and solving gives the solution
\begin{align}
    f(r)=&\frac{18 k^2}{r^2(k+r^6)^2}+c_4+2c_7+c_1\qty(2-\frac{1}{1+\frac{k}{r^6}})-\frac{c_3+2c_6}{6(k+r^6)},\nn\\
    g(r)=&0,\nn\\
    h(r)=&-\frac{18 k^2}{r^2(k+r^6)^2}-\frac{c_3}{6(k+r^6)}+c_4,\nn\\
    j(r)=&\frac{18 k^2}{r^2(k+r^6)^2}-\frac{c_1k^2}{2r^6(k+r^6)}-\frac{c_6}{6(k+r^6)},\nn\\
    \kay(r)=&\frac{(c_1-2c_2)k}{2r^6}+\frac{3c_1k-c_3-c_6}{6(k+r^6)}+c_2+c_4+c_7,\nn\\
    \ell(r)=&\frac{3c_1k-c_3-c_6}{12(k+r^6)}+c_5+\frac{c_7}{2},
\end{align}
where the $c_i$ are (as yet undetermined) constants of integration, along with the four-derivative corrected Killing spinor
\begin{equation}
    \epsilon=\qty[1+\frac{\alpha'}{2}\qty(\frac{9k^2}{r^2(k+r^6)^2}-\frac{c_3+2c_6-6c_1 k}{12(k+r^6)})]\epsilon^{(0)}.
\end{equation}
Now, we can further reduce the number of free constants. Our mass $M$, electric charge $Q^{(A)}$, $B$ charge $Q^{(B)}$, and scalar charge $\Sigma$ are given by
\begin{align}
    M&=2k+\alpha'\qty(-\qty(3c_1+2c_4+4c_7)k+\frac{c_3+2c_6}{6}),\nn\\
    Q^{(A)}&=-k-\alpha'\qty(\frac{c_6}{6}+c_7k),\nn\\
    Q^{(B)}&=k+\alpha'\qty(\frac{c_3+c_6}{6}+(c_4+c_7)k),\nn\\
    \Sigma&=-\frac{k}{2}+\frac{\alpha'}{12}\qty(3k\qty(c_1-2c_5-c_7)-c_3-c_6),
\end{align}
where the scalar charge is defined as the coefficient of $r^{-6}$ in the large-$r$ expansion of the dilaton
\begin{equation}
    \phi=\phi_{\infty}+\frac{\Sigma}{r^6}+\mathcal{O}(r^{-12}).
\end{equation}
If we fix the metric at infinity, as well as the asymptotic value of the scalar $\phi_\infty$ and the charges $Q^{(A)}$ and $Q^{(B)}$, then this requires that all the $c_i$ vanish. Upon doing so, we are left with the four-derivative corrected metric
\begin{align}
    \dd s_{10}^2=&-\qty(1+\frac{k}{r^6})^{-2}\qty(1+\frac{18\alpha' k^2}{r^2(k+r^6)^2})\dd t^2+\dd r^2+r^2\dd\Omega_7^2\nn\\
    &+\qty(1-\frac{18\alpha' k^2}{r^2(k+r^6)^2})\qty(\dd z-\frac{1}{1+\frac{k}{r^6}}\qty(1+\frac{18\alpha' k^2}{r^2(k+r^6)^2})\dd t)^2+\mathcal{O}(\alpha'^2),
\end{align}
with $B$ and $\phi$ left uncorrected.

We now compare this ten-dimensional solution with the form of the consistent truncation (\ref{eq:truncanz1}).  Making note of
\begin{equation}
    F^2=\frac{1}{3}H^2=-\frac{72k^2}{r^2(k+r^6)^2},
\end{equation}
we find
\begin{subequations}
   \begin{align}
    g_{zz}=& 1+\frac{\alpha'}{4}F^2,\label{eq:gzz}\\
    B =& \qty(-A+\frac{\alpha'}{2}\Omega^{\alpha\beta}F_{\alpha\beta})\land\dd z.\label{eq:B}
\end{align}
\end{subequations}
In particular, \eqref{eq:gzz} is exactly what we expect from the general expression \eqref{eq:truncanz1} when $n=1$. On the other hand, \eqref{eq:B} na\"ively seems to have an extra term compared with \eqref{eq:truncanz}, but the $\Tr\Omega F$ is precisely the term that we implicitly absorbed into $B$ to remove a total derivative from the Lorentz Chern-Simons form.

\section{Discussion}\label{sec:conclusion}

In this chapter, we have shown that it is consistent to truncate out the vector multiplets that arise in the toroidal reduction of heterotic supergravity in the presence of four-derivative corrections. In particular, this truncation does not ruin the supersymmetry of the reduced theory. We further verified our truncation by looking at the example of a four-derivative corrected black string solution. We view this work as a step towards more general non-trivial truncations of higher-derivative theories, such as what may arise from sphere reductions.

One may be tempted to interpret these results as a statement that a two-derivative truncation automatically implies the existence of a four-derivative one. However, this is not always the case. For example, one might consider further truncating the $T^5$ reduction to minimal $D=5$, $\mathcal{N}=2$ supergravity. For $n=5$, upon transforming to the Einstein frame $g=e^{4\varphi/3}\tilde g$ and dualizing $h=e^{-2\varphi}\star \mathcal G=e^{-2\varphi}\star\dd C$, the two-derivative Lagrangian reads\footnote{Here, we have chosen to use the notation $\star$ to refer to the Hodge star in the string frame and $\tilde\star$ the Hodge star in the Einstein frame.}
\begin{equation}
    \tilde\star\mathcal L^{(0)}=\tilde\star R-\frac{8}{3}\dd\varphi\land\tilde\star\dd\varphi-\frac{1}{2}e^{8\varphi/3}\mathcal G\land\tilde\star\mathcal G-\frac{1}{4}F^{(-)\,i}\land F^{(-)\,i}\land C-\frac{1}{4}e^{-4\varphi/3}F^{(-)\,i}\land\tilde\star F^{(-)\,i},
\end{equation}
where the Chern-Simons term arises from requiring that the $h$ Bianchi identity become the $\mathcal G$ equation of motion. This theory may be thought of, in bosonic $\mathcal{N}=2$ language, as a graviton multiplet $(g_{\mu\nu},C_{\mu})$ coupled to a gravitino multiplet $A^{(-)\,i\ne 1}_{\mu}$ and a vector multiplet $(A^{(-)\,1}_{\mu},\varphi)$. One may then check that it is consistent to truncate the additional multiplets
\begin{equation}
    F^{(-)\,1}=\pm 2\mathcal G,\qquad F^{(-)\,i\ne 1}=0,\qquad \varphi = 0,
\end{equation}
which, upon rescaling $\mathcal G=F/\sqrt{3}$, gives us pure minimal ungauged $\mathcal N=2$ supergravity
\begin{equation}
    \star\mathcal L^{(0)}=\star R-\frac{1}{2}F\land\star F-\frac{1}{3\sqrt{3}}F\land F\land A.
\end{equation}
Similarly, upon truncating, the two-derivative dilatino equation takes the form
\begin{equation}
    \delta_\epsilon\tilde\lambda^{(0)}=-\frac{i}{4\sqrt{3}}F_{\mu\nu}\gamma^{\mu\nu}\qty(1\pm i\Gamma^{\underline 1})\epsilon,
\end{equation}
where the presence of a projector is consistent with the fact that we are truncating $\mathcal N=4$ supersymmetry down to $\mathcal{N}=2$.

At the four-derivative level, we na\"ively expect the minimal $\mathcal N=2$ truncation to yield the Lagrangian \eqref{eq:param4d} with coefficients \eqref{eq:correctc} that we saw in Chapter \ref{chap:UE}. However, the dilaton equation of motion in the Einstein frame contains the term
\begin{equation}
    \mathcal E_\varphi^{(1)}\supset \frac{1}{3}e^{4\varphi/3}\qty(R_{\mu\nu\rho\sigma})^2.
\end{equation}
As is well-known, such a term cannot be removed by field redefinitions. This spells doom for our truncation. An alternative way of seeing the same issue is to look at the $\mathcal G$ equation of motion (or, equivalently, the $\tilde h$ Bianchi identity)
\begin{equation}
    \dd\qty(e^{-2\varphi}\star \mathcal G)\sim -\frac{\alpha'}{4}R(\omega_+)\land R(\omega_+)+...
\end{equation}
No such term appears in the $F^{(-)\,i}$ equation of motion and likewise cannot be removed by a field redefinition. The issue in both cases is that the two-derivative equations of motion have no Riemann tensors, and so field redefinitions cannot generate two Riemann tensors.%
\footnote{Although, if we are clever, we can generate one, as happened for the Maxwell equation truncation earlier.}
Hence, the truncation is likely inconsistent at the four-derivative level. We may also see this in the fermionic sector. If we now identify $\tilde h=e^{-2\varphi}\star\mathcal G$ and set $F^{(-)\,i\ne 0}=0$, then the four-derivative part of the dilatino variation becomes
\begin{equation}
    \delta_\epsilon\tilde\lambda^{(1)}=\frac{1}{32}\qty(R_{\mu\nu}{}^{\alpha\beta}(\omega_+)-\frac{1}{8}F^{(-)\,1}_{\mu\nu}F^{(-)\,1}_{\alpha\beta})F^{(-)\,1}_{\alpha\beta}\gamma^{\mu\nu}\Gamma^{\underline1}\epsilon,
\end{equation}
where the Lorentz Chern-Simons piece has been absorbed into $\mathcal{G}$ and the last term vanishes since $\Gamma^{\underline{11}}=0$. We can indeed remove the Riemann term via a field redefinition similar to what was done for $\tilde\psi$ but at the cost of a $\nabla^{(+)}F^{(-)}\nabla^{(+)}F^{(-)}$ term that cannot be removed.\footnote{Note that we are very restricted in what field redefinitions we may make since $\delta_\epsilon\psi_i^{(0)}$ and $\delta\tilde\lambda^{(0)}$ both vanish, which means we can only shift by $\psi_\mu$.}

What this example illustrates is that not every two-derivative consistent truncation necessarily leads to a four-derivative version, even in the case of torus reductions. It is interesting to note that the problem rests with the vector multiplet; it is perfectly consistent to truncate out just the gravitino multiplet (leaving us with a matter-coupled $\mathcal N=2$ supergravity), but the vector multiplet gets non-trivially sourced by the graviton multiplet at the four-derivative level.  Here, the best we can do is to truncate to minimal supergravity coupled to a universal vector multiplet.  An almost identical issue is present if one tries to truncate to pure $\mathcal{N}=1$ or $\mathcal{N}=2$ supergravity in $D=4$, which leads to difficulty recovering the expected result of \cite{Bobev:2020egg,Bobev:2020zov,Bobev:2021oku}.  It should also be noted that at $(\alpha')^3$ order in $D=4$, it is known that the pure $\mathcal N=1$ theory has to be coupled to at least an extra chiral multiplet \cite{Moura:2007ks}.\footnote{See also \cite{Moura:2007ac} for a related story in the torus reduction of type II supergravity.}

One might also consider a similar truncation of the theory in $D=6$. Here, our $\mathcal N=(1,1)$ Lagrangian may be thought of, in bosonic $\mathcal{N}=(1,0)$ language, as a graviton multiplet\footnote{We denote the self-dual part of $b$ as $b^{+}$ and the anti-self-dual part as $b^-$. This is unrelated to the $\pm$ notation associated with $F^{(\pm)}$.} $(g_{\mu\nu},b_{\mu\nu}^+)$ coupled to a gravitino multiplet $A^{(-)\,i}_\mu$ and a tensor multiplet $(b_{\mu\nu}^{-},\varphi)$. As before, it is perfectly consistent to truncate out the gravitino multiplet
\begin{equation}
    F^{(-)\,i}=0,
\end{equation}
at the four-derivative level; in particular, the $F^{(-)\, i}$ equations of motion trivially vanish when we set $F^{(-)\, i}=0$. This yields the truncated Lagrangian
\begin{equation}
    e^{-1}\mathcal L=e^{-2\varphi}\qty[R+4\qty(\partial\varphi)^2-\fft1{12}\tilde h_{\mu\nu\rho}^2+\frac{\alpha'}{8}\qty(R_{\mu\nu\rho\sigma}(\omega_+))^2].
\end{equation}
However, analogous to the vector multiplet in the $D=5$ case, the universal tensor multiplet cannot be truncated away. This can be seen from the dilaton equation of motion, which again contains a $\qty(R_{\mu\nu\rho\sigma}(\omega_+))^2$ that cannot be removed with field redefinitions, or from the dilatino variation, which again has a Riemann term as in the $D=5$ case.  Presumably, the truncation of \cite{Chang:2021tsj} to $\mathcal N=(1,0)$ supergravity coupled to a tensor multiplet and four hypermultiplets will suffer the same fate; while the hypermultiplets can be removed by a further truncation, the tensor multiplet cannot.

While we have focused on four-derivative corrected heterotic supergravity, more generally, the dilaton coupling to higher curvature couplings precludes it from being consistently truncated out of the lower-dimensional theory.%
\footnote{It should be emphasized that this is a \textit{perturbative} statement. In principle, one could consider a non-perturbative reduction with compact dimensions of size $\alpha'$, which could provide a loophole.}
One implication is that a top-down approach to higher-derivative holography will necessarily include, at a minimum, the dilaton multiplet in addition to the supergravity multiplet. It would be interesting to see how this fits with the many non-trivial consistency checks of bottom-up holography performed in the absence of the dilaton.

%% file: Chapters/chap6.tex
Life is simple on a torus $T^n$ since the truncation to the massless sector restricts the $U(1)^n$ singlets. Products of singlets can never yield non-singlets, so symmetry protects the truncation and there is little that can go wrong. However, there are many subtleties involved with sphere truncations (see \textit{e.g.} \cite{deWit:1986oxb,Nastase:1999kf,Lu:1999bc,Cvetic:1999un,Lu:1999bw,Cvetic:1999au,Cvetic:2000dm,Cvetic:2000nc,Lee:2014mla,Nicolai:2011cy,Samtleben:2019zrh,Bonetti:2022gsl,Ciceri:2023bul}) and more general coset reductions (see \textit{e.g.} \cite{Cvetic:2003jy,House:2005yc,Cassani:2009ck,Cassani:2010na,Bena:2010pr}). The issue is that there are fields in the massless sector that transform non-trivially under the isometry group; amongst these is the gauge field that arises from the metric. In such cases, products of the retained fields may act as sources for the very fields we wish to truncate, which spells disaster for consistency. Hence, such truncations are quite delicate.

On the other hand, we are now entering an age of precision holography, where higher-derivative corrections are becoming increasingly important. This is motivated partially by advancements in the construction of higher-derivative supergravities but also by advancements on the field theory side that allow for precise comparison. However, such higher-derivative corrections are subtle. As we saw in the previous chapter, not every two-derivative truncation necessarily leads to a consistent four-derivative one. Some higher-derivative gauged supergravities could not be reached by a consistent truncation, in particular, because of non-trivial couplings between the graviton and dilaton-containing multiplets. Hence, it is doubly subtle to check that a four-derivative sphere truncation is indeed consistent.

As in Chapter \ref{chap:torus}, we work in the context of heterotic supergravity with four-derivative corrections. Here, we will be interested in the Scherk-Schwarz reduction on $S^3$, which yields half-maximal 7D gauged supergravity coupled to three vector multiplets. The truncation to pure 7D supergravity is known to be consistent at the two-derivative level \cite{Chamseddine:1999uy}, and, as we will show, it is indeed consistent at the four-derivative level as well.

Our starting point is 10D heterotic supergravity with field content $(g_{MN},\psi_M,B_{MN},\lambda,\phi)$, and leading order bosonic Lagrangian
\begin{equation}
    e^{-1}\mathcal L=e^{-2\phi}\left[R+4(\partial_M\phi)^2-\fft1{12}H_{MNP}^2\right].
\end{equation}
Note that we do not include the heterotic gauge fields, as we aim to truncate to pure 7D supergravity. For the $S^3$ reduction, we take a metric ansatz
\begin{equation}
        \dd s^2_{10}=g_{\mu\nu}\dd x^\mu\dd x^\nu+g_{ij}\eta^i\eta^j,\qquad \eta^i=\sigma^i+A^i,
\end{equation}
where $x^\mu$ are coordinates on the 7D base space, $A^i$ is a principal $SU(2)$ connection, $\sigma^i$ is a set of left-invariant one-forms on $S^3$, and $g_{ij}$ is a symmetric matrix of scalars. We reduce the two-form as
\begin{equation}
        B=b+B_i\land\eta^i+\frac{1}{2}b_{ij}\,\eta^i\land\eta^j+m\omega_{(2)},
\end{equation}
such that
\begin{equation}
        \dd\omega_{(2)}=\frac{1}{3!}\epsilon_{ijk}\,\sigma^i\land\sigma^j\land\sigma^k,
\end{equation}
is the volume form on $S^3$.  Here, $m$ is the $H_3$ flux supporting the sphere reduction.

At the two-derivative level, this reduction was worked out in \cite{Chamseddine:1999uy}; the resulting theory is a seven-dimensional gauged supergravity with Lagrangian
\begin{equation}
    e^{-1}\mathcal L_7=e^{-2\varphi}\left[R+4(\partial_\mu\varphi)^2-\fft1{12}h_{\mu\nu\lambda}^2-\fft1{2g^2}(F_{\mu\nu}^i)^2+g^2\right],
\end{equation}
where the gauge coupling constant $g$ is related to the flux by $g^2=-1/m$.  We extend this reduction to the four-derivative level.  In particular, after analyzing the bosonic equations of motion, we find a consistent $\mathcal O(\alpha')$ truncation
\begin{align}
    g_{ij}=\fft1{\tilde g^2}\delta_{ij}+\frac{\alpha'}{4\tilde g^4}F^i_{\alpha\beta}F^j_{\alpha\beta},\qquad b_{ij}=0,\qquad B_i=-\fft1{\tilde g^2}A^i.
\end{align}
where the gauge coupling constant, $\tilde g$, receives an $\mathcal O(\alpha')$ shift
\begin{equation}
    \fft1{\tilde g^2}=\fft1{g^2}-\fft{\alpha'}2=-m-\fft{\alpha'}2,
\end{equation}
This truncation reduces to the torus case of Chapter \ref{chap:torus} in the $g\to0$ limit, assuming the fields are appropriately rescaled.

We also consider the fermionic sector. Here, the 10D gravitino $\psi_M$ splits into a 7D gravitino $\psi_\mu$ and three gaugini $\psi_i$. Note that the left-invariant one-form basis trivializes the spin bundle and hence the reduction preserves the full supersymmetry of our original theory. To truncate the gaugini, we require a field redefinition
\begin{equation}
    \tilde\psi_i=\psi_i-\frac{\alpha'}{2\tilde g}F^i_{\mu\nu}\mathcal D_\mu\psi_\nu,
\end{equation}
where we have denoted $\delta_\epsilon\psi_\mu=\mathcal D_\mu\epsilon$. This then sets $\delta_\epsilon\tilde\psi_i=0$ and leads to consistency with supersymmetry.

The rest of this chapter is organized as follows. In Section \ref{sec:reduction1}, we review the two-derivative group manifold reduction of heterotic supergravity on $S^3$. We then show, in Section \ref{sec:bosonic1}, that the truncation is consistent at the four-derivative level by analysis of the equations of motion, as well as summarize the remaining equations of motion and effective Lagrangian. In Section \ref{sec:fermionic1}, we check that this truncation is consistent with supersymmetry, and, finally, we conclude in Section \ref{sec:discussion1}.

\section{Group manifold reduction on $S^3$}\label{sec:reduction1}

The reduction of heterotic supergravity on $S^3$ can be viewed as a reduction on the group manifold of $SU(2)$. We first summarize the standard results of the group manifold reduction \cite{Chamseddine:1999uy,Cvetic:2003jy,Lu:2006ah}. As a matter of setting notation, we will use the same notation for four-derivative heterotic supergravity as in Chapter \ref{chap:torus}. In particular, we split our indices as
\begin{equation*}
    M\to\{\mu,i\},\qquad A\to\{\alpha,a\}.
\end{equation*}

\subsection{$S^3$ reduction}
We now proceed to reduce the heterotic theory on $S^3$ to arrive at half-maximal seven-dimensional gauged supergravity coupled to three vector multiplets \cite{Chamseddine:1999uy,Lu:2006ah}.  The vector multiplets will be truncated out in the next section, but we retain them here since we must use the full non-truncated lowest-order equations of motion to compute their four-derivative shifts.

The sphere $S^3$ is, as a manifold, isomorphic to $SU(2)$, which admits a basis of globally defined left-invariant one-forms $\sigma^i$ satisfying the Maurer-Cartan equation
\begin{equation}
        \dd \sigma^i=-\frac{1}{2}\epsilon^{ijk}\sigma^j\land\sigma^k.
\end{equation}
Such a global frame reduces the structure group to the identity, which is just the statement that $S^3$ is parallelizable. Moreover, $\sigma^i$ generates the right isometries of the metric. The $\epsilon^{ijk}$ are the structure constants of $\mathfrak{su}(2)$, and formally the indices should be raised and lowered via the Cartan-Killing metric $\kappa$. However, in this case, the Cartan-Killing metric is trivial
\begin{equation}
        \kappa_{ij}\equiv -\frac{1}{2}\epsilon^{k}{}_{\ell i}\epsilon^{\ell}{}_{kj}=\delta_{ij},
\end{equation}
and so we will not concern ourselves with the positions of the indices of $\epsilon$. 
    
We then have a metric ansatz in the form of a standard Scherk-Schwarz reduction \cite{Scherk:1979zr}
\begin{equation}
        \dd s^2_{10}=g_{\mu\nu}\dd x^\mu\dd x^\nu+g_{ij}\eta^i\eta^j,\qquad \eta^i=\sigma^i+A^i,\label{eq:metric}
\end{equation}
where $x^\mu$ are coordinates on the 7D base space, $g_{ij}$ is a symmetric matrix of scalars and $A^i$ forms a principal $SU(2)$ connection with curvature locally given by
\begin{equation}
        F^i=\dd A^i-\frac{1}{2}\epsilon^{ijk}A^j\land A^k.
\end{equation}
Being non-abelian, our gauge field naturally has an associated gauge-covariant derivative, which we will denote as $D$. Given an $\mathfrak{su}(2)$-valued form $t^i$, $D$ acts as
\begin{equation}
        Dt^i=\dd t^i-\epsilon^{ijk}A^j\land t^k.
\end{equation}
Considering the metric \eqref{eq:metric}, we choose a natural zehnbein
\begin{equation}
        E^\alpha=e^\alpha,\qquad E^a=e^a_i\eta^i.
\end{equation}
where $e^\alpha$ is a siebenbein for $g_{\mu\nu}$ and $e^a$ is a dreibein for $g_{ij}$, so that $\dd s_{10}^2=\eta_{\alpha\beta}E^\alpha E^\beta+\delta_{ab}E^aE^b$. We then compute
\begin{align}
        \dd E^\alpha&=-\omega^{\alpha\beta}E^\beta,\nn\\
        \dd E^a&=\frac{1}{2}e_i^aF^i_{\alpha\beta}E^\alpha\land E^\beta+e^i_b D_\alpha e_i^a E^\alpha\land E^b-\frac{1}{2}\epsilon^{ijk}e_i^ae_b^je_c^k E^b\land E^c.
\end{align}
from which one may extract the components of the spin connection
\begin{align}
        \Omega^{\alpha\beta}&=\omega^{\alpha\beta}-\frac{1}{2}e_i^a F_{\alpha\beta}^i E^a,\nn\\
        \Omega^{\alpha b}&=-P_\alpha^{bc}E^c-\frac{1}{2}e_i^b F^i_{\alpha\beta}E^\beta,\nn\\
        \Omega^{ab}&=Q_\alpha^{ab} E^\alpha+\frac{1}{2}\mathcal C_{c,ab}E^c.
\end{align}
Here, we have made the conventional definitions
\begin{align}
        P_\alpha^{ab}&=e^{i(a} D_\alpha e^{b)}_i,\nn\\
        Q_\alpha^{ab}&=e^{i[a} D_\alpha e^{b]}_i,\nn\\
        \mathcal C_{c,ab}&=\epsilon^{ijk}\qty[e^c_i e^j_a e^k_b+e^b_i e^j_a e^k_c-e_i^a e^j_b e^k_c],
\end{align}
where $P$ and $Q$ are the scalar kinetic term and composite $SU(2)$ connection, respectively.

We may then compute the relevant Riemann tensor components to be
\begin{align}
        R_{\gamma\delta}{}^{\alpha\beta}(\Omega)=&R^{\alpha\beta}{}_{\gamma\delta}(\omega)-\frac{1}{2}F^i_{\alpha\beta}F^j_{\gamma\delta}g_{ij}-\frac{1}{2}F^i_{\alpha\gamma}F^j_{\beta\delta}g_{ij},\nn\\
        R_{\gamma d}{}^{\alpha\beta}(\Omega)=&-\frac{1}{2}D_\gamma\qty(e^d_i F_{\alpha\beta}^i)-\frac{1}{2}e_i^a F_{\alpha\beta}^i\qty(e^j_d D_\gamma e^a_j)-e^c_i F_{\alpha\gamma}^iP_{\beta cd},\nn\\
        R_{c\delta}{}^{\alpha b}(\Omega)=&D_\delta P_{\alpha bc}+P_{\alpha bd}e^i_c D_\delta e_i^d+\frac{1}{4}e_i^c e_j^b F_{\alpha\gamma}^iF_{\gamma\delta}^j-P_{\alpha dc}Q_{\delta db}+\frac{1}{4}e_i^d F^i_{\alpha\delta}\mathcal C_{c,db},\nn\\
        R_{cd}{}^{\alpha b}(\Omega)=&P_{\alpha ba}\epsilon^{ijk}e_i^ae^j_ce^k_d+e_i^cF_{\alpha\gamma}^iP_{\gamma bd}-P_{\alpha ac}\mathcal C_{d,ab},\nn\\
        R_{cd}{}^{ab}(\Omega)=&-2P_{\gamma ac}P_{\gamma bd}-\frac{1}{4}\mathcal C_{f,ab}\epsilon^{ijk}e_i^fe_j^ce_k^d+\frac{1}{2}\mathcal C_{c,af}\mathcal C_{d,fb},
\end{align}
where there is implicit antisymmetrization as relevant. Contracting Riemann then gives the Ricci tensor components
\begin{align}
        R_{\alpha\beta}(\Omega)=&R_{\alpha\beta}(\omega)-\frac{1}{2}F_{\alpha\gamma}^iF_{\beta\gamma}^j g_{ij}-D_\beta P_{\alpha cc}-P_{\alpha cd}P_{\beta}{}^{cd},\nn\\
        R_{\alpha b}(\Omega)=&\frac{1}{2}D_{\gamma}\qty(e_i^b F_{\alpha\gamma}^i)+\frac{1}{2}e_i^aF_{\alpha\gamma}^ie^j_bD_\gamma e_j^a+\frac{1}{2}e_i^bF_{\alpha\gamma}^iP_{\gamma cc},\nn\\
        R_{a b}(\Omega)=&-D^\gamma P_{\gamma ab}-P_{\gamma ad}\qty(e^i_bD_\gamma e^d_i)+\frac{1}{4}e_i^ae_j^bF_{\gamma\delta}^iF_{\gamma\delta}^j+P_{\gamma db}Q_{\gamma da}-2P_{\gamma a[b|}P_{\gamma c|c]}\nonumber\\
        &-\frac{1}{4}\mathcal C_{f,ac}\epsilon^{ijk}e_i^fe_j^be_k^c,
\end{align}
and the Ricci scalar
\begin{align}
        R(\Omega)&=R(\omega)-\frac{1}{4}F^i_{\alpha\beta}F^j_{\alpha\beta}g_{ij}-2\nabla^\gamma P_{\gamma cc}-P_{\gamma cc}P^{\gamma dd}-P^2\nn\\
        &\qquad-\frac{1}{4}\qty(\epsilon^{ijk}\epsilon^{\ell mn}h_{i\ell}h^{jm}h^{kn}+2\epsilon^{ijk}\epsilon^{ji\ell}h^{k\ell}).
\end{align}

For the reduction, we also need to consider the dilaton and $B$-field.  For derivatives of the dilaton, we find
\begin{equation}
    \hat\nabla_\alpha\hat\nabla_\beta\phi=\nabla_\alpha\nabla_\beta\phi,\qquad\hat\nabla_\alpha\hat\nabla_b\phi=-\frac{1}{2}e_{ib}F^{i}_{\alpha\gamma}\partial^\gamma\phi,\qquad \hat\nabla_a\hat\nabla_b\phi=P_{\gamma ab}\partial^\gamma\phi.
\end{equation}
For the $B$-field, we make the sphere reduction ansatz
\begin{equation}
        B=b+B_i\land\eta^i+\frac{1}{2}b_{ij}\,\eta^i\land\eta^j+m\omega_{(2)},
\end{equation}
such that
\begin{equation}
        \dd\omega_{(2)}=\frac{1}{3!}\epsilon_{ijk}\,\sigma^i\land\sigma^j\land\sigma^k,
\end{equation}
is the volume form on $S^3$, and $m$ is the three-form flux supporting the reduction.  This then leads to an expression for the three-form field strength in terms of the seven-dimensional field strengths (three-form $h$, two-form $\tilde G_i$ and one-form $G_{ij}$) according to
\begin{equation}
        H=h+\tilde G_i\land\eta^i+\frac{1}{2}G_{ij}\land\eta^i\land\eta^j+\frac{m}{6}\epsilon_{ijk}\,\eta^i\land\eta^j\land\eta^k,
\end{equation}
where
\begin{align}
        h&=\dd b-B_i\land F^i-\frac{m}{6}\epsilon_{ijk}A^i\land A^j\land A^k,\nn\\
        \tilde G_i&=G_i-b_{ij}F^j+\frac{m}{2}\epsilon_{ijk}A^j\land A^k,\qquad G_i=D B_i,\nn\\
        G_{ij}&=Db_{ij}+\epsilon_{kij}B_k-m\epsilon_{kij}A^k.
\label{eq:hGiGij}
\end{align}
Note that these field strengths satisfy the Bianchi identities
\begin{align}
        \dd h&=-\tilde G_i\land F^i,\nn\\
        D\tilde G_i&=-G_{ij}\land F^j,\nn\\
        DG_{ij}&=-m\epsilon_{ijk}F^k+\epsilon_{kij}\tilde G_k.\label{eq:ogBianchi}
\end{align}

\subsection{The bosonic reduction at leading order}
At the leading two-derivative level, the above reduction ansatz leads to the reduced bosonic Lagrangian
\begin{align}
        e^{-1}\mathcal L^{(0)}&=e^{-2\varphi}\Bigl[R+4(\partial\varphi)^2-\frac{1}{12}h^2-\frac{1}{4}\qty(g_{ij}F^i_{\alpha\beta}F^j_{\alpha\beta}+g^{ij}\tilde G_{\mu\nu\,i}\tilde G_{\mu\nu\,j})\nn\\
        &\kern4em-(P_\mu{}^{ab})^2-\frac{1}{4}g^{ij}g^{k\ell}G_{\mu ik}G_{\mu j\ell}\nn\\
        &\kern4em-\frac{1}{4}\qty(\epsilon^{ijk}\epsilon^{\ell mn}g_{i\ell}g^{jm}g^{kn}+2\epsilon^{ijk}\epsilon^{ji\ell}g^{k\ell})-\frac{1}{2}m^2\Bigr].\label{eq:redL}
\end{align}
For comparison, note that the first two lines of \eqref{eq:redL} is a gauge covariantized version of a standard torus reduction \cite{Maharana:1992my,Liu:2023fqq}, while the last line is a scalar potential generated by the gauged geometry. Note also that there is a natural identification of the 7D dilaton as
\begin{equation}
        \varphi=\phi-\frac{1}{4}\log\det g_{ij}.\label{eq:dilatonshift}
\end{equation}
Along with the reduced Lagrangian, we have a set of reduced equations of motion at leading order.  The reduced Einstein equation becomes
\begin{align}\label{eq:redEinstein}
        \mathcal E^{(0)}_{g,\alpha\beta}=\,&R_{\alpha\beta}(\omega)-\frac{1}{2}\qty(F_{\alpha\gamma}^iF_{\beta\gamma}^jg_{ij}+\tilde G_{\alpha\gamma i}\tilde G_{\beta\gamma j}g^{ij})+2\nabla_\alpha\nabla_\beta\varphi-P_{\alpha cd}P_\beta{}^{cd}-\frac{1}{4}h_{\alpha\gamma\delta}h_\beta{}^{\gamma\delta}\nonumber\\
        &-\frac{1}{4}G_{\alpha ij}G_{\beta k\ell}g^{ik}g^{j\ell},\nn\\
        \mathcal E^{(0)}_{g,\alpha b}=\,&\frac{1}{2}e^{2\varphi}D^{\gamma}\qty(e^{-2\varphi}F_{\alpha\gamma}^i)e_i^b+F_{\alpha\gamma}^ie_i^a P_{\gamma ba}-\frac{1}{4}h_{\alpha\gamma\delta}\tilde G_i^{\gamma\delta}e^i_b+\frac{1}{2}\tilde G_{\alpha\gamma i}G^{\gamma}_{jk}g^{ik}e^j_b\nonumber\\
        &-\frac{m}{4}G_{\alpha ij}g^{i\ell}g^{jm}\epsilon_{k\ell m}e^k_b,\nn\\
        \mathcal E^{(0)}_{g,ab}=\,&-e^{2\varphi}\nabla^\gamma (e^{-2\varphi} P_{\gamma ab})+2P_{\gamma (a|d}Q_{\gamma d|b)}-\frac{1}{12}\mathcal C_{f,ac}\mathcal C_{f,bc}\nonumber\\
        &+\frac{1}{4}\qty(F_{\gamma\delta}^iF_{\gamma\delta}^je_i^ae_j^b-\tilde G_{\gamma\delta i}\tilde G_{\gamma\delta j}e^i_a e^j_b)-\frac{1}{2}G_{\delta ik}G_{\delta j\ell}h^{k\ell}e^i_a e^j_b\nonumber\\
        &-\frac{m^2}{4}\epsilon_{ik\ell}\epsilon_{jmn}g^{km}g^{\ell n}e^i_a e^j_b,
\end{align}
while the $H$ equation becomes
\begin{align}\label{eq:redH}
        \mathcal E^{(0)}_{H,\alpha\beta}=\,&e^{2\varphi}\nabla^\gamma\qty(e^{-2\varphi}h_{\gamma\alpha\beta}),\nn\\
        \mathcal E^{(0)}_{H,\alpha b}=\,&e^{2\varphi}D^\gamma\qty(e^{-2\varphi}\tilde G_{\gamma\alpha i})e^i_b-2P_{\gamma bd}\tilde G_{\gamma\alpha i}e^i_d+\frac{1}{2}h_{\alpha\gamma\delta}F^i_{\gamma\delta}e_i^b+\frac{1}{2}\mathcal C_{c,bd}G_{\alpha ij}e^i_d e^j_c,\nn\\
        \mathcal E^{(0)}_{H,ab}=\,&e^{2\varphi}\nabla^\mu\qty(e^{-2\varphi}G_{\mu ij}e^i_a e^j_b)-F_{\alpha\gamma}^i\tilde G_{\alpha\gamma j}e_i^{[a}e^{b]j}+2Q_{\alpha[a|c} G_{\alpha c|b]}\nonumber\\
        &-2P_{\gamma c[a|}G_{\gamma ij}e^i_ce^j_{|b]}-\mathcal C_{c,[a|d}\epsilon^{ijk}e^i_d e^j_c e^k_{|b]}.
\end{align}
Finally, the dilaton equation becomes
\begin{align}\label{eq:redPhi}
        \mathcal E^{(0)}_{\phi}&=R(\omega)-\frac{1}{4}\qty(F^i_{\alpha\beta}F^j_{\alpha\beta}g_{ij}+\tilde G_{\alpha\beta i}\tilde G_{\alpha\beta j}g^{ij})-\frac{1}{12}h^2-4(\partial\varphi)^2+4\Box\varphi-\qty(P_{\mu ab})^2\nonumber\\
        &-\frac{1}{4}G_{\alpha ij}G_{\alpha k\ell}g^{ik}g^{j\ell}-\frac{1}{12}\mathcal C^2-\frac{m^2}{12}\epsilon^{ijk}\epsilon^{\ell mn}g^{i\ell}g^{jm}g^{kn}.
\end{align}
It is straightforward to check that the above equations of motion follow from the Lagrangian \eqref{eq:redL}, which confirms that the reduction is indeed consistent.
    
\subsection{The fermionic reduction at leading order}
We may also reduce the variations of the fermion fields. Since the dilaton is shifted \eqref{eq:dilatonshift}, we must likewise shift the dilatino as
\begin{equation}
        \tilde\lambda=\lambda-\Gamma^i\psi_i.
\end{equation}
The leading order fermion transformations then reduce to
\begin{align}
        \delta_\epsilon\psi_\mu^{(0)}&=\qty[\nabla_\mu(\omega_-)+\frac{1}{4}Q_{\mu ab}\Gamma^{ab}+\frac{1}{4}\qty(g_{ij}F^j_{\mu\nu}-\tilde G_{\mu\nu i})\gamma^\nu\Gamma^i-\frac{1}{8}G_{\mu ij}\Gamma^{ij}]\epsilon,\nn\\
        \delta_\epsilon\psi_i^{(0)}&=\Bigg[-\frac{1}{8}\qty(g_{ij}F^j_{\mu\nu}+\tilde G_{\mu\nu i})\gamma^{\mu\nu}-\frac{1}{2}e_i^b\qty(P_{\mu ab}+\frac{1}{2}G_{\mu jk}e_a^je^k_b)\gamma^\mu \Gamma^a\nonumber\\
        &\qquad+\frac{1}{8}\qty(\mathcal C_{c,ab}e^c_i-m\epsilon_{ijk}e^j_a e^k_b)\Gamma^{ab}\Bigg]\epsilon,\nn\\
        \delta_\epsilon\tilde\lambda^{(0)}&=\left[\gamma^\mu\partial_\mu\varphi-\frac{1}{12}h_{\mu\nu\rho}\gamma^{\mu\nu\rho}+\frac{1}{8}\qty(g_{ij}F^j_{\mu\nu}-\tilde G_{\mu\nu i})\gamma^{\mu\nu}\Gamma^i\right.\nn\\
        &\left.\qquad-\frac{1}{8}\qty(\mathcal C_{c,ab}-\frac{m}{3}\epsilon_{ijk}e^i_ce^j_a e^k_b)\Gamma^{abc}\right]\epsilon.\label{eq:redVar}
    \end{align}
Notice that the composite connection $Q$ appears in the gravitino variation to make the derivative covariant with respect to this connection.  Since the reduced gravitino, $\psi_\mu$, and dilatino, $\tilde\lambda$, are in the gravity multiplet, while the internal gravitino components, $\psi_i$, are in the vector multiplets, we can identify the graviphoton and vector multiplet gauge field combinations as
\begin{align}
        F_{\mu\nu}^{a\,(-)}&=e^a_iF_{\mu\nu}^i-e_a^i\tilde G_{\mu\nu\,i},&&(\hbox{graviphoton})\nn\\
        F_{\mu\nu}^{a\,(+)}&=e^a_iF_{\mu\nu}^i+e_a^i\tilde G_{\mu\nu\,i}.&&(\hbox{vector})
    \label{eq:gpv1}
\end{align}
This matches the torus case of Chapter \ref{chap:torus} and will be used as a guide to truncating out the three vector multiplets below.

\section{The bosonic truncation}\label{sec:bosonic1}

While the $SU(2)$ reduction includes three additional vectors coming from the reduced $B$-field, it is possible to consistently truncate them away at the two-derivative level, leading to pure 7D gauged supergravity \cite{Chamseddine:1999uy,Lu:2006ah}.  In this section, we demonstrate that it remains consistent to truncate out the additional vector multiplets at the four-derivative level by analysis of the bosonic equations of motion.  Before doing so, however, we review how the truncation works at the two-derivative level to set the stage for the four-derivative truncation.

\subsection{The leading order truncation}
At the leading order, the natural choice of truncation is \cite{Chamseddine:1999uy,Lu:2006ah}
\begin{equation}
    g_{ij}=g^{-2}\delta_{ij},\qquad b_{ij}=0,\qquad B_{i}=-g_{ij}A^j=-g^{-2}A^i,
\label{eq:truncation}
\end{equation}
where we have introduced the 7D gauge coupling constant $g$.  Here the choice of minus sign in the relation between $B_i$ and $A^i$ is motivated by the desire to truncate away the vector multiplets as identified in \eqref{eq:gpv1}.  As expected for a sphere reduction, the gauge coupling $g$ is necessarily related to the three-form flux $m$ on $S^3$.  To fix the relation between $m$ and $g$, we note that, since $B_i\propto A^i$, we expect a similar relation with the field strengths, $\tilde G_i\propto F^i$.  Substituting (\ref{eq:truncation}) into (\ref{eq:hGiGij}), we obtain
\begin{equation}
    \tilde G_i=-g^{-2}\qty[\dd A^i-\qty(1+\frac{m g^2}{2})\epsilon^{ijk}A^j\land A^k],
\end{equation}
and so we must pick $m=-g^{-2}$ in order to get a properly covariant field strength, $\tilde G_i=-g_{ij}F^i$. This can also be seen from the truncated scalar field strength term
\begin{equation}
    G_{ij}=-\epsilon_{ijk}(m+g^{-2})A^k.
\end{equation}
We expect this expression to vanish since we are truncating away the scalars with $b_{ij}=0$.

Alternatively, we could have started by freezing out the scalars with
\begin{equation}
    g_{ij}=g^{-2}\delta_{ij},\qquad b_{ij}=0,
\end{equation}
in which case the scalar equation arising from the internal Einstein equation becomes
\begin{equation}
        \mathcal E^{(0)}_{g,ab}=\frac{1}{4}\qty(g^{-2}F^i_{\gamma\delta}F^j_{\gamma\delta}-g^2\tilde G_{i\gamma\delta}\tilde G_{j\gamma\delta})\delta^i_a\delta^j_b-\frac{g^6}{2}\qty(m^2-g^{-4})\delta_{ab}.
\end{equation}
This tells us that we must pick the truncation
\begin{equation}
    B_i=\pm g^{-2}A^i,\qquad m=\pm g^{-2},
\end{equation}
to consistently remove this scalar equation as a constraint.  At the bosonic two-derivative level, either sign choice is valid, suggesting that either the graviphotons or the vector multiplet vectors can be removed.  However, based on supersymmetry, we must choose the minus sign to truncate out the vector multiplets while preserving supersymmetry in the gravity multiplet.

After truncation, the two-derivative equations of motion become
\begin{align}
    \mathcal E^{(0)}_{g,\alpha\beta}&=R_{\alpha\beta}(\omega)-g^{-2}F^i_{\alpha\gamma}F^i_{\beta\gamma}+2\nabla_\alpha\nabla_\beta\varphi-\frac{1}{4}h_{\alpha\gamma\delta}h_\beta{}^{\gamma\delta},\nn\\
    \mathcal E^{(0)}_{g,\alpha b}&=\frac{1}{2g}\qty[e^{2\varphi}D^\gamma\qty(e^{-2\varphi}F^i_{\gamma\alpha})-\frac{1}{2}h_{\alpha\gamma\delta}F^i_{\gamma\delta}]\delta^i_b,\nn\\
    \mathcal E^{(0)}_{g,ab}&=0,\nn\\
    \mathcal E^{(0)}_{H,\alpha\beta}&=e^{2\varphi}\nabla^\gamma\qty(e^{-2\varphi}h_{\gamma\alpha\beta}),\nn\\
    \mathcal E^{(0)}_{H,\alpha b}&=-g^{-1}\qty[e^{2\varphi}D^\gamma\qty(e^{-2\varphi}F^i_{\gamma\alpha})-\frac{1}{2}h_{\alpha\gamma\delta}F^i_{\gamma\delta}]\delta^i_b,\nn\\
    \mathcal E^{(0)}_{H,ab}&=0,\nn\\
    \mathcal E^{(0)}_{\varphi}&=R(\omega)-\frac{1}{2g^2}F^i_{\alpha\beta}F^i_{\alpha\beta}-\frac{1}{12}h^2-4(\partial\varphi)^2+4\Box\varphi+g^2.\label{eq:twoderivEOMs}
\end{align}
In particular, the scalar equations vanish, and the gauge field equations are proportional; hence, the truncation is indeed consistent. Note also that, after truncation, the $h$ Bianchi identity becomes
\begin{align}
        \dd h&=g^{-2}F_i\land F^i.\label{eq:hbianchi}
    \end{align}
The above equations of motion, \eqref{eq:twoderivEOMs}, correspond to the reduced Lagrangian \cite{Chamseddine:1999uy,Lu:2006ah}
\begin{align}
        e^{-1}\mathcal L^{(0)}=&e^{-2\varphi}\qty[R+4(\partial\varphi)^2-\frac{1}{12}h^2-\frac{1}{2g^2}\qty(F^i_{\alpha\beta})^2+g^2],
\end{align}
which matches the bosonic sector of gauged half-maximal 7D supergravity \cite{Townsend:1983kk} with gauge coupling constant $g$ related to the flux on $S^3$ according to $g^2=-1/m$.  (In our conventions, this indicates that the flux parameter $m$ is negative.)

\subsection{The truncation at $\mathcal O(\alpha')$}

We now extend the truncation at the four-derivative level.  Here it is important to note that the two-derivative truncation, (\ref{eq:truncation}), may require $\mathcal O(\alpha')$ corrections.  We thus write
\begin{align}
    &B_{\mu i}=-g^{-2}A_\mu^{i}+\alpha'\delta B_{\mu i}\qquad
    g_{ij}=g^{-2}\delta_{ij}+\alpha'\delta g_{ij},\qquad b_{ij}=0+\alpha'\delta b_{ij}.
\label{eq:corr}
\end{align}
We also split the equations of motion as
\begin{equation}
        \mathcal E=\mathcal E^{(0)}+\alpha'\qty(\delta\mathcal E^{(0)}+\mathcal E^{(1)}),
\end{equation}
where $\delta \mathcal E^{(0)}$ is the shift of $\mathcal E^{(0)}$ generated by the field redefinitions \eqref{eq:corr}. At the order we are interested in, the four-derivative piece $\mathcal E^{(1)}$ will only depend on the two-derivative truncation. As in the torus reduction case, the necessary condition for consistency of the truncation is that the scalar equations vanish
\begin{equation}
        \delta\mathcal E^{(0)}_{g,ij}+\mathcal E^{(1)}_{g,ij}=0,\qquad 
        \delta\mathcal E^{(0)}_{H,ij}+\mathcal E^{(1)}_{H,ij}=0,
\end{equation}
and the vector equations are compatible
\begin{equation}
        \delta\mathcal E^{(0)}_{g,\alpha i}+\mathcal E^{(1)}_{g,\alpha i}=-\frac{1}{2}\qty(\delta\mathcal E^{(0)}_{H,\alpha i}+\mathcal E^{(1)}_{H,\alpha i}).
\end{equation}

In order to compute the $\mathcal O(\alpha')$ equations of motion, $\mathcal E^{(1)}$, we require the torsionful Riemann tensor $R_{MN}{}^{AB}(\Omega_+)$ and the Lorentz-Chern-Simons form $\omega_{3L}(\Omega_+)$.  These can be obtained using the lowest order torsionful spin connection components, which become, after truncation
\begin{equation}
        \Omega_+=\begin{pmatrix}
        \omega_+^{\alpha\beta}-g^{-2}F_{\alpha\beta}^i\eta^i&0\\
        0&-\epsilon^{ijk}\delta^i_a\delta^j_b\sigma^k
        \end{pmatrix}.
\end{equation}
This then leads to expressions for the torsionful Riemann tensor components
\begin{align}
        R_{\gamma\delta}{}^{\alpha\beta}(\Omega_+)=&R_{\gamma\delta}{}^{\alpha\beta}(\omega_+)-g^{-2} F^i_{\alpha\beta}F^i_{\gamma\delta},\nn\\
        R_{\gamma d}{}^{\alpha\beta}(\Omega_+)=&-g^{-1}D_\gamma^{(+)}F_{\alpha\beta}^i\delta_i^d,\nn\\
        R_{cd}{}^{\alpha\beta}(\Omega_+)=&2g^{-2} F_{\alpha\gamma}^iF_{\gamma\beta}^j\delta_i^{[c}\delta_j^{d]}+\epsilon^{ijk}F^i_{\alpha\beta}\delta^j_c\delta^k_d,&
\end{align}
with all other independent components vanishing. Here $D^{(+)}$ is taken to mean $D(\omega_+)$. The Lorentz-Chern-Simons form is given by
\begin{align}
    \omega_{3L,\alpha\beta\gamma}(\Omega_+)&=\omega_{3L,\alpha\beta\gamma}(\omega_+)+2\epsilon^{ijk}A^i_\alpha A^j_\beta A^k_\gamma,\nonumber\\
    \omega_{3L,\alpha\beta c}(\Omega_+)&=\delta^i_c\qty(2g^{-1} R_{\alpha\beta}{}^{\gamma\delta}(\omega_+)F_{\gamma\delta}^{i}-g^{-3} F_{\alpha\beta}^{j}F_{\gamma\delta}^{j}F_{\gamma\delta}^{i}-2\epsilon^{ijk}A^j_\alpha A^k_\beta),\nonumber\\
    \omega_{3L,\alpha bc}(\Omega_+)&=\delta^{[i}_b\delta^{j]}_c\qty(2g^{-2}F_{\gamma\delta}^{i}D_\alpha^{(+)}F_{\gamma\delta}^{j}+2\epsilon^{kij}A^k_\alpha),\nonumber\\
    \omega_{3L,abc}(\Omega_+)&=\delta_a^{[i}\delta_b^j\delta_c^{k]}\qty(-4g^{-3}F_{\alpha\beta}^{i}F_{\beta\gamma}^{j}F_{\gamma\alpha}^{k}-2\epsilon^{ijk}).
\end{align}
Given the shifted $H$-field, \eqref{eq:Htilde}, the additional terms proportional to $A^i$ in $\omega_{3L}(\Omega_+)$ hint that we should define modified field strengths
\begin{align}
        \bar h&=\dd b-B_i\land F^i+\frac{1}{6}\qty(g^{-2}-\frac{\alpha'}{2})\epsilon_{ijk}A^i\land A^j\land A^k,\nn\\
        \bar G_i&=G_i-b_{ij}F^j+\qty(-\frac{1}{2g^2}+\frac{\alpha'}{4})\epsilon_{ijk}A^j\land A^k,\nn\\
        \bar G_{ij}&=Db_{ij}+\epsilon_{kij}B_k+\qty(g^{-2}-\frac{\alpha'}{2})\epsilon_{kij}A^k.
\end{align}
In order for $\bar G_i$ and $\bar G_{ij}$ to be proper gauge-covariant field strengths for $B_i$ and $b_{ij}$, we must choose the four-derivative truncation
\begin{equation}
        B_i=\qty(-g^{-2}+\frac{\alpha'}{2})A^i.
\end{equation}
which results in the modified field strengths truncating to
\begin{equation}
        \bar h=\dd b+\qty(g^{-2}-\frac{\alpha'}{2})\omega_{3Y},\qquad\bar G_i=\qty(-g^{-2}+\frac{\alpha'}{2})F^i,\qquad \bar G_{ij}=0,
\end{equation}
where we have defined the Yang-Mills Chern-Simons term as usual
\begin{equation}
        \omega_{3Y}=A^i\land F^i+\frac{1}{6}\epsilon^{ijk}A^i\land A^j\land A^k.
\end{equation}
One may be tempted then to view this as a shift of $g$
\begin{equation}
        \tilde g^{-2}= g^{-2}-\frac{\alpha'}{2}.\label{eq:gshift}
\end{equation}
Comparing with the field redefinitions of Chapter \ref{chap:torus} and shifting $g$ according to \eqref{eq:gshift}, we naturally infer the truncation
\begin{equation}
        \delta B_i=\frac{1}{2}A^i,\qquad\delta g_{ij}=\frac{1}{4g^4}F^i_{\alpha\beta} F^j_{\alpha\beta}-\frac{1}{2}\delta_{ij},\qquad \delta b_{ij}=0.
\label{eq:trunc}
\end{equation}
We will show that \eqref{eq:trunc} is indeed a consistent truncation. While we omit some details, the steps parallel those of Chapter \ref{chap:torus}, with additional terms due to the gauging that must be taken care of.

\subsubsection{Consistency of the truncation}

We start with the internal components of the Einstein equation.  Given (\ref{eq:trunc}), the shift to the two-derivative equations of motion for $g_{ab}$ are
\begin{align}
        \delta\mathcal E^{(0)}_{ab}=\,&-e^{2\varphi}\nabla^\gamma (e^{-2\varphi} \delta P_{\gamma ab})+2\delta P_{\gamma ad}Q_{\gamma db}-\frac{1}{6}\mathcal C_{f,ac}\delta\mathcal C_{f,bc}\nonumber\\
        &+\frac{1}{4}\qty(F_{\gamma\delta}^iF_{\gamma\delta}^j\delta(e_i^ae_j^b)-2\tilde G_{\gamma\delta i}\delta\tilde G_{\gamma\delta j}e^i_a e^j_b-\tilde G_{\gamma\delta i}\tilde G_{\gamma\delta j}\delta(e^i_a e^j_b))\nonumber\\
        &-\frac{m^2}{4}\epsilon_{ik\ell}\epsilon_{jmn}\delta(g^{km}g^{\ell n}e^i_a e^j_b),
\label{eq:dEab0}
\end{align}
where the $(ab)$ indices are implicitly symmetrized. Substituting in the lowest order equations of motion and making use of the $h$ Bianchi identity \eqref{eq:hbianchi}, one can derive the useful formula
\begin{align}
        e^{2\varphi}\nabla^\alpha\qty(e^{-2\varphi}\delta P_{\alpha ab})=&\Bigl[ -\frac{1}{4g^2}R_{\alpha\beta\gamma\delta}(\omega_+)F^i_{\alpha\beta}F^j_{\gamma\delta}+\frac{1}{4g^2}D_\gamma F_{\alpha\beta}^i D_\gamma F_{\alpha\beta}^j+\frac{1}{4g^4}F^i_{\alpha\beta}F^{k}_{\alpha\beta}F^j_{\gamma\delta}F^k_{\gamma\delta}\nn\\
        &+\frac{1}{2g^4} F^i_{\alpha\beta}F^j_{\beta\gamma}F^k_{\gamma\delta}F^k_{\delta\alpha}-\frac{1}{2g^4} F^i_{\alpha\beta}F^k_{\beta\gamma}F^j_{\gamma\delta}F^k_{\delta\alpha}-\frac{1}{2g^2} \epsilon^{jk\ell}F^i_{\alpha\beta}F^{k}_{\beta\gamma}F^\ell_{\gamma\alpha}\nn\\
        &-\frac{1}{2g}D_\gamma\mathcal E^{(0)}_{H,\delta j}F^i_{\gamma\delta}-\frac{1}{2g^2} \mathcal E^{(0)}_{g,\alpha\beta}F^j_{\gamma\alpha}F^i_{\beta\gamma}\Bigr]\delta^i_{(a}\delta^j_{b)}.
\end{align}
We may also evaluate
\begin{align}
        \mathcal E^{(1)}_{g,ab}=&\Bigg[-\frac{1}{8g^2}F^i_{\gamma\delta}F^j_{\alpha\beta}\qty(2R_{\gamma\delta}{}^{\alpha\beta}(\omega_+)-g^{-2}F^k_{\gamma\delta}F^k_{\alpha\beta})+\frac{1}{2g^4}F^i_{\alpha\beta}F^j_{\beta\gamma}F^k_{\gamma\delta}F^k_{\delta\alpha}\nn\\
        &\quad-\frac{1}{2g^4}F^i_{\alpha\beta}F^k_{\beta\gamma}F^j_{\gamma\delta}F^k_{\delta\alpha}+\frac{1}{4g^2}D^{(+)}_\gamma F^i_{\alpha\beta}D^{(+)}_\gamma F^j_{\alpha\beta}+\frac{1}{2}\delta_{ij}-\frac{1}{2g^2}\epsilon^{ik\ell}F^k_{\alpha\beta}F^{\ell}_{\beta\gamma}F^j_{\gamma\alpha}\nn\\
        &\quad+\frac{1}{4}F^2\delta_{ij}-\frac{1}{4}F^i_{\alpha\beta}F^j_{\alpha\beta}-\frac{1}{4g}\epsilon^{ik\ell}A^k_{\gamma}A^\ell_\delta F^i_{\gamma\delta}\Bigg]\delta^i_{(a}\delta^j_{b)},
\label{eq:scalarEOM1}
\end{align}
which may then be used to obtain that
\begin{equation}
        \delta\mathcal E^{(0)}_{g,ij}+\mathcal E^{(1)}_{g,ij}=\frac{1}{2g}\qty(D_\gamma\mathcal E^{(0)}_{H,\delta j}F^i_{\gamma\delta}+g^{-1} \mathcal E^{(0)}_{g,\alpha\beta}F^j_{\gamma\alpha}F^i_{\beta\gamma}),
\label{eq:scalarEOM}
\end{equation}
which vanishes upon imposing the leading-order equations of motion. In particular, the last term in \eqref{eq:scalarEOM1} with the bare $A$'s is precisely canceled by the corresponding shift $\delta\tilde G_i$ in (\ref{eq:dEab0}).

Similarly, $\delta\mathcal E^{(0)}_{H,ij}$ almost vanishes except for the bare $A^i$ that show up to account for the shifts to the modified field strengths $\bar G_i$ and $\bar G_{ij}$, but these are precisely canceled by the terms appearing in $\mathcal E^{(1)}_{H,ij}$. Keeping careful track of terms, it is straightforward to work out that
\begin{equation}
        \delta\mathcal E^{(0)}_{H,ij}+\mathcal E^{(1)}_{H,ij}=-g^{-1}D_\alpha\mathcal E^{(0)}_{H,\beta j}F^i_{\alpha\beta}-g^{-2}\mathcal E^{(0)}_{H,\alpha\beta}F^i_{\alpha\gamma}F^j_{\beta\gamma},
\end{equation}
which also vanishes by the leading-order equations of motion. This thus confirms that it is consistent to truncate out the scalars $g_{ij}$ and $b_{ij}$.
    
Finally, we turn our attention to the compatibility of the two Yang-Mills equations. Again, we can derive a useful formula
\begin{align}
    e^{2\varphi}D^\beta\qty(e^{-2\varphi}R_{\beta\alpha}{}^{\gamma\delta}(\omega_+)F^i_{\gamma\delta})=&\Bigg[-g^{-2}\nabla_\gamma\qty(F^{j}_{\delta\epsilon}F^{j}_{\epsilon\alpha})-\frac{1}{4}\nabla_\gamma\qty(h_{\delta\beta\epsilon}h_{\alpha\beta\epsilon})+\frac{1}{2}R_{\gamma[\delta|\beta\epsilon}h_{|\alpha]\beta\epsilon}\nn\\
    &-\frac{1}{2g^2}h_{\delta\alpha\epsilon}F^{j}_{\gamma\beta}F^{j}_{\epsilon\beta}-\frac{1}{8}h_{\delta\alpha\epsilon}h_{\gamma\beta\omega}h_{\epsilon\beta\omega}-\frac{1}{4g^2}h_{\alpha\beta\epsilon}F^{j}_{\beta\epsilon}F^{j}_{\gamma\delta}\nn\\
    &+\frac{1}{2g^2}F^{j}_{\alpha\beta}D^\beta F^{j}_{\gamma\delta}-\frac{1}{2g^2}h_{\delta\beta\epsilon}F^{j}_{\beta\epsilon}F^{j}_{\alpha\gamma}-g^{-2}F^{j}_{\beta\delta}D^\beta F^{j}_{\alpha\gamma}\nn\\
    &+\frac{1}{4}h_{\beta\epsilon\delta}\nabla^\beta h_{\alpha\gamma\epsilon}-\nabla_{[\gamma}\mathcal E^{(0)}_{g,\delta]\alpha}-\frac{1}{2}\nabla_\gamma\mathcal E^{(0)}_{H,\delta\alpha}-\partial_\gamma\varphi\mathcal E^{(0)}_{H,\delta\alpha}\nn\\
    &-\frac{1}{2}h_{\delta\alpha\epsilon}\mathcal E^{(0)}_{H,\gamma\epsilon}+\frac{1}{4}h_{\alpha\gamma\epsilon}\mathcal E^{(0)}_{H,\epsilon\delta}\Bigg]F^{i}_{\gamma\delta}+R_{\alpha\beta}^{\ \ \ \gamma\delta}(\omega_+)D^\beta F^{i}_{\gamma\delta}.
\end{align}
which may then be used to show that
\begin{align}
    &\qty(\delta \mathcal E^{(0)}_{H,\alpha a}+\mathcal E^{(1)}_{H,\alpha a})+2\qty(\delta \mathcal E^{(0)}_{g,\alpha a}+\mathcal E^{(1)}_{g,\alpha a})\nn\\
    &\kern4em=-\frac{1}{2}\delta_i^aF^{i}_{\gamma\delta}\Bigg[-\nabla_{[\gamma}\mathcal E^{(0)}_{g,\delta]\alpha}-\frac{1}{2}\nabla_\gamma\mathcal E^{(0)}_{H,\delta\alpha}-\partial_\gamma\varphi\mathcal E^{(0)}_{H,\delta\alpha}-\frac{1}{2}h_{\delta\alpha\epsilon}\mathcal E^{(0)}_{H,\gamma\epsilon}+\frac{1}{4}h_{\alpha\gamma\epsilon}\mathcal E^{(0)}_{H,\epsilon\delta}\Bigg]\nn\\
    &\kern5em+\delta_i^a\qty(\delta e^a_i \mathcal E^{(0)}_{H,\alpha a}+2\delta e^i_a\mathcal E^{(0)}_{g,\alpha a}),
\end{align}
which demonstrates that the two equations are indeed consistent after imposing the two-derivative equations of motion.

\subsubsection{The surviving equations of motion}

Here we summarize the equations of motion for the remaining degrees of freedom, namely the 7D metric $g_{\mu\nu}$, the two-form $b$-field, the graviphoton $A^i$, and the dilaton $\varphi$.  The equations of motion for the metric become
\begin{align}
    \mathcal E_{g,\alpha\beta}&=R(\omega)_{\alpha\beta}-\tilde g^{-2}F_{\alpha\gamma}^{i}F_{\beta\gamma}^{i}-\fft14\tilde h_{\alpha\gamma\delta}\tilde h_{\beta\gamma\delta}+2\nabla_\alpha\nabla_\beta\varphi\nn\\
    &\quad+\frac{\alpha'}{4}\biggl(R_{\alpha\gamma\delta\epsilon}(\omega_+)R_\beta{}^{\gamma\delta\epsilon}(\omega_+)-4g^{-2}R_{\alpha}{}^{\gamma\delta\epsilon}(\omega_+)F^{i}_{\beta\gamma}F^{i}_{\delta\epsilon}\nn\\
    &\kern4em+2g^{-4}F^{i}_{\alpha\gamma}F_{\beta}{}^{\gamma\,j}F^{i}_{\delta\epsilon}F^{\delta\epsilon\,j}+g^{-2}D^{(+)}_\alpha F^{i}_{\gamma\delta}D^{(+)}_
    \beta F^{\gamma\delta\,i}\biggr),
\end{align}
where $(\alpha\beta)$ symmetrization is assumed implicitly. Here, we have conveniently defined
\begin{equation}
        \tilde h\equiv\bar h-\frac{\alpha'}{4}\omega_{3L}(\omega_+)=\dd b+\tilde g^{-2}\omega_{3Y}-\frac{\alpha'}{4}\omega_{3L}(\omega_+),
\end{equation}
such that the Bianchi identity becomes
\begin{equation}
        \dd\tilde h=\tilde g^{-2}F^i\land F^i-\frac{\alpha'}{4}\Tr R(\omega_+)\land R(\omega_+).
\end{equation}
This newly defined $\tilde h$ then has the equation of motion
\begin{equation}
        \mathcal E_{H,\alpha\beta}=e^{2\varphi}\nabla^\gamma\qty(e^{-2\varphi}\tilde h_{\alpha\beta\gamma}).
\end{equation}

The graviphoton equation becomes
\begin{align}
    \mathcal E_{A,\alpha i}&=\tilde g^{-1}e^{2\varphi}\nabla^\gamma\qty(e^{-2\varphi}F_{\gamma\alpha}^{i})-\frac{1}{2\tilde g}\tilde h_{\alpha\beta\gamma}F_{\beta\gamma}^{i}\nn\\
    &\quad+\fft{\alpha'}4\biggl(-g^{-1}\tilde h_{\alpha\beta\gamma}R_{\beta\gamma\delta\epsilon}(\omega_+)F_{\delta\epsilon}^{i}+g^{-3}\tilde h_{\alpha\beta\gamma}F_{\beta\gamma}^{j}F_{\delta\epsilon}^{i}F_{\delta\epsilon}^{j}-2g^{-1}R_{\alpha\beta\gamma\delta}(\omega_+)D_\beta^{(+)}F_{\gamma\delta}^{i}\nn\\
    &\kern4em+2g^{-3}F_{\alpha\gamma}^{j}F_{\delta\epsilon}^{j}D_\gamma^{(+)}F_{\delta\epsilon}^{i}-2g^{-3}F_{\alpha\gamma}^{j}F_{\delta\epsilon}^{i}D_\gamma^{(+)}F_{\delta\epsilon}^{j}+4g^{-3}F_{\beta\gamma}^{i}F_{\gamma\delta}^{j}D_\alpha^{(+)}F_{\beta\delta}^{j}\nn\\
    &\kern4em +\frac{1}{4g}\epsilon^{ijk}F_{\gamma\delta}^jD^{(+)}_\alpha F^k_{\gamma\delta}\biggr),
\label{eq:redA}
\end{align}
while the dilaton equation becomes
\begin{align}
    \mathcal E_\phi&=R(\omega)-\frac{1}{2\tilde g^2}F_{\alpha\beta}^{i}F_{\alpha\beta}^{i}-\fft1{12}\tilde h_{\alpha\beta\gamma}^2+4\Box\varphi-4(\partial\varphi)^2+\tilde g^2\nn\\
    &\quad+\fft{\alpha'}8\biggl((R_{\alpha\beta\gamma\delta}(\omega_+))^2-4g^{-2}R_{\alpha\beta\gamma\delta}(\omega_+)F_{\alpha\beta}^{i}F_{\gamma\delta}^{i}+2g^{-2}\qty(D_\alpha^{(+)}F_{\beta\gamma}^{i})^2\nn\\
    &\kern4em+2g^{-4}F_{\alpha\beta}^{i}F_{\beta\gamma}^{i}F_{\gamma\delta}^{j}F_{\delta\alpha}^{j}-2g^{-4}F_{\alpha\beta}^{i}F_{\beta\gamma}^{j}F_{\gamma\delta}^{i}F_{\delta\alpha}^{j}+2g^{-4}F_{\alpha\beta}^{i}F_{\alpha\beta}^{j}F_{\gamma\delta}^{i}F_{\gamma\delta}^{j}\nn\\
    &\kern4em-\frac{1}{3g^2}\epsilon^{ijk}F^i_{\alpha\beta}F^j_{\beta\gamma}F^k_{\gamma\alpha}\biggr).
    \label{eq:reddil}
\end{align}

Having shown that the truncation is consistent, we may also compute the truncated Lagrangian to be
\begin{align}
    e^{-1}\mathcal L=&e^{-2\varphi}\Biggl[R(\omega)+4(\partial\varphi)^2-\fft1{12}\tilde h_{\alpha\beta\gamma}^2-\frac{1}{2\tilde g^2}\qty(F_{\alpha\beta}^{i})^2+\tilde g^2\nn\\
    &\kern2em+\fft{\alpha'}8\biggl((R_{\alpha\beta\gamma\delta}(\omega_+))^2-4\tilde g^{-2}R_{\alpha\beta\gamma\delta}(\omega_+)F_{\alpha\beta}^{i}F_{\gamma\delta}^{i}+2\tilde g^{-2}\qty(D_\alpha^{(+)}F_{\beta\gamma}^{i})^2\nn\\
    &\kern5em+2\tilde g^{-4}F_{\alpha\beta}^{i}F_{\beta\gamma}^{i}F_{\gamma\delta}^{j}F_{\delta\alpha}^{j}-2\tilde g^{-4}F_{\alpha\beta}^{i}F_{\beta\gamma}^{j}F_{\gamma\delta}^{i}F_{\delta\alpha}^{j}+2\tilde g^{-4}F_{\alpha\beta}^{i}F_{\alpha\beta}^{j}F_{\gamma\delta}^{i}F_{\gamma\delta}^{j}\nn\\
    &\kern5em-\frac{1}{3\tilde g^2}\epsilon^{ijk}F^i_{\alpha\beta}F^j_{\beta\gamma}F^k_{\gamma\alpha}\biggr)\Biggr].\label{eq:effectiveL}
\end{align}
Notably, all of the coupling constant $g$'s at the two-derivative level combine in just the right way with $\alpha'$ to be consistent with the shift \eqref{eq:gshift}.  Since we are working only to first order in $\alpha'$, we have also replaced $g$ by $\tilde g$ in the $\mathcal O(\alpha')$ contribution to the Lagrangian.

\section{The fermionic truncation}\label{sec:fermionic1}

We now turn our attention to the fermion sector. The gravitino $\psi_M$ naturally splits into components along the spacetime directions $\psi_\mu$, which should be interpreted as the lower-dimensional gravitino, and components along the internal directions $\psi_i$, which should be interpreted as gaugini for the vectors $B_i$. Since we are truncating away the $B_i$, we expect to also truncate out the associated gaugini. 

Since we are reducing to seven dimensions, it is useful to decompose our gamma matrices as
\begin{align}
    \Gamma^\alpha&=\gamma^\alpha\otimes\mathbbm{1}\otimes\sigma^1,\nn\\
    \Gamma^a&=\mathbbm{1}\otimes\tau^a\otimes\sigma^2.
\end{align}
Here the $\gamma^\alpha$ form a seven-dimensional Clifford algebra $\mathrm{Cliff}(6,1)$, while the $\tau^a$ are the Pauli matrices of our three-dimensional Clifford algebra $\mathrm{Cliff}(3)$.\footnote{We have denoted Pauli matrices by both $\tau^a$ and $\sigma^i$. While they are the same matrices, this is done to clarify which spinor subspace they are acting on.} We take the convention that $\gamma^{0123456}=1$ and $\tau^{789}=i$. The chirality matrix then becomes
\begin{equation}
    \Gamma_{11}=\Gamma^{0123456789}=-\mathbbm{1}\otimes\mathbbm{1}\otimes\sigma^3.
\end{equation}
The choice of 10D chirality, which we take to be $\Gamma_{11}\epsilon=-\epsilon$, thus implies that $\sigma^3\epsilon=\epsilon$.  The ten-dimensional gravitino has the same chirality as $\epsilon$, while the ten-dimensional dilatino has the opposite.  Thus we can represent the heterotic Majorana-Weyl spinors as
\begin{equation}
    \epsilon\to\epsilon\otimes\begin{bmatrix}1\\0\end{bmatrix},\qquad
    \psi_M\to\psi_M\otimes\begin{bmatrix}1\\0\end{bmatrix},\qquad
    \lambda\to\lambda\otimes\begin{bmatrix}0\\1\end{bmatrix}.
\end{equation}
The spinors on the right-hand side of these expressions are 16 component spinors that further decompose into a pair of seven-dimensional spinors that are acted on by $\tau^a$.  This pair of spinors satisfies a Majorana condition that we do not concern ourselves with here.

After truncation, the leading order supersymmetry variations become
\begin{align}
        \delta_\epsilon\psi_\mu^{(0)}=&\qty[D_\mu(\omega_-)+\frac{i}{2g}F_{\mu\nu}^i\gamma^\nu\tau^{\underline i}]\epsilon,\nn\\
        \delta_\epsilon\psi_i^{(0)}=&\,0,\nn\\
        \delta_\epsilon\tilde\lambda^{(0)}=&\qty[\gamma^\mu\partial_\mu\varphi-\frac{1}{12}h_{\mu\nu\rho}\gamma^{\mu\nu\rho}+\frac{i}{4g}F^i_{\mu\nu}\gamma^{\mu\nu}\tau^{\underline i}-\frac{g}{2}]\epsilon.
\end{align}
It is also noteworthy that the composite $SU(2)$ connection, upon truncation, becomes the \textit{gauge} $SU(2)$ connection
\begin{equation}
        Q_{\alpha ab}=\epsilon^{ijk}\delta^i_a\delta^j_b A^k_\alpha,
\end{equation}
which is what promotes the covariant derivative to a gauge-covariant one in $\delta_\epsilon\psi_\mu$, which acts by
\begin{equation}
        D_\mu\epsilon=\nabla_\mu\epsilon+\frac{i}{2}A^i_\mu\tau^{\underline{i}}\epsilon.
\end{equation}
In analogy to the torus case, we make the definition
\begin{equation}
        \mathcal D_\mu\equiv D_\mu(\omega_-)+\frac{i}{2g}F_{\mu\nu}^i\gamma^\nu\tau^{\underline i},
\end{equation}
which will be useful for the fermionic field redefinitions.

The bosonic field redefinitions \eqref{eq:trunc} combined with the two-derivative truncation \eqref{eq:truncation} lead to four-derivative contributions to the supersymmetry variations
\begin{align}
        \delta\qty(\delta_\epsilon\psi_\mu^{(0)})+\delta_\epsilon\psi^{(1)}_\mu=&\frac{1}{32}\Bigg[\omega_{3L,\mu\nu\rho}(\omega_+)\gamma^{\nu\rho}+4ig^{-1}\qty(R_{\mu\nu}{}^{\alpha\beta}(\omega_+)-\frac{1}{2g^2}F^{j}_{\mu\nu}F^{j}_{\alpha\beta})F_{\alpha\beta}^{i}\gamma^\nu\tau^{\underline i}\nn\\
        &\qquad+2g^{-2}F^{i}_{\alpha\beta}D^{(+)}_\mu F^{j}_{\alpha\beta}\tau^{\underline{ij}}-4igF^i_{\mu\nu}\gamma^\nu\tau^{\underline i}\Bigg]\epsilon,\nn\\
        \delta\qty(\delta_\epsilon\psi_i^{(0)})+\delta_\epsilon\psi^{(1)}_i=&\fft1{16g}F_{\alpha\beta}^{i}\Big(R_{\gamma\delta}{}^{\alpha\beta}(\omega_+)\gamma^{\gamma\delta}-g^{-2}F_{\alpha\beta}^{j}F_{\gamma\delta}^{j}\gamma^{\gamma\delta}-2ig^{-1}D_\gamma^{(+)} F_{\alpha\beta}^{j}\gamma^\gamma\tau^{\underline j}\nn\\
        &\qquad\qquad-2g^{-2}F_{\beta\gamma}^{j}F_{\gamma\alpha}^{k}\tau^{\underline{jk}}-2iF^j_{\alpha\beta}\Gamma^{\underline j}\Big)\epsilon,\nn\\
        \delta\qty(\delta_\epsilon\tilde\lambda^{(0)})+\delta_\epsilon\tilde\lambda^{(1)}=&\frac{1}{48}\biggl[\omega_{3L,\mu\nu\rho}(\omega_+)\gamma^{\mu\nu\rho}+3ig^{-1}\left(R_{\mu\nu}{}^{\alpha\beta}(\omega_+)-\frac{1}{2g^2}F_{\mu\nu}^{j}F_{\alpha\beta}^{j}\right)F_{\alpha\beta}^{i}\gamma^{\mu\nu}\tau^{\underline{i}}\nn\\
        &\kern2em-2g^{-3}\epsilon^{ijk}F^{i}_{\alpha\beta}F^{j}_{\beta\gamma}F^{k}_{\gamma\alpha}-6g^3-3igF^i_{\mu\nu}\gamma^{\mu\nu}\tau^{\underline i}-\frac{6}{g}F^i_{\alpha\beta}F^i_{\alpha\beta}\biggr]\epsilon.
\label{eq:susyvar}
\end{align}
While the supersymmetry variation of the gaugino, $\delta_\epsilon\psi_i$, is undesirable, it has but a single extra term compared to the ungauged case. In particular, using the fact that
\begin{equation}
        [D_\mu,D_\nu]\epsilon=\frac{1}{4}\qty(R_{\mu\nu}{}^{\alpha\beta}\gamma^{\alpha\beta}+\epsilon^{ijk}F_{\mu\nu}^i\Gamma^{\underline{jk}})\epsilon,
\end{equation}
we see that an analogous field redefinition holds as that in Chapter \ref{chap:torus}
\begin{equation}
        \tilde\psi_i=\psi_i-\frac{\alpha'}{2g}F^i_{\mu\nu}\mathcal D_\mu\psi_\nu,
\end{equation}
such that
\begin{equation}
        \delta_\epsilon\tilde\psi_i=0.
\end{equation}
Thus truncating $\tilde\psi_i$ is indeed consistent with supersymmetry.

Interestingly, the higher-derivative corrections in the variations, (\ref{eq:susyvar}), appear in exactly the appropriate way to be consistent with the shifted gauge coupling \eqref{eq:gshift}. This is more readily seen in the combined expressions
\begin{align}
    \delta_\epsilon\psi_\mu=&\biggl[D_\mu(\tilde\omega_-)+\frac{i}{2\tilde g}F_{\mu\nu}^i\gamma^\nu\tau^{\underline i}+\fft{i\alpha'}{8\tilde g}\qty(R_{\mu\nu}{}^{\alpha\beta}(\omega_+)-\frac{1}{2\tilde g^2}F^{j}_{\mu\nu}F^{j}_{\alpha\beta})F_{\alpha\beta}^{i}\gamma^\nu\tau^{\underline i}\nn\\
    &\qquad+\fft{\alpha'}{16\tilde g^{2}}F^{i}_{\alpha\beta}D^{(+)}_\mu F^{j}_{\alpha\beta}\tau^{\underline{ij}}\biggr]\epsilon,\nn\\
    \delta_\epsilon\tilde\lambda=&\biggl[\gamma^\mu\partial_\mu\varphi-\frac{1}{12}h_{\mu\nu\rho}\gamma^{\mu\nu\rho}+\frac{i}{4\tilde g}F^i_{\mu\nu}\gamma^{\mu\nu}\tau^{\underline i}-\frac{\tilde g}{2}-\frac{\alpha'}{8\tilde g}F^i_{\alpha\beta}F^i_{\alpha\beta}\nn\\
    &\kern2em+\fft{i\alpha'}{16\tilde g}\left(R_{\mu\nu}{}^{\alpha\beta}(\omega_+)-\frac{1}{2\tilde g^2}F_{\mu\nu}^{j}F_{\alpha\beta}^{j}\right)F_{\alpha\beta}^{i}\gamma^{\mu\nu}\tau^{\underline{i}}-\fft{\alpha'}{24\tilde g^3}\epsilon^{ijk}F^{i}_{\alpha\beta}F^{j}_{\beta\gamma}F^{k}_{\gamma\alpha}\biggr]\epsilon,
\end{align}
where again we make no distinction between $g$ and $\tilde g$ in the $\mathcal{O}(\alpha')$ terms.

\section{Discussion}\label{sec:discussion1}
In this chapter, we have shown that the $SU(2)$ group manifold reduction of four-derivative heterotic supergravity on $S^3$ may be consistently truncated to pure half-maximal gauged 7D supergravity. This may be seen as supporting evidence that the Gauntlett-Varela conjecture \cite{Gauntlett:2007ma} extends to higher-derivative truncations. 

We may, of course, compare our results to those of the ungauged case. This requires that we rescale $A^i\to g A^i$ to recover the proper normalization and also rescale $\sigma^i\to g\sigma^i$ so that the Maurer-Cartan equation becomes
\begin{equation}
    \dd\sigma^i=-\frac{g}{2}\epsilon^{ijk}\sigma^j\land\sigma^k.
\end{equation}
Upon sending $g\to 0$, the field strength becomes abelian and the $\sigma^i$ become torus coordinates
\begin{equation}
   F^i=\dd A^i,\qquad \dd \sigma^i=0.
\end{equation}
This also has the effect of reducing the gauge-covariant derivatives $D$ to regular covariant derivatives $\nabla$. Writing $\sigma^i=\dd y^i$, we then recover the torus metric
\begin{equation}
    \dd s^2=g_{\mu\nu}\dd x^\mu\dd x^\nu+(\dd y^i+ A^i)^2,
\end{equation}
as well as the ungauged Lagrangian
\begin{align}
        e^{-1}\mathcal L=&e^{-2\varphi}\Biggl[R+4(\partial\varphi)^2-\fft1{12}\tilde h_{\alpha\beta\gamma}^2-\fft12\qty(F_{\alpha\beta}^{i})^2\nn\\
    &\kern2em+\fft{\alpha'}8\biggl((R_{\alpha\beta\gamma\delta}(\omega_+))^2-4R_{\alpha\beta\gamma\delta}(\omega_+)F_{\alpha\beta}^{i}F_{\gamma\delta}^{i}+2\qty(\nabla_\alpha^{(+)}F_{\beta\gamma}^{i})^2\nn\\
    &\kern5em+2F_{\alpha\beta}^{i}F_{\beta\gamma}^{i}F_{\gamma\delta}^{j}F_{\delta\alpha}^{j}-2F_{\alpha\beta}^{i}F_{\beta\gamma}^{j}F_{\gamma\delta}^{i}F_{\delta\alpha}^{j}+2F_{\alpha\beta}^{i}F_{\alpha\beta}^{j}F_{\gamma\delta}^{i}F_{\gamma\delta}^{j}\biggr)\Biggr].
    \end{align}
Moreover, since the shift $\sigma\to g\sigma$, $A^i\to g A^i$ effectively rescales $g_{ij}$ by $g^2$, the internal metric becomes that of the torus case
\begin{equation}
   g_{ij}=\delta_{ij}+\frac{\alpha'}{4}F^i_{\alpha\beta}F^j_{\alpha\beta}.
\end{equation}
We also recover the ungauged supersymmetry variations (in the original ten-dimensional Dirac matrix notation)
\begin{align}
    \delta_\epsilon\psi_\mu=&\Bigg[D_\mu(\tilde\omega_-)+\frac{1}{2}F_{\mu\nu}^i\Gamma^\nu\Gamma^{\underline i}+\frac{\alpha'}{8}\qty(R_{\mu\nu}{}^{\alpha\beta}(\omega_+)-\frac{1}{2}F^{j}_{\mu\nu}F^{j}_{\alpha\beta})F_{\alpha\beta}^{i}\Gamma^\nu\Gamma^{\underline i}\nn\\
    &\ +\frac{\alpha'}{16}F^{i}_{\alpha\beta}\nabla^{(+)}_\mu F^{j}_{\alpha\beta}\Gamma^{\underline{ij}}\Bigg]\epsilon,\nn\\
    \delta_\epsilon\tilde\lambda=&\Bigg[\Gamma^\mu\partial_\mu\varphi-\frac{1}{12}\tilde h_{\mu\nu\rho}\Gamma^{\mu\nu\rho}+\frac{1}{4}F^i_{\mu\nu}\Gamma^{\mu\nu}\Gamma^{\underline i}+\fft{\alpha'}{16}\left(R_{\mu\nu}{}^{\alpha\beta}(\omega_+)-\frac{1}{2}F_{\mu\nu}^{j}F_{\alpha\beta}^{j}\right)F_{\alpha\beta}^{i}\Gamma^{\mu\nu}\Gamma^{\underline{i}}\nn\\
    &\ +\fft{\alpha'}{24}F^{i}_{\alpha\beta}F^{j}_{\beta\gamma}F^{k}_{\gamma\alpha}\Gamma^{\underline{ijk}}\Bigg]\epsilon,
\end{align}
where
\begin{equation}
    \tilde\omega_-\equiv\omega-\frac{1}{2}\tilde h=\omega-\frac{1}{2}h+\frac{\alpha'}{8}\omega_{3L}.
\end{equation}
Hence, our results are consistent with the previous chapter, as they should be.

It is also interesting to note that the terms proportional to $g$ at the two-derivative level always have corresponding four-derivative terms in exactly the way to respect the shift \eqref{eq:gshift}.  While we have restricted our attention to $S^3$, there is no reason to suspect this is specific to that setup. In particular, assuming there are no other obstructions to truncation, we should generically expect \cite{Lu:2006ah}
\begin{align}
    H_{ijk}=mf_{ijk},
\end{align}
for a group manifold reduction on a unimodular Lie group  $G$, whose Lie algebra has structure constants $f^{ijk}$. In general, we expect the two-derivative truncation to be $m=-g^{-2}$ and $g_{ij}=g^{-2}\kappa_{ij}$, where $\kappa$ is the Cartan-Killing metric. Under the assumption that $G$ is compact and semi-simple, $\kappa$ is necessarily symmetric, non-degenerate, and positive-definite, so we write $\kappa_{ij}=\delta_{ab}k_i^a k_j^b$. The torsionful spin connection should, in general, be
\begin{equation}
    \Omega_+^{ab}=-f_{ijk}k^i_a k^j_b\sigma^k,
\end{equation}
which leads to the terms
\begin{equation}
    \omega_{3L,abc}\supset-2k_a^{i}k_b^j k_c^{k}f_{ijk},
\end{equation}
in the Lorentz-Chern-Simons form. This then \emph{suggests} that we ought always to get the effective coupling shift \eqref{eq:gshift} for a group manifold reduction of heterotic supergravity.

While we have truncated away the heterotic gauge fields from the start, one might wonder if they may be included in the reduction.  In the current context, since we were interested in truncating out all the vector multiplets, it was natural to truncate them in 10D.  As this is a truncation to gauge singlets, the initial removal of the heterotic gauge fields is guaranteed to be consistent, even at the higher derivative order.  Nevertheless, it would be interesting to see how these play into the story of higher-derivative consistent truncations and if they might obstruct truncation more generally. We leave this to future work.

It would be interesting to see how our results extend to more general group manifold or coset reductions. In particular, it is known that any unimodular Lie group $G$ may be used for a group manifold reduction of heterotic supergravity \cite{Lu:2006ah} and one is free (at least at the two-derivative level) to truncate out the vector multiplets that arise. One could then wonder if any new constraints on $G$ arise at the four-derivative level. It would also be interesting to see if consistency extends to more general coset reductions such as the $SO(4)/SO(3)$ coset reduction of heterotic supergravity \cite{Cvetic:2000dm}.

Finally, one may ask whether such higher derivative truncations may be done more systematically in the framework of \ac{DFT}, along the lines of \cite{Eloy:2020dko}. In particular, gauged DFT was used in \cite{Baron:2017dvb} to construct a large class of consistent truncations, including the $S^3$ group manifold reduction. Indeed, after truncating the result in \cite{Baron:2017dvb} and performing suitable field redefinitions, one finds that the effective action matches \eqref{eq:effectiveL} as expected. In the process, one must truncate the $O(3,3)$ covariant packaging to a subsector. For example, the truncation sets half of the components of the $O(3,3)$ field strength $\mathcal F^I=(F^{(+)\,i},F^{(-)\,i})$ to zero. More generally, while DFT can be extremely useful in constructing consistent truncations where all the fields in an $O(d,d)$ multiplet are kept, we still have to break apart the $O(d,d)$ covariant language to truncate away some of the multiplets and check the consistency with the equations of motion.

%% file: Chapters/chap7.tex
Throughout this thesis, we have studied precision holography and higher-derivative supergravity. In Chapter \ref{chap:GT}, we considered the $\mathcal N=3$ Gaiotto-Tomasiello theory and used resolvent techniques to compute subleading corrections to the free energy, which were then dual to the on-shell action of massive IIA supergravity on $\mathrm{AdS}_4\times \mathbb{CP}^3$. In particular, our results implied that the higher-derivative $D^{6n}R^4$ and $R^{4+3n}$ terms vanish on-shell, and it would be interesting to see if this can be verified directly in the gravitational dual.

In Chapter \ref{chap:UE}, we turned to the gravity side of the AdS/CFT correspondence, focusing on minimal gauged supergravity in five dimensions. At the four-derivative level, there are three independent supersymmetric invariants, the supersymmetrizations of the curvature tensors squared. We have shown that after appropriate field redefinitions, these three terms are all the same, up to six-derivative corrections that we have perturbatively ignored. It would be interesting to see how this applies more generally to theories in higher dimensions or with additional matter couplings.

In Chapter \ref{chap:RG}, we applied precision holography to study holographic RG flows across dimensions. There, we extended previous work on higher-derivative RG flows within the same dimension and two-derivative flows across dimensions. Adding a Gauss-Bonnet term allowed us to distinguish the central charges, and we indeed found that the monotonic behavior was of the $A$-type central charge. In particular, we constructed a local, monotonic $c$-function valid in arbitrary dimensions and for arbitrary internal Einstein metrics by directly using the Null Energy Condition, as well as a non-local, monotonic $c$-function valid for flows to $\mathrm{AdS}_3$ with no internal curvature by using the entanglement entropy. We further showed that these two $c$-functions were intimately related in their overlapping regime of validity. It is an open problem to incorporate non-zero internal curvature, as well as to extend the proof to higher dimensions.

Finally, in Chapters \ref{chap:torus} and \ref{chap:sphere}, we turned our attention to supergravity itself. There, we studied the consistency of truncations in the presence of four-derivative corrections. The lesson is that truncations protected at the two-derivative level by T-duality will automatically continue to be protected at the four-derivative level. However, our truncation may fail when we do not have this symmetry to protect us. We see this in the reduction of four-derivative heterotic supergravity on $T^5$: Attempting to truncate away the dilaton-containing multiplet is inconsistent due to the coupling to the Riemann squared term. It is an open question whether this statement applies only to truncations or also to the solutions. It would be interesting to see if such solutions with an uncorrected, vanishing dilaton exist.

%% file: Appendices/Appendix_GT.tex
\section{Endpoint computations}\label{appendix:endpoints}

After manipulating the $A$-cycle integrals for the endpoint relations, we have arrived at the expression
\begin{equation}
    t_1=\fft{\kappa_1}{2\pi^2}J_1+\fft{\kappa_2}{2\pi^2}J_2,
\label{eq:t1=J1J2}
\end{equation}
where
\begin{equation}
    J_1=\int_0^\beta \dd v\,\fft{v\,e^{-\fft12(\gamma+v)}I_1(v)}{2\sqrt{\cosh\fft{\alpha-v}2\sinh\fft{\beta-v}2}},
\label{eq:J1integ}
\end{equation}
with
\begin{equation}
    I_1(v)=\int_0^\beta \dd u\,2e^{\fft12(\gamma+u)}\sqrt{\textstyle\cosh\fft{\alpha-u}2\sinh\fft{\beta-u}2}\left(\coth\fft{v-u}2+\coth\fft{v+u}2\right),
\end{equation}
up to exponentially small corrections in the large $\gamma$ limit.  (The principal value of $I_1(v)$ has to be taken in the $J_1$ integral.)  Similar expressions for $J_2$ are given in (\ref{eq:J1intdef}) and (\ref{eq:I12intdef}).  Here we carry out the integration to obtain the endpoint relations (\ref{eq:tbD}).

We first work on the $I_1(v)$ integral.  As it turns out, this can be integrated in closed form, with the result
\begin{align}
    I_1(v)&=2\sinh v\left(\fft\pi2-\tan^{-1}\sinh\delta\right)\nn\\
    &\qquad+2e^{\fft12(\gamma+v)}\sqrt{\textstyle\cosh\fft{\alpha-v}2\sinh\fft{\beta-v}2}\Big[-\gamma+\log(2\cosh\delta)+\log(1-e^v)\nn\\
    &\kern16em-2\log\left(\sqrt{1-e^{v-\beta}}+\sqrt{1+e^{v-\alpha}}\right)\Big]\nn\\
    &\qquad-2e^{\fft12(\gamma-v)}\sqrt{\textstyle\cosh\fft{\alpha+v}2\sinh\fft{\beta+v}2}\Big[-\gamma+\log(2\cosh\delta)+\log(1-e^{-v})\nn\\
    &\kern16em-2\log\left(\sqrt{1-e^{-v-\beta}}+\sqrt{1+e^{-v-\alpha}}\right)\Big],
\end{align}
up to exponentially small terms in the large $\gamma$ limit.  Since $I_1(v)$ is only needed for $v\in[0,\beta]$, we can further drop exponentially small terms to get
\begin{align}
    I_1(v)&=e^v\left(\fft\pi2-\tan^{-1}\sinh\delta\right)+e^\gamma\Bigl(\gamma-\log(\ft12\cosh\delta)-\log(1-e^{-v})\Bigr)\nn\\
    &\qquad+2e^{\fft12(\gamma+v)}\sqrt{\textstyle\cosh\fft{\alpha-v}2\sinh\fft{\beta-v}2}\Big[-\gamma+\log(2\cosh\delta)+\log(1-e^v)\nn\\
    &\kern16em-2\log\left(\sqrt{1-e^{v-\beta}}+\sqrt{1+e^{v-\alpha}}\right)\Big].
\label{eq:I1asymp}
\end{align}
Note that the replacement $2\sinh v\to e^v$ in the first line of this expression is not strictly valid for $v\approx0$.  However, the rest of the integrand for $J_1$ in (\ref{eq:J1integ}) is exponentially suppressed in this limit, so there is no harm in making this substitution.

Substituting (\ref{eq:I1asymp}) into (\ref{eq:J1integ}) now gives
\begin{align}
    J_1&=2\int_0^\beta \dd v\Bigg[\fft{v}{\sqrt{(1+e^{v-\alpha})(1-e^{v-\beta})}}\Bigl(
    e^{v-\gamma}\left(\fft\pi2-\tan^{-1}\sinh\delta\right)\nn\\
    &\kern18em+\gamma-\log(\ft12\cosh\delta)-\log(1-e^{-v})\Bigr)\nn\\
    &\kern3em+v\left(-\gamma+\log(2\cosh\delta)+\log(e^v-1)-2\log\left(\sqrt{1-e^{v-\beta}}+\sqrt{1+e^{v+\alpha}}\right)\right)\Bigg],
\end{align}
where we flipped the sign of $1-e^v$ in the log on the second line to take the principal value into account.  Some of the integrals in the second line can be readily done.  We also integrate the final log term in the second line by parts, with the result
\begin{align}
    J_1&=-\ft13\beta^3-2\zeta(3)-\int_0^\beta dv\,v^2\left(\fft1{\sqrt{(1+e^{v-\alpha})(1-e^{v-\beta})}}-1\right)\nn\\
    &\qquad+2\int_0^\beta \dd v\fft{v}{\sqrt{(1+e^{v-\alpha})(1-e^{v-\beta})}}\Bigl(
    e^{v-\gamma}\left(\fft\pi2-\tan^{-1}\sinh\delta\right)\nn\\
    &\kern19em+\gamma-\log(\ft12\cosh\delta)-\log(1-e^{-v})\Bigr).
\end{align}
The final term proportional to $\log(1-e^{-v})$ is only important for $v$ close to zero.  Thus, for this term, we can replace the square root factor in the denominator by $1$ up to exponentially small terms and then integrate.  For the remaining terms, we define $x=v-\beta$ and extend the lower range of integration to $-\infty$ (which only introduces exponentially small corrections) to obtain
\begin{align}
    J_1&=-\ft13\beta^3+\beta^2\left(\gamma+\log2-\log\cosh\delta\right)\nn\\
    &\qquad+\int_{-\infty}^0 \dd x\left(\fft1{\sqrt{(1-e^x)(1+e^{x-2\delta})}}-1\right)\big(-(x+\beta)^2\nn\\
    &\kern20em+2(x+\beta)(\gamma+\log2-\log\cosh\delta)\big)\nn\\
    &\qquad+2\int_{-\infty}^0\dd x\fft{(x+\beta)e^{x-\delta}}{\sqrt{(1-e^x)(1+e^{x-2\delta})}}\left(\fft\pi2-\tan^{-1}\sinh\delta\right).
\label{eq:J1expanded}
\end{align}
Recalling that $\beta=\gamma-\delta$, the first line gives the leading order $\fft23\gamma^3$ factor we expect from \eqref{eq:endptIntermed}.

To proceed, we define a set of basis integrals
\begin{align}
    f_n(\delta)&\equiv\int_{-\infty}^0 \dd x\,x^n\left(\fft1{\sqrt{(1-e^x)(1+e^{x-2\delta})}}-1\right),\nn\\
    g_n(\delta)&\equiv\int_{-\infty}^0\dd x\,\fft{x^ne^{x-\delta}}{\sqrt{(1-e^x)(1+e^{x-2\delta})}}.
\end{align}
Some of the integrals can be performed without too much difficulty. In particular,
\begin{equation}
    f_0(\delta)=\delta-\log(\ft12\cosh\delta),\qquad
    g_0(\delta)=\fft\pi2-\tan^{-1}\sinh\delta.
\end{equation}
With some effort, it is also possible to obtain
\begin{subequations}
\begin{align}
    f_1(\delta)&=-\fft{\pi^2}3-\delta^2+2\delta\log2+\log^22+2(\cot^{-1}e^\delta)^2+(\log\cosh\delta)^2-\Li_2\left(-e^{-2\delta}\right)\nn\\
    &\qquad+\Li_2\left(\fft1{1+e^{-2\delta}}\right),\\
    g_1(\delta)&=2\cot^{-1}e^\delta(\delta+\log(\ft12\cosh\delta))-\Im\Li_2\left(\left(\fft{i+e^\delta}{-i+e^\delta}\right)^2\right).
\end{align}
\end{subequations}
This leaves just the $f_2(\delta)$ integral to be done to obtain a closed-form result for $J_1$.  While we have not managed to analytically find an exact form for $f_2$, it can nevertheless be expanded for $\delta\ll 1$ as
\begin{align}
    f_2(\delta)=&\frac{1}{12} \left(6 \zeta (3)+4 \log ^3(2)-\pi ^2 \log 2\right)+\qty(-\frac{\pi ^2}{12}+\log ^2(2)+\pi\log 2)\delta\nn\\
    &+\frac{1}{24}((\pi -24) \pi -12 (\log2-4) \log 2)\delta ^2-\frac{1}{6} (-8+\pi  (\log 2-1)+6\log 2)\delta ^3\nn\\
    &-\frac{1}{144} \left((\pi -24) \pi +96-12 \log ^2(2)+12\log2\right)\delta ^4+\frac{1}{480}(10 (4+\pi ) \log (4)\nn\\
    &-8 (5+2 \pi ))\delta ^5+\frac{\left(2 \pi ^2-45 \pi +123-6 \log ^2(4)+15 \log4\right)}{1080}\delta ^6+\mathcal{O}\qty(\delta^7).
\end{align}

Although we have focused on $J_1$, the second integral, $J_2$, can be worked out similarly, with the result
\begin{equation}
    J_2(\delta)=J_1(-\delta)+\gamma(-\pi^2-2\pi\tan^{-1}\sinh\delta)-2\pi(\delta g_0(-\delta)+g_1(-\delta)).
\end{equation}
An interesting feature of the integral (\ref{eq:J1expanded}) is that it is precisely a cubic function of $\gamma$ up to exponentially small terms.  In particular, we find
\begin{subequations}
\begin{align}
    J_1&=\ft23\gamma^3-2\gamma^2\log(\ft12\cosh\delta)+2\gamma\left(\log^2(\ft12\cosh\delta)+(\ft12\pi-\tan^{-1}\sinh\delta)^2\right)+j_1(\delta)\nn\\
    &\qquad+\mathcal O(e^{-\gamma}),\\
    J_2&=\ft23\gamma^3-2\gamma^2\log(\ft12\cosh\delta)+2\gamma\left(\log^2(\ft12\cosh\delta)-\ft14\pi^2+(\tan^{-1}\sinh\delta)^2\right)+j_2(\delta)\nn\\
    &\qquad+\mathcal O(e^{-\gamma}),
\end{align}
\end{subequations}
where most of the complication is only in the $\gamma$-independent terms $j_1(\delta)$ and $j_2(\delta)$, defined as
\begin{subequations}
\begin{align}
     j_1(\delta)&=-\ft23\delta^3+2\delta f_1(\delta)-f_2(\delta)+2g_1(\delta)(\ft12\pi-\tan^{-1}\sinh\delta)-2\delta(\ft12\pi-\tan^{-1}\sinh\delta)^2\nn\\
     &\qquad+2(\delta^2-f_1(\delta))\log(\ft12\cosh\delta)-2\delta\log^2(\ft12\cosh\delta),\\
     j_2(\delta)&=\ft23\delta^3-2\delta f_1(-\delta)-f_2(-\delta)-2g_1(-\delta)(\ft12\pi-\tan^{-1}\sinh\delta)\nn\\
     &\qquad+2\delta(-\ft14\pi^2+(\tan^{-1}\sinh\delta)^2)+2(\delta^2-f_1(-\delta))\log(\ft12\cosh\delta)\nn\\
     &\qquad+2\delta\log^2(\ft12\cosh\delta).
\end{align}
\end{subequations}
Note that
\begin{equation}
    j_1(0)=\ft23\log^32-\ft12\pi^2\log2-\ft12\zeta(3),\qquad
    j_2(0)=\ft23\log^32+\ft12\pi^2\log2-\ft12\zeta(3).
\end{equation}
Moreover, we can see numerically that $j_2(\delta)$ is an even function, \emph{i.e.}, $j_2(-\delta)=j_2(\delta)$. This then suggests an exact expression for the odd part of $f_2$
\begin{align}
    f_{2,o}(\delta)&=-\ft23\delta^3+2\delta f_{1,e}(\delta)-2g_{1,o}(\delta)\ft12\pi-2g_{1,e}(\delta)\tan^{-1}\sinh\delta\nn\\
     &\qquad-2\delta(-\ft14\pi^2+(\tan^{-1}\sinh\delta)^2)-2f_{1,o}(\delta)\log(\ft12\cosh\delta)-2\delta\log^2(\ft12\cosh\delta),
\end{align}
where subscripts $e$ and $o$ denote even and odd parts, respectively. This has been verified numerically. Moreover, playing around with $j_1$ and $j_2$ numerically allows one to get an empirical expression for the even part of $f_2$
\begin{align}
    f_{2,e}(\delta)=&2 \Cl_3(2\pi-\arctan\sinh\delta)+4 \delta ^2 \log (\cosh (\delta ))+\frac{\pi ^2 \delta }{6}-\frac{1}{3} 2 \log ^3(2 \text{sech}(\delta ))\nn\\
    &+\log (2 \text{sech}(\delta )) \left(2 \left(\delta ^2+\log ^2(\cosh (\delta ))+\log ^2(2)\right)+\text{gd}^2(\delta )-\frac{\pi ^2}{4}\right)\nn\\
    &-\text{gd}(\delta ) \left(\pi  \delta +\Im\left(\text{Li}_2\left(\frac{\left(i+e^{\delta }\right)^2}{\left(-i+e^{\delta }\right)^2}\right)\right)-\Im\left(\text{Li}_2\left(\frac{\left(-i+e^{\delta }\right)^2}{\left(i+e^{\delta }\right)^2}\right)\right)\right)\nn\\
    &+2 \delta  \text{Li}_2\left(-e^{2 \delta }\right)+\text{Li}_2\left(\frac{1}{1+e^{2 \delta }}\right) (-\delta +\log (\text{sech}(\delta ))+\log (2))\nn\\
    &+(\delta +\log (\text{sech}(\delta ))+\log (2)) \text{Li}_2\left(\frac{1}{2} (\tanh (\delta )+1)\right)+2 \left(\delta ^3+\zeta (3)\right),
\end{align}
where $\Cl_3$ is the Clausen function. This holds to very high precision numerically, but we have not managed to show this identity analytically.

Finally, combining $J_1$ and $J_2$ according to (\ref{eq:t1=J1J2}) gives
\begin{align}
    t_1\approx&~\fft{\kappa_1+\kappa_2}{2\pi^2}\Bigl[\ft23\gamma^3-2\gamma^2\log(\ft12\cosh\delta)\nn\\
    &\qquad\qquad+2\gamma\left(\log^2(\ft12\cosh\delta)-\ft12\pi\tan^{-1}\sinh\delta+(\tan^{-1}\sinh\delta)^2\right)\nn\\
    &\qquad\qquad+\ft12\qty(j_1(\delta)+j_2(\delta))\Bigr]\nn\\
    &+\fft{\kappa_1-\kappa_2}{2\pi^2}\Bigl[2\gamma\left(\ft14\pi^2-\ft12\pi\tan^{-1}\sinh\delta\right)+\ft12\qty(j_1(\delta)-j_2(\delta))\Bigr].
\end{align}
We can obtain a similar expression for $t_2$ by interchanging $\kappa_1\leftrightarrow\kappa_2$ and taking $\delta\to-\delta$. Taking sums and differences, and defining
\begin{equation}
    \bar t=\ft12(t_1+t_2),\qquad\Delta=\ft12(t_1-t_2),
\end{equation}
then gives the expressions (\ref{eq:tbD}) for $\bar t$ and $\Delta$ in terms of $\gamma$ and $\delta$.  Note that the odd combination $j_{2,o}$ vanishes, at least numerically.  Numerically, we also have
\begin{equation}
    j_{1,e}-j_{2,e}=\pi^2\log(\ft12\cosh\delta).
\end{equation}
However, the remaining functions $j_{1,o}$ and $j_{1,e}+j_{2,e}$ that show up in (\ref{eq:tbD}) do not seem to have similar compact expressions.

\section{Free energy calculations}\label{appendix:freeEnergy}

The integrals involved in evaluating the derivative of the free energy
\begin{equation}
    \fft{\partial F_0}{\partial t_1}=\fft{\kappa_1}2\beta^2-(t_1-t_2)(\Lambda-\log4)-\fft{\kappa_1}\pi K_1-\fft{\kappa_2}\pi K_2,
\end{equation}
are similar to those for determining the endpoints.  In particular, the integrals
\begin{subequations}
\begin{align}
    K_1=\int_c^d\dd x\fft{\log x}{\sqrt{(x-a)(x-b)(x-c)(d-x)}}I_B(x),\\ K_2=\int_a^b\dd x\fft{\log(-x)}{\sqrt{(x-a)(b-x)(c-x)(d-x)}}I_B(x),
\end{align}
\end{subequations}
correspond directly to the $J_1$ and $J_2$ integrals, (\ref{eq:J1J2ints}), except with $I(x)$ replaced by the $B$-cycle integral
\begin{equation}
    I_B(z)=\int_d^{e^\Lambda}\fft{\dd y}y\fft{\sqrt{(y-a)(y-b)(y-c)(y-d)}}{z-y}.
\end{equation}

After dropping exponentially small terms, we can write
\begin{equation}
    I_B(z)\approx\int_\beta^\Lambda \dd u\fft{\sqrt{(e^u+e^\alpha)(e^u-e^\beta)}}{ze^{-u}-1}.
\end{equation}
This can be integrated to give
\begin{align}
    I_B(z)&=-e^\Lambda-\ft12(e^\alpha-e^\beta)\left(\Lambda+\log4+1-\log(e^\alpha+e^\beta)\right)\nn\\
    &\quad-z\left(\Lambda+\log4-\log(e^\alpha+e^\beta)\right)+2\sqrt{(e^\beta-z)(e^\alpha+z)}\tan^{-1}\sqrt{\fft{e^\alpha+z}{e^\beta-z}}.
\end{align}
Note that the first line of this expression is independent of $z$.  We can also rewrite the $K_1$ and $K_2$ integrals over the half intervals and use reflection symmetry to write
\begin{equation}
    K_1\approx\int_0^\beta \dd v\fft{v(I_B(e^v)-I_B(e^{-v}))}{\sqrt{(e^\alpha+e^v)(e^\beta-e^v)}},\qquad
    K_2\approx\int_0^\alpha \dd v\fft{v(I_B(-e^v)-I_B(-e^{-v}))}{\sqrt{(e^\alpha-e^v)(e^\beta+e^v)}},
\end{equation}
where as usual we drop exponentially small terms.  In both integrals, we only need the antisymmetric combination $I_B(z)-I_B(1/z)$.  As a result, the $z$-independent part of $I_B(z)$ drops out, and we are left with
\begin{equation}
    K_1\approx\int_0^\beta \dd v\,\fft{2v\hat I_B(e^v)}{\sqrt{(1+e^{v-\alpha})(1-e^{v-\beta})}},\qquad
    K_2\approx\int_0^\alpha \dd v\,\fft{2v\hat I_B(-e^v)}{\sqrt{(1-e^{v-\alpha})(1+e^{v-\beta})}},
\end{equation}
where
\begin{equation}
    \hat I_B(z)=e^{-\gamma}\fft{I_B(z)-I_B(z^{-1})}2.
\end{equation}

Just as with the endpoint integrals, we can work these integrals out using the explicit form of $\hat I_B(z)$.  The arctan contribution can be integrated by parts, and after some manipulation, we find
\begin{subequations}
\begin{align}
    K_1&=\beta^2(\ft12\pi-\tan^{-1}e^\delta)-(\Lambda-\gamma-\log\ft12\cosh\delta)(\beta g_0(\delta)+g_1(\delta))\nn\\
    &\qquad-\ft12(\beta^2g_0(\delta)+2\beta g_1(\delta)+g_2(\delta))-2\tan^{-1}e^\delta(\beta f_0(\delta)+f_1(\delta)),\\
    K_2&=-\alpha^2\tan^{-1}e^\delta+(\Lambda-\gamma-\log\ft12\cosh\delta)(\alpha g_0(-\delta)+g_1(-\delta))\nn\\
    &\qquad+\ft12(\alpha^2g_0(-\delta)+2\alpha g_1(-\delta)+g_2(-\delta))-2\tan^{-1}e^\delta(\alpha f_0(-\delta)+f_1(-\delta)).
\end{align}
\end{subequations}
Replacing $\alpha$ and $\beta$ with $\gamma$ and $\delta$ gives
\begin{subequations}
\begin{align}
    K_1&=(\Lambda-\gamma-\log\ft12\cosh\delta)(\gamma(-\ft12\pi+\tan^{-1}\sinh\delta)+k_1^\Lambda)\nn\\
    &\qquad+\gamma(-g_1(\delta)-2\tan^{-1}e^\delta f_0(\delta))+k_1^0,\\
    K_2&=(\Lambda-\gamma-\log\ft12\cosh\delta)(\gamma(\ft12\pi+\tan^{-1}\sinh\delta)+k_2^\Lambda)\nn\\
    &\qquad+\gamma(g_1(-\delta)-2\tan^{-1}e^\delta f_0(-\delta))+k_2^0,
\end{align}
\end{subequations}
where
\begin{equation}
    k_1^\Lambda=\delta g_0(\delta)-g_1(\delta),\qquad k_2^\Lambda=\delta g_0(-\delta)+g_1(-\delta),
\end{equation}
and
\begin{subequations}
\begin{align}
    k_1^0&=\delta g_1(\delta)-\ft12g_2(\delta)+2\tan^{-1}e^\delta(\delta f_0(\delta)-f_1(\delta)),\\
    k_2^0&=\delta g_1(-\delta)+\ft12g_2(-\delta)-2\tan^{-1}e^\delta(\delta f_0(-\delta)+f_1(-\delta)).
\end{align}
\end{subequations}

The derivative of the free energy can then be written as
\begin{align}
    \fft{\partial F_0}{\partial t_1}&=\fft{\kappa_1}2\beta^2-(\Lambda-\log4)(t_1-t_2)\nn\\
    &\qquad+\qty(\Lambda-\gamma-\log\ft12\cosh\delta)\Bigg(\fft{\kappa_1+\kappa_2}\pi\qty(-\gamma\tan^{-1}\sinh\delta-\ft12\qty(k_1^\Lambda+k_2^\Lambda))\nn\\
    &\qquad+\fft{\kappa_1-\kappa_2}2\qty(\gamma-\ft1\pi\qty(k_1^\Lambda-k_2^\Lambda))\Bigg)\nn\\
    &\qquad-\fft{\kappa_1+\kappa_2}\pi(\gamma(-g_{1,o}-2\tan^{-1}e^\delta f_{0,e})+\ft12(k_1^0+k_2^0))\nn\\
    &\qquad-\fft{\kappa_1-\kappa_2}\pi(\gamma(-g_{1,e}-2\tan^{-1}e^\delta f_{0,o})+\ft12(k_1^0-k_2^0)),
\end{align}
where we have again used the $e$ and $o$ notation to denote the even and odd components of the function.  Note that the cutoff $\Lambda$ should drop out of this expression.  Comparison with the expression for $\Delta$ in (\ref{eq:tbD}) indicates that this requires the identities
\begin{equation}
    j_{1,o}=-2\pi(\delta g_{0,e}-g_{1,o}),\qquad j_{1,e}-j_{2,e}=-2\pi(\delta g_{0,o}-g_{1,e}).
\end{equation}
along with the assumed vanishing of $j_{2,o}$.  These identities do hold numerically.  The result is then
\begin{subequations}
\begin{align}
    \fft{\partial F_0}{\partial t_1}&=\fft{\kappa_1}2(\gamma-\delta)^2+(\log4-\gamma-\log\ft12\cosh\delta)(t_1-t_2)\nn\\
    &\qquad-\fft{\kappa_1+\kappa_2}\pi(\gamma(-g_{1,o}-2\tan^{-1}e^\delta f_{0,e})+\ft12(k_1^0(\delta)+k_2^0(\delta)))\nn\\
    &\qquad-\fft{\kappa_1-\kappa_2}\pi(\gamma(-g_{1,e}-2\tan^{-1}e^\delta f_{0,o})+\ft12(k_1^0(\delta)-k_2^0(\delta))),\\
    \fft{\partial F_0}{\partial t_2}&=\fft{\kappa_1}2(\gamma+\delta)^2-(\log4-\gamma-\log\ft12\cosh\delta)(t_1-t_2)\nn\\
    &\qquad-\fft{\kappa_1+\kappa_2}\pi(\gamma(g_{1,o}-2\tan^{-1}e^{-\delta}f_{0,e})+\ft12(k_1^0(-\delta)+k_2^0(-\delta)))\nn\\
    &\qquad-\fft{\kappa_1-\kappa_2}\pi(\gamma(g_{1,e}-2\tan^{-1}e^{-\delta}f_{0,o})-\ft12(k_1^0(-\delta)-k_2^0(-\delta))).
\end{align}
\end{subequations}

We now transform from $t_1$ and $t_2$ to $\bar t$ and $\Delta$. In particular, we have
\begin{equation}
   \fft{\partial F_0}{\partial\bar t}=\fft{\partial F_0}{\partial t_1}+\fft{\partial F_0}{\partial t_2},\qquad\fft{\partial F_0}{\partial\Delta}=\fft{\partial F_0}{\partial t_1}-\fft{\partial F_0}{\partial t_2}.
\end{equation}
After some simplification, we find
\begin{subequations}
\begin{align}
    \fft{\partial F_0}{\partial\bar t}&=\fft{\kappa_1+\kappa_2}2\left((\gamma-\log\ft12\cosh\delta)^2-(\log\ft12\cosh\delta)^2-\ft2\pi(k_{1,e}^0+k_{2,e}^0)+\delta^2\right)\nn\\
    &\qquad+\fft{\kappa_1-\kappa_2}2\left(-\ft2\pi(k_{1,o}^0-k_{2,o}^0)\right),\\
    \fft{\partial F_0}{\partial\Delta}&=4\Delta(\log4-\gamma-\log\ft12\cosh\delta)\nn\\
    &\qquad+\fft{\kappa_1+\kappa_2}2\left(\ft1\pi\gamma(-2\pi\delta+4g_{1,o}-4(\tan^{-1}\sinh\delta) \log\ft12\cosh\delta)-\ft2\pi(k_{1,o}^0+k_{2,o}^0)\right)\nn\\
    &\qquad+\fft{\kappa_1-\kappa_2}2\left(\gamma^2+\ft1\pi\gamma(4g_{1,e}+4\delta\tan^{-1}\sinh\delta )-\ft2\pi(k_{1,e}^0-k_{2,e}^0)+\delta^2\right).
\end{align}
\end{subequations}

Since we have an explicit expression for $f_1(\delta)$, we should be able to verify $f_1=f_{1,e}+f_{1,o}$ where
\begin{subequations}
\begin{align}
    f_{1,e}&=\ft12\delta^2-\ft1{24}\pi^2+\ft12(\log\ft12\cosh\delta)^2+\ft12(\tan^{-1}\sinh\delta)^2,\\
    f_{1,o}&=-\ft12\pi\tan^{-1}\sinh\delta-\delta\log\ft12\cosh\delta.
\end{align}
\end{subequations}
This leads to the identities
\begin{subequations}
\begin{align}
    k_{1,e}+k_{2,e}&=
    \ft\pi2\left(\delta^2+\ft1{12}\pi^2-(\log\ft12\cosh\delta)^2-(\tan^{-1}\sinh\delta)^2\right),\\
    k_{1,o}^0-k_{2,o}^0&=\ft12\pi^2\tan^{-1}\sinh\delta,
\end{align}
\end{subequations}
which results in the simple expression for the $\bar t$ derivative of the free energy given in (\ref{eq:dF0dtb}).

%% file: Appendices/Appendix_Unreasonable_Effectiveness.tex
\section{Equations of motion simplifications}\label{app:simplifications}

Here we show several helpful simplifications that we use throughout Section \ref{sec:sugra}. All these simplifications make use of the two-derivative equations of motion and can be considered as perturbative field redefinitions.  Some also use integration by parts and hence are only valid inside the action.

We start by evaluating $(\nabla_\nu F^{\nu\mu})^2$ using the two-derivative equations of motion.
\begin{align}
    (\nabla_\nu F^{\nu\mu})^2=&\frac{1}{24}\epsilon^{\mu\nu\rho\sigma\lambda}F_{\nu\rho}F_{\sigma\lambda}\epsilon_{\mu\alpha\beta\gamma\delta}F^{\alpha\beta}F^{\gamma\delta}\nn\\
    =&-\frac{1}{24}\delta^{\nu\rho\sigma\lambda}_{\alpha\beta\gamma\delta}F_{\nu\rho}F_{\sigma\lambda}F^{\alpha\beta}F^{\gamma\delta}\nn\\
    =&-F_{\nu\rho}F_{\sigma\lambda}F^{[\nu\rho}F^{\sigma\lambda]}\nn\\
    =&-\frac{1}{3}(F^2)^2+\frac{2}{3}F^4.
\label{eq:simp1}
\end{align}
We quickly remark that the two-derivative equations of motion \eqref{eq:eoms} imply that
\begin{subequations}
\begin{align}
    R_{\mu\nu}F^{\mu\sigma}F^\nu{}_\sigma&=F^4-\frac{1}{6}\qty(F^2)^2-4g^2F^2,\\
    R&=\frac{1}{6}F^2-20g^2.
\end{align}
\end{subequations}
Next, we evaluate
\begin{subequations}
\begin{align}
    F^{\nu\rho}[\nabla_\mu,\nabla_\nu]F^\mu{}_\rho&=F^{\nu\rho}\qty(R^{\delta\mu}{\nu\mu}F_{\delta\rho}+R^{\delta}{}_{\rho\nu\mu}F^\mu{}_\delta)\\
    &=F^{\nu\rho}\qty(R^\delta{}_\nu F_{\delta\rho}+R_{\delta\rho\nu\mu}F^{\mu\delta})\\
    &=R^{\delta\nu}F_{\delta\rho}F_\nu{}^\rho+\frac{1}{2}R_{\nu\rho\delta\mu}F_{\mu\delta}F_{\nu\rho}
\label{eq:Bianchi2}\\
    &=F^4-\frac{1}{6}\qty(F^2)^2-\frac{1}{2}R_{\mu\nu\rho\sigma}F^{\mu\nu}F^{\rho\sigma}-4g^2F^2.
\label{eq:simp3}
\end{align}
\end{subequations}
where \eqref{eq:Bianchi2} follows from the first Bianchi identity for the Riemann tensor
\begin{equation}
    R_{[\mu\nu\rho]\sigma}=0.
\end{equation}
Now, we evaluate $(\nabla F)^2$. We recall the field strength Bianchi identity
\begin{equation}
    \nabla_{[\mu}F_{\nu\rho]}=0.
\end{equation}
This allows us to rewrite
\begin{equation}
    \nabla_\mu F_{\nu\rho}=\nabla_\nu F_{\mu\rho}+\nabla_\rho F_{\nu\mu}.\label{eq:Bianchi}
\end{equation}
Using this, we have
\begin{align}
    \nabla_\mu F_{\nu\rho}\nabla^\mu F^{\nu\rho}=&(\nabla_\nu F_{\mu\rho}+\nabla_\rho F_{\nu\mu})\nabla^\mu F^{\nu\rho}\nn\\
    =&2\nabla_\nu F_{\mu\rho}\nabla^\mu F^{\nu\rho}\nn\\
    \to&-2(\nabla_\mu\nabla_\nu F^{\mu\rho})F^{\nu}{}_\rho\nn\\
    =&-2(\nabla_\nu\nabla_\mu F^{\mu\rho})F^{\nu}{}_\rho-2([\nabla_\mu,\nabla_\nu]F^{\mu\rho})F^{\nu}{}_\rho\nn\\
    \to&\nabla_\mu F^{\mu\rho}\nabla_\nu F^{\nu\rho}-2F^{\nu\rho}[\nabla_\mu,\nabla_\nu]F^{\mu}{}_{\rho}\nn\\
    =&-\frac{1}{3}(F^2)^2-\frac{2}{3}F^4+R_{\mu\nu\rho\sigma}F^{\mu\nu}F^{\rho\sigma}+8g^2F^2,
\end{align}
where the arrows denote integration by parts, which is valid as long as we are applying this formula inside an integral. The last line follows from the equations of motion and \eqref{eq:simp1} and \eqref{eq:simp3}.

Now, we wish to evaluate
\begin{align}
    \epsilon_{\mu\nu\rho\sigma\lambda}F^{\mu\nu}F^{\rho\sigma}\nabla_\tau F^{\tau\lambda}&=-\frac{1}{2\sqrt{6}}\epsilon_{\mu\nu\rho\sigma\lambda}F^{\mu\nu}F^{\rho\sigma}\epsilon^{\lambda\alpha\beta\gamma\delta}F_{\alpha\beta}F_{\gamma\delta}\nn\\
    &=\frac{1}{2\sqrt{6}}\delta_{\mu\nu\rho\sigma}^{\alpha\beta\gamma\delta}F^{\mu\nu}F^{\rho\sigma}F_{\alpha\beta}F_{\gamma\delta}\nn\\
    &=\frac{12}{\sqrt{6}}F_{\mu\nu}F_{\rho\sigma}F^{[\mu\nu}F^{\rho\sigma]}\nn\\
    &=\frac{4}{\sqrt{6}}\qty[\qty(F^2)^2-2F^4].
\label{eq:simpeFFdF}
\end{align}
Finally, we compute
\begin{subequations}
\begin{align}
    \epsilon_{\mu\nu\rho\lambda\delta}F^{\lambda\delta}F^{\rho}{}_\beta\nabla^\mu F^{\nu\beta}&=\frac{1}{2}\epsilon_{\mu\nu\rho\lambda\delta}F^{\lambda\delta}F^{\rho}{}_\beta\nabla^\beta F^{\nu\mu}\label{eq:simp2}\\
    &=-\frac{1}{2}\epsilon_{\mu\nu\rho\lambda\delta}F^{\lambda\delta}F^{\rho}{}_\beta\nabla^\beta F^{\mu\nu}\\
    &=-\frac{1}{4}\epsilon_{\mu\nu\rho\lambda\delta}F^{\rho}{}_\beta\nabla^\beta \qty(F^{\mu\nu}F^{\lambda\delta})\\
    &\to\frac{1}{4}\epsilon_{\mu\nu\rho\lambda\delta}\qty(\nabla^\beta F^{\rho}{}_\beta) F^{\mu\nu}F^{\lambda\delta}\\
    &=-\frac{1}{4}\epsilon_{\mu\nu\rho\lambda\delta}F^{\mu\nu}F^{\lambda\delta}\nabla_\beta F^{\beta\rho}\\
    &=-\frac{1}{\sqrt{6}}\qty[\qty(F^2)^2-2F^4],
\label{eq:simp4}
\end{align}
\end{subequations}
where \eqref{eq:simp2} follows from the Bianchi identity \eqref{eq:Bianchi}. The last line in the computation follows from \eqref{eq:simpeFFdF}.

We can also evaluate some curvature squared terms using \eqref{eq:eoms}
\begin{equation}
    R_{\mu\nu}R^{\mu\nu}=F^4-\frac{7}{36}\qty(F^2)^2-\frac{4}{3}g^2F^2+80g^4,
\end{equation}
and also
\begin{equation}
    R^2=\frac{1}{36}\qty(F^2)^2-\frac{20}{3}g^2F^2+200g^4.
\end{equation}

One last useful formula for us is
\begin{align}
    R_{\mu\nu\rho\sigma}F^{\mu\nu}F^{\rho\sigma}&=\qty(C_{\mu\nu\rho\sigma}+\frac{4}{3}g_{\mu\rho}R_{\sigma\nu}-\frac{1}{6}Rg_{\mu\rho}g_{\sigma\nu})F^{\mu\nu}F^{\rho\sigma}\nn\\
    &=C_{\mu\nu\rho\sigma}F^{\mu\nu}F^{\rho\sigma}+\frac{4}{3}R_{\sigma\nu}F^{\rho\nu}F_\rho{}^\sigma-\frac{1}{6}RF^2\nn\\
    &=C_{\mu\nu\rho\sigma}F^{\mu\nu}F^{\rho\sigma}+\frac{4}{3}F^4-\frac{1}{4}\qty(F^2)^2-2g^2F^2.
\end{align}
%

\section{The Gutowski-Reall black hole}\label{app:GutowskiReall}

Here we compute the on-shell value of the parametrized four-derivative corrected action, (\ref{eq:param4d}), for the Gutowski-Reall black hole \cite{Gutowski:2004r}.  When viewed as an asymptotically AdS$_5\times S^5$ solution to IIB supergravity, it is known that the first correction occurs at the eight-derivative level.  The corrected action with curvature and the Ramond-Ramond five-form was obtained in \cite{Gross:1986iv,Paulos:2008tn}, and it was shown in \cite{Melo:2007} that the on-shell eight-derivative correction vanishes for the supersymmetric Gutowski-Reall solution.  However, here we take more of a bottom-up view and introduce four-derivative corrections to the five-dimensional action as may occur in theories with reduced supersymmetry.

The Gutowski-Reall black hole \cite{Gutowski:2004r} is a solution of minimal gauged supergravity in 5D given by
\begin{subequations}
\begin{equation}
    \dd{s}^2=-U(r)\Lambda(r)^{-1}\dd{t}^2+U(r)^{-1}\dd{r}^2+\frac{r^2}{4}\qty[\qty(\sigma^{1'}_L)^2+\qty(\sigma^{2'}_L)^2+\Lambda(r)\qty(\sigma_L^{3'}-\Omega(r)\dd{t})^2],
\end{equation}
\begin{equation}
    A=\sqrt{3}\left[\left(1-\frac{R_{0}^{2}}{r^{2}}-\frac{R_{0}^{4}}{2 L^{2} r^{2}}\right) \dd{t}+\frac{\epsilon R_{0}^{4}}{4 Lr^{2}} \sigma_{L}^{3'}\right],
\end{equation}
\end{subequations}
where
\begin{subequations}
\begin{equation}
    U(r)=\qty(1-\frac{R_0^2}{r^2})\qty(1+\frac{2R_0^2}{L^2}+\frac{r^2}{L^2}),
\end{equation}
\begin{equation}
    \Lambda(r)=1+\frac{R_0^6}{L^2r^4}-\frac{R_0^8}{4L^2r^6},
\end{equation}
\begin{equation}
    \Omega(r)=\frac{2\epsilon}{L\Lambda(r)}\qty[\qty(\frac{3}{2}+\frac{R_0^2}{L^2})\frac{R_0^4}{r^4}-\qty(\frac{1}{2}+\frac{R_0^2}{4L^2})\frac{R_0^6}{r^6}],
\end{equation}
\begin{equation}
    \sigma_L^{1'}=\sin\phi\dd{\theta}-\cos\phi\sin\theta\dd{\psi},
\end{equation}
\begin{equation}
    \sigma_L^{2'}=\cos\phi\dd{\theta}+\sin\phi\sin\theta\dd{\psi},
\end{equation}
\begin{equation}
    \sigma_L^{3'}=\dd{\phi}+\cos\theta\dd{\psi},
\end{equation}
\end{subequations}
and $\epsilon^2=\pm1$. We also have that
\begin{equation}
    \theta\in[0,\pi),\ \ \phi\in\left[\frac{2\epsilon t}{L},4\pi+\frac{2\epsilon t}{L}\right),\ \ \psi\in[0,2\pi),
\end{equation}
in these coordinates. This solution corresponds to a charged spinning black hole with mass
\begin{equation}
    M=12 \pi^2 R_{0}^{2}\left(1+\frac{3 R_{0}^{2}}{2 L^{2}}+\frac{2 R_{0}^{4}}{3 L^{4}}\right),
\end{equation}
angular momenta
\begin{subequations}
\begin{align}
    J_\phi&=\frac{6 \epsilon \pi^2 R_{0}^{4}}{L}\left(1+\frac{2 R_{0}^{2}}{3 L^{2}}\right),\\
    J_\psi&=0,
\end{align}
\end{subequations}
and charge
\begin{equation}
    Q=8\sqrt{3} \pi^2 R_{0}^{2}\left(1+\frac{R_{0}^{2}}{2 L^{2}}\right).
\end{equation}
This is easily seen to satisfy the BPS equation
\begin{equation}
    M-\frac{2}{L}|J|=\frac{\sqrt{3}}{2}|Q|.
\end{equation}

In the coordinates induced from the bulk solution, the boundary metric becomes
\begin{equation}
    \dd{s}^2_\text{bdy}=-\dd{t}^2+\frac{L^2}{4}\qty(\qty(\sigma_L^{1'})^2+\qty(\sigma_L^{2'})^2+\qty(\sigma_L^{3'})^2).
\end{equation}
Hence, we see that the boundary topology is $\bR\times S^3$. As a result, we should not expect a conformal anomaly, i.e., we do not need to cancel any logarithmic divergences.

The two-derivative action \eqref{eq:2derivAction} is simple to compute. Using the standard Gibbons-Hawking term \cite{Gibbons:1977h}
\begin{equation}
    S_{2\partial}^\text{GH}=2\int\dd[4]{x}\sqrt{-h}\,K,
\end{equation}
and boundary counterterm
\begin{equation}
    S_{2\partial}^\text{CT}=\int\dd[4]{x}\sqrt{-h}\,\qty(\frac{6}{L}+\frac{L}{2}\cR),
\end{equation}
to cancel the divergences, we find the holographically renormalized two-derivative action is
\begin{equation}
    I_{2\partial}^{\text{Ren}}=\frac{\pi^2T}{2L^2}\qty(-3L^4+4R_0^4).
\end{equation}

Now, we must compute five pieces of the four-derivative action
\begin{subequations}
\begin{align}
    S_1&:=\int\dd[5]{x}\sqrt{-g}\,\hat{R}_\text{GB},\\
    S_2&:=\int\dd[5]{x}\sqrt{-g}\,C_{\mu\nu\rho\sigma}F^{\mu\nu}F^{\rho\sigma},\\
    S_3&:=\int\dd[5]{x}\sqrt{-g}\,\qty(F^2)^2,\\
    S_4&:=\int\dd[5]{x}\sqrt{-g}\,F^4,\\
    S_5&:=\int\dd[5]{x}\sqrt{-g}\,\epsilon^{\mu\nu\rho\sigma\lambda}R_{\mu\nu ab}R_{\rho\sigma}^{ab}A_\lambda.
\end{align}
\end{subequations}
The only term we expect to give rise to divergences is $S_1$. This can be cured with an appropriate Gibbons Hawking term \cite{Teitelboim:1987z,Myers:1987yn}
\begin{equation}
    S_1^{\text{GH}}=2\int\dd[4]{x}\sqrt{-h}\qty[-\frac{2}{3}K^3+2KK_{ab}K^{ab}-\frac{4}{3}K_{ab}K^{bc}K_c{}^a-4\qty(\cR_{ab}-\frac{1}{2}\cR h_{ab})K^{ab}],
\end{equation}
where $K_{ab}$ is the extrinsic curvature, $K=h^{ab}K_{ab}$ is its trace, $\cR_{ab}$ is the induced Ricci tensor on the boundary, and $\cR$ is the induced Ricci scalar on the boundary. We also have boundary counterterms \cite{Cremonini2009ls}
\begin{equation}
    S_1^{\text{CT}}=-\int\dd[4]x\sqrt{-h}\,\qty(-\frac{2}{L^2}+\frac{3}{2L}\cR),
\end{equation}
per the usual holographic renormalization procedure. With this in hand, it is straightforward to compute
\begin{subequations}
\begin{align}
    I_1^\text{Ren}&=\frac{\pi^2\mathrm{vol}(\bR)}{20L^4}\qty(210L^4+408L^2R_0^2+128R_0^4),\\
    I_2&=\frac{2\pi^2\mathrm{vol}(\bR)}{5L^4}\qty(-15L^4+66L^2R_0^2+55R_0^4),\\
    I_3&=\frac{24\pi^2\mathrm{vol}(\bR)}{5L^4}\qty(30L^4+36L^2R_0^2+11R_0^4),\\
    I_4&=\frac{6\pi^2\mathrm{vol}(\bR)}{5L^4}\qty(60L^4+96L^2R_0^2+41R_0^4),\\
    I_5&=\frac{9\pi^2\mathrm{vol}(\bR)}{2L^4}\qty(72L^2R_0^2-7R_0^4).
\end{align}
\end{subequations}
Of particular note is that
\begin{equation}
    I_1^\text{Ren}-\frac{1}{2}I_2+\frac{1}{8}S_I+\frac{1}{2\sqrt{3}}S_I=\frac{9\pi^2\mathrm{vol}(\bR)}{2L^2}\qty(5L^2+8R_0^2),
\end{equation}
which is neither zero nor the two-derivative result. If one writes the four-derivative action with generic coefficients
\begin{equation}
    S_{4\partial}=c_1S_1+c_2S_2+c_3S_3+c_4S_4+c_5S_5,
\end{equation}
then we see that $I_{4\partial}=0$ requires that
\begin{subequations}
\begin{align}
    -7c_1 + 4 c_2 - 96 c_3 - 48 c_4&=0,\\
    17c_1 + 22 c_2 + 144 c_3 + 96 c_4 + 12 \sqrt{3} c_5&=0,\\
    125c_1 + 440 c_2 + 1056 c_3 + 984 c_4 - 28 \sqrt{3} c_5&=0,
\end{align}
\end{subequations}
which can be solved by
\begin{subequations}
\begin{align}
    c_2&=-\frac{151c_1+512\sqrt{3}c_5}{96},\\
    c_3&=-\frac{707c_1+2368\sqrt{3}c_5}{768},\\
    c_4&=\frac{901c_1+3296\sqrt{3}c_5}{576}.
\end{align}
\end{subequations}
If we enforce that $c_5=\frac{1}{2\sqrt{3}}c_1$, we get that
\begin{equation}
    c_2=-\frac{407}{96}c_1,\qquad c_3 =-\frac{1891}{768}c_1,\qquad c_4=\frac{2549}{576}c_1.
\end{equation}
These coefficients do not agree with any known results in the literature. Alternatively, enforcing $c_2=-\frac{1}{2}c_1$ gives
\begin{equation}
    c_3=-\frac{615}{2048},\qquad c_4=\frac{423}{1024},\qquad c_5=-\frac{103}{512\sqrt{3}},
\end{equation}
which is likewise undesirable.

%% file: Appendices/Appendix_RG.tex
\section{Technical details}\label{App:Geom}
In this Appendix, we provide some supplementary technical details that were omitted from the main text.

\subsection{Riemann tensors}
\label{app:Riemann}
Here we collect Riemann tensors for the metric
\begin{equation}
    \dd s^2 = e^{2f(z)}(\eta_{\mu\nu}\dd x^\mu \dd x^\nu+ \dd z^2)+ e^{2g(z)}g_{ij}(y)\dd y^i\dd y^j.
\end{equation}
We will use $\mu,\nu,\rho,...$ for curved indices in the $d$-dimensional base space and $i,j,k,...$ for curved indices on $M_{D-d}$, as well as $\alpha,\beta,\gamma,...$ for rigid indices in the $d$-dimensional spacetime and $a,b,c,d,...$ for rigid indices in the compact directions. We will use $z$ to denote the curved $z$-direction index and $\ul z$ to denote the rigid $z$-direction index. We will use $M, N,...$ for curved indices and $A,B,C,...$ for rigid indices of the whole $(D+1)$-dimensional spacetime. We choose a vielbein
\begin{align}
    \hat{e}^\alpha=e^{f(z)} e^\alpha,\ \ \ \hat e^{\ul z}=e^{f(z)}\dd z,\ \ \ \hat{e}^a=e^{g(z)} \tilde e^a,
\end{align}
so that $\dd s^2=\eta_{\alpha\beta}\hat e^\alpha\hat e^\beta+\hat e^{\ul z}\hat e^{\ul z}+\delta_{ab}\hat e^a\hat e^b$. Here we have defined $e^\alpha$ to be a vielbein for the flat $d$-dimensional space with metric $\eta_{\mu\nu}$ and $\tilde e^a$ to be a vielbein on $M_{D-d}$. Imposing the torsion-free condition
\begin{equation}
    \dd\hat e^A+\hat{\omega}^{A}_{\ \, B}\hat e^B=0,
\end{equation}
gives a spin connection
\begin{subequations}
    \begin{align}
    \hat\omega^{\alpha\beta}&=\omega^{\alpha\beta},\\
    \hat\omega^{\alpha\ul z}&=e^{-f}\partial^{\ul z}f\hat e^{\alpha},\\
    \hat\omega^{a \beta}&=0,\\
    \hat\omega^{a \ul z}&=e^{-f}\partial_{\ul z} g\,\hat e^a,\\
    \hat\omega^{ab}&=\tilde\omega^{ab},
    \end{align}
\end{subequations}
where $\omega$ is the spin connection on the $d$-dimensional base space and $\tilde\omega$ is the spin connection on $M_{D-d}$. The Riemann curvature two-form is then given by
\begin{equation}
    \hat R^{AB}=\dd\hat\omega^{AB}+\hat\omega^A_{\ \, C}\land\hat\omega^{CB},
\end{equation}
which, in components, reads
\begin{subequations}
\begin{align}
    \hat R^{\alpha\beta}_{\ \ \ \gamma\delta}&=-2e^{-2f}(f')^2\delta^{\alpha}_{[\gamma}\delta^{\beta}_{\delta]},\\
    \hat R^{\alpha\beta}_{\ \ \ \gamma\ul z}&=0,\\
    \hat R^{\alpha\ul z}_{\ \ \ \gamma\ul z}&=-e^{-2f}f''\delta^{\alpha}_{\gamma},\\
    \hat R^{\alpha b}_{\ \ \ \gamma d}&=-e^{-2f}f'g'\delta^{\alpha}_{\gamma}\delta^b_d,\\
    \hat R^{\alpha b}_{\ \ \ c \ul z}&=0,\\
    \hat R^{a\ul z}_{\ \ \ c\ul z}&=-e^{-2f}\qty(g''-f'g'+(g')^2)\delta^a_c,\\
    \hat R^{a b}_{\ \ \ c d}&=e^{-2g}\tilde R^{a b}_{\ \ \ c d}-2e^{-2f}(g')^2\delta^a_{[c}\delta^b_{d]},
\end{align}
\end{subequations}
where we have denoted the Riemann tensor on $M_{D-d}$ by $\tilde R^{a b}_{\ \ \ c d}$. Note that in the above, we have used the fact that the $d$-dimensional base space is flat to remove all the corresponding curvature tensors, hence why there is no $R_{\alpha\beta\gamma\delta}$. From here, one can compute the Ricci tensor, $\hat R_{AB}=\hat R^C_{\ \, ACB}$, to be
\begin{subequations}
    \begin{align}
        \hat R_{\alpha\beta}&= -e^{-2f}\qty[f''+(d-1)(f')^2+(D-d)f'g']\eta_{\alpha\beta},\\
        \hat R_{\alpha\ul z}&=0,\\
        \hat R_{\ul z\ul z}&=-e^{-2f}\qty[df''+(D-d)\qty(g''-g'f'+(g')^2)],\\
        \hat R_{a\beta}&=0,\\
        \hat R_{a\ul z}&=0,\\
        \hat R_{ab}&=e^{-2g}\tilde R_{ab}-e^{-2f}\qty[g''+(D-d)(g')^2+(d-1)f'g']\delta_{ab},
    \end{align}
\end{subequations}
where $\tilde R_{ab}$ denotes the Ricci tensor on $M_{D-d}$. Finally, the Ricci scalar is given by
\begin{align}
    \hat R=&e^{-2g}\tilde R-e^{-2f}\Big[2d f''+2(D-d)g''+d(d-1)(f')^2+2(d-1)(D-d)f'g'\nonumber\\
    &+(D-d+1)(D-d)(g')^2\Big],
\end{align}
where $\tilde R$ denotes the Ricci scalar on $M_{D-d}$.

\subsection{The general expression for $c'(z)$}
\label{app:c'(z)}

In section~\ref{sec:GBad}, we made an ansatz for a candidate $c$-function, (\ref{eq:LHcfnct}), in terms of real parameters $\{a_i,b_j\}$.  Given this ansatz, we find
\begin{align}
    c'(z)=&\fft{e^{-\tilde f}(L_{\mathrm{eff}})^d}{G_N}\nn\\
    &\times\biggl\{-(d-1)(f''-(f')^2)-(D-d)(g''+g'(g'-2f'))+\fft{(D-1)(D-d)}{d-1}(g')^2\nonumber\\
    &\qquad+\alpha e^{-2f}\Bigl[f''\left(\xi_1(f')^2+\xi_2f'g'+\xi_3 (g')^2\right)+g''\left(\xi_4(f')^2+\xi_5f'g'+\xi_6(g')^2\right)\nonumber\\
    &\qquad\kern1.5cm+\xi_7(f')^4+\xi_8(f')^3g'+\xi_9(f')^2(g')^2+\xi_{10}f'(g')^3+\xi_{11}(g')^4\Bigr]\nonumber\\
    &\qquad+\alpha e^{-2g}\frac{\kappa}{\ell^2}\Bigl[\omega_1f''+\omega_2g''+\omega_3(f')^2+\omega_4f'g'+\omega_5(g')^2\Bigr]\biggr\},
\label{eq:cprime}
\end{align}
where, for brevity, we have defined coefficients
{\allowdisplaybreaks
\begin{align}
    \xi_1&=-(d-3)a_1+3(d-1)a_4,\nn\\
    \xi_2&=2\frac{D-d}{d-1}a_1-(d-2)a_2+2(d-1)a_5,\nn\\
    \xi_3&=(d-1)(a_6-a_3)+\frac{D-d}{d-1}a_2,\nn\\
    \xi_4&=-(D-d)a_1+a_2+(d-1)a_5,\nn\\
    \xi_5&=-\frac{(D-d)(d-2)}{d-1}a_2+2a_3+2(d-1)a_6,\nn\\
    \xi_6&=-\frac{(D-d)(d-3)}{d-1}a_3+3(d-1)a_7,\nn\\
    \xi_7&=(d-3)a_1-3(d-1)a_4,\nn\\
    \xi_8&=2\frac{(D-d)(d-2)}{d-1}a_1+(d-3)a_2-(D-d)a_4-3(d-1)a_5,\nn\\
    \xi_9&=\frac{(D-d)\qty((D-d)a_1+2(d-2)a_2)}{d-1}+(d-3)a_3-(D-d)a_5-3(d-1)a_6,\nn\\
    \xi_{10}&=\frac{(D-d)\qty((D-d)a_2+2(d-2)a_3)}{d-1}-(D-d)a_6-3(d-1)a_7,\nn\\
    \xi_{11}&=(D-d)\qty(\frac{D-d}{d-1}a_3-a_7),\nn\\
    \omega_1&=-(d-1)(b_1-b_2),\nn\\
    \omega_2&=(d-1)b_3-(D-d)b_1,\nn\\
    \omega_3&=(d-1)(b_1-b_2),\nn\\
    \omega_4&=2(D-d-1)b_1-(D+d-2)b_2-(d-1)b_3,\nn\\
    \omega_5&=\frac{(D-d)(D-d-2)}{d-1}b_1-(D+d-2)b_3.
\end{align}
}%
Note that the form of the ansatz, (\ref{eq:LHcfnct}), was chosen so that no higher than second derivatives of $f$ and $g$ appear in (\ref{eq:cprime}).


\subsection{Induced Ricci scalar}\label{app:inducedR}
In Section \ref{sec:EE}, we require an expression for the Ricci scalar of the induced metric on the entangling surface, which we compute here. The induced metric is given by
\begin{equation}
    \dd \sigma^2=e^{2f(z)}\qty(1+r'(z)^2)\dd z^2+e^{2f(z)}r(z)^2\dd\Omega_{d-2}^2+e^{2g(z)}\dd s_{M_{D-d}}^2.
\end{equation}
By slight abuse of notation, we will use $\alpha,\beta,\gamma,\delta,...$ to index the rigid indices along the unit $(d-2)$-sphere (for this section only, these indices will not run over $t$ or $r$). A natural choice of vielbein is then
\begin{equation}
    \hat e^{\bar z}=e^f\sqrt{1+(r')^2}\dd z,\ \ \hat e^\alpha=e^fre^\alpha,\ \ \hat e^a=e^g\tilde e^a,
\end{equation}
where $e^\alpha$ is a vielbein on the $(d-2)$-sphere and $\tilde e^a$ is a vielbein on $M_{D-d}$. Note that this notation differs from the previous subsection. As before, we make use of the torsion-free condition to compute the components of the spin connection
\begin{subequations}
    \begin{align}
        \hat\omega^{\alpha\beta}&=\omega^{\alpha\beta},\\
        \hat\omega^{\alpha\ul z}&=\frac{e^{-f}}{\sqrt{1+(r')^2}}\qty(f'+\frac{r'}{r})\hat e^\alpha,\\
        \hat\omega^{\alpha b}&=0,\\
        \hat\omega^{ab}&=\tilde\omega^{ab},\\
        \hat\omega^{a\ul z}&=\frac{g'\,e^{-f}}{\sqrt{1+(r')^2}},
    \end{align}
\end{subequations}
where $\omega$ is the spin connection on the $(d-2)$-sphere and $\tilde\omega$ is the spin connection on $M_{D-d}$. The  induced Riemann tensor components may then be computed to be
\begin{subequations}
    \begin{align}
        \cal{R}^{\alpha\beta}_{\ \ \ \gamma\delta}&=\frac{e^{-2f}}{r^2}\bar R^{\alpha\beta}_{\ \ \ \gamma\delta}-2\frac{e^{-2f}}{1+(r')^2}\qty(f'+\frac{r'}{r})^2\delta^{[\alpha}_\gamma\delta^{\beta]}_\delta,\\
        \mathcal{R}^{\alpha\ul z}_{\ \ \ \beta\ul z}&=-\qty[\dv{}{z}\qty(\frac{e^{-f}}{\sqrt{1+(r')^2}}\qty(f'+\frac{r'}{r}))\frac{e^{-f}}{\sqrt{1+(r')^2}}+\frac{e^{-2f}}{1+(r')^2}\qty(f'+\frac{r'}{r})^2]\delta^\alpha_\beta,\\
        \mathcal{R}^{a\ul z}_{\ \ \ b\ul z}&=-\qty[\dv{}{z}\qty(\frac{g'\,e^{-f}}{\sqrt{1+(r')^2}})\frac{e^{-f}}{\sqrt{1+(r')^2}}+\frac{e^{-2f}(g')^2}{1+(r')^2}]\delta^a_b,\\
        \mathcal{R}^{ab}_{\ \ \ cd}&=e^{-2g}\tilde R^{ab}_{\ \ \ cd}-2\frac{e^{-2f}(g')^2}{1+(r')^2}\delta^{[a}_c\delta^{b]}_d,\\
        \mathcal{R}^{\alpha b}_{\ \ \ \gamma d}&=-\frac{g'\, e^{-2f}}{1+(r')^2}\qty(f'+\frac{r'}{r})\delta^a_c\delta^\beta_\delta,
    \end{align}\end{subequations}
    where $\bar R^{\alpha\beta}_{\ \ \ \gamma\delta}$ denotes the Riemann tensor on the (unit) $(d-1)$-sphere and $\tilde R^{ab}_{\ \ \ cd}$ denotes the Riemann tensor on $M_{D-d}$. Computing the induced Ricci scalar as $\mathcal{R}=\mathcal{R}^{AB}_{\ \ \ \,AB}$, and using the identities for the Ricci scalars of the constituent metrics
    \begin{subequations}
        \begin{align}
            \bar R&=(d-2)(d-3),\\
            \tilde R&=(D-d)(D-d-1)\frac{\kappa}{\ell^2},
        \end{align}
    \end{subequations}
    we finally arrive at our expression for the induced Ricci scalar
    \begin{align}
    \mathcal{R}=&(d-2)(d-3)\frac{e^{-2f}}{r^2}+(D-d)(D-d-1)\frac{\kappa}{\ell^2}e^{-2g}\nonumber\\
    &+\frac{e^{-2f}}{r^2(1+(r')^2)^2}\Big[-\qty(1+(r')^2)\Big(2(d-2)\qty((d-2)f'+(D-d)g')rr'\nonumber\\
    &\kern8em+(d-2)(d-3)(r')^2+\big((d-2)(d-3)(f')^2\nonumber\\
    &\kern8em+2(d-3)(D-d)f'g'+(D-d)(D-d-1)(g')^2\nn\\
    &\kern8em+2(d-2)f''+2(D-d)g''\big)\Big)r^2\nn\\
    &\kern8em+2\qty(-(d-2)+\qty((d-2)f'+(D-d)g')rr')rr''\Big].
\end{align}

\section{Alternate coordinates}
Here we collect some of the results from the main text reexpressed in alternate coordinates, more akin to those used in \cite{Myers:2010xs,Myers:2010tj,Myers:2012ed}. These are not new results, but the reader might find them more useful for some purposes.

\subsection{NEC-motivated $c$-function}\label{app:altCoords1}
One may alternately parameterize the metric as
\begin{equation}
    \dd{s}^2=e^{2A(r)}\eta_{\mu\nu}\dd{x}^\mu \dd{x}^\nu+\dd{r}^2+e^{2B(r)}g_{ij}(y)\dd{y}^i\dd{y}^j.\label{eq:altCoords}
\end{equation}
These are the coordinates that are used in \cite{Myers:2010tj,Myers:2010xs}. Pure AdS corresponds to $A(r)=B(r)=r/L$, and so it is natural to identify $r=0$ with the IR and $r=\infty$ with the UV. We expect the asymptotic behavior of the metric functions to be
\begin{align}
    &r\to\infty:\ &A(r)\to \frac{r}{L_\text{UV}},&\qquad B(r)\to \frac{r}{L_\text{UV}},\nonumber\\
    &r\to0:\ &A(r)\to \frac{r}{L_\text{IR}},&\qquad B(r)\to B_\text{IR}.
\end{align}
We still assume that the internal manifold is maximally symmetric with Ricci scalar
\begin{equation}
    \tilde R=(D-d)(D-d-1)\frac{\kappa}{\ell^2}.
\end{equation}
One can take this metric and compute the resulting $t$-$z$ null energy condition NEC1 for arbitrary dimensions, which gives
\begin{align}\label{NEC1 arbitrary}
    0\le& -(d-1)A''-(D-d)B''+(D-d)A'B'-(D-d)(B')^2\nonumber\\
    &+\alpha\Big[2(d-1)(d-2)(d-3)(A')^2A''+4(d-1)(d-2)(D-d)A'B'A''\nonumber\\
    &+2(d-1)(D-d)(D-d-1)(B')^2A''+2(d-1)(d-2)(D-d)(A')^2B''\nonumber\\
    &+4(d-1)(D-d)(D-d-1)A'B'B''+2(D-d)(D-d-1)(D-d-2)(B')^2B''\nonumber\\
    &-2(d-1)(d-2)(D-d)(A')^3B'-2(d-1)(D-d)(2D-3d)(A')^2(B')^2\nonumber\\
    &-2(D-d)(D-d-1)(D-3d)A'(B')^3+2(D-d)(D-d-1)(D-d-2)(B')^4\Big]\nonumber\\
    &+2\alpha(D-d)(D-d-1)\frac{\kappa}{\ell^2}\qty[-(d-1)A''+(D-d-2)\qty(-B''+A'B'-(B')^2)].
\end{align}
One might then propose a generic candidate $c$-function
\begin{align}
    &c(r)=\nn\\
    &\frac{e^{(D-d)B}\qty(1+\alpha\qty(a_1 (A')^2+a_2A'B'+a_3(B')^2+b_1\frac{\kappa}{\ell^2}e^{-2B}))}{\qty(\tilde{A}'+\alpha\qty(a_4(A')^3+a_5(A')^2B'+a_6A'(B')^2+a_7(B')^3+b_2\frac{\kappa}{\ell^2}e^{-2B}A'+b_3\frac{\kappa}{\ell^2}e^{-2B}B'))^{d-1}},
\end{align}
where we have defined
\begin{equation}
    \tilde{A}=A+\frac{D-d}{d-1}B,
\end{equation}
in analogy to $\tilde f$. This $c$-function is the obvious generalization of the two-derivative case (when $\alpha=0$). As before, one computes
\begin{align}
    c'(z)=&\frac{e^{(D-d)B}}{\qty(\tilde{A}'+\alpha\qty(a_3(A')^2B'+a_4A'(B')^2+a_5(B')^3+b_2\frac{\kappa}{\ell^2}e^{-2B}A'+b_3\frac{\kappa}{\ell^2}e^{-2B}B'))^{d}}\nonumber\\
    &\times\Big\{-(d-1)A''-(D-d)B''+(D-d)A'B'+(D-d)^2(B')^2\nonumber\\
    &+\alpha\Big[\xi_1(A')^2A''+\xi_2A'B'A''+\xi_3(B')^2A''+\xi_4(A')^2B''+\xi_5A'B'B''+\xi_6(B')^2B''\nonumber\\
    &+\xi_7(A')^4+(A')^3B'+\xi_8(A')^2(B')^2+\xi_9A'(B')^3+\xi_{10}(B')^4\Big]\nonumber\\
    &+\alpha\frac{\kappa}{\ell^2}\qty[\omega_1A''+\omega_2B''+\omega_3A'B'+\omega_4(B')^2]\Big\},
\end{align}
where we have defined
\begin{align}
    \xi_1&=3(a_1+a_4)-(a_1+3a_4)d,\nonumber\\
    \xi_2&=a_2-(d-1)(a_2+2a_5)+2\frac{D-d}{d-1}a_1,\nonumber\\
    \xi_3&=-(d-1)a_6+\frac{D-d}{d-1}a_2,\nonumber\\
    \xi_4&=a_2-(d-1)a_5-(D-d)a_1,\nonumber\\
    \xi_5&=-2(d-1)a_6+\frac{(D-d)(d-2)}{d-1}a_2,\nonumber\\
    \xi_6&=-3(d-1)a_7,\nonumber\\
    \xi_7&=(D-d)(a_1+a_4),\nonumber\\
    \xi_8&=(D-d)\qty(a_2+a_5+\frac{D-d}{d-1}a_1),\nonumber\\
    \xi_9&=(D-d)\qty(a_6+\frac{D-d}{d-1}a_2),\nonumber\\
    \xi_{10}&=(D-d)a_7,\nonumber\\
    \omega_1&=-(d-1)b_2,\nonumber\\
    \omega_2&=-(d-1)b_3,\nonumber\\
    \omega_3&=(D-d-2)b_1+(D-d-2)b_2,\nonumber\\
    \omega_4&=(D+d-2)b_3+\frac{(D-d)(D-d-2)}{d-1}b_1.
\end{align}
With the particular choice of
\begin{subequations}
    \begin{align}
        a_1&=-2(d-1)(d-2),\\
        a_2&=-4(D-d)(d-2),\\
        a_4&=0,\\
        a_5&=-4\frac{(D-d)(d-2)}{(d-1)},\\
        a_6&=\frac{a_2}{d-1}+2\frac{D-d}{(d-1)^2}(1+d(-5+3d-2D)+3D),\\
        b_1&=\frac{2(D-d-1)((D+1)d-D-d^2+2)}{d},\\
        b_2&=\frac{2(D-d-1)(D-d-2)}{d},\\
        b_3&=\frac{2(D-d)(D-d-1)(D-3d-2)}{d(d-1)},
    \end{align}
\end{subequations}
we get that
\begin{equation}
    c'(r)=\frac{e^{(D-d)B}\qty(\text{NEC1}+\frac{(D-1)(D-d)}{d-1}(B')^2\qty(1+\mathcal{O}(\alpha)))}{\qty(\tilde{A}'+\alpha\qty(a_3(A')^2B'+a_4A'(B')^2+a_5(B')^3+b_2\frac{\kappa}{\ell^2}e^{-2B}A'+b_3\frac{\kappa}{\ell^2}e^{-2B}B'))^{d}}\ge 0,
\end{equation}
and hence the candidate $c$-function gives us a monotonic flow from the UV to the IR. 

As before, we never need to use the all-internal components of the Riemann tensor $\hat R_{ijkl}$ to obtain NEC1, and so the above results also trivially generalize to arbitrary Einstein internal manifolds, as in the $f$ and $g$ coordinates. 

\subsection{Entanglement entropy $c$-function}\label{app:altCoords2}
One may also repeat the arguments of Section \ref{sec:EEcfnct} in the alternate coordinates \eqref{eq:altCoords}. Here we focus on flows from AdS$_{D+1}$ to AdS$_3$, and so we specialize the metric \eqref{eq:altCoords} to
\begin{equation}
    \dd \sigma^2=e^{2A(r)}\qty(-\dd t^2+\dd \rho^2)+\dd r^2+e^{2B(r)}\dd s_{M_{D-2}}^2.
\end{equation}
In terms of a profile $\rho(r)$, the induced Ricci scalar is
\begin{align}
    \mathcal{R}=&-\frac{(D-2)}{\left(1+e^{2 A(r)} \rho '(r)^2\right)^2}\nn\\
    &\times\Big(e^{2 A(r)} \rho '(r) \left(\rho '(r) \left(2 A'(r) B'(r)-(D-1) B'(r)^2-2 B''(r)\right)+2 B'(r) \rho ''(r)\right)\nonumber\\
    &-(D-1) B'(r)^2-2 B''(r)\Big)+(D-2)(D-3)\frac{\kappa}{\ell^2}e^{-2B},
\end{align}
which leads to a JM functional whose first integral is
\begin{equation}
    C=\frac{\rho'(r) e^{2 A(r)+(D-2)B(r)} \left(\left(1+e^{2 A(r)} \rho '(r)^2\right) \left(1+2 \tilde\alpha \frac{\kappa}{\ell^2}e^{-2B} \right)-2\tilde\alpha B'(r)^2\right)}{\left(1+e^{2 A(r)} \rho '(r)^2\right)^{3/2}},
\end{equation}
which can be solved to give
\begin{equation}
    \rho'(r)=\frac{e^{-A}\mathcal{F}}{\sqrt{1-\mathcal{F}^2+4\tilde\alpha\qty(\frac{\kappa}{\ell^2}e^{-2B}-(B')^2(1-\mathcal{F}^2))}},\ \ \mathcal{F}(r)\equiv C e^{-A-(D-2)B}.
\end{equation}
To fix the value of $C$, we note that we should have $\rho'(r)\to-\infty$ as $r\to r_0$, where $r_0$ is the deepest point in the bulk that the minimal surface. This then requires that
\begin{equation}
    C=e^{A(r_0)+(D-2)B(r_0)}\qty(1+2\tilde\alpha\frac{\kappa}{\ell^2}e^{-2B(r_0)}).
\end{equation}
Then the radius of the entangling area is
\begin{align}
    R=&\int_{r_0}^\infty\dd{r}\rho'(r)=\int_{r_0}^\infty \dd{r}\frac{e^{-A}\mathcal{F}}{\sqrt{1-\mathcal{F}^2+4\tilde\alpha\qty(\frac{\kappa}{\ell^2}e^{-2B}-(B')^2(1-\mathcal{F}^2))}}\nonumber\\
    =&\int_{r_0}^\infty \dd{r}\qty[\frac{e^{-A}\mathcal{F}}{\sqrt{1-\mathcal{F}^2+4\tilde\alpha\frac{\kappa}{\ell^2}e^{-2B}}}+2\tilde\alpha \frac{e^{-A}(B')^2\mathcal{F}}{\sqrt{1-\mathcal{F}^2}}]+\mathcal{O}(\tilde\alpha^2)\nonumber\\
    =&\lim_{r_c\to\infty}\int_{r_0}^{r_c}\dd{r}\Big[\sqrt{1-\mathcal{F}^2+4\tilde\alpha\frac{\kappa}{\ell^2}e^{-2B}}\dv{}{r}\frac{e^{-A}}{\mathcal{F}'+\frac{4\tilde\alpha}{\mathcal{F}}\frac{\kappa}{\ell^2}e^{-2B}B'}\nonumber\\
    &+2\tilde\alpha \sqrt{1-\mathcal{F}^2}\dv{}{r}\qty(\frac{e^{-A}(B')^2}{\mathcal{F}'})\Big]+2\tilde\alpha\lim_{r_c\to\infty}\frac{e^{-A}(B')^2}{\mathcal{F}'}\Bigg\vert_{r=r_c}+\mathcal{O}(\tilde\alpha^2).
\end{align}
Equivalently, in terms of a profile $r(\rho)$, we may write
\begin{equation}
    r'(\rho)=e^{A}\frac{\sqrt{1-\mathcal{F}^2+4\tilde\alpha\qty(\frac{\kappa}{\ell^2}e^{-2B}-(B')^2(1-\mathcal{F}^2))}}{\mathcal{F}}.
\end{equation}
The JM functional may then be calculated as
\begin{align}
    S_\text{JM}=&\frac{2\text{Vol}(M_{D-2})}{4G_N}\nn\\
    &\times\int\dd{\rho}\qty[\sqrt{e^{2A}+(r')^2}e^{(D-2)B}\qty(1+2\tilde\alpha\frac{\kappa}{\ell^2}e^{-2B})+2\tilde\alpha\frac{e^{(D-2)B}(B')^2(r')^2}{\sqrt{e^{2A}+(r')^2}}]\nonumber\\
    &-4\tilde\alpha e^{(D-2)B}B'\Big\vert_{\rho=\rho_c}.
\end{align}
As before, the boundary term is independent of $R$ and so it will not affect the succeeding analysis. The monotonic central charge is then given by
\begin{equation}
    c_\text{EE}=R\partial_R S_\text{JM}=\frac{2\pi\text{Vol}(M_{D-2})}{\ell_P^{D-1}}e^{A_0+(D-2)B_0}\qty(1+2\tilde\alpha\frac{\kappa}{\ell^2}e^{-2B_0})R.
\end{equation}

As before, this generically leads to rather complicated terms
\begin{align*}
    \dv{c_\text{EE}}{r_0}=&\frac{2\text{Vol}\qty(M_{D-2})}{4G_N}\nn\\
    &\times\int\dd\rho\frac{e^{(D-2)B}\mathcal{F}^2\tilde A_0'}{\sqrt{1-\mathcal{F}^2}(\tilde A')^2}\Bigg\{-A''-(D-2)B''+(D-2)A'B'+(D-2)^2(B')^2\nonumber\\
    &+2\tilde\alpha B'\qty[A'((D-2)(B')^2+2B'')+B'(-A''+(D-2)((D-2)(B')^2+B'')]\nonumber\\
    &+\tilde\alpha\frac{\kappa}{\ell^2}\Bigg[e^{\tilde A-2B}\frac{A''+(D-2)(B''-\tilde A' B')}{e^{\tilde A}-e^{\tilde A_0}}+4e^{-2B_0}B_0'\frac{A''+(D-2)(B''-\tilde A' B')}{\tilde A_0'}\nonumber\\
    &-\frac{4e^{2(\tilde A-2\tilde A_0-2B-B_0)}\qty(e^{2\tilde A_0}-2e^{\tilde A})}{\tilde A'}\nonumber\\
    &\times\qty(A'(2\tilde A'B'+B'')+B'((3D-8)\tilde A'B'-2A''-(D-2)B''))\Bigg]\Bigg\},
\end{align*}
but, if one sets $\kappa=0$, then
\begin{align}
    \dv{c_\text{EE}}{r_0}&=\frac{2\text{Vol}\qty(M_{D-2})}{4G_N}\int\dd\rho\frac{e^{(D-2)B}\mathcal{F}^2\tilde A_0'}{\sqrt{1-\mathcal{F}^2}(\tilde A')^2}\qty[\text{NEC1}+(D-1)(D-2)(B')^2\qty(1+\mathcal{O}(\tilde\alpha))]\nn\\
    &\ge0,
\end{align}
which gives monotonicity along flows to the IR. This parallels the computation that was done in $f$ and $g$ coordinates.

%% file: Appendices/Appendix_torus.tex
\section{Torsionful Riemann tensor}\label{app:riemann}
The torsionful Riemann tensor appears at $\mathcal O(\alpha')$ in the heterotic Lagrangian and equations of motion.  Although we only need its truncated form, here we give the general frame components of the torsionful Riemann tensor.  This is computed from $R(\Omega_+)=\dd \Omega_++\Omega_+\wedge\Omega_+$, where the torsionful connection is given in (\ref{eq:Omega+}).
\begin{align}
    R(\Omega_+)_{\gamma\delta}{}^{\alpha\beta}=&R(\omega_+)_{\gamma\delta}{}^{\alpha\beta}-\fft12F_{\gamma\delta}^i\qty(\qty(g_{ij}+b_{ij})F_{\alpha\beta}^j-G_{\alpha\beta i})\nonumber\\
    &-\fft12\qty(\qty(g_{ij}-b_{ij})F_{\gamma\alpha}^j+G_{\gamma\alpha i})g^{ik}\qty(\qty(g_{kl}-b_{kl})F_{\delta\beta}^l+G_{\delta\beta k}),\nn\\
    R(\Omega_+)_{\gamma d}{}^{\alpha\beta}=&-\fft12e_d^i\qty(\partial_\gamma\qty(g_{ij}+b_{ij})F_{\alpha\beta}^j+\qty(\qty(g_{ij}+b_{ij})\nabla^{(+)}_\gamma F_{\alpha\beta}^j-\nabla^{(+)}_\gamma G_{\alpha\beta i}))\nn\\
    &+\fft14e^l_d\qty(\qty(g_{ij}-b_{ij})F_{\gamma\alpha}^j+G_{\gamma\alpha i})g^{ik}\partial_\beta\qty(g_{kl}-b_{kl})\nn\\
    &-\fft14e^l_d\qty(\qty(g_{ij}-b_{ij})F_{\gamma\beta}^j+G_{\gamma\beta i})g^{ik}\partial_\alpha\qty(g_{kl}-b_{kl}),\nn\\
    R(\Omega_+)_{cd}{}^{\alpha\beta}=&\fft12e^{ic}e^{kd}\qty(\qty(g_{ij}+b_{ij})F_{\alpha\gamma}^j-G_{\alpha\gamma i})\qty(\qty(g_{kl}+b_{kl})F_{\gamma\beta}^l-G_{\gamma\beta k})\nn\\
    &-\fft12e^{jc}e^{ld}\partial_\alpha\qty(g_{ij}-b_{ij})g^{ik}\partial_\beta\qty(g_{kl}-b_{kl}),
\end{align}
along with
\begin{align}
    R(\Omega_+)_{\gamma\delta}{}^{\alpha b}=&-\fft12g^{ik}e^{lb}\partial_\gamma\qty(g_{kl}+b_{kl})\qty(\qty(g_{ij}-b_{ij})F_{\delta\alpha}^j+G_{\delta\alpha i})+e^{ib}\partial_\gamma\qty(g_{ij}-b_{ij})F_{\delta\alpha}^j\nn\\
    &-\fft12e^{ib}\partial_\alpha\qty(g_{ij}-b_{ij})F_{\gamma\delta}^j+e^{ib}\qty(\qty(g_{ij}-b_{ij})\nabla_\gamma^{(+)}F_{\hat\delta\alpha}^j+\nabla^{(+)}_\gamma G_{\hat\delta\alpha i}),\nn\\
    R(\Omega_+)_{\gamma d}{}^{\alpha b}=&\fft14g^{ik}e^{jd}e^{lb}\partial_\gamma\qty(g_{kl}+b_{kl})\partial_\alpha\qty(g_{ij}-b_{ij})-\fft12e^{ib}e^{jd}\nabla_\gamma^{(+)}\partial_\alpha\qty(g_{ij}-b_{ij})\nn\\
    &+\fft14e^{kb}e^{id}\qty(\qty(g_{ij}+b_{ij})F_{\alpha\epsilon}^j-G_{\alpha\epsilon i})\qty(\qty(g_{kl}-b_{kl})F_{\gamma\epsilon}^l+G_{\gamma\epsilon k}),\nn\\
    R(\Omega_+)_{cd}{}^{\alpha b}=&\fft12e^{kb}e^{ic}e^{ld}\qty(\qty(g_{ij}+b_{ij})F_{\alpha\gamma}^j-G_{\alpha\gamma i})\partial_\gamma\qty(g_{kl}-b_{kl}),
\end{align}
and
\begin{align}
    R(\Omega_+)_{\gamma\delta}{}^{ab}=&-\fft12e^{ia}e^{lb}\partial_\gamma\qty(g_{ij}-b_{ij})g^{jk}\partial_\delta\qty(g_{kl}+b_{kl})\nn\\
    &-\fft12e^{ia}e^{kb}\qty(\qty(g_{ij}-b_{ij})F_{\gamma\epsilon}^j+G_{\gamma\epsilon i})\qty(\qty(g_{kl}-b_{kl})F_{\delta\epsilon}^l+G_{\delta\epsilon k}),\nn\\
    R(\Omega_+)_{\gamma d}{}^{ab}=&\fft14e^{ia}e^{kb}e^{ld}\qty(\qty(g_{ij}-b_{ij})F_{\gamma\epsilon}^j+G_{\gamma\epsilon i})\partial_\epsilon\qty(g_{kl}-b_{kl})\nn\\
    &-\fft14e^{ib}e^{ka}e^{ld}\qty(\qty(g_{ij}-b_{ij})F_{\gamma\epsilon}^j+G_{\gamma\epsilon i})\partial_\epsilon\qty(g_{kl}-b_{kl}),\nn\\
    R(\Omega_+)_{cd}{}^{ab}=&-\fft12e^{ia}e^{kb}e^{jc}e^{ld}\partial_\gamma\qty(g_{ij}-b_{ij})\partial_\gamma\qty(g_{kl}-b_{kl}).
\end{align}
In some cases, implicit antisymmetrization with weight one is needed on the two-form indices.  Note that the covariant derivative $\nabla^{(+)}$ is taken with respect to the torsionful connection $\Omega_+$ on frame indices, except in the $R(\Omega_+)_{\gamma\delta}{}^{\alpha b}$ term where the hat on the $\delta$ index indicates that it is corrected using the torsion-free connection $\Omega$.  (The $\alpha$ index is corrected using the $\Omega_+$ connection.)

Since the two-derivative Lagrangian, (\ref{eq:Lags}), and equations of motion, (\ref{eq:eom2d}), are written in terms of a torsion-free connection, it is useful to make note of the standard reduction of the torsion-free Ricci tensor
\begin{align}
    R_{\alpha\beta}(\Omega)&=R_{\alpha\beta}(\omega)-\fft12F_{\alpha\gamma}^ig_{ij}F_{\beta\gamma}^j-\fft12g^{ij}\nabla_\alpha\nabla_\beta g_{ij}+\fft14g^{ij}g^{kl}\partial_\alpha g_{ik}\partial_\beta g_{jl},\nn\\
    R_{\alpha b}(\Omega)&=e^{ib}\left(\fft12\nabla_\gamma(g_{ij}F_{\alpha\gamma}^j)+\fft14g_{il}F_{\alpha\gamma}^lg^{jk}\partial_\gamma g_{jk}\right),\nn\\
    R_{ab}(\Omega)&=e^{ia}e^{jb}\left(\fft14g_{ik}g_{jl}F_{\gamma\delta}^kF_{\gamma\delta}^l-\fft12\Box g_{ij}+\fft12g^{kl}\partial_\gamma g_{ik}\partial_\gamma g_{jl}-\fft14\partial_\gamma g_{ij}g^{kl}\partial_\gamma g_{kl}\right).
\end{align}
In addition, the reduction of $\hat\nabla_M\hat\nabla_N\phi$ yields
\begin{equation}
    \hat\nabla_\alpha\hat\nabla_\beta\phi=\nabla_\alpha\nabla_\beta\phi,\qquad
    \hat\nabla_\alpha\hat\nabla_b\phi=-\fft12e_{ib}F_{\alpha\gamma}^i\partial_\gamma\phi,\qquad
    \hat\nabla_a\hat\nabla_b\phi=\fft12e^i_ae^j_b\partial_\gamma g_{ij}\partial_\gamma\phi.
\end{equation}

\section{A note on the torsionful Lorentz Chern-Simons term}\label{sec:CSterm}

The Lorentz Chern-Simons form, (\ref{eq:LCS+}), is computed with the torsionful connection, $\Omega_+$.  If we were to expand it out with $\Omega_+=\Omega+\fft12\mathcal H$, we would get
\begin{equation}
    \omega_{3L}(\Omega_+)=\omega_{3L}(\Omega)+\Tr(R(\Omega)\wedge\mathcal H+\fft14\mathcal H\wedge D\mathcal H+\fft1{12}\mathcal H\wedge\mathcal H\wedge\mathcal H)-\fft12\dd(\Tr\Omega\wedge\mathcal H).
\end{equation}
The final term is not Lorentz covariant but is a total derivative.  Hence it can be removed by a shift of the $B$-field.  In particular, with
\begin{equation}
\tilde H=\dd B-\fft{\alpha'}4\omega_{3L}(\Omega_+),
\end{equation}
we can make the shift
\begin{equation}
    B\to B+\fft{\alpha'}8\Tr(\Omega\wedge\mathcal H),\qquad \omega_{3L}(\Omega_+)\to\omega_{3L}(\Omega_+)+\fft12\dd(\Tr\Omega\wedge\mathcal H),
\end{equation}
to remove the total derivative from the Lorentz Chern-Simons form.

A similar manipulation can be performed in the lower-dimensional theory.  In particular, in (\ref{eq:LCSf}), we have made the shift
\begin{equation}
    \omega_{3L}(\Omega_+)\to\omega_{3L}(\Omega_+)-\fft12\dd(\omega_+^{\alpha\beta}F_{\beta\alpha\,i}^{(-)}\eta^i),
\end{equation}
to remove a total derivative.  Note that removing this mixed component of $\omega_{3L}$ corresponds to a shift of the gauge fields $B_{\mu\,i}$.